\documentclass[11pt,a4paper]{article}
\usepackage{jheppub,amsmath,  amssymb,slashed,url,bm,textgreek,upgreek}
\usepackage{graphicx}
\usepackage{epstopdf}
\usepackage{amsfonts}
\usepackage{booktabs}
\usepackage{siunitx}
\usepackage{tikz}
\newcommand{\cN}{{\cal N}}

\newcommand\Sp{\text{Sp}}
\newcommand\PSU{\text{PSU}}
\newcommand\uU{\text{U}}
\newcommand\hq{\widehat{q}}
\newcommand\SL{\text{SL}}
\newcommand\SU{\text{SU}}
\newcommand\OSp{\text{OSp}}
\newcommand{\beq}{\begin{equation}}
\newcommand{\eeq}{\end{equation}}
\newcommand{\nn}{\nonumber\\} 
\newcommand{\bea}{\begin{eqnarray}}
\newcommand{\ea}{\end{eqnarray}}
\newcommand{\lb}{{\langle}}
\newcommand{\rb}{{\rangle}}

\def\t{{ \sf t}} 
\def\tt{{\mathfrak t}}

\def\Q{{\mathcal Q}}

\def\X{{\mathcal X}}

\def\w{w}

\def\UU{U}

\def\g{\text{{\teneurm g}}}

\def\be{\begin{equation}}
\def\ee{\end{equation}}

\def\Ber{{\mathrm{Ber}}}

\def\X{{\eusm X}}

\def\hat{\widehat}

\def\tilde{\widetilde}

\def\h{\widehat}

\def\RP{{\Bbb{RP}}}

\def\S{{\mathcal S}}

\def\O{{\mathcal O}}

\def\Bbb{\mathbb}

\def\d{{\mathrm d}}

\def\b{\overline}
\def\R{{\mathbb R}}
\def\C{{\mathbb C}}
\def\U{{\mathcal U}}

\def\[{\bigl [}

\def\]{\bigr ]}

\def\CP{{\mathbb{CP}}}

\def\T{{\mathcal T}}

\def\tr{{\mathrm {tr}}}
\def\Z{{\mathbb Z}}

\def\L{{\mathcal  L}}
\def\t{\widetilde }
\def\h{\widehat}

\def\M{{\mathcal M}}

\def\H{{\mathcal H}}

\def\vol{{\mathrm{vol}}}

\def\tilde{\widetilde}

\def\bar{\overline}

\def\Ber{{\mathrm {Ber}}}

\def\qQ{{\sf Q}}

\font\teneurm=eurm10 \font\seveneurm=eurm7  \font\fiveeurm=eurm5
\newfam\eurmfam
\textfont\eurmfam=\teneurm \scriptfont\eurmfam=\seveneurm
\scriptscriptfont\eurmfam=\fiveeurm

\font\teneusm=eusm10 \font\seveneusm=eusm7 \font\fiveeusm=eusm5
\newfam\eusmfam
\textfont\eusmfam=\teneusm \scriptfont\eusmfam=\seveneusm
\scriptscriptfont\eusmfam=\fiveeusm
\def\eusm#1{{\fam\eusmfam\relax#1}}
\font\tencmmib=cmmib10 \skewchar\tencmmib='177
\font\sevencmmib=cmmib7 \skewchar\sevencmmib='177
\font\fivecmmib=cmmib5 \skewchar\fivecmmib='177
\newfam\cmmibfam
\textfont\cmmibfam=\tencmmib \scriptfont\cmmibfam=\sevencmmib
\scriptscriptfont\cmmibfam=\fivecmmib

\def\h{\widehat}
\def\N{{\mathcal N}}
\def\Tr{{\mathrm{Tr}}}
\def\i{{\mathrm i}} 
\def\diag{{\mathrm{diag}}}
\def\rR{{\sf R}}
\def\cR{{\mathcal R}}
\def\tT{{\sf T}}
\def\fF{{\sf F}}
\def\CT{{\sf {CT}}}
\def\CR{{\sf {CR}}}
\def\cC{{\sf C}}
\def\CPT{{\sf {CPT}}}
\def\g{{\mathfrak g}}
\def\M{{\mathrm M}}
\def\mM{{\mathcal M}}
\def\BPS{{\mathrm{BPS}}}
\def\SL{{\mathrm{SL}}}
\def\uU{{\mathrm U}}
\def\SU{{\mathrm{SU}}}
\def\vol{{\mathrm{vol}}}
\def\intt{{\mathrm{int}}}
\def\ext{{\mathrm{ext}}}

\def\Sp{{\mathrm{Sp}}}
\def\SU{{\mathrm{SU}}}
\def\b{{\sf b}}
\def\la{\langle}
\def\ra{\rangle}
\title{$\N=2$ JT Supergravity and Matrix Models}

 \author{Gustavo J. Turiaci${}^{1,2}$ and Edward Witten${}^1$}
\affiliation{${}^1$ Institute for Advanced Study, Einstein Drive, Princeton, NJ, USA}
\affiliation{${}^2$ Physics Department, University of Washington, Seattle, WA, USA}%\emailAdd{author@inst.edu}
\abstract{Generalizing previous results for $\N=0$ and $\N=1$, we 
analyze $\N=2$ JT supergravity on asymptotically AdS${}_2$ spaces with arbitrary topology and show that this theory of gravity is dual, in a holographic sense, to a certain
random matrix ensemble in which supermultiplets of different $R$-charge are statistically independent and each is described by its own $\N=2$ random matrix ensemble.
We also analyze the case with a time-reversal symmetry, either commuting or anticommuting with the $R$-charge.
In order to compare supergravity to random matrix theory, we develop an $\N=2$ analog of the recursion relations for  Weil-Petersson 
volumes originally discovered by Mirzakhani in the bosonic
case.
}

\begin{document}\maketitle

\section{Introduction}
Spacetime wormholes have been important in understanding quantum aspects of black holes. For example, for an evaporating black hole, wormhole contributions are important in order to obtain an entropy of Hawking radiation consistent with unitarity \cite{Almheiri:2020cfm}. 

Wormholes also appear in the following context, closer to the present article. The spectrum of a black hole can be computed via the gravitational path integral as pioneered by Gibbons and Hawking \cite{Gibbons:1976ue}. The classical analysis predicts a continuous spectrum, contradicting
 the expectation that the black hole spectrum is discrete and has a finite Bekenstein-Hawking entropy. 
It is believed that  perturbative quantum effects do not fix this problem \cite{Maldacena:2001kr}, but spacetime wormholes have played a role in addressing  some aspects of it.

It was proposed in \cite{Cotler:2016fpe} that a quantum system that describes a black hole has a chaotic spectrum. 
Similarly to strongly interacting many-body systems, this means that the statistics of black hole energy levels can be modeled by a random 
matrix ensemble. This can be tested by the time dependence of the spectral form factor, introduced in \cite{Cotler:2016fpe} in the context 
of black holes. It was later realized in \cite{Saad:2018bqo} that in order to reproduce the features in the spectral form factor characteristic of 
random matrix  level repulsion, it was necessary to include contributions from spacetime wormholes. The wormhole 
found in \cite{Saad:2018bqo} is universal and can be constructed, given the original black hole metric as a starting point, in any dimension 
and any theory of gravity.   At late enough times, the contribution from this wormhole reproduces
random matrix behavior,  supporting the idea of black hole chaos. 

The connection between black holes and chaos was beautifully illustrated in \cite{SSS}. The authors studied a solvable theory 
of gravity: pure Jackiw-Teitelboim (JT) gravity \cite{Jackiw:1984je,teitelboim1983gravitation} in a two-dimensional world asymptotic to AdS$_2$ \cite{Almheiri:2014cka,Jensen:2016pah,Maldacena:2016upp,Engelsoy:2016xyb}. As a first observation, it is possible to perform 
the exact gravitational path integral around the black hole solution, and verify that perturbative quantum effects do not give 
access to the discrete nature of the spectrum. The authors of \cite{SSS} proceeded by evaluating the exact path integral on (oriented)
surfaces with arbitrary topology.   It turned out that the gravitational path integral in this theory is equivalent to a matrix integral; the sum over topologies
corresponds to the $1/L$ expansion of the matrix integral, where $L$ is the rank of the matrices.
The matrix is interpreted as the Hamiltonian of the dual quantum mechanics and therefore, in this particular model, the black hole Hamiltonian really is a random matrix.

The duality between JT gravity and random matrix models has been extended in multiple ways. 
Some of the generalizations have been reviewed in  \cite{Mertens:2022irh} and references therein.
The generalization that is most directly relevant to the present article is the extension from JT gravity to JT supergravity \cite{SW}. 
   In this case, depending on certain anomaly coefficients and on whether time-reversal symmetry is assumed, one encounters various 
 $(\upalpha,\upbeta)$-ensembles 
of Altland and Zirnbauer \cite{AZ}, as well as Dyson $\upbeta$ ensembles.

In the present paper, we will extend the comparison between gravity and matrix models to the case of $\N=2$ JT supergravity.  We will identify a matrix model 
that agrees with the gravitational path integral of $\N=2$ JT supergravity to all orders in the topological expansion.\footnote{The  $\N=0$ and $\N=1$
versions of the duality between JT gravity and a matrix model 
 are limits of previously conjectured dualities between the minimal (super)string models and matrix integrals \cite{SSS, Johnson:2019eik, Mertens:2020hbs}. Curiously an $\N=2$ analog of this does not seem to be known.}
We also make some preliminary observations on $\N=4$ JT supergravity.

Before summarizing our results, we would like to motivate the problem. String theory provides multiple concrete examples of quantum systems that at strong coupling describe  black holes, e.g. the D0-brane black hole \cite{Maldacena:2023acv}, four dimensional $1/2$-BPS black holes arising from type II string theory compactified on a Calabi-Yau \cite{Mohaupt:2000mj}, excitations of the $1/16$-BPS black hole in ${\rm AdS}_5\times {\rm S}^5$ \cite{Cabo-Bizet:2018ehj,Choi:2018hmj,Benini:2018ywd}, among others. All these systems display extended supersymmetry, which affects the corresponding random matrix ensemble. 
Most cases can be approximated by JT supergravity models with $\N>1$ supersymmetry coupled to matter.
The last example mentioned  is particularly relevant, since at low temperatures this black hole is described by $\N=2$ JT supergravity coupled to matter \cite{Boruch:2022tno}. The four dimensional black holes mentioned in the second example are described instead by $\N=4$ JT supergravity \cite{Heydeman:2020hhw}. We summarize the spectral properties of JT supergravity in fig. \ref{fig:enter-label}.

\begin{figure}
\centering
\hspace{-0cm}\begin{tabular}{S|SSSS} \toprule
    {\footnotesize  Supercharges} & {\footnotesize  (BPS) Ground States} & {\footnotesize  Gap} & {\footnotesize  Index} & {\footnotesize  $R$-symmetry}  \\ \midrule\midrule
  {\footnotesize  $\mathcal{N}=0$} & {\footnotesize  No} & {\footnotesize  No} & {--} & {--} \\ 
  {\footnotesize  $\mathcal{N}=1$} & {\footnotesize  No}  & {\footnotesize  No} & {\footnotesize $0$} & {--}  \\ 
   {\footnotesize  $\mathcal{N}=2$}  & {\footnotesize  Yes $\forall (\widehat{q},\delta)\neq (1,\frac{1}{2})$}  & {\footnotesize  Yes $\forall \delta\neq \frac{1}{2}$} & {\footnotesize  $0$ unless $\widehat{q}=1\,\&\,\delta \neq \frac{1}{2}$} & {\footnotesize ${\rm U}(1)$} \\ 
   {\footnotesize  $ \mathcal{N}=4$}  & {\footnotesize  Yes}  & {\footnotesize  Yes} & {\footnotesize  Same as degeneracy} & {\footnotesize  ${\rm SU}(2)$}  \\ \bottomrule
\end{tabular}
 \caption{\footnotesize Spectral properties of JT gravity in the large $S_0$ approximation, for different number of supercharges. The theory is unique for $\mathcal{N}=0,1,4$ but multiple theories exist for $\mathcal{N}=2$ depending on an odd integer $\widehat{q}$, the $R$-charge of the supercharge, and a number $\delta~{\rm mod}~\mathbb{Z}$, a non-integer background $R$-charge. For most values of these parameters, there is an order $e^{S_0}$ number of BPS states but the index vanishes due to cancellations. (A refinement of the index exists which does not vanish, and predicts the presence of BPS states.) One can deform $\mathcal{N}=1$ JT gravity with Ramond punctures, which produces a finite number of BPS states and a non-vanishing index, which does not necessarily grow with $S_0$. (This case is not shown in the table.) }
 \label{fig:enter-label}
\end{figure}

In the remainder of this introduction we will summarize the contents of this article. In section \ref{randommatrix}, we study random matrix ensembles with $\N=2$
supersymmetry (and no time-reversal symmetry) and attempt to motivate the class of random matrix ensembles that we will compare to the gravity calculation.
In the ensembles that we consider, supermultiplets with different $R$-charges are statistically independent and each one is described by an AZ ensemble with $(\upalpha,\upbeta)=(1,2)$.
BPS states can be present and their wavefunctions are random, a property that
 was proposed recently \cite{Lin:2022zxd, Lin:2022rzw} as an aspect of black hole chaos.
 
In section \ref{sec:N2JTG}, we begin our analysis of $\N=2$ JT supergravity. We review the exact calculation of the path integral around the black hole geometry and use it to extract the leading spectral curve of the matrix model. We find a large number of BPS states in a restricted range of $R$-charges. Non-BPS states have a continuous density of states which can be mapped to a choice of matrix potential. In section \ref{sec:Z02} we evaluate the path integral on a space with the topology of a cylinder, dual to the spectral form factor. We find the result from gravity to be consistent with supermultiplets being statistically independent and the number of BPS states being fixed (rather than fluctuating as a random variable).

In section \ref{sec:GPI}, we perform the gravitational path integral for $\N=2$ JT supergravity on orientable spaces with an arbitrary number of handles and boundaries. Similarly to \cite{SSS}, the calculation reduces to evaluating the volumes of moduli spaces of hyperbolic surfaces, in this case with $\N=2$ supersymmetry.\footnote{See \cite{CraneRabin, Cohn:1986wn, Ip:2016ojn} for
mathematical work on $\N=2$ hyperbolic geometry.}  The main result of this section is to determine the measure on this moduli space induced by the supergravity path integral, presented in eqn. \eqref{eq:finalmeasure}. To obtain this result we evaluate the one-loop determinant arising from JT gravity using the Reidemeister torsion (an alternative, on orientable manifolds,
 would be to use the symplectic structure). We compute the integral over the moduli space in the simplest case of a three-holed sphere. 

In section \ref{sec:RR}, we complete the evaluation of the $\N=2$ JT supergravity path integral. Having the measure over moduli space of hyperbolic surfaces is not enough, since one needs to efficiently divide by the mapping class group. In section \ref{reviewmirz} we review the recursion relation by which Mirzakhani resolved this problem for bosonic surfaces. In section \ref{sec:RRfinalform}, after introducing the necessary elements of $\N=2$ supergeometry, we derive a similar recursion relation for volumes of moduli spaces of $\N=2$ hyperbolic surfaces, presented in eqn. \eqref{eq:N2recrel}. We list some interesting properties of these volumes. Importantly, we show in section \ref{equivalence} that they satisfy the  loop equations of
random matrix theory and use this result to finalize our proof that $\N=2$ JT supergravity is dual to the matrix ensemble proposed in section \ref{randommatrix}. 
\begin{figure}
\centering
\hspace{-0cm}\begin{tabular}{S|SS} \toprule
    {\footnotesize  $\mathcal{N}=2$ Random Matrices} & {\footnotesize  Ensemble for most supermultiplets} & {\footnotesize  Exceptions}   \\ \midrule\midrule
  {\footnotesize  No time-reversal symmetry} & {\footnotesize  ${\rm U}(L)\times {\rm U}(L')$ bif.} & {\footnotesize  None}  \\ \midrule
  {\footnotesize  ${\sf T}$ such that ${\sf T}^2=1$} & {\footnotesize  ${\rm O}(L)\times {\rm O}(L')$ bif.}  & {\footnotesize  None}  \\ 
   {\footnotesize  ${\sf T}$ such that ${\sf T}^2=-1$}  & {\footnotesize  ${\rm Sp}(L)\times {\rm Sp}(L')$ bif.}  & {\footnotesize  None}   \\  \midrule
  {\footnotesize  ${\sf CT}$ such that ${\sf CT}^2=1$}  & {\footnotesize  ${\rm U}(L)\times {\rm U}(L')$ bif.}  & {\footnotesize  $(-\frac{\widehat{q}}{2},\frac{\widehat{q}}{2})$: ${\rm U}(L)$ symm.}   \\ 
    {\footnotesize  ${\sf CT}$ such that ${\sf CT}^2=-1$}  & {\footnotesize  ${\rm U}(L)\times {\rm U}(L')$ bif.}  & {\footnotesize  $(-\frac{\widehat{q}}{2},\frac{\widehat{q}}{2})$: ${\rm U}(L)$ anti.}   \\ \bottomrule
\end{tabular}
 \caption{\footnotesize Summary of $\mathcal{N}=2$ random matrix ensembles. One of our main results is that different supermultiplets are statistically independent and can therefore be described by their own ensemble. We label them by their symmetry group, with $L$ and $L'$ being the appropriate dimensions for each supermultiplet of $R$-charges $(k,k+\widehat{q}\,)$ with $k\in \mathbb{Z}+\delta$. ${\sf T}$ is a time-reversal symmetry that commutes with all bosonic generators (can exist for any $\delta$) while ${\sf CT}$ anticommutes with the $R$-charge generator (can only exist for $\delta=0$ or $1/2$) and is defined such that ${\sf CT} Q^\dagger {\sf CT}^{-1} = Q$. For this reason, in the presence of ${\sf CT}$ symmetry the multiplets $(k,k+\widehat{q}\,)$ and $(-k-\widehat{q},-k)$ are identified. The final two exceptions only are relevant when $\delta= 1/2$. In section \ref{timereversal} we explain in what sense these are the only cases to consider.}
 \label{fig:enter-label2}
\end{figure}

In section \ref{timereversal},  we generalize parts of the story to $\N=2$ quantum mechanics with a time-reversal symmetry, either commuting or anticommuting with the $R$-charge. A summary appears in fig.~\ref{fig:enter-label2}. We identify the correct matrix model ensemble for each case. We also compute the 
JT supergravity path integral in the presence of crosscaps and match with the loop equations of the proposed matrix models.  In one case, one has to assume the existence of wormhole
corrections to the spectral curve. 

In section \ref{N4JT}, we generalize some of our findings to $\N=4$ JT supergravity, or alternatively to matrix models with four supercharges and  $\SU(2)$ $R$-symmetry. We evaluate the gravitational path integral around the black hole solution and also on the cylinder that computes the spectral form factor, showing that supermultiplets are statistically independent.  We also
develop the beginnings of $\N=4$ random matrix theory.

In appendix \ref{sec:disktrumpet}, we provide a more detailed analysis of the boundary reparametrization mode for $\N=2$ and $\N=4$ JT supergravity. In appendix \ref{app:Loop} we derive the matrix model loop equations in the presence of a logarithmic term in the matrix potential, necessary to match $\N=2$ JT gravity. We also define a notion of ``volumes'' valid for any matrix model and compute the analog of Mirzakhani's recursion for any spectral curve. This streamlines the derivations in section \ref{equivalence}. Other technical details are left for appendices \ref{sec:constn2} and \ref{app:JTU1bos}.

\section{Random Matrices with $\N=2$ Supersymmetry}\label{randommatrix} 

\subsection{Preliminaries}

Our goal in this section is to understand how random matrix theory might apply to $\N=2$ JT supergravity.\footnote{Random matrix theory with $\N=2$ supersymmetry was first investigated in \cite{Kanazawa:2017dpd}. Our approach has some differences in detail.}   We will motivate the idea that multiplets
with different $R$-charges are statistically independent, and  suggest what  type of random matrix ensemble should be relevant.

We assume that the random matrix theory relevant to $\N=2$ JT supergravity is supposed to realize a global $\N=2$ supersymmetry algebra that includes  a Hamiltonian $H$ and 
a hermitian conjugate pair of supersymmetry generators $Q$ and $Q^\dagger$ satisfying
\be\label{supercharges} Q^2=Q^\dagger{}^2 =0,\hskip.5cm \{Q,Q^\dagger\}=H,~~~ [H,Q]=[H,Q^\dagger]=0.\ee
We also assume the existence of a $\uU(1)$ ``$R$-symmetry'' group of outer automorphisms of this algebra, generated by the ``$R$-charge'' $J$, satisfying
\be\label{rcharge}[J,Q]=\h q\, Q, ~~~~[J,Q^\dagger]=-\h q \, Q^\dagger, ~~~~[J,H]=0. \ee 
What motivates these assumptions is that this global supersymmetry algebra can be naturally preserved in a cutoff version of $\N=2$ JT supergravity, with a cutoff near
the conformal boundary.
The operator $(-1)^\fF$ that distinguishes bosons from fermions is supposed to be 
\be\label{bosferm}(-1)^\fF=e^{\pm \i\pi J}.   \ee with either choice of sign.
This implies that $\h q$ is an odd integer, and that $J$ has even (odd) integer eigenvalues for bosonic (fermionic) states.   After possibly exchanging $Q$ and $Q^\dagger$,
one can assume that $\h q>0$.

As explained in section \ref{restoring}, there is also an ``anomalous'' case in which eigenvalues of $J$ are valued not in $\Z$ but in $\delta+\Z$, with a constant $\delta$.
Then $e^{\pm \i \pi J}$ still commutes with bosonic operators, and anticommutes with fermionic ones, but its eigenvalues are not $\pm 1$ and there is not really a natural
notion of which states are ``bosonic'' and which are ``fermionic.''  

In the present section and most of the article, we  consider $\N=2$ random matrix models
without time-reversal symmetry.   Time-reversal symmetry will be included in section \ref{timereversal}.

\subsection{Review of $\N=1$}\label{randomone}  

We begin by reviewing  random matrix theory  with $\N=1$ supersymmetry,  relevant to $\N=1$ JT supergravity \cite{SW}.   Here actually we will only consider the ``non-anomalous''
case that arises, for example, as a low energy approximation to an SYK model with an even number $N$ of Majorana fermions.   
As will be apparent, the assumption of a $\uU(1)$ $R$-symmetry leads for $\N=2$ to a structure similar to what one finds for $\N=1$ supersymmetry with even $N$.
The odd $N$ case of an $\N=1$ SYK model leads to a very different random matrix theory.    

In $\N=1$ random matrix theory, we assume that the Hamiltonian is $H=Q^2$, where $Q$ is a self-adjoint supercharge,
 odd under the symmetry $(-1)^\fF$ that distinguishes
bosons from fermions.  In a basis in which this operator is block diagonal
\be\label{minf}(-1)^{\fF}=\left(\begin{array} {c|c}  1 & 0\cr \hline 0 & -1\end{array}\right),\ee
with diagonal blocks of sizes  $L_b$, $L_f$ (respectively the numbers of bosonic and fermionic states), $Q$ is off-diagonal,
\be\label{zinf} Q=\left(\begin{array} {c|c}  0 & \qQ\cr \hline \qQ^\dagger & 0\end{array}\right).\ee  Here $\qQ$ is a complex-valued $L_b\times L_f$ matrix, subject to no constraint;
hermiticity of $Q$ just says that the lower left block in eqn. (\ref{zinf}) is the hermitian adjoint of $\qQ$. 

This structure is invariant under a unitary group $G=\uU(L_b)\times \uU(L_f)$  with the two factors acting on the bosonic and fermionic states, respectively:
\be\label{pinf}U= \left(\begin{array} {c|c}  U_{L_b} & 0\cr \hline 0 & U_{L_f}\end{array}\right).\ee
$\qQ$ transforms as 
\be\label{qtrans}\qQ\to U_{L_b} \qQ U_{L_f}^{-1}, \ee
that is, as a bifundamental of $\uU(L_b)\times \uU(L_f).$
In \cite{SW}, $\N=1$ JT supergravity was related to a random matrix ensemble in which $\qQ$ is treated as a random matrix with a measure of the general form\footnote{In this article,  for 
any matrix $M$,  we denote by  $\d M$  the product of Euclidean measure for all independent matrix elements of $M$.} 
\be\label{genform}\d\mu = \d \qQ \,\exp(-\Tr\, F(\qQ^\dagger \qQ)),\ee
with a suitable function $F$.  
As usual in random matrix theory, an integral over a random matrix $\qQ$ with such a measure can be conveniently reduced to an integral over its eigenvalues, in an appropriate
sense.  For example, in the present case, if $L_b =L_f=L$, then $\uU(L_b)\times \uU(L_f)$ can be used to diagonalize $\qQ$,
\be\label{diagQ}\qQ=\begin{pmatrix}\lambda_1 & &&& \cr & \lambda_2 &&& \cr && \ddots && \cr
         &&&&\lambda_L \end{pmatrix} ,\ee
         with all $\lambda_i\geq 0$.   There remains a residual $\uU(1)^L$ symmetry that preserves this form.   
         
         It is then necessary to find the appropriate measure for integration over the $\lambda$'s.    This may be done as follows.  Let $\mathfrak g$ be the Lie algebra of
         $G$, $\mathfrak h$ the Lie algebra of the residual symmetry group $\uU(1)^L$, and $\mathfrak g^\perp$ the orthocomplement of $\mathfrak h$ in $\mathfrak g$.
         Let $\O_\qQ$ be the $G$ orbit of the matrix $\qQ$ in the space of $L_b\times L_f$ complex matrices, and let $T_\qQ$ be the tangent space to this
         orbit at the point corresponding to $\qQ$.   The measure factor that we want is the determinant of the natural map  $W:\mathfrak g^\perp\to T_\qQ$ which expresses the
         fact that $\mathfrak g^\perp$ generates the orbit.   In practice,
        we will find an orthonormal basis  $e_1,e_2,\cdots \in \mathfrak g^\perp$ that maps to an orthonormal basis $W(e_1),W(e_2),\cdots\in T_\qQ$, 
         implying that the desired volume factor is the product of ``eigenvalues'' of $W$, that is, it is the product $\prod_a|W(e_a)|/|e_a|$, where the norm of a matrix $m$ is defined by
         $|m|^2=\Tr\,m^\dagger m$.  In evaluating $\prod_a |W(e_a)|/|e_a|$, we will drop constant factors independent of the $\lambda$'s.  
                  
         First of all, for each eigenvalue $\lambda_i$, there is a generator $e$ of $\uU(L_b)$ (or $\uU(L_f)$; this choice does not matter because of the residual $\uU(1)^L$ symmetry) that rotates
         $\lambda_i$ in the complex plane.   The generator $e$, if normalized to have norm 1, acts on $\lambda_i$ by 
         $\delta_e\lambda_i=\i \lambda_i$, with $|\delta_e\lambda_i|=\lambda_i$.   So $|W(e)|/|e|=\lambda_i$.
         By itself  this would lead to a measure $\prod_i \lambda_i\d \lambda_i$.
         
        Now consider a  pair of eigenvalues $\lambda_i,$ $\lambda_j$, and the subgroup $\uU(2)_b\times \uU(2)_f$ of $\uU(L_b)\times \uU(L_f)$ that rotates the corresponding eigenvectors of
        $\qQ^\dagger \qQ$ among themselves.  The group $\uU(2)_b\times \uU(2)_f$ is eight-dimensional; we have accounted for four generators (two residual symmetries and two
        that rotates the phases of $\lambda_i,\lambda_j$), so there are four more to consider.  Two linear combinations of generators take the form
        \be\label{matform} \begin{pmatrix}0 & \w \cr \bar\w & 0\end{pmatrix}\in \mathfrak{u}(2)_b, \hskip1cm    \begin{pmatrix}0 & \w \cr \bar\w & 0\end{pmatrix}\in \mathfrak{u}(2)_f,
        \hskip1cm \w\in\C.\ee
        These generators transform $\qQ$ by
        \be\label{transQ} \delta\begin{pmatrix} \lambda_i & 0\\ 0 &\lambda_j\end{pmatrix} =\begin{pmatrix} 0 & -\w(\lambda_i-\lambda_j) \cr
         \bar\w(\lambda_i-\lambda_j) & 0 \end{pmatrix}.\ee   For these two generators, we have $\prod_{a=1,2}|W(e_a)|/|e_a|=
        (\lambda_i-\lambda_j)^2$.   The other two linear combinations
         take the form
           \be\label{matform2} \begin{pmatrix}0 & \w \cr \bar\w & 0\end{pmatrix}\in \mathfrak{u}(2)_b, \hskip1cm    -\begin{pmatrix}0 &\w \cr \bar\w & 0\end{pmatrix}\in \mathfrak{u}(2)_f,
        \hskip1cm \w\in\C, \ee
transforming $\qQ$ by
        \be\label{transQtwo} \delta\begin{pmatrix} \lambda_i & 0\\ 0 &\lambda_j\end{pmatrix} =\begin{pmatrix} 0 & \w(\lambda_i+\lambda_j) \cr
         \bar\w(\lambda_i+\lambda_j) & 0 \end{pmatrix},\ee
         and for these generators we get $\prod_a |W(e_a)|/|e_a|= (\lambda_i+\lambda_j)^2$.
        
        At this stage, we have accounted for a full basis of $\mathfrak g^\perp$.
         Therefore, for this example with $L_b=L_f$, the measure for the eigenvalues is
         \be\label{firstmeasure}\prod_{i=1}^{L_b} \lambda_i \d\lambda_i \prod_{1\leq i<j\leq L_b} (\lambda_i-\lambda_j)^2(\lambda_i+\lambda_j)^2.\ee
         In random matrix theory, an ensemble with nonnegative ``eigenvalues'' $\lambda_i$ and measure
         \be\label{azmeasure}\prod_i\d\lambda_i  \lambda_i^{\upalpha} \prod_{i<j}(\lambda_i^2-\lambda_j^2)^\upbeta \ee
         is usually called an Altland-Zirnbauer \cite{AZ} $(\upalpha,\upbeta)$ ensemble,\footnote{Special cases of such ensembles had certainly been considered previously, for example in
         \cite{VZ,Anderson:1990nw,Anderson:1991ku,Dalley:1991qg,Dalley:1991vr}. The last two references were in the context of applications to gravity.} so here we have such an ensemble with $(\upalpha,\upbeta)=(1,2)$.
         
         It remains to consider the case $L_b\not= L_f$.   It turns out that the resulting measure depends only on $|L_b-L_f|$, so we lose nothing essential if we assume $L_b>L_f$,
         say $L_b=L_f+\nu$.   $\qQ$ is now a rectangular $L_b\times L_f$ matrix.   In its canonical form, it has $\nu$ rows of zeroes:
         \be\label{diagQnew}\qQ=\begin{pmatrix}\lambda_1 & &&& \cr & \lambda_2 &&& \cr && \ddots && \cr
         &&&&\lambda_{L_f}\cr 0 &&&&\cr &0&&&\cr &&\ddots && \end{pmatrix} ,\ee   
         The residual symmetry group is now $\uU(1)^{L_f}\times \uU(\nu)$, where the second factor acts on the last  $\nu$ rows of $\qQ$.   
    In addition, for each  $i$, there are $2\nu$ generators of $\mathfrak g^\perp$ that at linearized order map the $i^{th}$ (left) eigenvector of $\qQ$ into the space of null
         vectors.   For each of these generators, $|W(e)|/|e|=\lambda_i$.     Overall, for each $i$, we now get $1+2\nu$ factors of $\lambda_i$, and  the measure becomes 
           \be\label{azmeasuretwo}\prod_i\d\lambda_i  \lambda_i^{1+2\nu} \prod_{i<j}(\lambda_i^2-\lambda_j^2)^2, \ee
       corresponding to   an AZ ensemble with $(\upalpha,\upbeta)=(1+2\nu,2)$.   It will be convenient for later discussions to introduce a parameter $\upalpha_0$ and to write 
       \be\label{conven}\upalpha=\upalpha_0+\nu\upbeta,\hskip1cm (\upalpha_0,\upbeta)=(1,2). \ee

\subsection{Random Matrices for $\N=2$}\label{randomtwo}

Now we will try to generalize this discussion to $\N=2$ supersymmetry.   An immediate difference is that the $\Z_2$-valued symmetry $(-1)^\fF$ of the previous discussion
is now promoted to an additive $R$-charge symmetry, with generator $J$.   For our purposes in this section, it is convenient to assume that $J$ has integer eigenvalues and
that $\h q=1$.   The results we get are straightforwardly adapted to a more general case described in section \ref{restoring}.

Let $\H_k$, of dimension $L_k$, be the space of quantum states of $R$-charge $k$.   The full quantum Hilbert space is then $\H=\oplus_k \H_k$.
The supercharge $Q$, since it has $R$-charge 1,  takes the form $Q=\sum_k Q_k$ with $Q_k:\H_k\to \H_{k+1}$ and therefore $Q_k^\dagger:\H_{k+1}\to \H_k$.
 Similarly $Q^\dagger=\sum_k Q_k^\dagger$.    

The supersymmetry algebra has two types of irreducible multiplet.   First, a single zero-energy state of any $R$-charge, annihilated by $Q$ and $Q^\dagger$, is a multiplet by
itself.  These are the supersymmetric or BPS states.   Second, one can have a pair of states with adjacent values of the $R$-charge, say $\psi_k\in\H_k$, $\psi_{k+1}\in\H_{k+1}$,
with $Q\psi_k=\lambda\psi_{k+1}$, $Q^\dagger\psi_{k+1}=\bar\lambda \psi_k$.   These two states make up what we will call an $(k,k+1)$ supermultiplet.\footnote{It will be useful, starting from section \ref{sec:N2JTG}, to label a non-BPS multiplet
of charges $(k,k+\h q\, )$ (where in our present discussion $\hq=1$) by the average charge $q=k+\h q/2$.}

  Each $\H_k$
can be decomposed as the direct sum of three pieces:
\be\label{hn}\H_k=\H_k^0\oplus \H_k^+\oplus \H_k^-. \ee
Here $\H_k^0$ consists of supersymmetric states of $R$-charge $k$, $\H_k^+$ consists of states of $R$-charge $k$ that are in $(k,k+1)$ multiplets, and $\H_k^-$ consists
of states of $R$-charge $k$ that are in $(k-1,k)$ multiplets.  Letting $L_k^0$, $L_k^+$, and $L_k^-$ denote the dimensions of $\H_k^0$, $\H_k^+$, and $\H_k^-$, respectively,
we have clearly
\be\label{zn} L_k=L_k^0+L_k^++L_k^-. \ee
Another obvious implication is that $L_k^{+}=L_{k+1}^-$. 

Consider first a simplified model in which the $R$-charge takes only two adjacent values say $\H=\H_{k-1}\oplus \H_{k}$.   We then have simply $Q:\H_{k-1}\to \H_k$,
$Q^\dagger:\H_k\to\H_{k-1}$.   The structure is the same as in the $\N=1$ case discussed in section \ref{randomone}, with the nonhermitian supercharge $Q$ playing the
role that was played previously by $\qQ$.    A natural random matrix ensemble would then be an AZ $(\upalpha,\upbeta)=(1+2\nu,2)$ ensemble, with $\nu=|L_k-L_{k-1}|$, as in section \ref{randomone}.

However, as soon as the $R$-charge takes more than two adjacent values, we run into apparent difficulty.  For any integer $k$, let $Q_k$ be the restriction of $Q$ to a map from $\H_k$
to $\H_{k+1}$.
Suppose that there are three adjacent values of the $R$-charge, 
say $\H=\H_{k-1}\oplus \H_k\oplus \H_{k+1}$.   Now
we have $Q=Q_k+Q_{k-1}$, where $Q_k$ and $Q_{k-1}$ are related by 
\be\label{constrained} Q_k Q_{k-1}=0, \ee
which follows from the supersymmetry relation $Q^2=0$.  This constraint means that $Q_{k-1}:\H_{k-1}\to \H_k$ and $Q_k:\H_k\to \H_{k+1}$ cannot be chosen independently.
However, we will argue that nonetheless, it is natural in random matrix theory to expect that the $(k-1,k)$ and $(k,k+1)$ supermultiplets are statistically independent.  

As a preliminary, we have to think about what  a generic solution of the constraint $Q_k Q_{k-1}=0$ looks like.   In fact, the space of generic solutions of this constraint
has multiple components, differing in the values of $L_k^0$ and $L_k^\pm$.   There are only two general constraints.  One constraint 
 comes from the supersymmetric index. If $\H$ has finite dimension with $\sum_k L_k<\infty$ (implying in particular that $L_k=0$ for all but finitely many $k$), then
 \be\label{index} \sum_k (-1)^k L_k^0 = \sum_k (-1)^k L_k, \ee because the $(k,k+1)$ multiplets make no net contribution to the right hand side.   
 This condition is satisfied for all solutions of $Q^2=0$, not just generic ones.  
 In JT supergravity, the individual $L_k$ are infinite, for all $k$,
  but a regularized version of this condition makes sense, with $L_k$ replaced by $\Tr_{\H_k} e^{-\beta H}$.  
  A second constraint holds only for generic solutions of $Q^2=0$.   In a generic solution, for any $k$, either $L_k^0$ or $L_{k+1}^0$ vanishes, since otherwise
  $Q$ could be perturbed slightly to combine a pair of supersymmetric states of $R$-charges $k$ and $k+1$ into a single $(k,k+1)$ multiplet.
  
  These constraints and eqn. (\ref{zn}) are not enough to determine the $L_k^0, L_k^\pm$ in terms of the $L_k$.  
  For example, consider a simple case with $L_{k-1}=L_k=L_{k+1}=2$ and other $L_k$ vanishing.
  The spectrum could consist of two supersymmetric states of $R$-charge $k-1$ and two $(k,k+1)$ multiplets; but it could also consist of two supersymmetric
  states of $R$-charge $k+1$ and two $(k-1,k)$ multiplets, or one supersymmetric state of $R$-charge $k-1$ and one of $R$-charge $k+1$, along with
  an $(k-1,k)$ multiplet and an $(k,k+1)$ multiplet.

If $\sum_k L_k<\infty$, then the $L_k^\pm $ can be uniquely determined from eqn. (\ref{zn})  if the $L^0_k$ are given. 
In a matrix model relevant to JT supergravity, we expect that the $L_k$ and $L_k^\pm$ are all infinite in a double-scaled limit.
  But it is reasonable to imagine
starting with a cutoff matrix model with $\sum_k L_k<\infty$, in which the $L^\pm_k$ are known in terms of the $L_k$ and the $L_k^0$.  Our strategy will be to understand
what is a reasonable random matrix model in such a cutoff situation, and then to extrapolate to a double-scaled limit. 

But what should we assume for the $L_k^0$?
In an appropriate context, one might be interested in a probability distribution governing all possible solutions of $Q^2=0$.  In such a case, the $L^0_k$ would be random variables.
In $\N=2$ JT supergravity, however, one can just
calculate the $L_k^0$ from the disk partition function (see  \cite{Stanford:2017thb} or section \ref{sec:N2JTG}), and they appear to have no fluctuations associated to wormholes.

Since we want to assume that the $L_k^0$ are known, this means that in a cutoff version of the theory with $\sum_k L_k<\infty$, we should also consider the $L_k^\pm$ to be known.
Thus we want  to consider a particular component of the space of solutions of $Q^2=0$, characterized by given values of the $L_k^0$ and the $L_k^\pm$,
and understand what is a reasonable random matrix model in that context.   Hopefully, we will get an answer that makes sense in a double-scaled limit 
in which the Hilbert space dimensions become infinite.

Roughly speaking, we want to consider  a cutoff model with $\sum_k L_k<\infty$ and a measure of the general form
\be\label{roughmeasure} \prod_k \d Q_k   \prod_s \delta(Q_{s+1} Q_s) \prod_r \exp(-\Tr_{\H_r} F_r(Q_r^\dagger Q_r)),\ee
with suitable functions $F_r$. The delta functions are meant to impose the vanishing of the real and imaginary parts of each matrix element of $Q_{s+1}Q_s$.  The solutions of the delta function constraints consist of a finite number of families, as just explained, and we want to focus on just one family.   There is not
a convenient way to enforce the restriction to one family by including an extra factor in eqn. (\ref{roughmeasure}), and we will just impose that restriction by hand.
For the desired component of solutions of $Q^2=0$,  
after taking account of the delta functions and the symmetry group $G=\prod_k \uU(L_k)$, we want to reduce the model to an integral over a suitable set of
``eigenvalues.''    

There are two reasons that the output of this procedure might lead to a model in which the representations of $R$-charges $(k,k+1)$ for different values of $k$ are not statistically
independent.  The $(k-1,k)$ and $(k,k+1)$ representations are potentially coupled both by the fact that the symmetry group $\uU(L_k)$ acts on both $Q_{k-1}$ and $Q_k$ and by
the constraint $Q_k Q_{k-1}=0$.   However, it turns out that these two effects cancel, and the multiplets of different charge are statistically independent.  

Let us go back to our simplified model with three consecutive charges and first consider the case that all $L_k^0$ vanish, so there are generically no supersymmetric states.  
Hence $\H_{k+1}=\H_{k+1}^-$, $\H_k=\H_k^+\oplus \H_k^-$, $\H_{k-1}=\H_{k-1}^+$.    To see the essential cancellation, let us look at the case $L_{k+1}=L_{k-1}=1$, $L_k=2$.
By the action of $G$, if we arrange $\H=\H_{k+1}\oplus \H_k\oplus \H_{k-1}$ as a sort of column vector
 $\begin{pmatrix}\H_{k+1}\cr \hline \H_k\cr \hline \H_{k-1}\end{pmatrix}$, 
a  generic $Q$ with $Q^2=0$ and leaving no supersymmetric states can be put in the form
\be\label{canform}Q=\left(\begin{array}{c|cc|c} 0 & \lambda & 0 & 0\\  \hline 0&0&0&0 \\ 0&0&0&\t\lambda\\ \hline 0&0&0&0 \end{array}\right).\ee
We want to consider small fluctuations around this canonical form, and the measure that results from removing these small fluctuations by a combination
of delta functions and symmetries.  
A general ansatz for the small fluctuations is 
\be\label{cnform}Q=\left(\begin{array}{c|cc|c} 0 & \lambda &u  & 0\\  \hline 0&0&0&v \\ 0&0&0&\t\lambda\\ \hline 0&0&0&0 \end{array}\right),\ee
where the perturbation is parametrized by $u,v$.
A delta function constraint imposing $Q^2=0$ thus reduces in this case to
\be\label{deltafn}\delta^2(\t\lambda u+\lambda v), \ee where, for a complex number $b=b_1+\i b_2$, we define $\delta^2(b)=\delta(b_1)\delta(b_2)$.    
We should also analyze the action of $G=\uU(1)_{k+1}\times \uU(2)_{k} \times \uU(1)_{k-1}$.   There is a residual $\uU(1)\times \uU(1)$ subgroup of $G$ that leaves fixed the canonical
form (\ref{canform}) of $Q$. Also, another $\uU(1)\times \uU(1)$ subgroup -- say $\uU(1)_{k+1}\times \uU(1)_{k-1}$ -- can be used to set $\lambda,\t\lambda\geq 0$. An argument similar to the one presented in section \ref{randomone} implies that the contribution to the measure from these generators is a factor of $\lambda \, \tilde{\lambda}$. The symmetry
generators not yet accounted for are the off-diagonal generators of $\uU(2)_k$, say
\be\label{symgens}\begin{pmatrix} 0 & \w\cr \bar\w & 0 \end{pmatrix}. \ee
The transformation of $(u,v)$ that this generates is
\be\label{transgen} \delta_\w(u,v) =(-\w \lambda,\w\t\lambda). \ee   
Transforming to a new  basis by 
\be\label{stuv} (u,v)=\frac{s(\t\lambda,\lambda)+ t(-\lambda,\t\lambda)}{\sqrt{\lambda^2+\t\lambda^2}},\ee
normalized such that the Jacobian is trivial, the delta function is
\be\label{deltafn2}\delta^2\left(s\sqrt{\lambda^2+\t\lambda^2}\right)=\frac{1}{\lambda^2+\t\lambda^2}\delta^2(s),\ee
and the transformation generated by the off-diagonal symmetries is
\be\label{residtrans}\delta_\w s=0,~~\delta_\w t = (\lambda^2+\t\lambda^2)^{1/2}\w. \ee
Taking account the real and imaginary parts of $\w$, the volume factor associated to the off-diagonal symmetries is $\lambda^2+\t\lambda^2$,
and this precisely cancels the dependence on $\lambda,\t\lambda$  in (\ref{deltafn2}).   Hence in this example the measure is the product of a measure for $\lambda$
and a measure for $\t\lambda$, so the $(k-1,k)$ and $(k,k+1)$ multiplets are statistically independent.

Now consider the case of general $L_{k+1},L_k,L_{k-1}$, still with all $L_k^0=0$.   The constraint $Q^2=0$ and the symmetries imply that 
 $Q_k:\begin{pmatrix} \H_k^+\\ \hline \H_k^-\end{pmatrix}\to \H_{k+1}$
and $Q_{k-1}:\H_{k-1}\to \begin{pmatrix} \H_k^+\\ \hline \H_k^-\end{pmatrix}$
can be put in the canonical form 
\be\label{anotherform} Q_k=\left(\begin{array} {ccc|cc} \lambda_1 & 0 & 0 & \,0 & 0\\ 0& \lambda_2& 0 &\, 0 & 0 \cr 0 & 0&\lambda_3&\,0&0\end{array}\right),\hskip1cm
Q_{k-1}=\left(\begin{array}{cc} 0&0\cr 0&0\cr 0&0  \cr \hline\t\lambda_1&0\cr 0&\t\lambda_2\end{array}\right)\ee with all $\lambda_i,\t\lambda_j\geq 0$
(here we illustrate the case $L_{k+1}=3,\, L_k=5,\, L_{k-1}=2$).   The $\lambda$'s determine the energies of $(k,k+1)$ multiplets, and the $\t\lambda$'s determine the
energies of $(k-1,k)$ multiplets.   The effects that could potentially couple the two types of multiplet are the delta functions and symmetries that have been used to set to 0
the right block of $Q_{k}$ and the upper block of $Q_{k-1}$.   Those blocks are $L_{k+1}\times L_{k-1}$ matrices, with one matrix element for each pair $(\lambda_i,\t\lambda_j)$.  
For each  $i,j$, the vanishing of the $i,j$ matrix elements in the right block  of $Q_k$ and the upper block of $Q_{k-1}$ is determined by a complex delta function
that comes from the constraint $Q_k Q_{k-1}=0$, as well as a pair of off-diagonal generators of $\uU(L_k)$, generalizing eqn. (\ref{symgens}).   For each $i,j$,
the effects on the measure of the off-diagonal symmetry generators and the   delta function cancel in exactly the same fashion as in the previous paragraph.  The delta function
and symmetry generators that are relevant for the vanishing of the $i,j$ matrix elements are not sensitive to the existence or non-existence of other matrix elements,
so a cancellation occurs precisely as described previously.

Hence the $(k,k+1)$ and $(k-1,k)$ multiplets are statistically independent.  The measure for each multiplet is the same as if the other multiplet did not exist,
and therefore, as in section \ref{randomone}, each multiplet is governed by an AZ  ensemble with $(\upalpha,\upbeta)=(1,2)$.  

What happens if the $L_k^0$ do not all vanish?   For example, suppose that $L_k^0\not=0$ (and therefore $L_{k+1}^0=L_{k-1}^0=0$).   The canonical forms
of $Q_k$ and $Q_{k-1}$ then have extra columns and rows of zeroes.   This does not lead to the existence of any new delta functions.\footnote{The rows and columns of
zeroes can be enforced by gauge-fixing the symmetries, and this should be done before trying to analyze the delta functions associated to the $Q^2=0$ condition.}   It does lead, 
for each $\lambda_i$, to the existence of $2L_k^0$ generators  of $\uU(L_k)$ that rotate the column of $Q_k$ containing $\lambda_i$ into the columns of zeroes, and
also, for each $\t\lambda_j$, to an orthogonal set of $2L_k^0$  generators of $\uU(L_k)$  that rotate the row 
of $Q_{k-1}$ containing $\t\lambda_j$ into the $L_k^0$ rows of zeroes. 
 As in section \ref{randomone},  the effect of these generators is to supply a factor $\prod_i\lambda_i^{2 L_k^0}$
in the measure for the $\lambda$'s, and likewise a factor $\prod_j\t\lambda_j^{2 L_k^0}$ in the measure for the $\t\lambda$'s.   No coupling is induced between the
$\lambda$'s and $\t\lambda$'s because one set of new generators acts only on the $\lambda$'s and an orthogonal set of new generators acts only on the $\t\lambda$'s.    
Hence the $(k,k+1)$ and $(k-1,k)$ multiplets are statistically independent AZ ensembles with $(\upalpha,\upbeta)=(1+2 L_k^0,2)$.

If instead we assume that $L_k^0=0$ while $L_{k+1}^0$ and $L_{k-1}^0$ are potentially both nonzero, then there are $2 L_{k+1}^0$ new symmetry generators
rotating the column of $Q_k$ containing $\lambda_i$ for each $i$, and $2L_{k-1}^0$ new symmetry generators rotating the column of $Q_{k-1}$ containing $\t\lambda_j$ for each $j$.  
Now the $(k,k+1)$ multiplets are described by an AZ ensemble with $(\upalpha,\upbeta)=(1+2L_{k+1}^0, 2)$, and the $(k-1,k)$ multiplets are described by an AZ ensemble
with $(\upalpha,\upbeta)=(1+2 L_{k-1}^0,2)$.   

We can summarize what we have learned by saying that if there are just three adjacent values of the $R$-charge, then each multiplet, say of $R$-charges $r,r+1$ (where $r$ could be $k-1$ or $k$)
 is described by an AZ multiplet with
\be\label{bonven}\upalpha=\upalpha_0+\upbeta\nu, ~~~~(\upalpha_0,\upbeta)=(1,2), \ee
where $\nu$ is the number of BPS states with $R$-charge $r$ or $r+1$.

Perhaps surprisingly, something essentially new happens with four consecutive values of the $R$-charge, say $k-1,k,k+1,k+2$.   To explain the basic issue,
let us assume that each of the four Hilbert spaces is of dimension 1.   Then each of $Q_{k-1}, Q_k$, and $Q_{k+1}$ is a $1\times 1$ matrix, say $Q_{k-1}=a$, $Q_{k}=b$,
$Q_{k+1}=c$.   The constraints $Q_{k+1}Q_k=Q_k Q_{k-1}=0$ become $ab=bc=0$.    Let us look at the branch of the moduli space of solutions with $a,c\not=0$, $b=0$.
This branch describes a $(k-1,k)$ supermultiplet and a $(k+1,k+2)$ supermultiplet.   The delta functions would appear to be
\be\label{zimbo} \delta^2(ab)\delta^2(bc)=\frac{1}{|a|^2}\frac{1}{|c|^2} \delta^2(b) \delta^2(b) =\frac{1}{|a|^2}\frac{1}{|c|^2} \delta^2(b)\delta^2(0).\ee
Clearly this does not make sense as written, because of the $\delta^2(0)$.  

Concretely, the reason that this happened is that with four consecutive values of the $R$-charge, the constraints are not independent.  For any $R$-charge $k$, the constraint operators
$Z_k =Q_{k+1} Q_k$ satisfy $Z_k Q_{k-1}=Q_{k+1}Z_{k-1}$.   In the case under discussion, the constraints are $ab$ and $bc$, and the relation between the constraints
is $(ab)c=a(bc)$.   

There is actually a natural procedure to define a measure in such a problem with redundant constraints.   This procedure is known in the context of BV quantization and also in algebraic
geometry, but rather than giving a general explanation, we will just explain how the procedure works in the case at hand.   In doing so, to slightly simplify the story, we will
treat $a,b,c$ as real variables.  In the context of JT supergravity with time-reversal symmetry, there is actually a natural problem in which all the matrices are real and in particular
$a,b,c$ are real variables.  This will be explained in section \ref{boundaryview}.  But it will probably be obvious to the reader that by complexifying all of the objects that
we will consider, including the Lagrange multipliers and the ghosts, one could adapt what we will describe to the case of complex variables.

First of all, to impose a constraint $bc=0$ and also a constraint $ab=0$, we could add two Lagrange multipliers $\lambda$ and $\lambda'$ and include a term
$\i(\lambda bc+\lambda' ab)$ in the exponent of the matrix integral.   Thus we would define an extended version of the matrix integral that contains a factor
\be\label{zolbox}\int \frac{\d\lambda \d\lambda'}{2\pi}\,\:\exp\left(\i(\lambda bc+\lambda' ab)\right). \ee
One can indeed view eqn. (\ref{zolbox}) (and its obvious analog in more general cases) as a more microscopic description of the origin of the delta functions in eqn.
(\ref{roughmeasure}).
Eqn. (\ref{zolbox}) is  fine if the constraints are independent, but here we are interested precisely in a case in which they  are not independent:  they obey the identity $a(bc)-c(ab)=0$.
This means that the extended matrix model with the Lagrange multipliers included has a gauge invariance
\be\label{olbox}\delta \lambda= -\epsilon a,\hskip1cm \delta\lambda'=\epsilon c, \ee with infinitesimal parameter $\epsilon$.
Let us treat the extended theory as a gauge theory with this gauge invariance.   It could be gauge-fixed with a  gauge condition $\Lambda=0$ for a rather general function $\Lambda$; let us
consider the simple choice $\Lambda=\lambda'$.
Following standard reasoning, the ghost determinant for this gauge condition is\footnote{Faddeev-Popov quantization is often described without specifying that one
should take the absolute value of the ghost determinant, mainly because in many standard applications this determinant is anyway positive.   
But in fact, since one is trying to define a gauge-fixed measure, and measures are positive, one does want
the absolute value of the ghost determinant.}  $|\frac{\d \Lambda}{\d\epsilon}|=|c|$.  So in this gauge, the additional factor that we introduced to
deal with the constraints reduces to 
\be\label{nolbox} \frac{|c|}{2\pi}\int\d \lambda \exp(\i\lambda bc)=|c|\delta(bc).\ee
On the locus with $c\not=0$, we can scale $c$ out of the delta function, with $|c|\delta(bc)=\delta(b)$.
  With this measure, the offending variable $b$ can be eliminated
and forgotten, and we reduce to the case that the two multiplets of charges $(k-1,k)$ and $(k+1,k+2)$ are statistically independent, and described by the ensembles
that one finds by studying them separately.

We will leave our study of random $\N=2$ supersymmetry here, remarking only that it appears that no essentially new issues arise if one considers Hilbert spaces of higher
dimension or more than four consecutive values of the $R$-charge.    Hopefully, what we have said is sufficient to convince the reader that statistically independent AZ ensembles
with the parameters in eqn. (\ref{bonven}) are a natural candidate for random $\N=2$ supersymmetry.   In the rest of this article, we will confirm that this kind of ensemble, with the 
appropriate spectral curve, does match predictions of $\N=2$ JT supergravity.

\subsection{Restoring $\h q$}\label{restoring}

In this discussion, we have arbitrarily assumed that the $R$-charge generator $J$ has integer eigenvalues, and that the supercharges $Q,Q^\dagger$ have $R$-charges 
1, $-1$.

To see why one would want to relax these assumptions, we can consider $\N=2$ SYK models \cite{FGMS}.   The simplest such models are
 constructed from $N$ complex fermions $\psi_a$, each
of $R$-charge 1, with a supercharge $Q$ that is a polynomial in the $\psi_a$ of some odd degree $\h q\geq 3$.  $Q$ is then of $R$-charge $\h q$.
   By including complex fermions with different $R$-charges, one can construct an $\N=2$ SYK model
with $\h q=1$ \cite{Heydeman:2022lse}, and a model in which $Q$ has negative odd $R$-charge can be constructed by just exchanging $Q$ and 
$Q^\dagger$.  So in general $\h q$ is any odd integer.

In addition, quantizing a complex fermion of $R$-charge $n$ gives a pair of states of $R$-charges\footnote{This statement
assumes that the $R$-charge generator $J$ is defined to be odd under a charge conjugation operation that exchanges $\psi_a$ with $\psi_a^\dagger$.  Without that
assumption, one is free to add an arbitrary constant in the definition of $J$.} $\pm n/2$.   As a result,
depending on the numbers and $R$-charges of the elementary fermions, the spectrum of an $\N=2$ SYK model may consist of states of integer $R$-charge,
or states with $R$-charge valued in $1/2+\Z$.  The nonintegralilty of the $R$-charge is a sort of anomaly.
 
Though these values for the $R$-charges of the operator $Q$ and of the physical states
differ from what was assumed in section \ref{randomtwo}, the analysis in that section carries over in an obvious way 
to the family of all states whose $R$-charges differ from some given value by integer multiples of $\h q$.   The states of a supersymmetric SYK model come in $\h q$ families of that type,
and as we will see, such a spectrum is also natural in JT supergravity.

If one is going to consider anomalous models in which the $R$-charges of physical states are not integer-valued,
 it is natural to consider a general case in which the $R$-charges take values in $\delta+\Z$ (and $\h q$ is any positive odd integer), for any $\delta\in [0,1)$.
From a boundary point of view, arbitrary $\delta$ can be achieved by just adding a constant to the $R$-charge generator.
   Roughly speaking, the 
 parameter $\delta$ can be interpreted from a bulk point of view as $\theta/2\pi$, where
$\theta$ is the theta-angle of the bulk $\uU(1)$ gauge field $A$, with field strength $F=\d A$,  that is related to the $R$-charge on the boundary.   This statement reflects the identity
\be\label{welid} \exp\left(\i\frac{\theta}{2\pi}\int_D F\right) =\exp\left(\i\frac{\theta}{2\pi} \oint_{\partial D}A  \right) \ee for an oriented two-manifold $D$, 
showing that a bulk topological angle $\theta$  induces a charge $\theta/2\pi$ on the boundary of $D$.   However, in the context of $\N=2$ JT supergravity, one has to take into account
the fact that the super Schwarzian mode, which we discuss in more detail in section \ref{sec:N2JTG}, contains a complex fermion of charge $\h q$.  Quantization of this fermion gives two
boundary states of charge $\pm \h q/2$.  Adding this to the boundary charge associated to a bulk theta-angle, we find that the relation between the charge defect $\delta$
in a dual quantum mechanics model that lives on the boundary and the bulk theta-angle is really
\be\label{delfo} \delta =\frac{1}{2} +\frac{\theta}{2\pi}~{\mathrm{mod}}~\Z.\ee
This formula is analogous to the relation between bulk and boundary anomalies that was found for $\N=1$ JT supergravity in \cite{SW}. We will comment more on this later when we analyze the path integral of $\cN=2$ JT supergravity.

\subsection{Check By Simulation of $\N=2$ SYK}\label{check}
     \begin{figure}[t!]
      \begin{center}
   \includegraphics[width=6in]{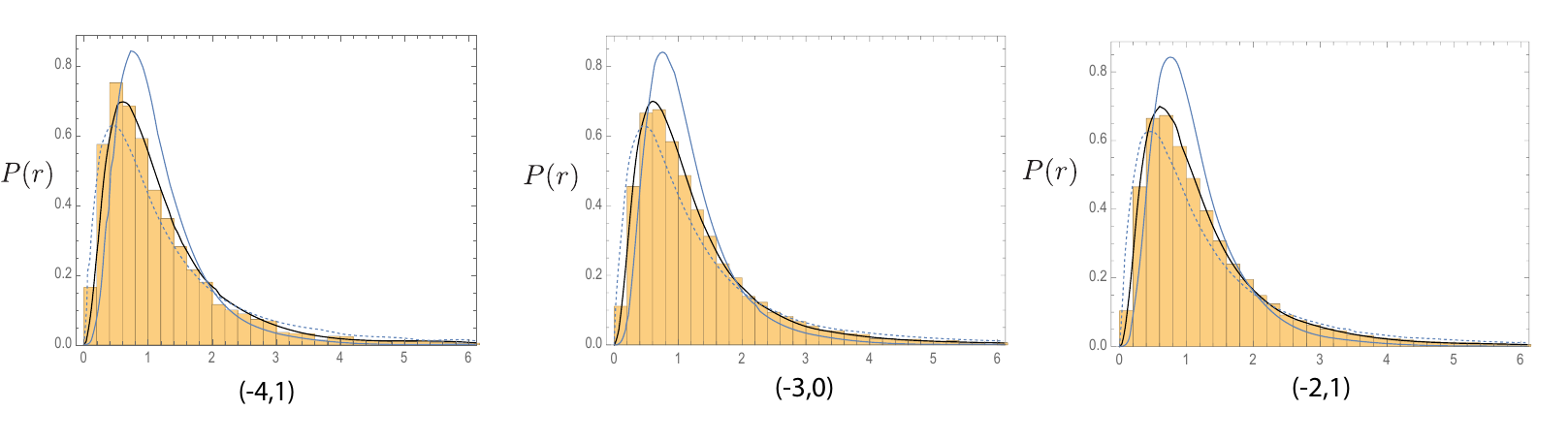}
 \end{center}
 \vspace{-0.8cm}
\caption{\footnotesize  Level spacing statistics for $\cN=2$ SYK with $N=10$ fermions, averaging over $400$ instances of the couplings. The plots show the probability distribution of $r=(\lambda_{j+2}-\lambda_{j+1})/(\lambda_{j+1}-\lambda_j)$ for the multiplets of charges $(-4,-1)$, $(-3,0)$, and $(-2,1)$.  We omit multiplets related to these by $\CT$ symmetry
or with only one state. The smooth curves denote the surmise $P_\upbeta(r)$ with $\upbeta=1$ (dashed blue), $\upbeta=2$ (black) and $\upbeta=4$ (blue). The distribution of all multiplets is consistent with $\upbeta=2$. The case $(-4,-1)$ has larger  deviations since the Hilbert space dimension is smaller. \label{fig:N10}}
\end{figure} 

In this section, we describe the results of a simulation comparing the local level statistics of $\N=2$ SYK models \cite{FGMS} to predictions of the
 $\cN=2$ random matrix ensemble that was proposed in section \ref{randomtwo}. The simplest $\N=2$ SYK models are constructed from 
 $N$ complex fermions $\psi_a$, $a=1,\ldots,N$, of $R$-charge 1,
 satisfying $\{ \psi_a , \psi^\dagger_b\} = \delta_{ab}$, with $\{\psi_a , \psi_b\} = \{ \psi^\dagger_a , \psi^\dagger_b\} = 0$. 
For simplicity, we take $\h q=3$ and thus a cubic supercharge:
\bea\label{eq:N2SYKQQ}
Q = \i \sum_{1\leq a<b<c\leq N} C_{abc}\,  \psi_{a} \psi_{b} \psi_{c}.
\ea
The $C_{abc}$ are random complex numbers with Gaussian statistics and we measure time in units such that $\langle C_{abc} \overline{C}_{abc}\rangle = 2/N^2$. This supercharge automatically satisfies the algebra $Q^2=Q^\dagger{}^2=0$ for all values of the couplings (this would not be true if $Q$ depended on both $\psi$ and $\psi^\dagger$). The Hamiltonian is given by $H = \{ Q, Q^\dagger \}$ and $[H,Q]=[H,Q^\dagger]=0$.   The $R$-charge is defined explicitly as $J = \sum_a \psi^\dagger_a \psi_a  - N/2$.

 We represent the fermions $\psi_a$ in the Hilbert space $\H$ of $N$ qubits
 as $\psi_a=Z_1 \otimes \ldots Z_{a-1}\otimes (X_a+\i Y_a)/2$, where $X_a$,$Y_a$ and $Z_a$ are the Pauli matrices acting on the $a^{th}$ qubit. 
 Then we construct $Q$ using \eqref{eq:N2SYKQQ}. In this representation $J$ is already diagonal.  The subspace $\H_n$ of $\H$ consisting of states with $J=n$ has dimension
${\rm dim}\,\mathcal{H}_n = {N \choose N/2+n}$. 
For moderate values of $N$, exact diagonalization of the model is possible and the level statistics can be compared to expectations.
In more detail, we  compute the ``eigenvalues'' (or singular values) of $Q_k: \mathcal{H}_k \to \mathcal{H}_{k+3}$. 
Here  $k$ ranges from  from $-N/2$ to $N/2-\hq=N/2-3$. We can test the value of $\upbeta$ by computing the ratio 
$r=(\lambda_{j+2}-\lambda_{j+1})/(\lambda_{j+1}-\lambda_j)$, evaluated after arranging the spectrum in increasing order. This quantity was introduced in 
\cite{OganesyanHuse} and does not require ``unfolding'' (rescaling according to the local density of states). In an AZ $(\upalpha,\upbeta)$ ensemble, this quantity
depends only on $\upbeta$, not on $\upalpha$ (which affects the local behavior only near zero energy).
The analog of the Wigner surmise is given by 
$P_\upbeta(r) = Z_{\upbeta}^{-1} \frac{(r+r^2)^{\upbeta}}{(1+r+r^2)^{1+3\upbeta/2}} $; see \cite{ABGR}. 
The coefficient $Z_{\upbeta}$ is found by normalizing the probability distribution to $\int_0^\infty \d r P_\upbeta(r) = 1$.  
It would be interesting but possibly more difficult to verify the random matrix prediction for $\upalpha$.
   \begin{figure}[t!]
 \begin{center}
     \begin{center}
   \includegraphics[width=4.5in]{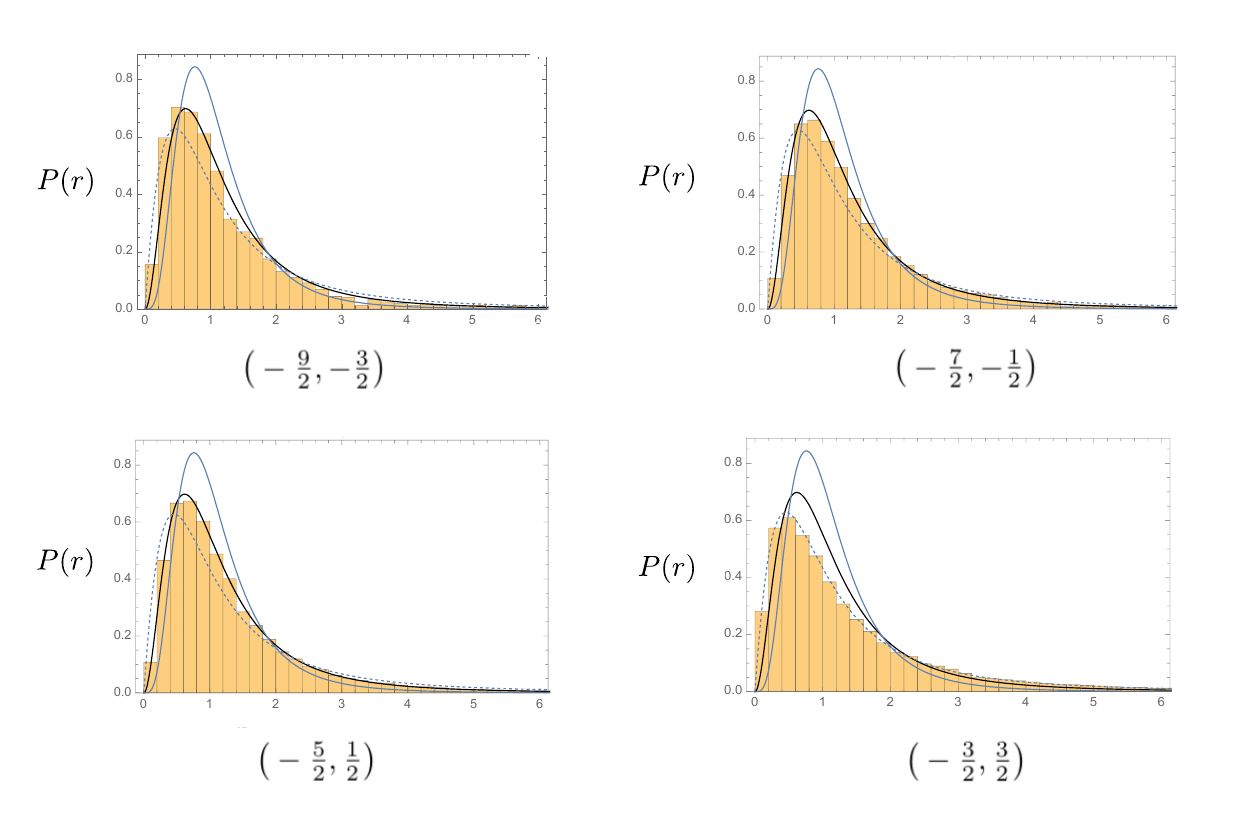}
 \end{center}
\end{center}
\vspace{-0.6cm}
\caption{\footnotesize  Level spacing statistics for $\cN=2$ SYK with $N=11$ fermions, averaging over $200$ instances of the couplings. The  plots show the probability distribution of $r$ for
multiplets with the indicated $R$-charges. The smooth curves denote the surmise $P_\upbeta(r)$ with $\upbeta=1$ (dashed blue), $\upbeta=2$ (black) and $\upbeta=4$ (blue). The distribution of all multiplets is consistent with $\upbeta=2$, except for the case of charges $(-3/2,3/2)$, which is matched by $\upbeta=1$. \label{fig:N11}}
\end{figure} 

We show the results of a simulation  for $N=10$ in fig. \ref{fig:N10} and for $N=11$ in fig. \ref{fig:N11}.  The model has a $\CT$ symmetry (charge conjugation plus
time-reversal)   that acts by $\psi_a \leftrightarrow \psi_a^\dagger$ and $J \leftrightarrow - J$. 
Multiplets related by $\CT$ symmetry  to the ones shown are omitted in the figures. We also omit the multiplets $(-N/2,-N/2+\h q \,)$, $(N/2-\h q,N/2)$, which always occur with multiplicity 1.
For  $N=10$, all cases are described by an ensemble with $\upbeta=2$. For $N=11$, 
this is the case except for the $\CT$ self-conjugate $(-3/2,3/2)$ multiplet, for which $\upbeta=1$.
Some of these features are anticipated in \cite{Kanazawa:2017dpd}.

The results sketched in the figures can be understood as follows.  For generic $k$, the $\CT$  symmetry exchanges the $(k,k+\hq \,)$  and $(-k-\hq,-k)$ multiplets,
implying that they have the same spectrum and level statistics, 
 without changing the random matrix theory prediction $\upbeta=2$.  
However,
  when $N$ is odd  there is a $\CT$-invariant multiplet  $(-\h q/2,\h q/2)$, and for this multiplet the random matrix prediction is modified in a way we explain in section \ref{boundaryview}.
 In the case $N=11$, illustrated in the figure, if we define $\CT$ so that $\CT Q\CT^{-1}=Q^\dagger$, then $\CT^2=1$, and this leads to $\upbeta=1$ for the $\CT$-invariant
 multiplet, as explained
 in section \ref{boundaryview} and as illustrated in the figure.   In a similar model with $\CT Q \CT^{-1}=Q^\dagger$ and $\CT^2=-1$, one expects $\upbeta=4$, as explained in
 section \ref{boundaryview}.

The fact that different supermultiplets have the level spacing statistics characteristic of random matrix universality supports the proposal that the supermultiplets are statistically independent. A more precise verification of this claim in an SYK simulation is an open problem.

\section{$\mathcal{N}=2$ JT Gravity Black Hole}\label{sec:N2JTG}
Our next goal is to analyze $\cN=2$ JT gravity in the simplest spaces, the disk and the cylinder. In section \ref{sec:Z01}, we study the disk, which
 determines the black hole spectrum to leading order in the topological expansion and provides us with the spectral curve of the random matrix ensemble outlined in section \ref{randomtwo}. 
 In section \ref{sec:Z02}, we turn to the cylinder, which in random matrix theory  is uniquely fixed by the ensemble and is independent of the details of the spectral curve. By 
 analyzing the cylinder,  we verify the prediction that different supermultiplets are statistically independent. We conclude in  section 
\ref{sec:3bwh} with some preliminary calculations for the three-boundary wormhole, which we explore   more fully in section \ref{sec:GPI}.

\subsection{The Leading Order Black Hole Spectrum}\label{sec:Z01}
 We will be brief in describing $\N=2$ JT supergravity, since this theory has been  studied extensively in the past \cite{FGMS,Stanford:2017thb, Forste:2017kwy, Forste:2017apw,  Mertens:2017mtv, Cardenas:2018krd, Forste:2020xwx, Lin:2022zxd, Lin:2022rzw}.  
 The action of this theory consists of two terms. The first is given by a constant $-S_0$ times the Euler characteristic of the spacetime. Since this term is a topological
 invariant,  it does not require a supersymmetric completion. In the sum over spacetime topologies,  this term weights surfaces by $(e^{S_0})^{2-2g-n}$, where $g$ is the number of handles, or  spacetime wormholes, and $n$ is the number of boundaries.  The expansion is  thus an asymptotic expansion in $e^{-S_0}$.

The second term consists of a bulk piece that can be written, in a first order formulation of gravity, as a topological $BF$ theory with gauge supergroup $G=\SU(1,1|1)$, or equivalently\footnote{\label{osptwotwo} Actually, $\OSp(2|2)$ has a center $\Z_2$, while $\SU(1,1|1)$ has trivial center.   The precise relation between the two
supergroups is that $\SU(1,1|1)=\OSp(2|2)/\Z_2$.   We write $\OSp'(2|2)=\OSp(2|2)/\Z_2$.  Similarly a maximal bosonic subgroup of $\SU(1,1|1)$ is really $(\SL(2,\R)\times \uU(1))/\Z_2$
(where $\Z_2$ is generated by the product of  $-1\in \SL(2,\R)$ and a $\pi$ rotation in $ \uU(1)$), not quite $\SL(2,\R)\times \uU(1)$ as stated in the text.  For brevity, we will
omit details such as these when not essential.}  $\OSp(2|2)$. This group has ${\rm SL}(2,\mathbb{R})\times \uU(1)$ as a maximal bosonic subgroup, together with four fermionic generators. An explicit construction of this action using the first order formulation can be found in \cite{Cardenas:2018krd}, while a construction using two-dimensional $\cN=2$ superspace is developed in \cite{Forste:2017kwy, Forste:2017apw}. In the second order formulation, the theory describes a two-dimensional spacetime metric together with a complex gravitino, the JT gravity dilaton together with a complex dilatino, and a 
$BF$ theory of the $\uU(1)$ $R$-symmetry. 
The dual quantum mechanical system lives on a nearly AdS (NAdS) boundary, defined by specifying the proper length of the boundary and the boundary value of the dilaton,
and taking them to both diverge with a fixed ratio.
In the presence of NAdS boundaries,  one needs to supplement the bulk term with boundary counterterms 
including the Gibbons-Hawking-York boundary term. 
 Depending on the boundary conditions,
boundary terms might be required for the $\uU(1)$ gauge field as well. 

We normalize the generator $J$ of the $\uU(1)$ $R$-symmetry such that its spectrum consists of the integers $\Z$. Fermionic operators have odd charge and bosonic 
operators  have even charge. In the simplest theory,  the gravitino and dilatino have unit charge, but one can consider more general theories where their charge is any odd integer $\widehat{q}$, where we follow the notation of section \ref{randommatrix}. In the $BF$ formulation of the theory, this is equivalent to taking the gauge group to be not
$\SU(1,1|1)$ but its  $\hq$-fold cover obtained  by ``unwrapping'' the $\uU(1)$ subgroup $\hq$ times. 

After these preliminaries, let us begin with the evaluation of the $\cN=2$ JT gravity path integral on the disk with NAdS boundary conditions.  We denote
this path integral as\footnote{The reader
will hopefully not confuse the inverse temperature $\beta$ and the $R$-charge fugacity $\alpha$ with the Altland-Zirnbauer parameters $(\upalpha,\upbeta)$.}  $Z_{\rm disk}(\beta,\alpha)$.  Here $\beta$ is the  renormalized circumference of the boundary circle, and $\alpha$ is a fugacity: fermions are twisted in phase by $e^{\i \alpha J}$ as they go around the thermal circle. Therefore $\mu = \i \alpha/\beta $ has the interpretation of a  chemical potential for the $R$ symmetry.

The general rules of the gravitational path integral imply that the disk path integral defined in the previous paragraph computes the following quantity for the quantum system describing the $\cN=2$ JT gravity black hole:
\bea\label{eq:zdiskrmtp}
Z_{\rm disk}(\beta,\alpha)=\langle {\rm Tr}\big( e^{-\beta H} e^{\i \alpha J}\big)\rangle_{g=0}.
\ea
$H$ is the Hamiltonian acting on the Hilbert space of the boundary theory, and $J$ generates  the $\uU(1)$ $R$-symmetry. The brackets on the right hand side anticipate that for pure $\cN=2$ JT gravity the quantum mechanical dual is an ensemble average over boundary theories. The label $g=0$ on the right means this is evaluated to leading order in the genus expansion of the matrix integral describing this ensemble.   The trace that appears in eqn. (\ref{eq:zdiskrmtp}) is a trace in the Hilbert space of the boundary theory.

In general, the $BF$ theory partition function reduces to an integral over the moduli space of flat connections modulo gauge transformations. In the case of gravity, corresponding to ${\rm SL}(2,\mathbb{R})\subset G$, one further restricts to the component with a geometric description and divides by the mapping class group; more details can be found in section \ref{sec:GPI}. This gauge theory procedure is  equivalent to integrating over the moduli space of hyperbolic surfaces, in our case with extended $\cN=2$ supersymmetry (see \cite{Ip:2016ojn} for mathematical
theory of $\N=2$ hyperbolic surfaces). The case of the disk is particularly simple, since the moduli space consists of a single configuration up to a boundary reparametrization mode. 
This mode is governed by the $\cN=2$ Schwarzian theory introduced in \cite{FGMS}. Its partition function was computed in \cite{Stanford:2017thb}, and we now give a quick overview of the calculation. 

We begin with an outline of how the $\cN=2$ Schwarzian mode arises. Parametrize the hyperbolic disk with a complex coordinate $e^{-\i x}$, such that $x \cong x+2\pi $. (The precise form of the metric will not be important for the following discussion.) The center of the disk is at $e^{-\i x} = 0$ or  ${\mathrm{Im}}\,x=-\infty$, while the NAdS boundary is located at ${\rm Im}\, x = T$ for some large $T$. Including the fermions, the disk is described by $x,\psi,\bar{\psi}$ with equivalence 
\bea
(x,\psi,\bar{\psi})\cong (x+ 2\pi ,- \psi, -\bar{\psi}).
\ea
In principle there could be a non-trivial holonomy $\psi \to- e^{\i \hq \phi} \psi$ as we go around the thermal circle, but smoothness at
the center of the disk requires $\phi=0$. We set $x= \i T + y$ and the boundary is parametrized  by $y,\psi,\bar{\psi}$ with real $y$ and identification 
\bea
(y,\psi,\bar{\psi})\cong (y+ 2\pi ,- \psi, -\bar{\psi}).
\ea
Now we pick the observable that we want to insert at the boundary, namely the grand canonical partition function ${\rm Tr}(e^{-\beta H}e^{\i \alpha J})$. To compute this observable, the boundary quantum mechanical system is formulated on a supercircle parametrized by $t,\theta,\bar{\theta}$ with an equivalence relation 
\bea\label{eq:equivreltttb}
(t,\theta,\bar{\theta}) \cong (t+\beta, -e^{\i \hq \alpha} \theta, - e^{-\i \hq\alpha} \bar{\theta}).
\ea
After localizing the gravity path integral on the hyperbolic disk, the $\cN=2$ Schwarzian theory is the only leftover gravitational mode. It involves an integral over all $\cN=2$ superconformal identifications between the boundary of the disk, parametrized by $(y,\psi)$ and the worldline of the quantum mechanical system, parametrized by $(t,\theta)$, that are compatible with the above equivalences, and such that 
the map from $t$ to $y$ has winding number one.  
Any such identification can be obtained from a standard one by an invertible superconformal reparametrization of $(y,\psi)$.   Here we are 
not just being careless in omitting $\bar\psi$ and $\bar\theta$.
The identification from $(y,\psi,\bar\psi)$ to $(t,\theta,\bar\theta)$ and therefore the reparametrization of $(y,\psi,\bar\psi)$ 
is required to be $\N=2$ superconformal.  For a general number $\N$ of supersymmetries, this condition is  subtle,
but in the special case $\N=2$, there is an easy way to implement it: the group of $\N=2$ superconformal reparametrizations of $(y,\psi,\bar\psi)$ is the same as the group of
all reparametrizations of $(y,\psi)$. (This is related to the fact that the $\N=2$ superconformal algebra can act on a chiral superspace that is parametrized by $y,\psi$ with $\bar\psi$ forgotten.)
So we will implement the superconformal condition by writing formulas relating $(t,\theta)$ to $(y,\psi)$ with no mention of $\bar\psi$, $\bar\theta$.
As for the restriction to maps of winding number 1, this a simple illustration of the fact that to go from $BF$ theory to gravity we need to restrict the class of flat connections by hand.  Finally,
to  not overcount geometries, it is important to remove the reparametrizations that correspond to the isometries of the $\cN=2$ hyperbolic disk, which form a group $\SU(1,1|1)$. 

The integration space ${\mathcal W}$ of the Schwarzian theory is therefore the group of superconformal reparametrizations of $(y,\psi)$ divided by the right action of
 $\SU(1,1|1)$.   This  is a symplectic supermanifold and the action arising from $\cN=2$ JT gravity is the generator, via Poisson brackets, of a global $\uU(1)$ symmetry of ${\mathcal W}$,
 with generator $\partial/\partial t$; see \cite{Stanford:2017thb}.  For these reasons, the integral over the Schwarzian mode can be computed by localization on fixed points of $\partial/\partial t$. However, in this context, because of the quotient by $\SU(1|1)$, ``fixed point'' means ``fixed point up to a superconformal automorphism of the disk.'' The relevant automorphisms of the disk are generated by $\partial/\partial y$ and $\i (\psi \partial_{\psi} - \bar{\psi} \partial_{\bar{\psi}})$ (we will include $\bar\psi$ in 
some formulas, but it can be consistently set to zero). 
The coefficients of each generator can be different at different fixed points. Taking into account the winding number constraint, the fixed points take the form 
\begin{align}\notag
y(t,\theta) &= \frac{2\pi}{\beta} t,\\
\psi(t,\theta) &= e^{-\i \hq \alpha\, t/\beta} \theta.\label{fixedpt}
\end{align}
These are annihilated by 
\bea
D = \frac{\partial}{\partial t} - \frac{2\pi}{\beta} \frac{\partial}{\partial y} + \frac{\hq \alpha}{\beta}\, \i (\psi \partial_\psi - \bar{\psi} \partial_{\bar{\psi}}),
\ea
so they do represent fixed points.   Since the boundary condition only depends on the value of $\alpha$ mod\footnote{The identification on $(t,\theta,\bar{\theta})$ depends on the value 
of $\alpha$ mod $2\pi n/\h q$.   But when we introduce the parameter $\h q$, we presume that the theory can be probed by operators of $R$-charge 1.   Taking this into
account, the indeterminacy in $\alpha$ is $\alpha\to \alpha+2\pi \Z$.}  $2\pi$, eqn. (\ref{fixedpt}) really describes an infinite family of fixed
points, one for each real-valued lift of $\alpha$.

Now we want to study the one-loop integral that arises in expanding around these fixed points. For this, one needs to know the value of the classical Schwarzian action at a fixed point, and also the ``rotation eigenvalues,'' that is, the eigenvalues of the operator $D$ acting on the tangent space of a fixed point. These are evaluated  in appendix \ref{sec:disktrumpet}. The final answer for the disk partition function, after summing over fixed points weighted by a product of functional determinants, is given by
\beq
Z_{\rm disk}(\beta,\alpha) =e^{S_0} \sum_{n\in \mathbb{Z}} \exp(2\pi \i n \delta)\frac{\hq \,\cos\left( \pi \widehat{q}\, (\frac{\alpha}{2\pi}+ n)\right)}{2\pi^3 \left(1-4 \hq^{\, 2}(\frac{\alpha}{2\pi}+ n)^2\right)} e^{\frac{\pi^2}{\beta}(1-4\widehat{q}^{\, 2} (\frac{\alpha}{2\pi}+  n)^2)}.
\eeq
We have weighted the $n^{th}$ fixed point with a factor  $\exp(2\pi \i n \delta)$.   As we will see in a moment, $\delta$ is the charge defect parameter that was introduced in section \ref{restoring} and related to the bulk theta-angle in eqn. (\ref{delfo}). 

 As we explain in section \ref{randommatrix}, the Hilbert space decomposes into supermultiplets with charges $(k,k+\hq\,)$, where $k\in \Z+\delta$ and $k,k+\hq$ are the $R$-charges of the states
in a two-dimensional supermultiplet. It will be convenient, in the gravity analysis, to label a supermultiplet by the average charge $q=k+\hq/2$, such that the $R$-charges of the multiplets are $(q-\frac{\hq}{2},q+\frac{\hq}{2})$. Since the fermions of the boundary super-reparametrization mode carries $R$-charge $+\hq/2$ and $-\hq/2$, the parameter $q$ can be interpreted as the bulk contribution to the microscopic $R$-charge. In this presentation, the bulk charge takes values $q \in \mathbb{Z}+\delta - \frac{1}{2}$, since $\hq$ is always odd.

The partition function on the boundary side can be expanded into  BPS states and two-dimensional supermultiplets  as
\bea\label{eq:spectrumcomm}
\langle {\rm Tr}\big( e^{-\beta H} e^{\i \alpha J}\big)\rangle_{g=0} &=&\sum_{k\in \mathbb{Z}+\delta} e^{\i \alpha k} N_{\rm BPS}(k) \nonumber\\
&&+  \sum_{q\in \mathbb{Z}+\delta-\frac{1}{2}} \big(e^{\i \alpha (q-\frac{\hq}{2})} + e^{\i \alpha (q+\frac{\hq}{2})}\big) \langle {\rm Tr}_{\M,q}\big( e^{-\beta H}\big)\rangle_{g=0}.
\ea
On the left hand side of this formula, the symbol $\Tr$ represents a trace over the whole Hilbert space of the boundary theory (including BPS states and both bose and fermi states of 
two-dimensional supermultiplets).  On the right hand side, the sum over BPS states of $R$-charge $k$ has been separated out. In the second line, we included a sum over non-BPS multiplets,  labeled by the average $R$-charge $q$, 
accounted for by the explicit factor $e^{\i \alpha (q-\frac{\hq}{2})} + e^{\i \alpha (q+\frac{\hq}{2})}$. Therefore $\Tr_{\M,q}$ is a matrix model trace in which each  supermultiplet with average charge $q$ is counted once;
in other words, what are counted are the independent ``eigenvalues'' of the matrix model that governs $q\pm \hq/2$ supermultiplets.  At the end of section \ref{sec:Z02}, we will
justify the claim that the number of BPS states of charge $k$, which we denote
 $N_{\rm BPS}(k)$, does not fluctuate as a random variable and instead can be regarded as a fixed number.
 
It is convenient to define $ \langle {\rm Tr}_{\M,q}\big( e^{-\beta H}\big)\rangle_{g=0} =  \int_0^\infty \d E e^{-\beta E } \rho_q(E)$, 
where $\rho_q(E)$ is the spectral density for a supermultiplet of charge $(q-\frac{\hq}{2},q+\frac{\hq}{2})$. The disk amplitude encodes both the number of BPS states of a given $R$-charge and the non-BPS density of states. The $\cN=2$ Schwarzian partition function can be expressed
 in the following way \cite{Mertens:2017mtv}, via an inverse Laplace transform and Poisson resummation:
\bea \label{handyexp}
Z_{\rm disk}(\beta,\alpha)&=& \sum_{k\in \mathbb{Z}+\delta, |k|<\frac{\hat{q}}{2} } e^{ \i \alpha k} ~ \frac{e^{S_0}}{4\pi^2}\cos \Big(\frac{\pi k}{\hq}\Big) \label{diskdos}\\ 
&+&  \sum_{q\in \mathbb{Z} + \delta-\frac{1}{2}} \big( e^{\i \alpha (q-\frac{\hq}{2})} + e^{\i \alpha (q+\frac{\hq}{2})}\big) \int_{E_0(q)}^\infty dE e^{-\beta E}\frac{e^{S_0}}{2\pi}\frac{\sinh{\big(2\pi \sqrt{E-E_0(q)}\big)}}{4\pi^2 E}.\nonumber
\ea 
where $E_0(q) \equiv \frac{q^2}{4\hq^{\, 2}}$ corresponds to the energy gap for a non-BPS supermultiplet of $R$-charges $q\pm\hq/2$. This is explained in appendix \ref{sec:disktrumpet}.  Since $E_0(q)$ is monotonically increasing in $|q|$,  only finitely many of the $\rho_q(E)$ are nonzero at any given $E$.    Given
this, the formula (\ref{handyexp}) is uniquely determined.

We can now read off the necessary input to the random matrix ensemble that governs $\N=2$ JT supergravity:
\bea
N_{\rm BPS}(k) =&\begin{cases} \frac{e^{S_0}}{4\pi^2} \cos \big( \frac{\pi k}{\hq} \big) &\hskip.5cm |k|<\frac{\hq}{2} \cr
   0 & \hskip.5cm|k|\geq \frac{\h q}{2}\end{cases}\\
\rho_q(E)=&\begin{cases}\frac{e^{S_0}}{2\pi}\frac{\sinh{\big(2\pi \sqrt{E-E_0(q)}\big)}}{4\pi^2 E}& E>E_0(q)\\
0 & E\leq E_0(q).\end{cases}
\ea

 We conclude  with several observations.  The formulas that we have written make sense for any value of $\delta$, but
 are consistent with a charge conjugation symmetry $\cC$ or a charge conjugation plus time-reversal symmetry $\CT$ precisely if $\delta=0$ or $1/2$.  (At $\delta=0,1/2$, JT gravity
 and the boundary matrix model potentially have both of these symmetries, though  in an $\N=2$ SYK model, $\CT$ is more natural.) 
 When $\delta=0$, the charges that appear in the spectrum are integers and the theory is anomaly-free.  When $\delta=1/2$, the charges are  half-integers 
 and there is a mixed anomaly between $J$ and $\cC$ or $\CT$, since the expectation that $e^{2\pi \i J}=1$ clashes with 
  the expectation that $J$ is odd under $\cC$ or $\CT$: if $J$ is defined to be odd, then $e^{2\pi \i J}=-1$. 
 For generic $\delta$, there is no $\cC$ or $\CT$ symmetry, and
 one can preserve the statement $e^{2\pi \i J}=1$ by shifting $J$ by $\delta$.

Another observation is that since BPS states only occur for  charges $-\hq/2 < k < \hq/2$,
 there are never two BPS states differing in charge by $\hq$.   This is expected, since  two BPS states with charges separated by $\hq$ could combine into a non-BPS
multiplet.

Finally, we note that the energy $E_0(q)$ of the lightest non-BPS state is positive except for $q=0$, corresponding to a multiplet with $R$-charges $(- \hq/2,+\hq/2)$ (a value that is available precisely if
$\delta=1/2$).
The behavior of the density of states near threshold is quite different in the two cases.  For generic $q$, 
 $\rho_{q}(E_0(q) +\varepsilon) \sim \sqrt{\varepsilon}$, while for $q=0$, the threshold is at $E=0$ and   $\rho_{q=0}(\varepsilon) \sim 1/\sqrt{\varepsilon}$. 
This difference is actually expected based on the random matrix analysis of section \ref{randomtwo}.   There, we claimed that the non-BPS multiplets are governed by an AZ ensemble with $(\upalpha,\upbeta)=(1+2\nu,2)$ (where $\nu$ is the number of BPS states).   In leading order at large $e^{S_0}$, the difference
between an AZ ensemble $(\upalpha,\upbeta)$ ensemble and a Dyson $\upbeta$ ensemble with a suitable potential is unimportant except near zero energy.
So as long as  $E_0(q)>0$, we expect the typical threshold behavior  $\rho_{q}(E_0(q) +\varepsilon) \sim \sqrt{\varepsilon}$ of a Dyson ensemble.
However, when the threshold is at zero energy, an AZ $(\upalpha,\upbeta)$ ensemble has the threshold behavior $\rho(\varepsilon)\sim 1/\sqrt \varepsilon$, 
as long as $\upalpha$ is of order 1
(not of order $e^{S_0}$).   This condition is satisfied here, since $N_\BPS(\pm \hq/2)=0$ and the expected ensemble therefore has $(\upalpha,\upbeta)=(1,2)$.

We define the spectral curve $y_q(x)$, for a multiplet with $R$-charges $q\pm \hq/2$, by the condition that $y_q(x\pm \i \varepsilon) = \mp \i \pi e^{-S_0}\rho_q(x)$.  So
\bea\label{eq:N2sc}
y_q(x) =\frac{1}{2\pi} \frac{\sin \big( 2\pi \sqrt{-x + E_0(q )}\big)}{4\pi x}.
\ea
This function, by definition, has a branch cut along the support of $\rho_q(x)$.  It also has, for $q \neq  0$, a pole at $x=0$.  
The interpretation of this pole in random matrix theory is explored in appendix \ref{app:Loop}; here we summarize a few facts.
 For $|q|<\hq$, the residue of the pole, namely
\beq
{\rm Res}_{x=0}~ y_{q}(x) = \frac{\sin \big( \frac{\pi |q|}{\hq}\big)}{8\pi^2},
\eeq
is determined by the number of BPS states in the multiplet of charges $q\pm \hq/2$; it is 
$\frac{1}{2} N_\BPS(k)$ where $k=q\pm \hq/2$ with the sign determined such that $|k|<\hq/2$ (except that a factor $e^{S_0}$ present in $N_\BPS$ was removed in defining $y_q$).  
Random matrix
analysis predicts a pole in $y_q(x)$ with this residue, assuming that the matrix potential is regular at zero energy.
For $|q|\geq\hq$, there are no BPS states, but there is still a pole.  To account for this in random matrix theory, one has to assume that the matrix potential
$F(\qQ^\dagger \qQ)$ defined in eqn. (\ref{genform}) has a logarithmic behavior near the origin, $F(\qQ^\dagger \qQ)\sim \log \qQ^\dagger\qQ$.   For $|q|>\hq$, the residue of the pole
and therefore the coefficient of the logarithm can have either sign, depending on $q$; this coefficient is large, of order $e^{S_0}$.   
When the logarithm has a positive coefficient, the matrix integral is divergent.   This divergence does not affect the loop expansion of the matrix model, because the loop
expansion is only sensitive to the matrix potential at energies of $ E_0(q)$ or greater, and $E_0(q)>0$ whenever the matrix potential has a wrong sign coefficient.   However,
the wrong sign of the logarithm does appear to imply 
that the matrix model has a nonperturbative instability, somewhat analogous to the familiar nonperturbative instability of many purely bosonic matrix
models.   It is surprising to run into such an instability in a supersymmetric model.\footnote{We mentioned earlier that some higher dimensional black holes can be described in the near-BPS limit by $\mathcal{N}=2$ JT gravity coupled to matter. The instability found here does not necessarily implies an instability of these UV-complete models since both the random matrix ensemble and pure $\mathcal{N}=2$ JT gravity are only approximations to the full theory.}

\subsection{The Two-Boundary Wormhole}\label{sec:Z02}

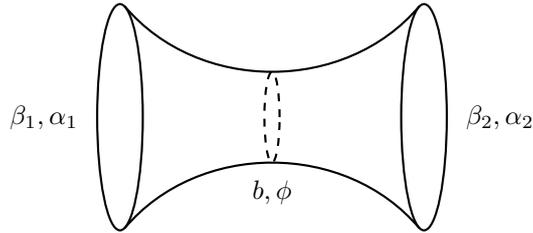
\begin{figure}
\begin{center}
\begin{tikzpicture}[scale=1, baseline={([yshift=0.1cm]current bounding box.center)}]
\node at (-3,0) {\small $\beta_1,\alpha_1$};
\node at (3,0) {\small $\beta_2,\alpha_2$};
\draw[thick] (-2,0) ellipse (0.3 and 1.5);
\draw[thick,dashed] (0,0) ellipse (0.1 and .6);
\node at (0,-1) {\small $b,\phi$};
\draw[thick] (2,0) ellipse (0.3 and 1.5);
\draw[thick] (-1.937,1.47) to [bend right=50] (1.937,1.47);
\draw[thick] (-1.937,-1.47) to [bend left=50] (1.937,-1.47);
\node at (0,-1.985) {};
\end{tikzpicture}
\end{center}
\caption{\footnotesize A cylinder with two NAdS boundaries -- the leading connected  two-boundary wormhole. We label each NAdS boundary by their inverse temperature $\beta$ and $\uU(1)$ $R$-symmetry chemical potential $\alpha$. Each $\N=2$ hyperbolic surface has an interior geodesic, indicated by the dashed line, with length $b$ and $\uU(1)$ holonomy $\phi$, that should be integrated over.}
\label{fig:2bdywh}
\end{figure}

We now begin our study of the gravitational path integral of $\cN=2$ JT 
gravity in geometries with spacetime wormholes. We start with the simplest case: the double trumpet or two-boundary 
wormhole, corresponding to geometries of the type displayed in fig. \ref{fig:2bdywh} with the topology of a cylinder with 
two boundaries. On each boundary we impose the usual NAdS${}_2$ boundary conditions labeled by an inverse temperature $\beta$ and the $\uU(1)$ fugacity $e^{\i\alpha}$.  

We can anticipate which quantity this gravity path integral corresponds to, in the random matrix ensemble. Each asymptotically NAdS boundary is associated to an insertion of ${\rm Tr}( e^{-\beta H} e^{\i \alpha J})$. The parameters $\alpha,\beta$ can be chosen independently for the different boundaries.   The path integral over a cylinder with two NAdS boundaries computes the connected genus $0$ contribution to
\beq\label{eq:02rmt}
Z_{0,2}(\beta_1,\alpha_1;\beta_2,\alpha_2)=\big\langle {\rm Tr}\big( e^{-\beta_1 H} e^{\i \alpha_1 J}\big)~{\rm Tr}\big( e^{-\beta_2 H} e^{\i \alpha_2 J}\big)\big\rangle_{{\rm conn.}, g=0},
\eeq
In the total path integral with two boundaries, this contribution is combined with the leading disconnected answer made out of a product of two disconnected disks of order $e^{2S_0}$. The trace in \eqref{eq:02rmt} can be decomposed in supermultiplets\footnote{In principle there could be contributions from the BPS states as well. In writing \eqref{eq:Z02rmt2} we anticipated that such contributions will turn out to be absent.}
\bea\label{eq:Z02rmt2}
Z_{0,2}(\beta_1,\alpha_1;\beta_2,\alpha_2)&=&\sum_{q_1,q_2\in\mathbb{Z}+\delta-\frac{1}{2}} \big(e^{\i \alpha_1 (q_1-\frac{\hq}{2})} + e^{\i \alpha_1  (q_1+\frac{\hq}{2})}\big)\big(e^{\i \alpha_2  (q_2-\frac{\hq}{2})} + e^{\i \alpha_2  (q_2+\frac{\hq}{2})}\big)\nn
&&~~~\times \big\langle {\rm Tr}_{\M,q_1}( e^{-\beta_1 H})~{\rm Tr}_{\M,q_2}( e^{-\beta_2 H})\big\rangle_{{\rm conn.}, g=0}.
\ea
The expected random ensemble of $\cN=2$ quantum mechanics defined in section \ref{randommatrix} has the property that different supermultiplets are statistically independent. Then, the gravity calculation should reproduce 
\beq
\big\langle {\rm Tr}_{\M,q_1}( e^{-\beta_1 H})~{\rm Tr}_{\M,q_2}( e^{-\beta_2 H})\big\rangle_{{\rm conn.}, g=0} \overset{?}{\propto} \delta_{q_1,q_2}
\eeq
In the rest of this section, we verify this expectation. Moreover, we also will see that the function of $\beta_1$ and $\beta_2$ on the right hand side also matches with the result from the ensemble proposed in section \ref{randommatrix}.

After these preliminaries, let us go back to gravity and compute the partition function, following \cite{SSS}. The path integral localizes to an integral over the moduli space of $\cN=2$ hyperbolic spaces. 
If we ignore the gluing on the boundaries, the only moduli of the cylinder are $b,\phi$.   
 From the point of view of $BF$ theory, these moduli arise as follows.
 A flat connection on the cylinder can be described by its holonomy around the interior geodesic.  We denote this holonomy by  $U\in G$. Two holonomies related by conjugation $U\to R U R^{-1}$, $R\in G$, are gauge-equivalent, so we only care about the conjugacy class of $U$.   In order for a flat connection with holonomy $U$ to have a geometric interpretation in terms
 of a hyperbolic metric, the conjugacy class of $U$ must be hyperbolic (meaning that, after reducing modulo the odd variables, the ${\rm SL}(2,\R)$ component of $U$ is
 hyperbolic).
A hyperbolic element $U$ can be conjugated to a diagonal form:
\bea\label{hyperbolicconj}
U=\left(\begin{array}{cc|c}
-e^{\i\hq\phi}e^{\frac{b}{2}}&0&0 \\
0&-e^{\i\hq\phi}e^{-\frac{b}{2}}&0 \\
\hline
0&0&e^{2\i\hq\phi}  
\end{array}\right).\label{eq:Ugeo}
\ea\\ 
Here  $b$ is the length of the interior geodesic in fig.  \ref{fig:2bdywh} and  $e^{\i \hq \phi}$ is the $R$-symmetry holonomy around this  geodesic.
The minus sign appearing in the $\SL(2,\mathbb{R})$ block indicates that when $\phi=0$, fermions are antiperiodic around this geodesic.  At $\phi=\pi$, the fermions are periodic.
As opposed to the case of 
$\cN=1$ supergravity, we do not need to consider spin structures separately, because changing the
spin structure would be equivalent to shifting $\phi$ by $\pi$.

For convenience, we  have represented $U$as a $3 \times 3$ supermatrix, 
but we remind the reader that the gauge group is really a $\hq$-fold cover of $\SU(1,1|1)$.   This cover has no convenient matrix representation, so it is often convenient to write formulas
projected to $\SU(1,1|1)$.   
In the $\hq$-fold cover, the parameter $\phi$ runs from $0$ to $2\pi$, although in $\SU(1,1|1)$ itself, a shift of $\phi$ by $2\pi/\hq$ would give the same group element.

One can separate the cylinder into two ``trumpets'' by cutting along the interior geodesic in 
fig.  \ref{fig:2bdywh}.   Each trumpet is a cylinder with one asymptotically NAdS boundary and one geodesic boundary. An $\N=2$ Schwarzian mode propagates on each
NAdS boundary.  Taking this into account, we now  have additional moduli that arise from the way we glue the two trumpets along the interior geodesic. 
These moduli are present because a global rotation or $R$-symmetry rotation of either trumpet would act non-trivially on the Schwarzian modes of the other one.  Hence
one can make a  rotation or $R$-symmetry transformation of one trumpet relative to the other before gluing.
Again, the $BF$ theory perspective is useful.  Gauge transformations are constrained to be trivial along an NAdS boundary, so with two such boundaries, one can define
 a gauge-invariant 
``holonomy'' $V$ by parallel transport from one boundary to the other.   Flatness of the connection implies that $V$ must commute with $U$, so they are diagonal in the same
basis, $V={\rm diag} (-e^{\varrho/2 + \i \hq\varphi},-e^{-\varrho/2 + \i \hq \varphi}, e^{2 \i\hq \varphi})$, with ``gluing parameters'' $\varrho,\varphi$.    In order not to overcount geometries, we restrict  to $0\leq \varrho \leq b$,  $0\leq \varphi \leq 2\pi$. To summarize, the interior moduli of the cylinder are $b$, $\phi$ together with the geometrical twist parameter
 $\varrho$ and the $R$-symmetry twist parameter $\varphi$.

In standard approaches to the path integral, as in \cite{SSS}, the integration measure 
for these  parameters  is $\d b \, \d \phi \, \d\varrho \, \d \varphi$.   (The torsion, which we use in section \ref{sec:GPI}, suggests a slightly different though ultimately equivalent
gluing procedure.)   
The path integral over the bulk is independent of $\varrho$ and $\varphi$ and therefore they can be integrated out  from the beginning, producing an effective measure $2\pi b \,\d b\, \d \phi$. 

In this example of the trumpet, all interior moduli are bosonic, with fermionic moduli coming only from the Schwarzian modes that we study next. As soon as we 
increase the number of boundaries or handles, there are also  bulk  fermionic moduli.

The final ingredient we need to complete the evaluation of $Z_{0,2}$ is  the path integral of the $\N=2$ Schwarzian modes.   The Schwarzian action arises in gluing the bulk geometry
to the worldline of the boundary quantum mechanics, so it depends on the parameters $b,\phi$ of the bulk geometry.   We will focus on the Schwarzian path integral on one NAdS$_2$
boundary,

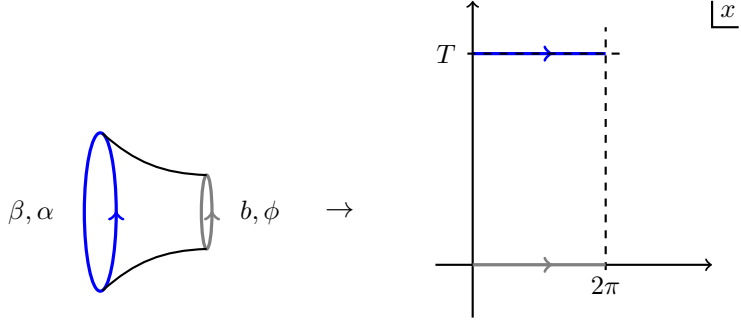
\begin{figure}
\begin{center}
\begin{tikzpicture}[scale=0.7, baseline={([yshift=-0.1cm]current bounding box.center)}]
    \draw[very thick, blue] (6.5,0) ellipse (0.3 and 1.5);
    \draw[very thick,blue,->] (6.8,0) -- (6.8,0.01); 
\draw[very thick,gray] (8.5,0) ellipse (0.1 and 0.7);
\draw[very thick,gray,->] (8.6,0) -- (8.6,0.01); 
\draw[thick] (6.54,1.49) to [bend right=20] (8.5,0.7);
\draw[thick] (6.54,-1.49) to [bend left=20] (8.5,-0.7);
\node at (5.2,0) {\small $\beta,\alpha $};
\node at (9.5,0) {\small $b,\phi $};
\node at (11,0) {$\to$};
\draw[thick,->] (13.5,-2) to (13.5,4);
\draw[thick,->] (12.8,-1) to (18,-1);
\draw[thick,dashed] (16,-1.1) to (16,3.5);
\draw[very thick, gray,->] (13.5,-1) to  (15,-1);
\draw[very thick, gray] (14.9,-1) to  (16,-1);
\draw[very thick, blue,->] (13.5,3) to  (15,3);
\draw[very thick, blue] (14.9,3) to  (16,3);
\draw[thick,dashed] (13.4,3) to (16.3,3);
\node at (18.3,3.8) {\small $x$};
\draw[thick] (18,4) -- (18,3.5) -- (18.5,3.5);
\node at (16,-1.4) {\small $2\pi$};
\node at (13,3) {\small $T$};
    \end{tikzpicture}
    \end{center}
\caption{\footnotesize On the left panel we represent the trumpet geometry which is a cylinder with one NAdS boundary (left) with inverse temperature $\beta$ and chemical potential $\alpha$, and one geodesic boundary (right) with length $b$ and $\uU(1)$ holonomy $\phi$. On the right panel we represent the trumpet in the complex $x$ plane with $x\cong x+2\pi$. The NAdS boundary is at $x=\i T + y$, for $T\gg1$, while the geodesic boundary is along the real axis. }
\label{fig:trumpet}
\end{figure}

The calculation is very similar to the one that computes the disk partition function. We view the trumpet as the complex $x$ plane with the equivalence $x\cong x+2\pi$, and a metric of scalar curvature $-2$. Let us put the geodesic at ${\rm Im}~x=0$ and the outer NAdS boundary at ${\rm Im}~x=T$, for some large $T$. Including the fermions, the trumpet is described by coordinates $x,\psi,\bar{\psi}$ with equivalence 
\beq
(x,\psi,\bar{\psi} ) \cong (x+2\pi, -e^{\i \hq \phi} \psi, -e^{-\i \hq \phi} \bar{\psi}).
\eeq
On the upper boundary, write $x=\i T+ y$ with real $y$. Thus the upper boundary is described by $y,\psi,\bar{\psi}$ with 
\beq
(y,\psi,\bar{\psi})\cong (y+2\pi, -e^{\i\hq  \phi} \psi,- e^{-\i\hq \phi} \bar{\psi}).
\eeq
Just like in the case of the disk, the boundary quantum mechanics system is formulated on a supercircle with coordinates $t,\theta,\bar{\theta}$ and an equivalence relation \eqref{eq:equivreltttb}.
For the case of the trumpet, the fixed points take the form
\bea
y(t,\theta) &=& \frac{2\pi}{\beta} t,\nn
\psi(t,\theta) &=& e^{\i\hq ( \phi-\alpha) t/\beta} \theta,
\ea
with $\alpha \to \alpha + 2\pi n$, and $n\in\mathbb{Z}$. These fixed points are annihilated by an operator similar to the one appearing in the case of the disk 
\beq
D= \frac{\partial}{\partial t} - \frac{2\pi}{\beta} \frac{\partial}{\partial y} - \hq \frac{\phi-\alpha}{\beta} \i (\psi \partial_\psi - \bar{\psi}\partial_{\bar{\psi}}).
\eeq
The action on these fixed points can be computed from the bosonic sector only. The rotation eigenvalues from the bosonic sector are also straightforward. The fermionic rotation eigenvalues 
are independent of $\phi$. The calculation is explained in appendix \ref{sec:disktrumpet}. The trumpet partition function with bulk parameters
$b,\phi$  is given by 
\bea\label{trumpet}
Z_{\rm tr}(\beta,\alpha; b,\phi)&=& \sum_{n\in \mathbb{Z}} \exp(2\pi\i n\delta)  \frac{\hq \, \cos\left( \pi \hq \,(n+\frac{\alpha}{2\pi})\right)}{\pi \beta} e^{-\frac{b^2}{4\beta}-\frac{4\pi^2\, \hq^{\, 2}}{\beta}(n+\frac{\alpha}{2\pi}-\frac{\phi}{2\pi})^2},
\ea
The exponential term is precisely the sum of the bosonic Schwarzian action for the trumpet and the bosonic $\uU(1)$ mode action, with the same boundary conditions for bosonic fields as in $\cN=2$ JT gravity. Using Poisson resummation, the expression \eqref{trumpet} can be written as a sum over supermultiplets
\bea
Z_{\rm tr}(\beta,\alpha; b,\phi)=\sum_{q\in \mathbb{Z} + \delta-\frac{1}{2}} \big( e^{\i \alpha (q-\frac{\hq}{2} )} + e^{\i \alpha (q+\frac{\hq}{2})}\big)  e^{- \i \phi q} e^{-\beta E_0(q)} \frac{1}{2\pi} \frac{ e^{-\frac{b^2}{4\beta}}}{\sqrt{4\pi \beta}}.\label{trumpet2}
\ea
This expression is particularly simple. Other than the chemical potential dependence, which reproduces the expectation for the charge spectrum of a supermultiplet, the rest is almost the same as if one considered bosonic JT gravity coupled to a $\uU(1)$ gauge field. We remind the reader that the bosonic JT gravity trumpet is indeed $Z^{\rm bos.}_{\rm tr}(\beta, b) = e^{-b^2/4\beta}/\sqrt{4\pi\beta}$ in our conventions.

Formally, we expect the trumpet path integral to be invariant under $\phi\to\phi+2\pi$, but eqn. (\ref{delfo}) generically does not have this property; under $\phi\to\phi+2\pi$,
it is multiplied by $\exp(-2\pi \i(\delta-1/2))$.
 We have introduced a charge $\delta$ mod $\Z$
in the boundary quantum mechanics.   The bulk theory must describe the same charge propagating on the outer boundary.
 The Schwarzian mode on the outer boundary carries a charge $\pm \h q/2=1/2$ mod $\Z$.   So it accounts for a half-unit of boundary charge.
To reproduce the boundary quantum mechanics, an
 additional charge $\delta-1/2$ on the outer boundary is needed; the bulk theory must have a gauge theory theta-angle $2\pi(\delta-1/2)$, as claimed in eqn. (\ref{delfo}).   This
 induces a charge $\delta-1/2$ mod $\Z$ on the outer boundary, which together with the contribution of the Schwarzian mode reproduces the charge in the boundary
 quantum mechanics.   But this theta-angle will also place a charge $1/2-\delta$ on the inner boundary.   We have not included the corresponding Wilson operator
 on the inner boundary; doing so would give a factor $\exp(\i\phi(\delta-1/2))$ in the path integral, canceling the anomaly.    
 When we combine two trumpets to make the double trumpet,  the charges on the inner boundaries  cancel
and the anomaly under $\phi\to\phi+2\pi$ disappears.

We are ready now to assemble the pieces by gluing together the two trumpets along their common interior geodesic:
\vspace{-0.3cm}\\
\bea
\begin{tikzpicture}[scale=0.8, baseline={([yshift=-0.1cm]current bounding box.center)}]
\draw[thick] (5.5,0) ellipse (0.3 and 1.5);
\draw[thick] (7.5,0) ellipse (0.1 and 0.7);
\draw[thick] (5.54,1.49) to [bend right=20] (7.5,0.7);
\draw[thick] (5.54,-1.49) to [bend left=20] (7.5,-0.7);
\draw[thick] (11.5,0) ellipse (0.3 and 1.5);
\draw[thick] (9.5,0) ellipse (0.1 and 0.7);
\draw[thick] (11.46,1.49) to [bend left=20] (9.5,0.7);
\draw[thick] (11.46,-1.49) to [bend right=20] (9.5,-0.7);
\draw[thick,red,->] (8.6,0) to [bend right=20] (8.5,0.5)  ;
\node at (8.5,1) {\textcolor{red}{\small $\varrho,\varphi$}};
\node at (7.5,-1.3) {\small $b,\phi$};
\node at (9.5,-1.3) {\small $b,-\phi$};
\end{tikzpicture}
\ea
\vspace{-0cm}\\
Twist or gluing parameters are shown in red.
After integrating over the twist parameters, the path integral on the cylinder is
\bea
Z_{0,2}(\beta_1,\alpha_1 ; \beta_2, \alpha_2 ) &=& (2\pi) \int_0^{2\pi} \d\phi \int_0^\infty b\d b ~Z_{\rm tr}(\beta_1, \alpha_1; b,\phi)Z_{\rm tr}(\beta_2, \alpha_2;b,-\phi).\nonumber
\ea
When we glue together the two trumpets, we reverse the orientation of the inner boundary prior to gluing (differently put, the orientation of the geodesic that comes by viewing it as the boundary of one trumpet is opposite to the orientation that comes from viewing it as the boundary of the other trumpet). This has the effect of reversing the sign of $\phi$ on one side. In appendix \ref{app:JTU1bos} we recall the analogous result for bosonic JT gravity coupled to a $\uU(1)$ gauge field \cite{Kapec:2019ecr}.

To compare with the matrix model, it is convenient to use the representation of the trumpet given in \eqref{trumpet2}. After integrating over the geodesic length, we obtain
\bea
Z_{0,2}(\beta_1,\alpha_1 ; \beta_2, \alpha_2 )  &=& \int_0^{2\pi} \frac{\d\phi}{2\pi} \sum_{q_1,q_2 \in \mathbb{Z}+\delta-\frac{1}{2}}\big( e^{\i \alpha_1 (q_1-\frac{\hq}{2}) } + e^{\i \alpha_1  (q_1+\frac{\hq}{2})}\big) \big( e^{ \i \alpha_2  (q_2-\frac{\hq}{2}) } + e^{\i \alpha_2  (q_2+\frac{\hq}{2})}\big)   \nonumber\\
&&\times \, e^{ \i \phi (q_2-q_1)}\, \frac{\sqrt{\beta_1 \beta_2}}{2\pi(\beta_1+\beta_2)} e^{-\beta_1 E_0(q_1)-\beta_2 E_0(q_2)}.
\ea
The anomaly under $\phi\to\phi+2\pi$ has disappeared, as expected.

We now integrate over the internal holonomy. This produces a delta function imposing that $q_1 = q_2 \equiv q$. The final answer is  
\beq\label{eq:Z0202}
Z_{0,2}=\sum_{q\in \mathbb{Z}+\delta-\frac{1}{2}}  \big( e^{\i \alpha_1 (q-\frac{\hq}{2}) } + e^{\i \alpha_1  (q+\frac{\hq}{2}) }\big) \big( e^{ \i \alpha_2 (q-\frac{\hq}{2}) } + e^{ \i \alpha_2  (q+\frac{\hq}{2}) }\big)   \frac{\sqrt{\beta_1 \beta_2}}{2\pi (\beta_1+\beta_2)}  e^{-(\beta_1+\beta_2) E_0(q)}.
\eeq
This is consistent with the statistical independence of the Hamiltonian on each supermultiplet, a feature present in the ensemble described in section \ref{randommatrix}. 
As in the discussion of the disk path integral, the factors $ e^{\i \alpha (q-\frac{\hq}{2}) } + e^{\i \alpha  (q+\frac{\hq}{2}) } $ represent a sum over the two states in a supermultiplet, labeled by the average charge $q$.   
Going back to the random matrix interpretation in eqn. \eqref{eq:Z02rmt2}, the result \eqref{eq:Z0202} implies 
\bea\label{trumpresult}
\langle {\rm Tr}_{\M,q_1}( e^{-\beta_1 H})~{\rm Tr}_{\M,q_2}( e^{-\beta_2 H})\rangle_{{\rm conn.}, g=0} = \frac{\sqrt{\beta_1 \beta_2}}{2\pi(\beta_1+\beta_2)} e^{-(\beta_1+\beta_2)E_0(q_1)} \delta_{q_1,q_2}.
\ea 
This agrees with the prediction based on the random matrix theory analysis of section \ref{randomtwo}, since eqn. (\ref{trumpresult}) is the prediction of a $\upbeta=2$  AZ ensemble  if
either $E_0(q)>0$ (in which case the value of $\upalpha$ does not matter) or $E_0(q)=0$ and 
$\upalpha$ is of order 1 (not of order $e^{S_0}$).   In $\N=2$ JT supergravity,
generically $E_0(q)>0$, and when $E_0(q)=0$, $\upalpha=1$.

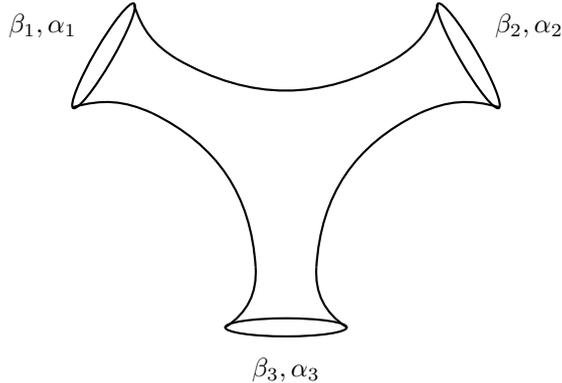
\begin{figure}[t!]
\begin{center}
\begin{tikzpicture}[scale=0.8, baseline={([yshift=-0.1cm]current bounding box.center)}]
\draw[thick] (0,-2) ellipse (1 and .15);%Boundary
\draw[thick, rotate around={-30:(-3,2.5)}] (-3,2.5) ellipse (.15 and 1);%Boundary
\draw[thick, rotate around={30:(3,2.5)}] (3,2.5) ellipse (.15 and 1);%Boundary
 \draw[thick, bend right=30] (-1.8,2.45) to (1.8,2.45);
  \draw[thick, bend left=20] (-1.8,2.45) to (-2.5,3.35);
  \draw[thick, bend right=20] (1.8,2.45) to (2.5,3.35);
  \draw[thick, bend right=20] (-2.2,1.55) to (-3.5,1.65);
  \draw[thick, bend left=20] (2.2,1.55) to (3.5,1.65);
 \draw[thick, bend left=30] (-2.2,1.55) to (-0.5,-1);
  \draw[thick, bend right=30] (2.2,1.55) to (0.5,-1);
   \draw[thick, bend left=30] (-0.5,-1) to (-1,-2);
  \draw[thick, bend right=30] (0.5,-1) to (1,-2);
  \node at (-4,3) {\small $\beta_1,\alpha_1$};
   \node at (4,3) {\small $\beta_2,\alpha_2$};
   \node at (0,-2.7) {\small $\beta_3,\alpha_3$};
 %   \node at (-6,0) {\LARGE $Z_{0,3}~~=$};
    \end{tikzpicture} 
    \end{center}
    \caption{\footnotesize 
    The leading connected three-boundary wormhole: a three-holed sphere with three NAdS boundaries, each of which is labeled by its 
inverse temperature $\beta$ and $\uU(1)$ $R$-charge chemical potential $\alpha$. }
    \label{fig:3bdywh}
    \end{figure}
    
   From \eqref{eq:Z0202}, we see that the cylinder  path integral receives contributions only from the non-BPS multiplets.   Therefore, the numbers 
   $N_{\rm BPS}(k)$ of BPS states have trivial dispersion, at least at leading order in $e^{S_0}$.   This justifies a claim made 
 in section \ref{sec:Z01}, following eqn. \eqref{eq:spectrumcomm}.   The computations in the rest of this article will show that the absence of dispersion in
 $N_{\rm BPS}(k)$ remains true to all orders in the topological expansion; the path integral for every topology except the disk receives contributions only from the non-BPS 
 multiplets (see footnote \ref{footnoteBPSfluct}).

\subsection{Preliminaries on the Three-Boundary Wormhole}\label{sec:3bwh}

So far we have considered the simple cases of the disk and the cylinder, which did not require integrals over complicated moduli space of $\cN=2$ hyperbolic surfaces. We will study the general case for an arbitrary number of handles and boundaries in the next two sections. 

As an introduction, to familiarize the reader with what the final answers look like, we anticipate the result for the leading 
three-boundary wormhole. This corresponds to the gravitational path integral $Z_{0,3}(\beta_1,\alpha_1;\beta_2,\alpha_2;\beta_3,\alpha_3)$ over a spacetime that is a three-holed sphere with three asymptotically NAdS boundaries labeled by  inverse temperatures $\beta_1,\beta_2,\beta_3$ and chemical potentials $\alpha_1,\alpha_2,\alpha_3$; see fig. \ref{fig:3bdywh}. The goal is to study to what extent the gravity path integral can be identified with 
\beq\label{eq:03rmt}
Z_{0,3}(\beta_1,\alpha_1;\beta_2,\alpha_2;\beta_3,\alpha_3)=\big\langle {\rm Tr}( e^{-\beta_1 H+\i \alpha_1 J})~{\rm Tr}( e^{-\beta_2 H+\i \alpha_2 J})~{\rm Tr}( e^{-\beta_3 H+\i \alpha_3 J})\big\rangle_{{\rm conn.}, g=0}.
\eeq
Similar to the case with two boundaries, if supermultiplets are statistically independent, we expect to obtain that the fixed multiplet connected correlator is zero unless the three multiplets coincide
\beq
\big\langle {\rm Tr}_{\M,q_1}( e^{-\beta_1 H})~{\rm Tr}_{\M,q_2}( e^{-\beta_2 H})~{\rm Tr}_{\M,q_3}( e^{-\beta_3 H})\big\rangle_{{\rm conn.}, g=0} \propto \delta_{q_1,q_2} \delta_{q_2,q_3},
\eeq
In the rest of this section we reproduce this result from gravity. Moreover, we also match the function of multiplet and temperature appearing as a prefactor in the equation above with the prediction from the topological recursion of the AZ matrix ensemble.

We begin as in the previous cases.  The first step consists, in $BF$ language, in localizing the path integral on flat $\SU(1,1|1)$ connections that correspond to
hyperbolic geometries. This guarantees the presence of a geodesic homotopic to each asymptotically NAdS boundary. As in  \cite{SSS}, we
 can cut along those geodesics and divide the geometry into three trumpets and an internal manifold that is a hyperbolic three-holed sphere with geodesic boundaries. 
The path integral on the trumpets was already computed in  \eqref{trumpet2}. We denote the interior or bulk contribution by\\ 
\beq
~~~~V_{0,3}({\sf b}_1,{\sf b}_2, {\sf b}_3)~~~~  \to~~~~ \begin{tikzpicture}[scale=0.6, baseline={([yshift=-0.1cm]current bounding box.center)}]
\draw[thick, rotate around={-30:(-2,2)}] (-2,2) ellipse (.15 and .5);%Geodesic
\draw[thick, rotate around={30:(2,2)}] (2,2) ellipse (.15 and .5);%Geodesic
\draw[thick] (0,-1) ellipse (.5 and .15); %Geodesic
 \draw[thick, bend right=30] (-1.8,2.45) to (1.8,2.45);
 \draw[thick, bend left=30] (-2.2,1.55) to (-0.5,-1);
  \draw[thick, bend right=30] (2.2,1.55) to (0.5,-1);
  \node at (-4.2,2) {\small ${\sf b}_1=(b_1,\phi_1)$};
   \node at (4.2,2) {\small ${\sf b}_2=(b_2,\phi_2)$};
   \node at (1.6-1.5,-1.8) {\small ${\sf b}_3=(b_3,\phi_3)$};
    \end{tikzpicture} 
\eeq\\
We introduced a notation we will use throughout, where ${\sf b} = (b,\phi)$ combines the length $b$ of a boundary with the $\uU(1)$ holonomy $e^{\i \hq\phi}$ around the boundary. The final answer for $Z_{0,3}$ is given by gluing this quantity $V_{0,3}({\sf b}_1,{\sf b}_2, {\sf b}_3)$ to the three trumpets, including the integral over the twist parameters,
\bea
Z_{0,3}= \left[ \prod_{j=1,2,3} \int_0^\infty b_j \d b_j \int_0^{2\pi}2\pi \d \phi_j\, Z_{\rm tr} (\beta_j,\alpha_j; \overline{{\sf b}}_j)\right] V_{0,3}({\sf b}_1,{\sf b}_2, {\sf b}_3),\nonumber
\ea
where $\overline{{\sf b}} = (b, - \phi)$ is the orientation reversed version of ${\sf b}=(b,\phi)$. As explained before, one needs to reverse the orientation of the $\uU(1)$ holonomy around one of the geodesics involved in the gluing; in this expression we chose to apply this reversal on the trumpet. To avoid clutter, we omitted the arguments of $Z_{0,3}$. The factors of $b_j$ come from the integral over the length-twist parameter while the factors of $2\pi$ come from the $\uU(1)$-twist parameter. 

We leave a detailed derivation of the gravity path integral on spaces with geodesic boundaries and handles for the following two sections. For now we quote the result derived in section \ref{sec:3hole} for the particular case of $V_{0,3}$;
\beq\label{ansatz}
V_{0,3}({\sf b}_1,{\sf b}_2, {\sf b}_3)= - \frac{1}{2\pi} \frac{1}{4\hq^{\, 2}}\delta''(\phi_1+\phi_2+\phi_3).
\eeq
The delta function imposes the constraint $\phi_1 + \phi_2 + \phi_3 = 0$.  A flat $\uU(1)$ connection on the three-holed sphere with boundary holonomies $\phi_1,\phi_2,\phi_3$
exists if and only if $\phi_1+\phi_2+\phi_3=0$.   Accordingly, in $BF$ theory with gauge group $\uU(1)$, we would get a delta function $\delta(\phi_1+\phi_2+\phi_3)$.
In $BF$ theory of $\SU(1,1|1)$, the three-holed sphere has fermionic moduli which appear as corrections to the argument of the delta function.  
 Integrating  over those fermionic moduli produces the two derivatives acting on the delta function. Details of this calculation are spelled out in section \ref{sec:GPI}. 

The overall prefactor  in \eqref{ansatz} depends on the normalization of the $\cN=2$ JT gravity path integral measure. The calculation in the next section also clarifies our choice of 
normalization.

Accepting for now this result for $V_{0,3}({\sf b}_1,{\sf b}_2,{\sf b}_3)$, we are ready to put the pieces together. The answer is actually quite simple. For concreteness, integrate first over the third holonomy $\phi_3$. Using the delta function appearing in \eqref{ansatz}, the only remaining dependence on the other holonomies is \be\label{output}e^{\i\phi_1(q_1-q_3) + \i \phi_2 (q_2-q_3)}.\ee We observe again that although the individual $q_i$ are valued in $\delta+\Z$ for some $\delta$,  the differences $q_1-q_3$ and $q_2-q_3$ are always integers.  So the
expression that results from integration over $\phi_3$ is periodic under $\phi_1 \sim \phi_1 + 2\pi$ and $\phi_2 \sim \phi_2 + 2\pi$. Next we perform the integrals over $\phi_1$ and $\phi_2$, which sets $q_1=q_3$ and $q_2=q_3$. So the three supermultiplets propagating in the three trumpets are all the same. The final answer after including the trumpet
partition functions and integrating over the three geodesic lengths is given by
\bea
Z_{0,3}&=& \sum_{q\in \mathbb{Z}+\delta-\frac{1}{2}} \big(e^{\i\alpha_1 (q-\frac{\hq}{2})} + e^{\i \alpha_1 (q+\frac{\hq}{2})} \big)\big(e^{\i\alpha_2 (q-\frac{\hq}{2})} + e^{\i \alpha_2  (q+\frac{\hq}{2})}\big)\big(e^{\i\alpha_3 (q-\frac{\hq}{2})} + e^{\i \alpha_3  (q+\frac{\hq}{2})}\big)\nonumber\\
&&~~~~\times (2\pi) \frac{q^2}{4\hq^{\, 2}} \frac{\sqrt{\beta_1 \beta_2\beta_3}}{\pi^{3/2}} e^{-(\beta_1+\beta_2+\beta_3)E_0(q)}.
\ea
The first line is completely fixed by the structure of the $\cN=2$ supermultiplet and the constraint of the three of 
them being the same. The second line has more information regarding the details of the matrix model spectral curve. 
The factor of $2\pi$ is the answer one would obtain from a $\uU(1)$ gauge theory on the three-holed sphere. 
The factor of $q^2/4 \hq^{\, 2}= E_0(q)$ comes from the details inherent to the $\cN=2$ supergravity. The temperature 
dependence is similar to the one appearing in bosonic JT gravity. We would like to emphasize that these features are   
highly non-trivial consequences of the explicit evaluation. This will become clear from the derivation in the following sections.

The gravity prediction for $\langle {\rm Tr}_{\M,q_1}( e^{-\beta_1 H})~{\rm Tr}_{\M,q_2}( e^{-\beta_2 H})~{\rm Tr}_{\M,q_3}( e^{-\beta_3 H})\rangle_{{\rm conn.}, g=0} $, 
the random matrix partition function correlator  for  fixed supermultiplet, is:
\bea
 (2\pi)E_0(q_3) \frac{\sqrt{\beta_1 \beta_2\beta_3}}{\pi^{3/2}} e^{-(\beta_1 + \beta_2 + \beta_3)E_0(q_3)} \delta_{q_1,q_2} \delta_{q_2,q_3}.
\ea
Consistent with the discussion in section \ref{randommatrix}, the different supermultiplets are statistically independent. 
The temperature dependence reproduces precisely the matrix model topological recursion for an $(\upalpha_0,\upbeta)=(1,2)$ 
ensemble, with $\nu$ either zero or proportional to $e^{S_0}$, applied to the spectral curve \eqref{eq:N2sc}. We leave it  for the reader to verify this, by a 
straightforward application of the matrix model loop equation presented in appendix \ref{app:Loop}. 
In the next two sections we will provide a procedure to compute the gravitational path integral of $\cN=2$ JT gravity for any topology and 
prove to all orders in the genus expansion the match with the random matrix topological recursion. This check will therefore become a particular case of a more general analysis.

\section{Gravity Path Integral for Spaces with Wormholes}\label{sec:GPI}

\subsection{The Path Integral of $BF$ Theory}\label{pbf}
With the exception of the disk, which we have already studied, 
any oriented NAdS two-manifold can be built by gluing trumpets onto an ``interior'' manifold, a hyperbolic  two-manifolds with geodesic boundary.
This decomposition leads to a convenient evaluation of the path integral of JT gravity \cite{SSS}.   This analysis has been  to $\N=1$ JT supergravity \cite{SW},
and here we will describe the generalization to $\N=2$.

The  path integral on the trumpet requires special treatment, but in any event we have already discussed this, in arriving at  eqn. \eqref{trumpet}.
The remaining JT gravity path integral on the interior manifold  $C$ is naturally formulated in $BF$ theory.   The action of $BF$ theory with group $G$, 
on a surface $Y$, is $I= - \i \int_Y {\rm Tr} \, BF$. $B$ is an adjoint-valued spin zero field and  $F=\d A + A\wedge A$ is the field strength of a gauge connection  $A$. 
Integrating over $B$ produces a delta function $\delta (F)$, localizing the path integral to the moduli space\footnote{We use the same letter $\T$ for the full moduli 
space of flat connections, and for the geometric component relevant for gravity. We hope this will not cause confusion. The motivation for the name $\T$ is that in the case of
gauge group $\SU(1,1|1)$, $\T$ is an $\N=2$ version of Teichm\"{u}ller space.}  $\T$ of flat connections on $C$ modulo gauge transformations. 
In defining $\T$, one wants to specify the conjugacy classes of the geodesics around the boundaries of $C$; then in gluing onto the external trumpets, one integrates
over those holonomies. 
In the $BF$ description, the boundary holonomy is specified by fixing the gauge connection $A$ along the boundary.
In gravity one has to further reduce to a component of $\T$ that is related to hyperbolic metrics, and also mod out by the mapping class group. 
 In the present section, we do not consider such global questions, and merely aim to
determine the appropriate measure on $\T$, deferring global questions to section \ref{sec:RR}.

To derive the measure on the moduli space, one can expand the path integral around a chosen flat connection $A_0$, which represents a  point in $\mathcal{T}$. After
integrating over $B$ to get the delta function $\delta(F)=\delta(\d A+A\wedge A)$, one integrates over fluctuations in $A$ and over the ghosts that are introduced
for gauge-fixing.   With a standard gauge choice, this leads to a measure $\mu$ on $\T$ that differs from a classical measure $\mu_0$ by a certain ratio of determinants:
\bea\label{bfrmeasure}
 \mu =  \mu_0 \cdot \frac{\sqrt{{\rm det}'\Delta_0}}{\sqrt{{\rm det}'\Delta_2}}.
\ea   
In eqn. (\ref{bfrmeasure}),  $\Delta_q = D^* D  + D D^*$ is the Laplacian acting on adjoint-valued $q$-forms, constructed from the exterior derivative $D= \d + [ A_0, \cdot]$ and its adjoint $D^*$. 
The numerator arises from the ghost path integral, while the denominator arises from the integral over $A$.  The primes denote the removal of zero-modes.  
 One way to describe the classical measure $\mu_0$ is the following: $\T$ is naturally a Riemannian manifold,\footnote{For compact $G$, if
 one imposes the harmonic gauge condition $\d\star \delta A=0$ on a zero-mode $\delta A$ of $A$ (where $\star$ is the Hodge star), 
  then the  metric on the space of zero-modes of $A$ and the Riemannian structure of $\T$ are defined
  by  $|\delta A|^2=\int_C\Tr\,\delta A\wedge \star\delta A$.  For non-compact $G$, this definition needs to be slightly modified.}  
  and $\mu_0$ is the Riemannian measure on $\T$;
physically, $\mu_0$ is the natural measure on the space of zero-modes of $A$.

On an oriented two-manifold,  an alternative description of $\mu_0$ is generally more useful.  
 If $C$ is oriented, then $\T$ is a symplectic manifold and $\mu_0$ is its natural symplectic measure.
In addition, if $C$ is oriented, then  
the ratio $ \frac{\sqrt{{\rm det}'\Delta_0}}{\sqrt{{\rm det}'\Delta_2}}$ is equal to 1 because Hodge duality between 0-forms and 2-forms implies
that $\Delta_0$ and $\Delta_2$ have the same spectrum.   Therefore, in this case,  the quantum measure $\mu$ coincides with the classical measure $\mu_0$.

Yet another way to look at the measure $\mu$ that comes from $BF$ theory is as follows.
On any two-manifold, orientable or not, this measure  is the analytic torsion of Ray and Singer  \cite{RaySinger}.  This observation, or more precisely its generalization to
any spacetime dimension, was made by Schwarz \cite{Schwarz:1978cn}.
In any dimension,  analytic torsion is a certain ratio of determinants times a classical measure $\mu_0$ on the space of zero-modes; in two dimensions, 
the definition reduces to precisely what we have written
in eqn. (\ref{bfrmeasure}).   
Analytic torsion is equivalent to combinatorial torsion,   originally defined by Reidemeister \cite{Reidemeister}, as was proved in 
 \cite{Cheeger,Muller1,Muller2,BismutZhang}. In physical terms, this means that $BF$ theory has the unusual property that it can be defined by a lattice theory with
 a result that is completely independent of the choice of lattice.   By taking a very fine lattice, one can motivate the idea that the lattice theory has a continuum limit (namely $BF$ theory).
 On the other hand, one can choose a very crude lattice so that the path integral can be directly calculated, leading to explicit formulas.

Although the measure $\mu_0$ in eqn. (\ref{bfrmeasure}) 
can be understood as a Riemannian measure, as far as we know this does not lead to an effective method of calculation.  On an orientable two-manifold,
the symplectic interpretation of $\mu_0$ definitely does lead to an effective method of calculation.  However, as far as we know, such calculations are not  simpler than
calculations based on the torsion.  We have chosen to base our calculations on the relationship of the measure to the torsion, because this is directly related to the physics,
as the derivation of eqn. (\ref{bfrmeasure}) showed, and because this approach is equally valid on orientable or unorientable two-manifolds.

In this section and the next one, we consider orientable two-manifolds only.   Unorientable two-manifolds are considered in section \ref{timereversal}.
Any orientable two-manifold can be constructed by gluing three-holed spheres along their boundaries.   By determining the moduli space measure for the three-holed sphere
as well as the procedure for gluing, we can determine the moduli space measure for an arbitrary oriented two-manifold.  

The computation of  the torsion of a three-holed sphere
and the gluing procedure were described in \cite{SW}, section 3, in a way that is valid for any possibly non-compact gauge group or supergroup.   (The examples considered in that
paper were $\SL(2,\R)$ and ${\mathrm{OSp}}(1|2)$.)    Here we will not repeat those general explanations and will just apply the procedure for the gauge group  $\SU(1,1|1)$ that
is relevant to $\N=2$ JT supergravity. 

There is a subtlety in the statement that the symplectic structure and the torsion define the same measure.   On an oriented two-manifold $C$ without boundary, this is true as stated:
if $\T_{g}$ is the moduli space of flat connections with gauge group $G$ on a closed surface of genus $g$, then the symplectic structure and the torsion define the same measure
$\mu$ on $\T_g$ (assuming the symplectic structure is properly normalized).  On the other hand, if $C$ is an oriented two-manifold of genus $g$ with $n$ boundary circles, then the symplectic structure and the torsion define measures on two closely related but slightly different spaces.
The symplectic structure determines a measure $\mu$ on what we will call $\T_{g,n}$ (or $\T_{g,n}(\vec w)$ if we wish to be more precise), the moduli space of flat bundles with
prescribed conjugacy classes $\vec w=(w_1,\cdots,w_n)$ of the holonomies 
around the boundaries.   The torsion defines a measure $\tau$ on what we will call $\cR_{g,n}$, the moduli space of flat bundles
over $C$ without a restriction on the boundary holonomies.   (In the case of gravity, we restrict the holonomies to be hyperbolic, but $\tau$  can be defined without this restriction.)
The two spaces and measures are closely related, and in the case of $\N=2$ JT gravity, we will ultimately determine the precise relationship in eqn. (\ref{comparison}).

\subsection{Measure On Moduli Space of $\cN=2$ Hyperbolic Surfaces}\label{sec:3hole}
In this section, after some preliminaries, we determine the torsion for a  three-holed sphere, the torsion for a circle and the associated gluing procedure, and the final answer for 
the measure that comes from
 the $BF$ path integral on an arbitrary orientable surface.

\subsubsection{JT Gravity as a $BF$ Theory}\label{JTBF}

We start by reviewing some aspects of $BF$ theory with gauge group ${\rm SL}(2,\R)$  on an oriented  two-manifold $C$. We study flat connections on  $C$,
with the proviso that if $C$ has a non-trivial boundary, then 
the holonomy  of the flat connection around any boundary component is a hyperbolic 
element of $\SL(2,\R)$, that is, an element that is conjugate to 
\bea\label{eq:Ul}
U_b^\pm = \pm \left(\hspace{-1mm}\begin{array}{cc}
 e^{b/2} & 0 \\
0& e^{-b/2}
\end{array}\hspace{-1mm}\right),
\ea
with some $b>0$.   With this specification, the
 moduli space of flat ${\rm SL}(2,\mathbb{R})$ connections on $C$ has several disconnected components, one of which parametrizes hyperbolic metrics on $C$ that have
  curvature $R=-2$ and such that, for any given boundary, the corresponding parameter $b$ is the circumference of that boundary.\footnote{The group $\SL(2,\R)$ is contractible onto
its maximal compact subgroup $\uU(1)$, so topologically any $\SL(2,\R)$ bundle over $C$
is equivalent to a $\uU(1)$ bundle and therefore has a first Chern class.  (This first Chern class can be expressed as an integral over $C$, once one
 picks a reduction of the structure group of the bundle from $\SL(2,\R)$ to $\uU(1)$.)   If $C$ is a closed oriented surface of genus $g$, then the values of the first Chern class of a flat
$\SL(2,\R)$ bundle over $C$ 
 range from $g-1$ to $-(g-1)$, and hyperbolic metrics on $C$ are related to the component of first Chern class $g-1$.   The first Chern class can still be defined if $C$
is an oriented two-manifold with boundary and the flat connection on $C$ is constrained to have hyperbolic conjugacy around the boundary, because hyperbolic holonomy around
a given boundary component reduces
the structure group of the flat bundle along that boundary  to a contractible subgroup (the one-parameter subgroup of $\SL(2,\R)$ that contains the holonomy around
the given boundary).   Again the component of $\T$
with maximum possible first Chern class is the one related to hyperbolic metrics on $C$.} That is the component that we will study.

In purely bosonic JT gravity, the sign in eqn. (\ref{eq:Ul})  is irrelevant.   One can take the gauge group to be ${\mathrm{PSL}}(2,\R)$ and disregard this sign.  
In a theory with fermions, one wants to endow $C$ with a spin structure as well as a hyperbolic metric.    In that case, one must take the gauge group to be $\SL(2,\R)$
and the sign in eqn. (\ref{eq:Ul}) is meaningful: it determines the spin structure on the given boundary component of $C$.

\begin{figure}
    \centering
\includegraphics[scale=0.3]{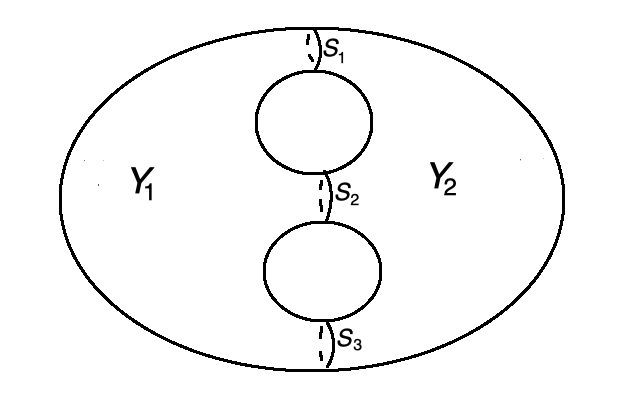}
    \caption{\footnotesize Decomposition of a genus 2 surface as a union of three-holed spheres $Y_1,$ $Y_2$, glued on circles $S_1,S_2,S_3$.}
    \label{fig:pair of pants}
\end{figure}
The flat connections on $C$ that are related to hyperbolic metrics on $C$ have hyperbolic  holonomy around any homotopically nontrivial embedded
 loop in $C$ (not just the boundaries
of $C$).   As a result, if we ``cut'' $C$ along some embedded circles to decompose it into a union of three-holed spheres, there will be hyperbolic
holonomy around each of the boundaries of these three-holed spheres.   The loops along which we cut can always be chosen to be geodesics, so the cutting decomposes $C$ into
a union of three-holed spheres with geodesic boundaries.  In brief, we will call these geodesic three-holed spheres. In fig. \ref{fig:pair of pants}, we give an example of
building a genus 2 surface by gluing together two three-holed spheres.

In $\N=2$ JT supergravity,  the gauge group is $\SU(1,1|1)$ (or more precisely a $\hq$-fold cover thereof).   Since a maximal  
 bosonic subgroup of $\SU(1,1|1)$ is $\SU(1,1)\times\uU(1)\cong \SL(2,\R)\times \uU(1)$, we will briefly discuss $BF$ theory of $\SL(2,\R)\times \uU(1)$ before
 studying $\SU(1,1|1)$.
$BF$ theories based on  $\SL(2,\R)\times \widetilde G$, with any compact $\widetilde G$, were studied in \cite{Kapec:2019ecr}, the main result being a match
with a random matrix theory in which the different representations of $\widetilde G$ are statistically independent.   
The case $\widetilde G=\uU(1)$, relevant to $\N=2$ JT supergravity, is important for the present article and has special properties because $\uU(1)$ is abelian.
In $\uU(1)$ gauge theory, the holonomy around a given boundary can be an arbitrary element $e^{\i \phi}\in \uU(1)$.   If $C$ has $n$ boundary components, then
the holonomies $e^{\i\phi_1},\cdots,e^{\i\phi_n}$ around these boundaries cannot be chosen arbitrarily.   A flat connection on $C$ with these holonomies
exists if and only if 
\be\label{holconstraint} \sum_{i=1}^n \phi_i = 0 ~{\mathrm {mod}}~2\pi \mathbb{Z}.\ee
This constraint, which is slightly deformed in the case of $\SU(1,1|1)$,  will be important in what follows.

$\SU(1,1|1)$, the gauge group relevant to $\N=2$ JT supergravity, 
 can be  presented\footnote{$\SU(1,1|1)$ is isomorphic to ${\mathrm{OSp}}(2|2)/\Z_2$, so it is also related to the group of linear transformations of
a vector space of dimension $2|2$ that preserves a certain symplectic form.    This will be important in section \ref{sec:RR} when we derive the $\N=2$ analog of
Mirzakhani's recursion  relation for volumes. But except in that application, we use the $\SU(1,1|1)$ description, since it is easier to calculate with $3\times 3$ matrices.} as the group of linear transformations of a vector space $V$ of dimension $2|1$
that preserves a certain hermitian inner product.   This inner product can be described by saying that for
$a,b|\eta\in V$, 
\bea\label{eq:innprod}
|(a,b|\eta)|^2 =\i( a \bar{b} - b \bar{a})+ 2 \i \bar{\eta} \eta.
\ea
For  bosonic generators of $\mathfrak{su}(1,1|1)$, we can choose
\bea\label{bosegen}
{\sf e}=\left(\begin{array}{cc|c}
 0&1\hspace{.5mm}&\hspace{.5mm}0 \\
0&0\hspace{.5mm}&\hspace{.5mm}0 \\
\hline 0&0\hspace{.5mm}&\hspace{.5mm}0 
\end{array}\right),
~{\sf f}=\left(\begin{array}{cc|c}
 0&0\hspace{.5mm}&\hspace{.5mm}0 \\
 1&0\hspace{.5mm}&\hspace{.5mm}0 \\
\hline 0&0\hspace{.5mm}&\hspace{.5mm}0 
\end{array}\right),
~{\sf h}=\left(\begin{array}{cc|c}
 1&0\hspace{.5mm}&\hspace{.5mm}0 \\
0&-1\hspace{.5mm}&\hspace{.5mm}0 \\
\hline 0&0\hspace{.5mm}&\hspace{.5mm}0 
\end{array}\right),
~{\sf z}=\left(\begin{array}{cc|c}
 \i&0\hspace{.5mm}&\hspace{.5mm}0 \\
 0& \i\hspace{.5mm}&\hspace{.5mm}0 \\
\hline 0&0\hspace{.5mm}&\hspace{.5mm}2\i 
\end{array}\right).\label{slgenerators}
\ea
The first three generators satisfy the $\mathfrak{sl}(2)$ algebra $[{\sf h},{\sf e}]=2{\sf e}$, $[{\sf h},{\sf f}]=-2{\sf f}$ and $[{\sf e},{\sf f]}={\sf h}$, while ${\sf z}$ is the $\mathfrak{u}(1)$ generator  and 
accordingly commutes with ${\sf e},{\sf f}$, and ${\sf h}$.   The $R$-charge generator $J$ that was introduced in section \ref{randomtwo} is represented in this basis by $J=-\i {\sf z}$ (the
following formulas will be slightly simplified by using $\sf z$ rather than $J$).  We have chosen ${\sf e},{\sf f},{\sf h}, $ and ${\sf z}$ so that they generate symmetries of the hermitian form 
(\ref{eq:innprod}).   Therefore, they are generators of $\SU(1,1|1)$; in Lorentz signature, they are represented by anti-hermitian operators, whose exponentials are unitary.

The four fermionic generators of $\SU(1,1|1)$ are complex linear combinations of 
\bea\label{fermigen}
{\sf q}_1=\left(\begin{array}{cc|c}
0&0\hspace{.5mm}&\hspace{.5mm}0 \\
0&0\hspace{.5mm}&\hspace{.5mm}0 \\
\hline 0&1\hspace{.5mm}&\hspace{.5mm}0 
\end{array}\right),
~{\sf q}_2=\left(\begin{array}{cc|c}
0&0\hspace{.5mm}&\hspace{.5mm}1 \\
0&0\hspace{.5mm}&\hspace{.5mm}0 \\
\hline 0&0\hspace{.5mm}&\hspace{.5mm}0 
\end{array}\right),
~{\sf q}_3=\left(\begin{array}{cc|c}
0&0\hspace{.5mm}&\hspace{.5mm}0 \\
0&0\hspace{.5mm}&\hspace{.5mm}0 \\
\hline 1&0\hspace{.5mm}&\hspace{.5mm}0 
\end{array}\right),~
{\sf q}_4=\left(\begin{array}{cc|c}
0&0\hspace{.5mm}&\hspace{.5mm}0 \\
0&0\hspace{.5mm}&\hspace{.5mm}1 \\
\hline 0&0\hspace{.5mm}&\hspace{.5mm}0 
\end{array}\right).
\ea   One can work out the commutators of the ${\sf q}_i$ with the bosonic generators defined in eqn. (\ref{bosegen}) and their anticommutators with each other, but we omit
these formulas.

In eqn. (\ref{fermigen}),  we have simply picked a convenient basis for the four-dimensional space of fermionic linear transformations of the vector space $V$.   Certain complex
linear combinations of the ${\sf q}_i$ preserve the quadratic form (\ref{eq:innprod}) and therefore are generators of $\SU(1,1|1)$.   However, for the Euclidean signature calculation
that we will perform, the existence of such a real basis for the fermionic generators is not important, though the real structure of the maximal bosonic subgroup $\SL(2,\R)\times \uU(1)$ is 
important.    We will pause to explain this point, which sometimes causes  confusion.

First of all, in any quantum field theory with or without gravity, 
in Lorentz signature, any group or supergroup $G$ of symmetries always has a real structure. The reason for this is that elements of $G$ act on the quantum
Hilbert space $\H$ of the theory as unitary transformations.  This statement only makes sense if $G$ has a real structure.   One can (with some restrictions on the states
a given group element can act on) 
analytically continue the  action of $G$ on $\H$  to an action of the complexification $G_\C$ of $G$, but only elements of $G$ are unitary; elements of $G_\C$ that are not
in $G$ act on $\H$ as (densely defined) unbounded and nonunitary operators.  

This remark applies to any theory, supersymmetric or not.   In the case of a supersymmetric theory, after Wick rotation to Euclidean signature, generically the fermionic
generators do not carry any real structure.   For example, consider $\N=1$ supersymmetry in four dimensions.   This superalgebra has four fermionic generators, and, 
although they carry a real structure in Lorentz signature, they do not admit any real structure in Euclidean structure  (they take values in a pseudoreal rather than real
representation of the rotation group ${\mathrm{Spin}}(4)$).   

More broadly, even without supersymmetry, in Euclidean signature, fermionic integration variables in general do not carry any real structure.   For example, in the standard model of particle physics -- because of the chiral nature of the weak interactions -- after Wick rotation to Euclidean signature, there is no way to define a real structure for the fermionic
variables.   This causes no difficulty, because fermionic integration is a purely algebraic operation.   When one says that $\int \d \psi \,\psi = 1$, $\int \d\psi\,1=0$,
there is no need to know whether $\psi$ can be considered ``real'' (and if the answer to that question is ``no,'' as it is in the standard model or many four-dimensional
models with $\N=1$
supersymmetry, there is no need to apologize and add a complex conjugate variable $\psi^*$).      

By contrast, for the bosonic variables, one does need a real structure in Euclidean signature in order to define the appropriate integration cycle.   This statement generalizes
the following fact about 1-dimensional integrals: if $F(z)$ is a holomorphic function on $\C$ (possibly with some singularities), then to make sense of an integral
$\int_\gamma F(z)\d z$, one needs to pick an integration cycle $\gamma\subset \C$.   For example, in  general, the integral of $F(z)\d z$ on the real $z$-axis gives a different result
from the ``same'' integral on the imaginary $z$-axis, even if both of these integrals converge.  

The measure  that we will compute using the relation of $BF$ theory to torsion makes sense as a holomorphic measure on the moduli space of flat  connections
on $C$ with structure group $\SL(1,1|1)_\C$, the complexification of $\SU(1,1|1)$.   
To integrate this measure, we need a real form for the bosonic integration variables but we do not need a real form for the fermionic integration variables.
When we say that the gauge group we work with is $\SU(1,1|1)$, we are really specifying the complex Lie algebra of the gauge group and the real form
that we will use in integrating over the bosons, which will be  a maximal bosonic subgroup $\SU(1,1)\times \uU(1)\cong \SL(2,\R)\times \uU(1)$ of $\SU(1,1|1)$.

At this stage, the reader may wonder why the supergroup $\SU(1,1|1)$ that is appropriate for $\N=2$ JT supergravity in Euclidean signature admits a real structure
for the fermionic generators, given that this does not happen in a generic supersymmetric theory.   This fact can be viewed as an accident
of two dimensions.   The connected component of the symmetry group of $\mathrm{AdS}_2$ in Lorentz signature is ${\mathrm{SO}}(2,1)$; Wick rotation converts this
to $\mathrm{SO}(1,2)$, which happens to be isomorphic to $\mathrm{SO}(2,1)$. As a result of this equivalence, after including spin and fermionic generators,
one gets the same supergroup $\SU(1,1|1)$ in either Lorentz or Euclidean signature.  Since it is always possible to define a real structure in Lorentz signature,
it follows that in this theory one can also do so in Euclidean signature.    In dimension $d$, the two groups would be $\mathrm{SO}(2,d-1)$ and $\mathrm{SO}(1,d)$,
which are no longer isomorphic.  For some values of $d$, a superextension of this symmetry group will not have any real structure in Euclidean signature.

Going back to $\SU(1,1|1)$, although this supergroup does admit a real structure for the fermionic generators, it will be more convenient in our calculations to ignore
this fact and make no use of this real structure.   

In this discussion, we have  implicitly assumed that $\hq=1$.  To incorporate $\hq$, we want to take a $\hq$-fold cover of  $\SU(1,1|1)$ and of its maximal
bosonic subgroup $\SL(2,\R)\times \uU(1)$.  The cover is defined by taking a $\hq$-fold cover of the second factor $\uU(1)$.  Taking this cover means that the fermionic generators
of $\SU(1,1|1)$ have $R$-charge that is greater than the minimum possible $R$-charge of any operator by a factor $\pm \hq$.

\subsubsection{The Three-Holed Sphere}\label{sec:3hs}

The ultimate goal of this section is to compute the path integral measure for the component of the moduli of flat $\SU(1,1|1)$ connections on an orientable surface $C$
associated to hyperbolic surfaces with $\cN=2$ supersymmetry.   Any orientable surface can be constructed by gluing together three-holed spheres along some of their boundaries, which are circles.   So to obtain a general result, we will
need to know the torsion of a flat connection on a three-holed sphere or a circle, and the procedure for gluing.

\begin{figure}
\begin{center}
\begin{tikzpicture}[scale=0.8, baseline={([yshift=-0.1cm]current bounding box.center)}]
\draw[thick, rotate around={-30:(-2,2)}] (-2,2) ellipse (.15 and .5);%Geodesic
\draw[thick, rotate around={30:(2,2)}] (2,2) ellipse (.15 and .5);%Geodesic
\draw[thick] (0,-1) ellipse (.5 and .15); %Geodesic
 \draw[thick, bend right=30] (-1.8,2.45) to (1.8,2.45);
 \draw[thick, bend left=30] (-2.2,1.55) to (-0.5,-1);
  \draw[thick, bend right=30] (2.2,1.55) to (0.5,-1);
  \node at (-2.6,2.2) {\small $U$};
   \node at (2.6,2.2) {\small $V$};
   \node at (0,-1.6) {\small $W$};
       \node at (0,1) {\small $Y$};
    \end{tikzpicture}
    \hspace{1cm}\begin{tikzpicture}[scale=0.9, baseline={([yshift=-0mm]current bounding box.center)}]
\draw[thick,fill=gray!60] (0,0) ellipse (2.5 and 1.5); %W
 \draw[thick,fill=white] (0,-1.5) to [out=150, in=-90] (-1.3,-1.5+1.5) to [out=90, in=90] (0,-1.5) ;
  \draw[thick,fill=white] (0,-1.5) to [out=30, in=-90] (1.3,-1.5+1.5) to [out=90, in=90] (0,-1.5) ;
  \draw[fill,black] (0,-1.5) circle (0.1);
  \draw[thick,->] (1.85,-1.01) to [bend right=5] (1.85+0.1,-1.01+0.09)  ;
  \draw[thick,->] (-1,-.8) to [bend right = 30] (-1,-.8+0.01)  ;
  \draw[thick,->] (1,-.8) to [bend left= 20] (1,-.8-0.01)  ;
  \node at (-5,0) {\small $\to$};
  \node at (2.5,-1) {\small $W$};
  \node at (-1.7,0) {\small $U$};
  \node at (1.7,0) {\small $V$};
  \node at (0,-2) {\small $P$};
      \node at (0,0.5) {\small $Y$};
    \end{tikzpicture}
    \end{center}
    \caption{\footnotesize On the left we show the hyperbolic three-holed sphere $Y$ with geodesic boundaries labeled by $\SU(1,1|1)$ holonomies $U$, $V$ and $W$. On the right we show a topological space given by the interior of the curves labeled by the same three holonomies. The combinatorial torsion is conveniently calculated using the picture on the right.}
    \label{fig:torsion3hs}
    \end{figure}
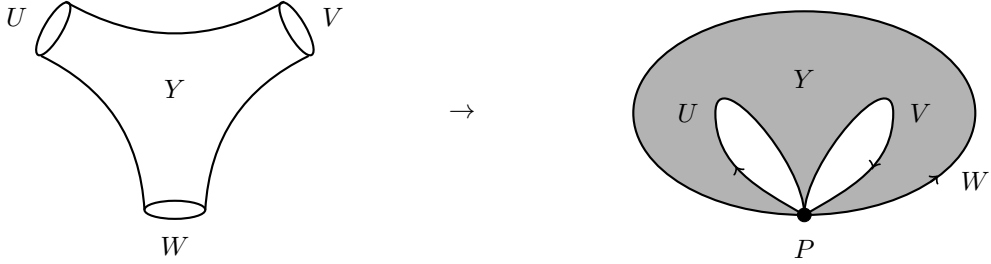
    
 We denote as $\mathcal{R}$ the moduli ``space'' of flat $\SU(1,1|1)$ connections on a three-holed sphere $Y$, with the boundary holonomies required to be hyperbolic
 but not otherwise restricted.   As we will see, $\cR$  is a rather subtle sort of space, not a supermanifold.
 We denote the holonomies of a flat $\SU(1,1|1)$ connection on $Y$ around the boundaries as $U,V,W$; see fig. \ref{fig:torsion3hs}. 
 Actually, to define the holonomies as group elements (not just conjugacy classes), one needs to pick a base-point in the definition of the fundamental group and to choose, for each boundary circle, a corresponding element of the fundamental group.   A way to do this is given in the right panel of the figure.   From this picture, it is hopefully obvious that, as $UVW$ is the holonomy around
 a contractible loop in $Y$, any flat connection on $Y$ will satisfy
\bea\label{eq:smooth3hs}
U V W = 1.
\ea
We can solve this relation for $W$: $W=V^{-1}U^{-1}$.   $U$ and $V$ are not subject to any constraint, so we can pick an arbitrary pair $U,V\in \SU(1,1|1)$, and this
will determine a flat $\SU(1,1|1)$ connection on $Y$.   This parametrization is redundant, however: 
a conjugation  $(U,V,W) \cong (R U R^{-1}, R V R^{-1}, R W R^{-1})$, which one can think of as resulting from a gauge transformation that equals $R$ at the base-point $P$
that was used in defining the fundamental group,  preserves constraint \eqref{eq:smooth3hs} and is associated to  a gauge-equivalent flat connection.

We now present the parametrization of $\mathcal{R}$ which we use in the torsion computation.   Given a pair $(U,V)\in \SU(1,1|1)$, we cannot conjugate them
to be simultaneously diagonal, since they do not commute, but
we can conjugate them to a pair $(U_0,V_0)$ such that $U_0$ is upper triangular and $V_0$ is lower triangular:
\beq\label{eq:U0V0}
U_0 = \hspace{-1mm} \left(\hspace{-1mm}\begin{array}{cc|c}
-e^{\i\hq\phi_a}e^{\frac{a}{2}}&-e^{\i\hq\phi_a}\kappa&0 \\
0&-e^{\i\hq\phi_a}e^{-\frac{a}{2}}&0 \\
\hline
0&0&e^{2\i\hq\phi_a}  
\end{array}\hspace{-1mm}\right) e^{\psi_2 {\sf q}_2} e^{\psi_4 {\sf q}_4},
~V_0 =\hspace{-1mm}  \left(\hspace{-1mm}\begin{array}{cc|c}
-e^{\i\hq\phi_b}e^{-\frac{b}{2}}&0&0 \\
-e^{\i\hq\phi_b}&-e^{\i\hq\phi_b}e^{\frac{b}{2}}&0 \\
\hline
0&0&e^{2\i\hq\phi_b}  
\end{array}\hspace{-1mm}\right)e^{\psi_1 {\sf q}_1} e^{\psi_3 {\sf q}_3}.
\eeq\\
Every pair $(U,V)$ is conjugate to a unique such pair $(U_0,V_0)$.   The triangular nature of $U_0,V_0$ will simplify the calculation considerably.\footnote{It is to achieve this simplification
that, as discussed in section \ref{JTBF}, we ignore the real structure that could be imposed on the fermionic generators.   $\SU(1,1|1)$ actually has a real Borel
subgroup that could be used to get somewhat similar simplifications in a real basis. (For this, one would define ``upper triangular'' or ``lower triangular'' generators to be those with
a positive or negative eigenvalue of ${\sf h}$.)    It does not appear that this would lead to simpler calculations.}   
The boundary parameters $(a,b,\phi_a,\phi_b,\kappa)$  are the bosonic moduli of $\cR$.   $\cR$ also has four ``internal''  fermionic moduli $\psi_1,\cdots,\psi_4$.  The torsion $\tau_Y$ will be holomorphic in all of these
parameters.  When ultimately we integrate over moduli (usually after gluing $Y$ onto other three-holed spheres and/or external trumpets)  $a,b,\phi_a,\phi_b,$ and $\kappa$ are all  taken to be real,\footnote{It is actually more natural to say that the parameters  $a,b,\cdots$ are real modulo nilpotents.   The reason is that in general the same oriented two-manifold $C$
can be decomposed in three-holed spheres in many different ways.   The different sets of variables $a,b,\cdots$ and $a',b',\cdots$ associated to these
decompositions are related by a change of variables
that is real modulo nilpotents  but not actually real.
  So we cannot
in an invariant way say that the bosonic parameters are real, but we can say that they are real modulo nilpotents, which is good enough so that the moduli
space integrals make sense.  See for example  \cite{Witten:2012bg}  for discussion of such issues.} but we do not invoke a real structure for the fermionic parameters $\psi_1,\cdots,\psi_4$.
As usual, we actually  work with a $\hq$-fold cover of $\SU(1,1|1)$, so $\phi_a$ and $\phi_b$ are defined
mod $2\pi$  (not $2\pi/\hq$ as the formulas in $3\times 3$ matrices suggest).

If two boundary holonomies are $U_0$ and $V_0$, then the third is $W_0=V_0^{-1}U_0^{-1}$.
In the parametrization (\ref{eq:U0V0}) of the moduli space, 
$a$ and $b$ are the geodesic lengths of the $U_0$ and $V_0$ boundaries while $\phi_a$ and $\phi_b$ parametrize their $\uU(1)$ holonomies. But what is the interpretation of $\kappa$? We would like to replace it instead by the geodesic length of the boundary with holonomy $W_0$; we denote this length by $c$. The length and $\uU(1)$ holonomy of $W_0$
can be determined by conjugating $W_0$ to be diagonal.  Thus, if there is $D\in \SU(1,1|1)$ such that
\bea\label{conjform}
D W_0 D^{-1} =  \left(\begin{array}{cc|c}
-e^{\i\hq\phi_c}e^{-\frac{c}{2}}&0&0 \\
0&-e^{\i\hq\phi_c}e^{\frac{c}{2}}&0 \\
\hline
0&0&e^{2\i\hq\phi_c}  
\end{array}\right)
\ea\\
then the $W_0$ boundary has circumference $c$ and $\uU(1)$ twist parameter $\phi_c$.
The straightforward but tedious calculation to determine $c$ and $\phi_c$ is presented 
in appendix \ref{sec:constn2}. Before quoting the final result, it is instructive to consider the simpler case with vanishing fermionic moduli $\psi_1 = \cdots= \psi_4 =0$. In this case the matrix $W_0=V_0^{-1}U_0^{-1}$ is\\ 
\bea\label{wform}
W_0 \Big|_{\psi = 0} =  \left(\begin{array}{cc|c}
e^{-\i\hq(\phi_a + \phi_b)}e^{\frac{b-a}{2}}&-e^{-\i\hq(\phi_a + \phi_b)}e^{\frac{b}{2}}\kappa&0 \\
-e^{-\i\hq(\phi_a+ \phi_b)}e^{-\frac{a}{2}}&e^{-\i\hq(\phi_a + \phi_b)}(e^{\frac{a-b}{2}}+\kappa)&0 \\
\hline
0&0&e^{-2 \i\hq(\phi_a + \phi_b)}  
\end{array}\right).
\ea\\
This case simplifies since $W_0$ is block diagonal.   From the bottom right matrix element of eqn. (\ref{wform}), we can read off that $\phi_c=-\phi_a-\phi_b$.  Equivalently,
\bea\label{eq:bosconstphi}
\phi_a + \phi_b + \phi_c = 0~~~~{\rm for}~ \psi_i =0,~i=1,\ldots,4.
\ea
Thus the boundary holonomies cannot be specified independently, or the moduli space will be empty.
We essentially already met this constraint in eqn. (\ref{holconstraint}); it reflects the fact that when we set the odd moduli to zero, the structure group of a flat
$\SU(1,1|1) $ connection reduces to the maximal bosonic subgroup $\SL(2,\R)\times \uU(1)$, in which the second factor is abelian.
The $2\times 2$ block in $W_0$ is
\bea
\left(\begin{array}{cc}
e^{-\i\hq(\phi_a + \phi_b)}e^{\frac{b-a}{2}}&~-e^{-\i\hq(\phi_a + \phi_b)}e^{\frac{b}{2}}\kappa \\
 -e^{-\i\hq(\phi_a + \phi_b)}e^{-\frac{a}{2}}&~e^{-\i\hq(\phi_a + \phi_b)}(e^{\frac{a-b}{2}}+\kappa)  
\end{array}\right) =  -e^{\i \hq\phi_c} \left(\begin{array}{cc}
-e^{\frac{b-a}{2}}&~e^{\frac{b}{2}}\kappa \\
e^{-\frac{a}{2}}&~-(e^{\frac{a-b}{2}}+\kappa  )
\end{array}\right).
\ea\\
The matrix on the right, after removing a factor of $-e^{\i \hq\phi_c}$, has eigenvalues $e^{\pm c/2}$, so its trace is $2\cosh\frac{c}{2}$.
Computing the trace,  we learn  that with all fermionic moduli set to zero, $\kappa=\kappa_0$ with
\bea\label{eq:bosconstk}
\kappa_0=- 2 \cosh \Big(\frac{c}{2}\Big) - 2 \cosh \Big( \frac{b-a}{2}\Big),~~~{\rm for}~ \psi_i =0,~i=1,\ldots,4.
\ea
When the odd moduli vanish, these relations determine the parameters $c,\phi_c$ that parametrize the conjugacy class of $W_0$.

Conceptually, with non-vanishing fermionic moduli, one proceeds  in the same way. The only modification is that now the right hand sides of \eqref{eq:bosconstphi} and \eqref{eq:bosconstk} get nilpotent contributions proportional to the fermionic moduli.
This will modify both the result for $\kappa$ and the constraint $\phi_a+\phi_b+\phi_c=0$.
The most general possible form of this modification is
\bea\label{kapform}
\kappa &=& \kappa_0 + \sum_{i<j} \kappa_{ij} \psi_i \psi_j + \kappa_4 \psi_1 \psi_2 \psi_3 \psi_4,\\ \label{phiform}
\phi_a +\phi_b + \phi_c&=&\sum_{i<j} f_{ij} \psi_i \psi_j + f_4 \psi_1\psi_2\psi_3\psi_4.
\ea
In appendix \ref{sec:constn2}, we derive explicit expressions for the  coefficients $\kappa_{ij}$ and $f_{ij}$, $i,j=1,\ldots,4$, that appear in these relations. The non-zero $f$'s are 
\bea
&& f_{12} = -\frac{\i e^{-\i\hq \phi_a}}{4\hq(\cosh(\frac{c}{2}) + \cos\hq(\phi_a+\phi_b))},~~~ f_{14}=\frac{\i (e^{-\frac{b}{2}-\i\hq \phi_a}-e^{-\frac{a}{2}+\i \hq\phi_b})}{4\hq(\cosh(\frac{c}{2}) + \cos\hq(\phi_a+\phi_b))},\nonumber\\
&&  f_{23} =-\frac{\i (e^{\frac{b}{2}-\i\hq \phi_a}-e^{\frac{a}{2}+\i \hq\phi_b})}{4\hq(\cosh(\frac{c}{2}) + \cos\hq(\phi_a+\phi_b))},~~~ f_{34}=\frac{\i  e^{\i \hq\phi_b} (\cosh(\frac{c}{2}) + \cosh(\frac{a-b}{2}))}{2\hq(\cosh(\frac{c}{2}) + \cos\hq(\phi_a+\phi_b))},\nonumber\\
&&  f_4=-\frac{\i  e^{-2\i\hq \phi_a} (1+e^{\i\hq(\phi_a+\phi_b)}\cosh(\frac{c}{2}) )}{8\hq(\cosh(\frac{c}{2}) + \cos\hq(\phi_a+\phi_b))^2}.\label{eq:Fexp}
\ea
and the nonzero $\kappa$'s are
\bea
&& \kappa_{12} =\frac{e^{\i \hq\phi_b}+e^{-\i \hq\phi_a}\cosh(\frac{c}{2})}{2(\cosh(\frac{c}{2}) + \cos\hq(\phi_a+\phi_b))},\nn
&&\kappa_{14}=-\frac{e^{-\frac{a}{2}-\i \hq\phi_a}+e^{-\frac{b}{2}+\i \hq\phi_b}+(e^{-\frac{b}{2}-\i \hq\phi_a}+e^{-\frac{a}{2}+\i\hq \phi_b})\cosh(\frac{c}{2})}{2(\cosh(\frac{c}{2}) + \cos\hq(\phi_a+\phi_b))}\nn
&& \kappa_{23} =\frac{e^{\frac{a}{2}-\i \hq\phi_a}+e^{\frac{b}{2}+\i \hq\phi_b}+(e^{\frac{b}{2}-\i \hq\phi_a}+e^{\frac{a}{2}+\i \hq\phi_b})\cosh(\frac{c}{2})}{2(\cosh(\frac{c}{2}) + \cos\hq(\phi_a+\phi_b))} , \nn
&& \kappa_{34} =\frac{(\cosh(\frac{c}{2})+\cosh(\frac{a-b}{2}))(e^{\i\hq \phi_b}\cosh(\frac{c}{2})+e^{-\i \hq\phi_a})}{\cosh(\frac{c}{2}) + \cos\hq(\phi_a+\phi_b)}.\nn
&&\kappa_4 = \frac{e^{\i\hq(\phi_b-\phi_a)}(1+\cos\hq(\phi_a+\phi_b)\cosh(\frac{c}{2}))}{4(\cosh(\frac{c}{2}) + \cos\hq(\phi_a+\phi_b))^2}.\label{eq:Kexp}
\ea
It will be convenient for future reference to define the nilpotent function 
\bea
K= \sum_{i<j} k_{ij} \psi_i \psi_j+k_4 \psi_1 \psi_2\psi_3\psi_4,~~~~k_{ij} = -\frac{\partial_c \kappa_{ij}}{\sinh(\frac{c}{2})},~~k_4 = - \frac{\partial_c \kappa_4}{\sinh(\frac{c}{2})}.
\ea
The coefficients on the right hand side can be easily extracted from \eqref{eq:Kexp}. We define this function $K$ since it will appear in the final expression for the torsion.

At this point, however, we should point out why $\cR$ is not a supermanifold and instead is a rather strange kind of ``space.''   Because $\uU(1)$ is abelian,
a  flat $\SU(1,1|1)$ connection on $Y$ with
$\psi_1=\cdots \psi_4=0$ is always invariant under the $R$-symmetry group $\uU(1)\subset \SU(1,1|1)$.   Accordingly, let $\cR'$ be the supermanifold of dimension $5|4$
parametrized by $a,b,\phi_a,\phi_b,\kappa|\psi_1,\psi_2,\psi_3,\psi_4$.   Then, formally, $\cR$ is the quotient $\cR'/\uU(1)$, with $\uU(1)$ acting only on $\psi_1,\cdots,\psi_4$.
The trouble with this is that dividing a supermanifold, in this case $\cR'$, by a continuous group, in this case $\uU(1)$, that acts only on odd variables is not a nice operation,
and the quotient $\cR=\cR'/\uU(1)$ is definitely not a supermanifold.  We do not want to define directly what should be meant by a volume form on $\cR$ or by integration on $\cR$.
Pragmatically, we want to define the volume of $\cR$ as $\vol(\cR')/\vol(\uU(1))$. For this purpose, we need volume forms on $\cR'$ and $\uU(1)$, or at least we need the
ratio of a volume-form on $\cR'$ and a volume form on $\uU(1)$.   We will see that the torsion gives precisely such a ratio.

\subsubsection{Torsion of the Three-Holed Sphere} \label{torthree}
We are now ready to compute the torsion $\tau_Y$ of the three-holed sphere $Y$.    We will follow the general procedure described in section 3.4.2 of \cite{SW}.   We will
review a few steps, emphasizing the points that are different for $\SU(1,1|1)$, but we will not give a self-contained explanation.
 
Since the torsion can be computed using any triangulation (or more generally any cell decomposition) of
a manifold, we can use the very simple description of a three-holed sphere shown in the right panel of fig. \ref{fig:torsion3hs}, with one two-cell, three boundary one-cells, and 
one vertex $P$.   Let us discuss
the situation with an arbitrary gauge group $G$ before specializing to $G=\SU(1,1|1)$.   
If we divide the space of flat connections on $Y$
not by all gauge transformations but only by gauge transformations that are trivial at $P$, we get a supermanifold $\h\cR\cong G\times G$ that is parametrized
by $U,V$ without any constraint or equivalence.   Then to get the moduli space $\cR$, we divide by $G$ acting on $\h\cR$ by conjugation:
\be\label{quomod}\cR=\h\cR/G. \ee

The definition of the torsion requires a choice of a left- and right-invariant measure $\vol_G$ on the $G$ manifold; this is equivalent to a choice of invariant measure $\vol_\g$
on the Lie algebra $\g$ of $G$.   Such a choice determines a measure on any copy of the group manifold,
so in particular it determines measures $\vol_G(U)$ or $\vol_G(V)$ on the copies of $G$ parametrized by $U$ or $V$.  
Similarly, if we define $U=R U_0 R^{-1}$, $V=R V_0 R^{-1}$, with some choices of $U_0$, $V_0$ (which in our application will be given by eqn.  (\ref{eq:U0V0})), one
has a measure $\vol_G(R)$ on the copy of $G$ parametrized by $R$.
 The torsion can be used to define
measures $\tau_Y$ and $\hat\tau_Y$ on $\cR$ and $\h\cR$, respectively.   One finds
\be\label{firstres} \h\tau_Y=\vol_G(U)\vol_G(V) \ee
and 
\be\label{secondres}\tau_Y=\frac{\h\tau_Y}{\vol_G(R)} =\frac{\vol_G(U)\vol_G(V)}{\vol_G(R)}. \ee 
It suffices to evaluate this formula for $\tau_Y$ at $R=1$, since every flat connection on $Y$ can be described by some pair $(U_0,V_0)$.  

When the quotient $\cR =\h \cR/G$ is sufficiently nice, which happens for the examples $\SL(2,\R)$ and $\OSp(1|2)$ studied in \cite{SW}, the quotient $\vol_G(U)\vol_G(V)/\vol_G(R)$
is a measure on $\cR$ and this is the appropriate moduli space measure.   For the example of $\SU(1,1|1)$ that we aim to understand here, it turns out instead
that $\vol_G(U)\vol_G(V)/{\vol_G(R)}$ is the ratio of a measure on $\cR'$, defined at the end of section \ref{sec:3hs}, and a measure on the $\uU(1)$ $R$-symmetry group.
We interpret this to mean that, as essentially stated in section \ref{sec:3hs}, integration over $\cR$ should be defined as integration over $\cR'$ divided by the volume of $\uU(1)$.

To evaluate the torsion, we first describe the volume form that we will use on $\mathfrak{su}(1,1|1)$.
 Consider an arbitrary Lie algebra element $x \in \mathfrak{su}(1,1|1)$ with components $x= x_e {\sf e} + x_f {\sf f} + x_h {\sf h} + x_z \hq \,{\sf z} + \sum_{j=1}^4 x_j {\sf q}_j$. (Notice the factor of $\hq$ in the normalization in $x_z$, such that the range of $x_z$ is $2\pi$ in the $\hq$-fold of $\SU(1,1|1)$ we are working with.)
 We choose the volume form on the Lie algebra to be
 \beq
\vol_{\g} = 4 [ \d x_e \d x_f \d x_h \d x _z | \d x_1  \d x_2  \d x_3 \d x_4],
\eeq 
and this also determines the volume form on the group.     As explained in \cite{SW}, a symbol such as $[\d y|\d\eta]$ (where $y$ is bosonic and $\eta$ is fermionic) 
represents the measure associated to a  basis
$\d y|\d\eta$ of 1-forms on a $\Z_2$-graded vector space.  
 A crucial fact is that in contrast to bosons, measures for fermions transform  oppositely to differential forms, in the sense that for\footnote{The reason
for the absolute value $|\lambda|$ in eqn (\ref{wellsense}) in the bosonic case is that for a bosonic variable $y$, the measure $[\d y]$, as opposed to the 1-form $\d y$, is invariant under
$y\to -y$.}  $\lambda
\in\C$,
\be\label{wellsense} [\lambda\d y|\d \eta]=|\lambda|\,[\d y|\d\eta],~~~~[\d y|\lambda\d\eta]=\lambda^{-1}[\d y|\d\eta].\ee

Evaluating ${\rm vol}_G(R)$ is immediate, because it suffices to evaluate this measure at $R=1$.  If we write $R=1+r$ then, at $R=1$,  $R^{-1} \d R = \d r$ and
$\vol_G(R)$ reduces to
\bea\label{eq:RHSTI} {\rm vol}_G(R)  =4 [ \d r_e \d r_f \d r_h \d r _z | \d r_1  \d r_2  \d r_3 \d r_4] .   \ea
Next we evaluate ${\rm vol}_G(U)$ for $U=RU_0 R^{-1}$, following the procedure outlined in  \cite{SW}. 
The quantity $U^{-1} \d U$ will get contributions both from the differential of $R$, resulting in terms proportional to components of $\d r$ which will cancel similar factors in 
$\vol_G(R)$ when evaluating eqn. (\ref{secondres}), and contributions from the differential of $U_0$.   When
we repeat this for $V=R V_0 R^{-1}$,  we will be left finally with  $\tau=F[\d a \d b \d \kappa \d \phi_a \d \phi_b | \d\psi_1\d \psi_2 \d\psi_3\d \psi_4]/[\d r_z]$ for some function $F$.  
As anticipated at the end of section 
\ref{sec:3hs}, we interpret the result as the ratio of a volume form on $\cR'$ and a volume form on $\uU(1)$.

The identity  (\ref{secondres}) can be greatly simplified by exploiting the fact that
the matrices $U_0$ and $V_0$ are respectively upper and lower triangular. Begin by evaluating the $q_3$ component of $U^{-1} \d U$ and the $q_2$ component of $V^{-1} \d V$:
\bea
(U^{-1} \d U )_3 = -(e^{\frac{a}{2}-\i\hq\phi_a}+1)\d r_3,~~~~(V^{-1} \d V )_2 = -(e^{\frac{b}{2}+\i\hq\phi_b}+1)\d r_2.
\ea
Since these expressions are proportional to $\d r_2$ and $\d r_3$, we can in evaluating the numerator of eqn. (\ref{secondres}) set to zero any other occurrences
of $\d r_2$ or $\d r_3$.  Then, we can cancel these factors of $\d r_2$ and $\d r_3$ with the factors
of $\d r_2$ and $\d r_3$ that appear in $\vol_G(R)$ in the denominator in eqn. (\ref{secondres}).   After this cancellation, we can forget about $\d r_2 $ and $\d r_3$ and set
them to zero.   We are left with a factor $(e^{\frac{a}{2}-\i\hq\phi_a}+1)^{-1}(e^{\frac{b}{2}+\i\hq\phi_b}+1)^{-1}$.   Note that these factors are inverted because for fermions,
measures transform oppositely to differential forms, as noted in eqn. (\ref{wellsense}).
 In the next step, evaluate the bosonic components
\bea
(U^{-1} \d U )_f = (e^a-1)\d r_f ,~~~~(V^{-1} \d V )_e = (e^b-1)\d r_e.
\ea
Because these expressions are proportional to $\d r_f$ and $\d r_e$, we can set to zero any other occurrences of $\d r_f$ and $\d r_e$ in the numerator
of eqn. (\ref{secondres}).   Then we can cancel $\d r_e$ and $\d r_f$  with similar occurrences in the denominator and set them to zero.
We are left this time with a factor $(e^a-1)(e^b-1)$.    Next evaluate
\bea
(U^{-1} \d U )_1 = -(e^{-\frac{a}{2}-\i\hq\phi_a}+1)\d r_1,~~~~
(V^{-1} \d V )_4 = -(e^{-\frac{b}{2}+\i\hq\phi_b}+1)\d r_4.
\ea
We can set to zero any other occurrences of $\d r_1$ and $\d r_4$ in the numerator in eqn. (\ref{secondres}), and  cancel $\d r_1$ and $\d r_4$ with similar factors in the denominator.
We are left with a factor $(e^{-\frac{a}{2}-\i\hq\phi_a}+1)^{-1}(e^{-\frac{b}{2}+\i\hq\phi_b}+1)^{-1}$.
 Next evaluate 
\bea
(U^{-1} \d U )_h = \frac{\d a}{2},~~~~(V^{-1} \d V )_h = -\frac{\d b}{2}.
\ea
Since these expressions are proportional to $\d a$ and $\d b$, we can set to zero any occurrences of $\d a$ and $\d b$ that we find in the rest of the evaluation.  We are left with a factor
$-\frac{1}{4}\d a\d b$.
 After dropping $\d a$ and $\d b$ where they appear elsewhere, we  obtain $(U^{-1} \d U )_z =\d\phi_a$ and $(V^{-1} \d V )_z =   \d\phi_b$, so we can set to zero any further occurrences of $\d\phi_a$ and $\d\phi_b$
 that we may encounter, and we remain with a factor $\d\phi_a\d\phi_b$.   Then $(V^{-1} \d V )_f = -2 e^{-\frac{b}{2}} \d r_h$, allowing us to cancel $\d r_h$ 
 with a similar factor in the denominator.
 The final bosonic component is  $(U^{-1} \d U)_e =  e^{-\frac{a}{2}} \d\kappa$.
 Now we can evaluate the remaining fermionic components
\bea\label{somefactors}
(U^{-1}  \d U )_4 =\d\psi_4 - \i \psi_4  \d r_z ,~~~~(V^{-1} \d V )_1 =\d\psi_1 +  \i \psi_1 \d r_z.
\ea
and
\bea\label{morefactors}
(U^{-1} \d U )_2 =\d\psi_2 - \i  \psi_2 \d r_z ,~~~~(V^{-1} \d V )_3 =\d\psi_3 +\i  \psi_3 \d r_z.
\ea 
The measure coming from \eqref{somefactors} and \eqref{morefactors} simplifies via $[\d\psi_1 +  \i \psi_1 \d r_z,\d\psi_2 - \i  \psi_2 \d r_z,\d\psi_3 +\i  \psi_3 \d r_z,\d\psi_4 - \i \psi_4  \d r_z] =[\d\psi_1\d\psi_2\d\psi_3\d\psi_4]$, since the Berezinian of the indicated change of basis
is 1.   We canceled almost all factors in the denominator of (\ref{secondres}), leaving a leftover factor of $[\d r_z]$ in the denominator.

After all these simplifications, keeping track of all the factors generated along the way, the identity (\ref{secondres}) becomes
\bea \label{answer} \tau_Y=\frac{1}{[\d r_z]}
\frac{8\sinh \frac{a}{2} \sinh \frac{b}{2} ~[\d a \, \d b\,  \d\kappa\, \d\phi_a\,  \d\phi_b | \d^4\psi] }{(e^{\frac{a}{2}-\i\hq\phi_a}+1)(e^{-\frac{a}{2}-\i\hq\phi_a}+1)(e^{\frac{b}{2}+\i\hq\phi_b}+1)(e^{-\frac{b}{2}+\i\hq\phi_b}+1)} 
\ea
where to shorten the notation we defined $[\d^4 \psi ] \equiv [\d \psi_1 \d \psi_2 \d \psi_3 \d \psi_4]$. 
As promised, $\tau$ is the ratio of a measure on $\cR'$ and a measure on $\uU(1)$.   

The result can be put in a more useful form by eliminating $\kappa$ in favor of the other moduli via eqn. (\ref{kapform}).
We have  $\d \kappa = \partial_c \kappa \, \d c + \ldots$, where the omitted terms involve derivatives with respect to other arguments and
do not contribute when this substitution is made for $\d\kappa$ in eqn. (\ref{answer}).   So 
\bea\label{eq:torsiondrz}\tau_Y=\frac{1}{[\d r_z]}
 \frac{8 (\partial_c \kappa)\sinh \frac{a}{2} \sinh \frac{b}{2}~[\d a \, \d b\,  \d c\, \d\phi_a\,  \d\phi_b | \d^4\psi] }{(e^{\frac{a}{2}-\i\hq\phi_a}+1)(e^{-\frac{a}{2}-\i\hq\phi_a}+1)(e^{\frac{b}{2}+\i\hq\phi_b}+1)(e^{-\frac{b}{2}+\i\hq\phi_b}+1)} .
\ea
We are going to make two further adjustments of this formula to put it in its final form.   

First, we have to decide what to do with the factor $1/[\d r_z]$.  The most
obvious thing to do, suggested by the discussion at the end of section \ref{sec:3hs}, is to interpret the factor $1/[\d r_z]$ to mean that we should divide by
the volume of $\uU(1)$, computed using the measure $[\d r_z]$.  This will
 replace $1/[\d r_z]$ with $1/2\pi$.  In
 appendix \ref{app:JTU1bos}, we consider the simpler case  of JT gravity coupled to a $\uU(1)$ gauge field, described by a $BF$ theory of
 ${\rm PSL}(2,\mathbb{R})\times \uU(1)$, and verify that this procedure is correct.

Second, the formulas that we have presented so far 
have  been written in terms of $\phi_a$ and $\phi_b$, with no mention of $\phi_c$.
We can write the result  somewhat more symmetrically by ``integrating in'' $\phi_c$ with the identity,
 \begin{equation}
 1 = \int \d\phi_c~ \delta( \phi_a + \phi_b + \phi_c - F),
 \end{equation}
where the coefficients in $F= f_{ij} \psi_{i} \psi_j + f_4 \psi_1 \psi_2 \psi_3 \psi_4$ were presented in eqn. (\ref{eq:Fexp}).
Making use of this, and replacing $1/[\d r_z]$ with $1/2\pi$, the final result for the torsion, which we will use from now on, is 
\beq\label{eq:torsionN2}
\tau_Y =-\frac{1}{2\pi} \frac{2e^{\i\hq\phi_a-\i \hq\phi_b}\sinh \frac{a}{2} \sinh \frac{b}{2} \sinh \frac{c}{2} (1+K)}{(\cosh \frac{a}{2} + \cos \hq\phi_a )(\cosh \frac{b}{2} + \cos \hq\phi_b)}\delta(\phi_a + \phi_b + \phi_c- F) [\d a \d \phi_a  \d b \d \phi_b \d c \d \phi_c  | \d^4\psi].
\eeq
We remind the reader that to derive this expression we singled out a boundary with holonomy $W$. This was used in the definition of the fermionic moduli, which is why the formula
(\ref{eq:torsionN2}) is not symmetrical in $a,b,c$.
 Nevertheless, we show in sections \ref{explicit} and \ref{sec:RR}  that the permutation symmetry is restored, as it should be, after integrating over the fermionic moduli.
 The formula (\ref{eq:torsionN2}) is actually symmetrical in $a$ and $b$, up to a charge conjugation operation that reverses the signs of $\phi_a,\phi_b$,
 because our definition of the fermionic moduli was symmetrical in $U$ and $V$, similarly
 to what was explained in eqns. (3.96-7) of \cite{SW}.

\subsubsection{The Torsion of a Circle}\label{torcircle}

Let us now discuss what happens when we glue together two manifolds with boundary  $Y_1$ and $Y_2$ along a common boundary circle $S$ to make a connected manifold $Y'$.
In quantum field theory in general, the gluing would be implemented by multiplying the path integrals on $Y_1$ and $Y_2$ and
summing over physical states propagating through $S$.   In standard approaches to $BF$ theory,
the sum can be carried out by integrating over the conjugacy class of the holonomy of a flat connection around $S$.   For gauge group $\SU(1,1|1)$, this
would mean an integration over a pair of parameters $a,\phi_a$ that appear as eigenvalues of the holonomy.

In the approach via the torsion, the appropriate gluing procedure is to multiply the torsions of $Y_1$ and $Y_2$ and divide by the torsion of $S$:
\be\label{gluetor}\tau_{Y'}=\tau_{Y_1}\frac{1}{\tau_S}\tau_{Y_2}. \ee
This procedure is explained heuristically in \cite{SW} (see eqn. (3.18)) and one can find a more detailed explanation in, for example, section 4 of \cite{Witten:1991we}.
The rough idea is that  $\tau_{Y_1}$ and $\tau_{Y_2}$ are both measures for the holonomy along $S$ (as well as other variables), and to avoid overcounting,
we have to divide by $\tau_S$, which is also such a measure.

 The general procedure to compute $\tau_S$ for any gauge group is outlined in section 3.4.3 of \cite{SW}.  
 For any $G$, let $U$ be the holonomy of a flat connection around $S$ and let $\h U$ be the linear transformation of the Lie algebra $\g$ of $G$ given
 by $\h U(s)=U s U^{-1}$.  Then $\tau_S=\Ber'(\h U-1)$, where $\Ber'$ is a sort of superanalog of the determinant.   To make this more concrete in the case of $G=\SU(1,1|1)$,
 up to conjugation we can assume that 
\beq\label{uansatz}
U = {\rm diag} (-e^{\frac{a}{2} + \i \hq\phi},-e^{-\frac{a}{2} + \i\hq \phi}, e^{2 \i \hq\phi}).
\eeq 
$\h U$ commutes with a decomposition of $\g=\mathfrak{su}(1,1|1)$ as $\g=\g^0\oplus \g^\perp$, where $\g^0$ is the subalgebra of diagonal matrices and $\g^\perp$ is the space
of strictly off-diagonal matrices.   The torsion has a corresponding factorization
\be\label{facttors} \tau_S=\tau_S^\perp\cdot \tau_S^0.\ee
Let $\h U^0$, $\h U^\perp$ be the restrictions of $\h U$ to $\g^0$, $\g^\perp$.    Then $\h U^\perp$ is invertible and $\tau_S^\perp =\Ber'(\h U^\perp-1)$.   The definition
of $\Ber'$ is such that  if $M=\left(\begin{array}{c|c} M_b& 0 \\ \hline 0 & M_f\end{array}\right) $ is a linear transformation of a $\Z_2$-graded vector space with invertible bosonic
and fermionic blocks $M_b$, $M_f$, then $\Ber'(M)=|\det M_b|/\det M_f$.     
In the present case, this leads to  \be\label{tauperp} \tau_S^\perp=  {\rm Ber}'(\h U^\perp-1) = \frac{\sinh^2 \frac{a}{2}}{(\cosh \frac{a}{2}+ \cos\hq \phi)^2}.\ee
On the other hand, $\h U^0-1=0$, which leads to $\tau_S^0$ being a simple quotient of measures, 
\be\label{taudiag}\tau_S^0=\d a \,\d\phi \, \d\varrho^{-1} \, \d\varphi^{-1}. \ee
Here $a,\phi$ are the parameters of $U$ introduced in eqn. (\ref{uansatz}), and $\varrho$, $\varphi$ are length-twist and $\uU(1)$-twist parameters that we already encountered in 
discussing the double trumpet.    This formula generalizes eqn. (3.42) of \cite{SW}.

The final answer for the torsion is
\begin{equation}\label{eq:torsioncirc}
\tau_S= \frac{\sinh^2 \frac{a}{2}}{(\cosh \frac{a}{2} + \cos \hq\phi)^2}\, \d a \, \d \phi \, (\d \varrho)^{-1} \, (\d \varphi)^{-1}.
\end{equation}
It is perhaps more to the point to write this as a formula for $1/\tau_S$, since that is what appears in the gluing law (\ref{gluetor}):
\be\label{invtor}
\frac{1}{\tau_S} =  \frac{(\cosh \frac{a}{2} + \cos \hq\phi)^2}{\sinh^2 \frac{a}{2}} \d a^{-1}\d\phi^{-1}\d\varrho \, \d\varphi. \ee
This formula can be interpreted, in part, as follows.   Both $\tau_{Y_1}$ and $\tau_{Y_2}$ are proportional to $\d a\d\phi$, and $1/\tau_S$ removes an unwanted extra factor
of $\d a\d\phi$ that would be present in the product $\tau_{Y_1}\tau_{Y_2}$.   On the other hand, gluing introduces new parameters 
$\varrho$ and $\varphi$, and $1/\tau_S$ generates a measure for integration over these parameters.

\subsubsection{General Case: Measure on Moduli Space}
Having computed the torsion of a three-holed sphere and a circle, we can assemble the pieces and compute the  moduli space measure of $\cN=2$ JT gravity on any oriented 
NAdS$_2$ manifold.

For simplicity, we begin with a discussion of  surfaces without boundary. Any closed oriented surface $C$ of genus $g$ can be assembled by gluing together a set $T$  of $2g-2$ three-holed spheres $Y_t$, $t\in T$. These three-holed spheres have to be glued along a set $U$ of $3g-3$ circles $S_u$, $u\in U$.
Two three-holed spheres (or two boundaries of the same three-holed sphere) are glued along each $S_u$.

We denote as $\T_g$ the moduli space of flat connections modulo gauge transformations on a genus $g$ surface $C$ without boundary, and we write $\mu$
for the corresponding moduli space measure.
The measure $\mu$ is simply the tensor product of the $\tau_{Y_t}$ for all $t\in T$ times the tensor product of the $\frac{1}{\tau_{S_u}}$ for all $u\in U$.  This follows inductively
from the gluing formula.
Hence
\be\label{closedmeas}\mu_{g,n=0}=\prod_{t\in T}\tau_{Y_t}\prod_{u\in U}\frac{1}{\tau_{S_u}}. \ee
Looking at eqn. (\ref{eq:torsionN2}) for $\tau_{Y_t}$ and eqn. (\ref{invtor}) for $1/\tau_{S_u}$, we see that significant cancellation occurs in this product between factors
coming from three-holed spheres and factors coming from gluing along circles.

However, 
eqn. (\ref{closedmeas}) is actually only a formal expression.   As seen in eqn. (\ref{eq:torsionN2}), $\tau_Y$ is proportional to a delta function of a sum $\phi_a+\phi_b+\phi_c$
of holonomy angles, modulo nilpotent corrections that lead to derivatives of the delta function.   If we actually evaluate $\mu_{g,n=0}$ using eqn. (\ref{closedmeas}), then,
since a closed surface without boundary has no external holonomies, after multiplying together all these delta functions and integrating over the $\phi$'s,  we will be left
over with a factor of $\delta(0)$ (or possibly a derivative thereof, $\delta^{[n]}(0)$ for some $n$).   It is quite likely that in $\N=2$ JT supergravity, the partition function of a closed
surface is not well-defined.   What are well-defined, though subtle, and what we need for understanding  
holographic duality, are partition functions for surfaces with NAdS external boundaries.
Also well-defined, and closely related, are volumes for surfaces with geodesic boundaries and with specified holonomy parameters $b,\phi$ on each boundary.

To describe the appropriate measure in the case of an NAdS boundary, we have to know how to glue a trumpet onto a three-holed sphere.
We have described in section \ref{sec:N2JTG}
the trumpet path integral $Z_{\rm tr}(b,\phi)$,
with the Schwarzian mode on the asymptotic boundary and standard boundary conditions on the interior geodesic;
we also have computed the path integral on a three-holed sphere using torsion boundary conditions.   We need to know how to compare or combine results computed with the two
different types of boundary condition.
It is slightly tricky to do this because  a convenient way to combine the Schwarzian path integral and the torsion in one calculation is not known.   Instead we proceed indirectly
as in \cite{SW}.   In the standard approach, the measure for the double trumpet is\footnote{For brevity, we do not show the external parameters $\alpha_i,\beta_i$ characterizing the
trumpets.}
\be\label{doubletrumpet} Z_{\rm tr}(b,\phi) \,\d b\d\phi \d\varrho\d\varphi \,Z_{\rm tr}(b,\phi). \ee
If we knew how to calculate a trumpet path integral with the Schwarzian mode on the outer boundary and torsion boundary conditions on the inner boundary, we would
get a measure $\t Z_{\rm tr}(b,\phi)\,\d b\d\phi. $    Then to get the appropriate measure for the double trumpet, we would glue together two trumpets using the gluing
law for the torsion, meaning that we would multiply the torsions of the two trumpets and divide by $\tau_S$, the torsion of a circle where the gluing occurs.   This would give
\be\label{torsdouble}\t Z_{\rm tr}(b,\phi)\d b\d \phi\frac{1}{\tau_S}\t Z_{\rm tr}(b,\phi) \d b\d\phi. \ee
Using eqn. (\ref{invtor})  for $1/\tau_S$, this becomes
\be\label{simptorsdouble} \t Z_{\rm tr}(b,\phi) \frac{(\cosh \frac{b}{2}+\cos\h q \phi)^2}{\sinh^2\frac{b}{2}}  \d b \d\phi \d\varrho \d\varphi   \, \t Z_{\rm tr}(b,\phi). \ee 
Comparing these formulas, we get
\be\label{tortrumpet} \t Z_{\rm tr}(b,\phi) = Z_{\rm tr}(b,\phi) \frac{\sinh\frac{b}{2}}{\cosh\frac{b}{2}+\cos \h q \phi}. \ee

Now consider a Riemann surface $C$ of genus $g$ with $n$ boundaries, all of NAdS type.   Thus $C$ is made by gluing $n$ trumpets onto an internal manifold $C_\intt$ along
$n$ circles $S_u$, $u\in U_\ext$, where $U_\ext$ is the set of boundaries of $C_\intt$.   Write $\tau_{C_\intt}$ for the torsion measure of the moduli space $\cR_{g,n}$ of flat connections
on $C_\intt$.  The gluing law for the torsion tells us
that the $\N=2$ JT supergravity path integral on $C$ is the integral over $\cR_{g,n}$ of 
\be\label{integralc}\prod_{u\in U_\ext}\left(  \t Z_{\rm tr}(b_u,\phi_u) \d b_u\d\phi_u \frac{1}{\tau_{S_u}} \right) \cdot  \tau_{C_\intt}.\ee
Using eqn. (\ref{tortrumpet}) for $\t Z_{\rm tr}(b,\phi)$ and also eqn. (\ref{invtor}) for $1/\tau_{S}$, this becomes
\be\label{integrald} \prod_{u\in U_\ext} Z_{\rm tr}(b_u,\phi_u) \frac{\cosh\frac{b_u}{2}+\cos\hq \phi_u}{\sinh\frac{b_u}{2}} \d\varrho_u \d\varphi_u \cdot \tau_{C_\intt}. \ee 
In the approach of \cite{SSS}, the expected answer is 
\be\label{expected} \prod_{u\in U_\ext} Z_{\rm tr}(b_u,\phi_u) \d b_u \d \phi_u \d\varrho_u \d\varphi_u \cdot \mu_{g,n}(b_1,\phi_1;\cdots;b_n,\phi_n) ,\ee
where $\mu_{g,n}(b_1,\phi_1;\cdots; b_n,\phi_n)$ is the Riemannian or symplectic measure on the moduli space $\T_{g,n}(b_1,\phi_1;\cdots;b_n,\phi_n)$
of flat connections on $C_\intt$ with the holonomy
around $S_u$ specified to be $b_u,\phi_u$.   
Comparing the two formulas, we find a formula for $\mu_{g,n}(b_1,\phi_1;\cdots;b_n,\phi_n)$:
\be\label{comparison}\mu_{g,n}(b_1,\phi_1;\cdots;b_n,\phi_n)=\prod_{u\in U_\ext}      \frac{\cosh\frac{b_u}{2}+\cos\hq \phi_u}{\sinh\frac{b_u}{2}}\frac{1}{\d b_u\d\phi_u}
\tau_{C_\intt}. \ee
Here the factor of $1/\d b_u\d\phi_u$ is an instruction to remove the factor $\d b_u\d\phi_u$ that is present\footnote{This factor is present in $\tau_{C_\intt}$ because
$\tau_{C_\intt}$ is a measure on $\cR_{g,n}$,  the moduli space of flat connections on $C_\intt$ with no restriction on the boundary holonomies except that they are hyperbolic;
it is absent in $\mu_{g,n}$, because $\mu_{g,n}$ is a measure on the moduli space $\T_{g,n}$ of flat connections on $C_\intt$ with specified conjugacy classes of the boundary
holonomies.}
 in $\tau_{C_\intt}$.   
$C_\intt$ can be built by gluing together $2g-2+n$ three-holed spheres $Y_t$, $t\in T$, along $3g-3+n $ circles $S_u$, $u\in U_\intt$.   Then we can write a formula
for $\tau_{C_\intt}$ that is precisely analogous to eqn. (\ref{closedmeas}):
\be\label{taubound}\tau_{C_\intt}=\prod_{t\in T} \tau_{Y_t}\prod_{u\in U_\intt}\frac{1}{\tau_{S_u}}. \ee
Finally, using (\ref{eq:torsionN2}) for $\tau_{Y_t}$ and eqn. (\ref{invtor}) for $1/\tau_{S_u}$, and inserting eqn. (\ref{taubound}) in eqn. (\ref{comparison}), 
we get a more explicit formula for $\mu_{g,n}$:
\bea\label{eq:finalmeasure}
\mu_{g,n}(b_1,\phi_1;\cdots;b_n,\phi_n) &=& \prod_{u\in U_{\rm int}} [\d a_u \d \phi_u] \cdot [\d \varrho_u \d \varphi_u]\prod_{t\in T} ~ 
\frac{-2e^{\i\hq\phi_{a_t} - \i\hq\phi_{b_t}}(\cosh \frac{c_t}{2} + \cos \hq\phi_{c_t}) }{(2\pi)}\nn
&&~~~~\times (1+K_t) \delta\left(\phi_{a_t}+\phi_{b_t}+\phi_{c_t}-F_t\right)  [\d^4 \psi_t] .
\ea
In writing this expression we used the same labels for each $t\in Y$ that we used in section \ref{sec:3hs}. The three lengths are 
$a_t$, $b_t$ and $c_t$, the three $\uU(1)$ holonomies are $\phi_{a_t}$, $\phi_{b_t}$ and $\phi_{c_t}$ and the fermionic moduli are  $\psi_t$. 
The nilpotent terms $K_t$ and $F_t$ are implicitly given by the functions appearing in \eqref{eq:Fexp} and \eqref{eq:Kexp} involving only the parameters of $t\in Y$. 

Since $C$ is assumed to be oriented, the measure \eqref{eq:finalmeasure} will equal  the Weil-Petersson symplectic measure  of the moduli space $\mathcal{T}_{g,n}$ (assuming that the symplectic
structure is normalized properly).   $\T_{g,n}$ as we have defined it so far is a version of $\N=2$ Teichm\"{u}ller space.\footnote{For an exploration
of  $\cN=2$ Teichm\"{u}ller theory and the associated Weil-Petersson structure, see \cite{Ip:2016ojn}.}  To get the corresponding moduli space $\mM_{g,n}$ of $\N=2$
super Riemann surfaces, we have to divide $\T_{g,n}$ by the mapping class group.   Since $BF$ theory -- and its partition function, the torsion -- is diffeomorphism-invariant,
$\mu_{g,n}$ descends to a measure on $\mM_{g,n}$.   
Finally, we can define formally the volume of $\mM_{g,n}$:
\beq
V_{g,n} = \int_{{\mM}_{g,n}} \mu_{g,n}.
\eeq
Although we will be able to compute these integrals and in some sense they certainly deserve to be called volumes, actually their interpretation is subtle.
That is because $\mM_{g,n}$ is not really a supermanifold, partly because of the constraint on the external $R$-symmetry holonomies, and partly because dividing by global
$R$-symmetries is not a very nice operation, as
was explained at the end of section \ref{sec:3hs}.  
We will see in actual computations, for example eqn. (\ref{eqn:V03final}) and generalizations calculated later, that these ``volumes'' in their dependence on the $R$-symmetry
holonomies are derivatives of a delta function.   So for given external holonomies they are either zero or infinite; they cannot be interpreted literally as volumes in any usual sense.  
To get actual numbers that would be volumes in a conventional sense, one has to integrate over one or more of the external holonomies to eliminate the delta function.
In fact, we will do just that in a particular way when we use these volumes as inputs to compute the partition function $Z_{g,n}$ via eqn. (\ref{eq:Zgnnn}) below.

Geometrically, integrating over external holonomies is similar to treating them as additional moduli.   When one does this, one replaces $\mM_{g,n}$ with a space $\X_{g,n}$
that is still not a supermanifold because the operation of dividing by global $R$-symmetries is not nice, but that at least locally can be interpreted as $\X'_{g,n}/\uU(1)$, where $\X'_{g,n}$ is
a supermanifold.   Then the ``volume'' of $\X_{g,n}$ is plausibly a quotient ${\rm vol}(\X'_{g,n})/{\rm vol}(\uU(1))$.

In this paper, it will not be important to understand the precise sense in which $V_{g,n}$ is a volume.   More relevant is that we can calculate $V_{g,n}$ (after learning in
section \ref{sec:RR} how to deal with the mapping class group), and that we can express the JT gravity partition function, which in turn can be compared to the matrix
model predictions, in terms of $V_{g,n}$.
To get the formula  for $Z_{g,n}$ that is analogous to that of \cite{SSS} in the bosonic case, 
we have to integrate the measure (\ref{expected}) over ${\mM}_{g,n}$.   This replaces $\mu_{g,n}$ by $V_{g,n}$.
In addition, since $V_{g,n}$ does not depend on the gluing parameters,
we can explicitly integrate over $\varrho_u$ and $\varphi_u$, giving for each $u\in U_\ext$ a factor $2\pi b_u$.   
Finally we get
\bea\label{eq:Zgnnn}
Z_{g,n}= \left[ \prod_{u=1}^n \int_0^\infty b_u\d b_u \int_0^{2\pi}2\pi \d \phi_u\,Z_{\rm tr} (\beta_u,\alpha_u;\overline{{\sf b}}_u)\right] V_{g,n}({\sf b}_1,\ldots, {\sf b}_n),
\ea
where ${\sf b}=(b,\phi)$ and $\overline{{\sf b}}=(b,-\phi)$. 

We
will now do a direct calculation of $V_{0,3}$,  enabling us to justify the formula \eqref{ansatz}  that was used in section \ref{sec:3bwh} to match with the matrix model prediction for
$Z_{0,3}$.
To compute the volumes in general, we will in section \ref{sec:RR} use the preceding formulas as input to develop a Mirzakhani-style recursion relation.
This will give general results for the volumes and the partition functions that can be compared to random matrix theory.

\subsection{Example: Computation of $V_{0,3}$}\label{explicit}

In bosonic JT gravity, $V_{0,3}=1$, because $\mM_{0,3}$, the moduli space of geodesic three-holed spheres with specified boundary lengths,  consists of a single point.  In $\N=1$ JT supergravity, $V_{0,3}=0$,
because of fermionic moduli.  The case of $\cN=2$ is more involved and more interesting.

$\mM_{0,3}$  has dimension $0|4$ with four fermionic moduli $\psi_1,\ldots, \psi_4$. We denote the three boundary lengths now by $b_1$, $b_2$ and $b_3$ and the corresponding $\uU(1)$ holonomies by $\phi_1$, $\phi_2$ and $\phi_3$. The measure $\mu_{0,3}$ on $\mM_{0,3}$ is given by eqns.  (\ref{eq:torsionN2}) and (\ref{comparison}).   Explicitly
\bea
V_{0,3} &=& \int_{\mM_{0,3}} \mu_{0,3} \nn
&=& -\frac{2e^{\i\hq\phi_{1} - \i\hq\phi_{2}}(\cosh \frac{b_3}{2} + \cos \phi_{3}) }{(2\pi)}\int (1+K) \delta\left(\phi_{1}+\phi_{2}+\phi_{3}-F\right)  [\d^4 \psi_t].\label{eq:V03intmu03}
\ea
This particular calculation is  relatively simple because the mapping class group of a three-holed sphere is trivial. If we did not have the nilpotent contributions $K= \sum_{i<j} k_{ij} \psi_i \psi_j  +k_4 \psi_1\psi_2\psi_3\psi_4$ and $F=\sum_{i<j} f_{ij} \psi_i \psi_j + f_4 \psi_1\psi_2\psi_3\psi_4$ that
were presented in eqns. (\ref{eq:Fexp}) and (\ref{eq:Kexp}), $V_{0,3}$ would vanish due to the rule of fermionic integration $\int \d \psi =0$. To obtain a non-zero answer requires a term proportional to the product $\psi_1\psi_2\psi_3\psi_4$. 

 We define the fermionic measure by $\int [\d^4 \psi] \psi_1 \psi_2 \psi_3 \psi_4 = 1$, and therefore
\begin{align}
\int [\d^4 \psi]  ~\big(1 + K\big)~\delta\Big(\sum_{i=1}^3\phi_i - F\Big) = & k_4 \,\delta\Big(\sum_{i=1}^3\phi_i \Big) \nonumber \\
&- (f_{12} k_{34} + f_{14} k_{23} + f_{23} k_{14} + f_{34} k_{12} + f_4  )  \delta'\Big(\sum_{i=1}^3\phi_i\Big)\nonumber\\
& + (f_{12} f_{34}+ f_{14} f_{23}) \delta''\Big(\sum_{i=1}^3\phi_i\Big).    \label{eqn:fermint2}
\end{align}
Using \eqref{eq:Fexp} and \eqref{eq:Kexp} for the coefficients, the formula for $V_{0,3}$  simplifies considerably:\footnote{We would like to thank D. Stanford for finding a mistake in an earlier version of this calculation.}
\begin{eqnarray}
V_{0,3} &=& -\frac{1}{2\pi}\frac{1}{4\hq^{\,2}} \Big[ G(\phi_1+\phi_2,\phi_3) \delta''(\phi_1 + \phi_2 + \phi_3) +2\partial_{\phi_1}G(\phi_1+\phi_2,\phi_3) \delta'(\phi_1 + \phi_2 + \phi_3) \nn
&&~~~+ \partial^2_{\phi_1}G(\phi_1+\phi_2,\phi_3) \delta(\phi_1 + \phi_2 + \phi_3)\Big],\label{eqn:V03final222}
\end{eqnarray}
where we define
\bea
G(\phi,\phi_3) =\frac{\cosh \frac{b_3}{2} + \cos \hq\phi_3}{\cosh \frac{b_3}{2} + \cos \hq\phi}.
\ea
Surprisingly, the three terms in eqn. (\ref{eqn:V03final222})   can be neatly combined to a single delta function:
\begin{equation}\label{surprising}
G(\phi,\phi_3) \delta''(\phi_3 + \phi) + 2 \partial_\phi  G(\phi,\phi_3) \delta'(\phi_3 + \phi)+\partial^2_\phi  G(\phi,\phi_3) \delta(\phi_3 + \phi)= \delta''(\phi_3 + \phi).
\end{equation}
One can prove this identity easily by multiplying by  an arbitrary test function and integrating, and then integrating by parts.  The only property of $G(\phi,\phi_3)$ that is required
is that $G(\phi_3,\phi_3)=1$.  By virtue of eqn. (\ref{surprising}), the Weil-Petersson volume of the $\cN=2$ hyperbolic three-holed sphere is equal to 
\beq
\label{eqn:V03final}
V_{0,3}({\sf b}_1, {\sf b}_2, {\sf b}_3) = - \frac{1}{2\pi} \frac{1}{4\hq^{\, 2}} \delta''(\phi_1+\phi_2+\phi_3).
\eeq
This precisely justifies eqn.  \eqref{ansatz} and therefore justifies claims made in section \ref{sec:3bwh} that follow from this formula. 
  It might be possible to define the fermionic moduli in a way that enables one to get this simple result without  the complicated intermediate steps, but it is not immediately clear how to do that.

In the next section we will find a recursion relation for $V_{g,n}$, and the formula for $V_{0,3}$  will be an initial condition in the recursion. This will extend the match with the random matrix ensemble from the three-holed sphere to an
arbitrary oriented surface.

\section{Topological Recursion from Gravity and Random Matrices}\label{sec:RR}

In the previous section, we determined  the path integral measure of $\cN=2$ JT supergravity on an oriented manifold with geodesic boundaries.   In order to calculate the partition
function, we need to integrate this measure over the appropriate moduli space.

Even though the measure is extremely simple, how to integrate it over  a fundamental domain of the mapping class group  is not obvious.   For ordinary Riemann surfaces, Mirzakhani   \cite{mirzakhani2007simple, Mirzakhani:2006eta} solved this problem by developing a recursion relation that can be used
to effectively compute the volumes.  This recursion relation was extended to $\N=1$ super Riemann surfaces in \cite{SW}, see also \cite{norbury2023enumerative}, and here we will extend this story to $\N=2$ super Riemann
surfaces. With the help of this result, we will be able to verify the duality between $\N=2$ JT supergravity and an $(\upalpha_0,\upbeta)=(1,2)$ AZ ensemble, to all orders in
perturbation theory.

\subsection{Review of Mirzakhani's Approach}\label{reviewmirz}

We begin with a brief 
review of Mirzakhani's method, and then in section \ref{ntwohyper}, we  explain the modifications needed to extend this approach to $\cN=2$ hyperbolic surfaces.  The
reader might find it helpful to consult appendix D of \cite{SW}, which contains some more detailed explanations.   In this discussion, for brevity we refer to a three-holed
sphere with a hyperbolic metric and geodesic boundaries as a geodesic three-holed sphere.

Let $\Sigma$ be a hyperbolic surface of genus  $g$ with $n+1$ geodesic boundaries.  We single out one boundary of length $b$, 
and denote the other $n$ boundary lengths as  $B=\{ b_1,\ldots,b_n\}$. It will be useful to label the boundary 
geodesics  by their lengths as $\gamma_{b}, \gamma_{b_1},\ldots, \gamma_{b_n}$.

     \begin{figure}
 \begin{center}
     \begin{center}
   \includegraphics[width=4.5in]{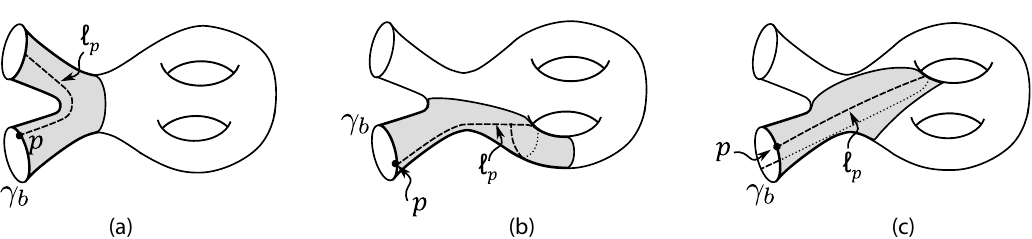}
 \end{center}
\end{center}
\caption{\footnotesize  Generically, as explained in the text, a geodesic $\ell_p$ orthogonal to the boundary of $\Sigma$ at $p\in\gamma_b$ determines a hyperbolic three-holed
sphere $Y\subset \Sigma$ that has $\gamma_b$  as one of its boundaries.  Three possibilities are sketched here.  \label{geodesics}}\end{figure}

The first step is a simple decomposition of $\gamma_b$. Consider a point $p\in \gamma_b$, and let $\ell_p$ be the geodesic orthogonal to $\gamma_b$ at $p$.  
The set of $p$ such that $\ell_p$ continues forever without intersecting itself or returning to the boundary has measure zero, according to a theorem of
Birman and Series.   
After throwing away this set of measure zero, we can assume that $\ell_p$ eventually intersects itself or returns to the boundary of $\Sigma$.
When that happens, we terminate $\ell_p$.   We take the union of $\ell_p$ and the boundary component or 
components it ends on, and thicken this set
slightly to get what topologically is a three-holed sphere whose boundaries are either boundaries of $\Sigma$ and therefore geodesics, or 
else embedded circles in $\Sigma$.   The lengths of such embedded circles can be minimized in their homotopy classes to get geodesics.   
In this way, the choice of a generic point $p\in \gamma_b$
determines an embedded geodesic three-holed sphere\footnote{The moduli space $\mM_{0,3}$ of a three-holed sphere is a point, of volume 1.  So we can assume
that $\Sigma$ is not itself a three-holed sphere, ensuring that 
$Y$ is a proper subset of $\Sigma$ and always has at least one boundary in the interior of $\Sigma$.
In the extension to $\N=2$ super Riemann surfaces, the volume of $\mM_{0,3}$ is not simply 1, but we have already computed it in  
section \ref{explicit}, so we will again assume that $\Sigma$ is not a three-holed sphere.}
 $Y\subset \Sigma$.   All this is pictured in fig. \ref{geodesics}.

We can now decompose $\gamma_b$ (with a set of measure zero discarded) as the union of three subsets:
\begin{itemize}
\item Set $A$: $\ell_p$  intersects itself or returns to the original boundary $\gamma_b$, and as depicted in figs.  \ref{geodesics}(b,c), the three-holed sphere $Y$
has one boundary $\gamma_b$ and  two boundaries in the interior of $\Sigma$.     The complement of $Y$ is
connected and is a surface $\Sigma'$ of genus $g-1$ with $n+2$ boundaries, as schematically shown here:
\bea
\begin{tikzpicture}[scale=0.5, baseline={([yshift=-0.1cm]current bounding box.center)}]
 %pants
 \node at (-5.8,0) {\small $b$};
  \draw[thick] (-5.2,0) ellipse (.15 and .5);
  \draw[thick] (-3.2,1.5) ellipse (.15 and .5);
  \draw[thick] (-3.2,-1.5) ellipse (.15 and .5);
   \draw[thick, bend right=45] (-5.2,.5) to (-4.2,1);
   \draw[thick, bend left=30] (-4.2,1) to (-3.2,2);
   \draw[thick, bend left=45] (-5.2,-.5) to (-4.2,-1);
   \draw[thick, bend right=30] (-4.2,-1) to (-3.2,-2);
   \draw[thick, bend right=95] (-3.2,1) to (-3.2,-1);
   \node at (-2.7,1.5) {\small $b'$};
   \node at (-2.6,-1.5) {\small $b''$};
   %g-1 surface
\draw[thick] (-2,1.5) ellipse (.15 and .5);
 \draw[thick] (-2,2) to (-.2,2);
  \draw[thick] (-2,1) to (-.2,1);
  \draw[thick] (-2,-1.5) ellipse (.15 and .5);
 \draw[thick] (-2,-2) to (-.2,-2);
  \draw[thick] (-2,-1) to (-.2,-1);
  \draw [gray!60,fill=gray!60] (1,0) ellipse (2 and 2.5);
  \draw[thick] (1,0) ellipse (2 and 2.5);
  \node at (1,0) {\small $g-1$};
\end{tikzpicture}
\ea
We denote the lengths of the new boundaries as $b'$ and $b''$.
\item Set $B$: again $\ell_p$ intersects itself or returns to $\gamma_b$, and $Y$ has two interior boundaries,  
but now $\Sigma'$ is disconnected and is the union of surfaces $\Sigma_1,$ $\Sigma_2$ of genera
$h_1,$ $h_2$ such that $h_1+h_2=g$, and with boundary lengths $b',B_1$ and $b'',B_2$ 
 such that $B_1 \cup B_2 = B$:
\bea
\begin{tikzpicture}[scale=0.5, baseline={([yshift=-0.1cm]current bounding box.center)}]
%h1 surface upper
\draw[thick] (-2,1.5) ellipse (.15 and .5);
 \draw[thick] (-2,2) to (-.2,2);
  \draw[thick] (-2,1) to (-.2,1);
  \draw [gray!60,fill=gray!60] (1,1.5) ellipse (1.2 and 1.2);
  \draw[thick] (1,1.5) ellipse (1.2 and 1.2);
  %h2 surface lower
  \draw[thick] (-2,-1.5) ellipse (.15 and .5);
 \draw[thick] (-2,-2) to (-.2,-2);
  \draw[thick] (-2,-1) to (-.2,-1);
  \draw [gray!60,fill=gray!60] (1,-1.5) ellipse (1.2 and 1.2);
  \draw[thick] (1,-1.5) ellipse (1.2 and 1.2);
  %pants
   \node at (-5.8,0) {\small $b$};
  \draw[thick] (-5.2,0) ellipse (.15 and .5);
  \draw[thick] (-3.2,1.5) ellipse (.15 and .5);
  \draw[thick] (-3.2,-1.5) ellipse (.15 and .5);
   \draw[thick, bend right=45] (-5.2,.5) to (-4.2,1);
   \draw[thick, bend left=30] (-4.2,1) to (-3.2,2);
   \draw[thick, bend left=45] (-5.2,-.5) to (-4.2,-1);
   \draw[thick, bend right=30] (-4.2,-1) to (-3.2,-2);
   \draw[thick, bend right=95] (-3.2,1) to (-3.2,-1);
     \node at (-2.7,1.5) {\small $b'$};
   \node at (-2.6,-1.5) {\small $b''$};
     \node at (1,1.5) {\small $h_1$};
   \node at (1,-1.5) {\small $h_2$};
\end{tikzpicture}
\ea
\item Set $C$: The geodesic  reaches another boundary $\gamma_{b_k}$, or it self-intersects and $\gamma_{b_k}$ is one of the boundaries of $Y$.
Either way,  $Y$ has the two boundaries just specified and one internal boundary, say of length $b'$. 
The remaining surface $\Sigma'$ is of genus $g$ with $n$ boundaries:
\bea
\begin{tikzpicture}[scale=0.5, baseline={([yshift=-0.1cm]current bounding box.center)}]
 %pants
  \draw[thick] (-3.2,0) ellipse (.15 and .5);
  \draw[thick] (-5.2,1.5) ellipse (.15 and .5);
  \draw[thick] (-5.2,-1.5) ellipse (.15 and .5);
   \draw[thick, bend left=45] (-3.2,.5) to (-4.2,1);
   \draw[thick, bend right=30] (-4.2,1) to (-5.2,2);
   \draw[thick, bend right=45] (-3.2,-.5) to (-4.2,-1);
   \draw[thick, bend left=30] (-4.2,-1) to (-5.2,-2);
   \draw[thick, bend left=95] (-5.2,1) to (-5.2,-1);
   \node at (-5.8,1.5) {\small $b$};
    \node at (-5.8,-1.5) {\small $b_k$};
   \node at (-2.6,0) {\small $b'$};
    %g surface lower
  \draw[thick] (-2,0) ellipse (.15 and .5);
 \draw[thick] (-2,-.5) to (-.2,-.5);
  \draw[thick] (-2,.5) to (-.2,.5);
  \draw [gray!60,fill=gray!60] (1,0) ellipse (1.2 and 1.2);
  \draw[thick] (1,0) ellipse (1.2 and 1.2);
   \node at (1,0) {\small $g$};
   \end{tikzpicture}
\ea
\end{itemize}

Keeping $\Sigma$ fixed, the sum of the lengths of the three sets $A,B,C$ is $b$, since $A\cup B\cup C=\gamma_b$, except for a set of measure 0.  If we integrate over all moduli,
including the moduli of $\Sigma$ and also the choice of the point $p\in\gamma_b$, the result will be simply $b V_{g,n+1}(b,B)$.  
However, we can make use of the fact that the choice of $p$ generically determines a decomposition of $\Sigma$ as the union of a three-holed sphere $Y$ with some other
surface $\Sigma'$ (possibly disconnected).  In general, $Y$ can be any geodesic three-holed sphere in $\Sigma$ one of whose boundaries is $\gamma_b$.
So we can organize the integral over the moduli of the pair $\Sigma,p$ as a sum over all choices of $Y$, and for each $Y$ an integral over $p$ and an integral over the moduli
of the remaining surface $\Sigma'$.   Since the Euler characteristic of $\Sigma'$ is always less negative than the Euler characteristic of $\Sigma$, this leads to a recursion
relation that ultimately completely determines the volumes.

To carry out this program, we need to know how to make the decomposition of $\gamma_b$ in the three sets $A,B,C$ if $Y$ is given.
Let $Y$ be a hyperbolic three-holed sphere 
 with geodesic boundaries $\gamma$, $\gamma'$ and $\gamma''$ of lengths $b$, $b'$ and $b''$ respectively. In the bosonic case, $Y$ has
 no moduli apart from these lengths (in the generalization to super Riemann surfaces, $Y$ does have fermionic moduli and we will have to integrate over them). 
  We denote by ${\sf T}(b,b',b'')$ the length of the set of points $p\in\gamma$  such that $\ell_p$ ends on $\gamma'$. Of course the set of points such
  that $\ell_p$ ends on $\gamma''$ then has length ${\sf T}(b,b'',b')$.   We also denote by ${\sf D}(b,b',b'')$ the length of the set of points $p\in\gamma$ such that $\ell_p$  self-intersects or 
  returns to $\gamma$. Since these are all the options (apart from a set of measure zero), we have 
\bea\label{eq:bDTT}
b = {\sf D}(b,b',b'') + {\sf T}(b,b',b'')+{\sf T}(b,b'',b').
\ea
In terms of these functions, which we compute shortly, the recursion relation becomes
\bea
b V^{\rm bos.}_{g,n+1} (b, B) &=& \frac{1}{2}\int_0^{\infty}  b'\text{d}b' \,b'' \text{d}b''\,{\sf D}_{\rm bos.} (b,b',b'') ~V^{\rm bos.}_{g-1,n+2}(b',b'',B) \nn
&& + \frac{1}{2}\int_0^{\infty}  b'\text{d}b' \,b''\text{d}b''\,{\sf D}_{\rm bos.} (b,b',b'') \sum_{\rm stable} V^{\rm bos.}_{h_1,|B_1|+1} (b',B_1) V^{\rm bos}_{h_2,|B_2|+1} (b'',B_2)\nn
&&+\sum_{k=1}^{|B|} \int_0^{\infty} b'\text{d}b' \,\big(b -{\sf T}_{\rm bos.}(b,b',b_k) \big) V^{\rm bos.}_{g,n} (b',B /b_k) \label{eq:Mirzakhanirec}
\ea
The three terms on each line of the right hand side correspond to each of the three possible behaviors of $\ell_p$. We have denoted  the bosonic volumes by $V_{g,n}^{\rm bos.}(b,B)$ to distinguish them from their supersymmetric counterpart which we analyze later, and similarly for the kernels appearing in the integrands. The qualification ``stable'' means that one
excludes the case that one of the components is of genus 0 with less than 3 boundaries, and hence does not have a hyperbolic metric.

In eqn. (\ref{eq:Mirzakhanirec}), the reason for the coefficient ${\sf D}_{\rm bos.}(b,b',b'')$ in the first two lines is relatively easy to understand.  
The outcomes that correspond to these lines arise if and only if  $\ell_p$, which is initially perpendicular to $\gamma_b$ at $p$, self-intersects or returns to $\gamma_b$
while remaining in a three-holed sphere $Y$ whose other boundaries are internal geodesics $\gamma_{b'}$, $\gamma_{b''}$ of $\Sigma$.   
In that case, the thickening of $\gamma_b\cup\ell_p$ to a geodesic three-holed sphere  will always be $Y$.
The measure of the subset of $\gamma_b$ for which $\ell_p$ has this behavior with this particular $Y$ is ${\sf D}_{\rm bos.}(b,b',b'')$,
so that is the coefficient in the first two lines of eqn. (\ref{eq:Mirzakhanirec}).
The coefficient in the last line is less obvious.   There are two ways that the thickening process can lead to a three-holed sphere $Y$ two of whose boundaries are $\gamma_b$
and another external boundary $\gamma_{b_k}$ of $\Sigma$.  First and perhaps more obvious, $\ell_p$ might terminate on $\gamma_{b_k}$ while remaining in $Y$.   This
corresponds to a set of measure ${\sf T}_{\rm bos.}(b,b_k,b')$.   Second and  possibly less obvious, $\ell_p$ might self-intersect or return to $\gamma_b$, in such a way that the
thickening process leads to a three-holed sphere with $\gamma_{b_k}$ as one of its boundaries.   This corresponds to a subset of $\gamma_b$ of measure ${\sf D}_{\rm bos.}(b,b',b_k)$.
The combined measure of the two subsets of $\gamma_b$ is ${\sf D}_{\rm bos.}(b,b',b_k)+{\sf T}_{\rm bos.}(b,b_k,b')=b-{\sf T}_{\rm bos.}(b,b',b_k)$, which is the
coefficient claimed in (\ref{eq:Mirzakhanirec}).

To complete the derivation, we need an explicit expression for the functions ${\sf D}_{\rm bos.}(b,b',b'')$ and ${\sf T}_{\rm bos.}(b,b',b'')$. Thanks to \eqref{eq:bDTT}, we only need to determine ${\sf T}_{\rm bos.}(b,b',b'')$, the length of the segment $S$ of points along $\gamma$ that end up in $\gamma'$. We explain this calculation in a way that  straightforwardly  generalizes 
to the $\cN=2$ case.   It is convenient to do the calculation by first ``unwrapping'' $Y$ in the upper half plane $H$,
which we parametrize by a complex coordinate $z = x + \i y $, $y>0$,  with metric \be\label{hypmet} \d s^2 = \frac{\d x^2+\d y^2}{y^2} .\ee 
As suggested in the right panel of fig. \ref{fig:torsion3hs}(b), the fundamental group of $Y$  is generated by loops that wind once around $\gamma$ or once around $\gamma'$, starting and
ending at the base-point $P$.   Taking the bosonic limit of the formulas that we used in $\N=2$ supergravity, we can assume that the hyperbolic
flat connection on $Y$ has  holonomies $U,V$ around $\gamma$ and $\gamma'$, respectively:
\be\label{holonomies}
U = \pm \Big(\begin{array}{cc}
 e^{b/2} & \kappa \\
0& e^{-b/2}
\end{array}\Big),\hskip2cm V=\pm \Big(\begin{array}{cc}
 e^{-b'/2} & 0 \\
1& e^{b'/2}
\end{array}\Big).\ee

We can view $Y$ as $\h Y/\Gamma$, where $\Gamma$ is the fundamental group of $Y$, and $\h Y$ is a region in $H$ that is bounded by geodesics that are lifts to $H$ of the boundaries 
$\gamma,\gamma',\gamma''$ of $Y$.   In fact, $\h Y$ has infinitely many
geodesic boundaries that project in $Y$ to $\gamma$, $\gamma'$, or $\gamma''$.   For example, $\gamma$ has a lift that is invariant under the action of $U$ on $H$; its other
lifts in $\h Y$ are invariant under elements of $\Gamma$ that are conjugate to $U$.   There is a unique $U$-invariant geodesic $\h\gamma$ in $H$; it connects the two fixed points of $U$
on the conformal boundary of $H$.   The conformal boundary of $H$ is at $y=0$; it is a copy of $\RP^1$, which can be parametrized by  homogeneous coordinates $u,v$ 
with $x=u/v$.
 As a $2\times 2$ matrix acting on $\begin{pmatrix}u\\[-4pt] v\end{pmatrix}$, $U$ has two eigenvectors, which correspond to the
two fixed points of $U$ on $ \RP^1$. The eigenvectors of $U$ are  $\Big(\begin{array}{c}
  1 \\[-4pt]
 0\end{array}\Big)$ and $\Big(\begin{array}{c}
  -\frac{\kappa}{2\sinh \frac{b}{2}} \\[-4pt] 1\end{array}\Big)$, so the end-points of the geodesic $\widehat{\gamma}$ are located at $x_0= - \kappa/(2\sinh \frac{b}{2})$ and $x_3=\infty$.
  Likewise, $\gamma'$ has a $V$-invariant lift $\h \gamma'$ to $\h Y$.   The endpoints of $\h\gamma'$ correspond to the eigenvectors of $V$, namely
   $\Big(\begin{array}{c}
  0 \\[-4pt]
 1
\end{array}\Big)$ and $\Big(\begin{array}{c}
  -2\sinh \frac{b'}{2} \\[-4pt]
1
\end{array}\Big)$, corresponding to $x_1=0$ and $x_2=-2\sinh \frac{b'}{2}$.  

 Let $\ell$ be a non-self-intersecting geodesic  in $Y$ orthogonal to $\gamma$ at a point $p$ and ending at   $p'\in\gamma'$.   
 Keeping $p'$ fixed, $\ell$ is the shortest path from $p'$ to $\gamma$ in its homotopy class.   A homotopy class of non-self-intersecting paths
from $p'$ to $\gamma$ is unique except for the possibility of adjoining at the beginning of the path a loop that winds around $\gamma'$ any integer number  of times.
As we vary $p'$, $\ell$ will continue to exist, being always the shortest path from $\gamma'$ to $\gamma$ in its homotopy class; $p$ will change, of course.  This continues
as $p'$ winds around $\ell'$ any number of times, and since in that way we explore all possible homotopy classes of non-intersecting paths from $\gamma'$ to $\gamma$,
it follows that the space of non-self-intersecting geodesics  in $H$ that are orthogonal to $\gamma$ and end at $\gamma'$ is connected.   Since that is the case, in lifting to $H$,
we can pick a specific
pair of lifts $\h\gamma$ and $\h\gamma'$ to $H$ and consider geodesics in $\h Y$ orthogonal to $\h\gamma$ and ending on $\h\gamma'$.  There is no need to consider
different lifts of $\gamma$ and $\gamma'$.    We can simply consider the lifts described in the last paragraph, such that $\h\gamma$ connects boundary points $x_0$ and $x_3$,
while $\h\gamma'$ connects $x_1$ and $x_2$.

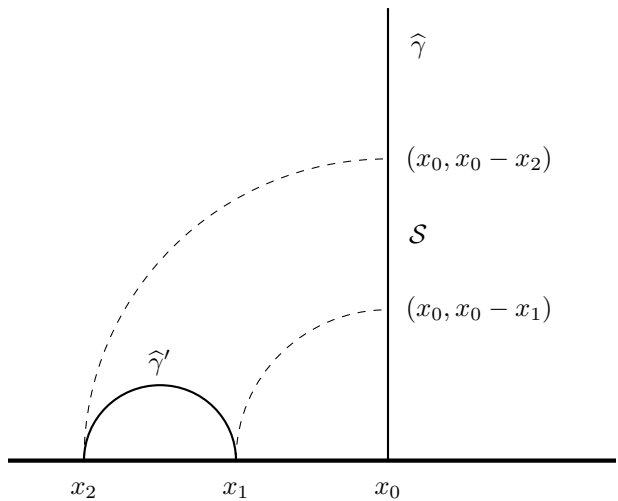
\begin{figure}
\begin{center}
\begin{tikzpicture}[scale=1, baseline={([yshift=-0.1cm]current bounding box.center)}]
\draw[ultra thick] (-5,0) -- (3,0);
\draw[ thick ] (0,0) -- (0,6);
\draw[thick] (-4,0) to [out=90,in=180] (-3,1) to [out =0,in=90] (-2,0);
\draw[dashed] (-2,0) to [out = 90, in =180] (0,2);
\draw[dashed] (-4,0) to [out = 90, in =180] (0,4);
\node at (0,-0.4) {\small $x_0$};
\node at (-2,-0.4) {\small $x_1$};
\node at (-4,-0.4) {\small $x_2$};
\node at (1.2,2) {\small $(x_0,x_0-x_1)$};
\node at (1.2,4) {\small $(x_0,x_0-x_2)$};
\node at (.4,3) {\small $\mathcal{S}$};
\node at (0.4,5.5) {\small $\widehat{\gamma}$};
\node at (-3,1.3) {\small $\widehat{\gamma}'$};
    \end{tikzpicture}
    \end{center}
\caption{\footnotesize The picture in the upper half-plane $H$ with its conformal boundary shown as a horizontal line.    $\widehat{\gamma}$ is the solid vertical line and $\widehat{\gamma}'$ is the solid semicircle.  The dashed lines, which are portions of circles that are orthogonal to both $\h\gamma$ and the conformal boundary, denote the  limiting cases of geodesics that are orthogonal to $\widehat{\gamma}$ and reach $\widehat{\gamma}'$. The interval between them represents the region $\mathcal{S}$ of $\gamma$.  Its length can be straightforwardly determined from
the picture.}
\label{fig:recursionkernel}
\end{figure}

To work out the length of the desired subset of $\gamma$ is now straightforward.   We just need to know that geodesics in $H$ are either vertical straight lines
or semicircles (in the Euclidean metric $\d x^2+\d y^2$) that are orthogonal to the line $x=0$.  So the geodesic $\h\gamma$ from $x_0$ to $x_3$ is 
the vertical line $x=x_0$ and the geodesic $\h\gamma'$ from $x_1$ to $x_2$ is the semi-circle shown in 
 fig. \ref{fig:recursionkernel}. We denote by $x_1$ and $x_2<x_1$ the end-points of $\widehat{\gamma}'$.
Points $p\in\h\gamma$ such that the geodesic $\ell_p$ orthogonal to $\h\gamma$ at $p$ reaches $\h\gamma'$ make up an interval $\mathcal S$,
and the desired function ${\sf T}_{\rm bos.}(b,b',b'')$ is the length of $\mathcal S$.   The endpoints of $\mathcal S$ are easily determined as in the figure,
and computing the length of $\mathcal S$ in the hyperbolic metric (\ref{hypmet}), we get $\log\frac{(x_0-x_2)}{(x_0-x_1)}$.    Equivalently (as $x_3=\infty$), 
\bea\label{crossr}
{\sf T}_{\rm bos.}(b,b',b'') = \log \frac{(x_0-x_2)(x_1-x_3)}{(x_0-x_1)(x_2-x_3)}.
\ea
In order words, it is the logarithm of a cross-ratio between the four endpoints of the two geodesics $\widehat{\gamma}$ and $\widehat{\gamma}'$. 

The final step is to find an expression for $x_0,\ldots, x_3$ in terms of the lengths $b$, $b'$ and $b''$.  All we need to know is that $\kappa=- 2 \cosh \frac{b-b'}{2} - 2 \cosh \frac{b''}{2}$
(eqn. \ref{kapform}).  With this information, we can compute 
\bea
{\sf T}_{\rm bos.}(b,b',b'') = \log \left[\frac{\cosh \frac{b''}{2} + \cosh \frac{b+b'}{2}}{\cosh \frac{b''}{2} + \cosh \frac{b-b'}{2}}\right].
\ea
This completes the derivation of Mirzakhani's recursion relation, since we can easily determine ${\sf D}_{\rm bos.}$ from \eqref{eq:bDTT} and replace it in \eqref{eq:Mirzakhanirec}.

\subsection{Geometry of $\cN=2$ Hyperbolic Surfaces}\label{ntwohyper}

In this section we present some important facts about $\cN=2$ hyperbolic surfaces, study the geometry of the $\cN=2$ three-holed sphere, and derive the kernels that will appear in the recursion, which is written in its final form in the next section \ref{sec:RRfinalform}.

\subsubsection{Chiral vs. Non-Chiral Formalism}\label{nonchiral}

Up to this point, we have not had to be very specific about what we mean by an $\N=2$ super Riemann surface.  We only cared about the moduli of $\N=2$ super Riemann
surfaces.   However, to obtain the recursion relation, we will have to use some geometric facts about $\N=2$ super Riemann surfaces themselves, as opposed to their moduli.

 There are actually two essentially different kinds of $\N=2$ super Riemann surface, which share the same moduli space.  The difference shows up when one writes
 an action for a theory with $\N=2$ supersymmetry.  There are ``$D$-terms,'' which are written as an integral $\int \d^2z \d^4\theta K$ over a supermanifold of  complex
 dimension $1|2$ or real dimension $2|4$, with some generic operator $K$.   And there are ``$F$-terms,'' which are written as an integral $\int \d^2z \d^2\theta W$, 
 over a supermanifold  of complex dimension $1|1$ or real 
 dimension $2|2$,
 with some chiral operator $W$. 
 
 To illustrate the difference between the two, we can describe what is a genus 0 super Riemann surface of one type or the other.
 The chiral version of a genus 0 super Riemann surface is $\CP^{1|1}$, with homogeneous coordinates $u,v|\theta$.  We observe that $\CP^{1|1}$ does admit
 an action of the supergroup $\SL(2|1)$, acting on $\left(\begin{array}{c}u\\[-4pt] v    \\[-1pt] \hline \theta\end{array}\right)$.  We can introduce a hermitian metric
 on $\CP^{1|1}$ and reduce to a real form $\SU(1,1|1)$ of $\SL(2|1)$. 
 The non-chiral version of a genus 0 super Riemann surface is actually $\CP^{1|2}$, with homogeneous coordinates $u,v|\uptheta,\uptheta'$, endowed with a symplectic
 form that is homogeneous of degree 2:
\bea\label{eq:ospsf}
\widehat{\omega}= \d u \d v +\frac{1}{2}(\d \uptheta^2+\d \uptheta'{}^2).
\ea
The group of linear transformations of these homogeneous coordinates that preserves the symplectic form is the complex Lie group 
$\OSp(2|2)_\C$.   Again, we can introduce a hermitian metric
and a real form which we will simply call $\OSp(2|2)$.

In fact, the group $\OSp(2|2)_\C$ is equivalent to a double cover\footnote{See footnote \ref{osptwotwo} for an explanation of this double cover, which however
will not play a major role in the following.} of $\SL(2|1)$, and, with the right choice of the real forms, $\OSp(2|2)$ is similarly a double cover of $\SU(1,1|1)$.
There is a natural one-to-one correspondence between super Riemann surfaces of the two types; see, for example,  section 9 of \cite{Witten:2012ga}.    So they have the
same moduli spaces and that is why up to this point, there was no need to make a choice.   

In this article, we will need very little systematic theory of super Riemann surfaces.   For either of the two types of super Riemann surface, we just define
$z=u/v$, define the analog $\h H$ of the upper half plane by ${\mathrm{Im}}\,z>0$, and define a hyperbolic super Riemann surface to be a quotient
$\h H/\Gamma$, where $\Gamma$ is a discrete subgroup of $\SU(1,1|1)$ or $\OSp(2|2)$, with the property that if we reduce modulo the fermionic variables 
and forget the $R$-symmetries, then $H/\Gamma$ is a hyperbolic Riemann surface in the ordinary sense. 

But should we use the chiral or non-chiral version of a super Riemann surface?   In fact, it turns out that if we work with non-chiral $\N=2$ super Riemann surfaces,
then Mirzakhani's derivation in the bosonic case and the derivation in \cite{SW} for $\N=1$ have straightforward generalizations to $\N=2$.   It appears to be less straightforward
to make similar arguments in terms of chiral $\N=2$ super Riemann surfaces.  An indication that the generalization to $\N=2$ is more direct if the arguments are expressed
in terms of non-chiral super Riemann
surfaces is the following.   Mirzakhani's derivation in the bosonic case leads to a cross-ratio of four points (eqn. (\ref{crossr})), and similarly, the $\N=1$ derivation in \cite{SW}
leads to an $\N=1$ version of the cross-ratio.   So we can guess that we will need an $\N=2$ version of the cross ratio.
In the non-chiral formalism, there is no problem with this.  
   If  a point $P_1\in \CP^{1|2}$ has homogeneous coordinates $u_1, v_1|\uptheta_1,\uptheta'_1$  (where we denote the odd variables as 
   $\uptheta$ and $\uptheta'$ and use the numerical subscript to label the point) and similarly for $P_2, P_3, P_4$, then one defines 
\bea\label{n2symbol}
\langle 1,2\rangle = u_1 v_2- u_2 v_1 - \uptheta_1 \uptheta_2- \uptheta'_1 \uptheta'_2,
\ea
 and the cross ratio is as usual $\langle1,2\rangle \langle3,4\rangle /\langle1,3\rangle \langle2,4\rangle $.   (In addition to this complex cross ratio, 
 four points in $\CP^{1|2}$ have four odd complex moduli, which will not play an important role.)  By contrast, there is no simple definition of a cross ratio for four points in $\CP^{1|1}$, 
 because there is no appropriate invariant in the $2|1$-dimensional representation of $\SL(2|1)$.   
 
Other facts about $\N=1$ super Riemann surfaces that were important in  \cite{SW}, appendix D, also have simple analogs for $\N=2$ in the non-chiral formulation but not in the 
chiral formulation.   In particular, the notions of geodesics, perpendicular geodesics, intersecting geodesics, and the like will be spelled out in section \ref{susysetup}. 

It is not clear that these notions have equally satisfactory analogs if we work with chiral super Riemann surfaces.
 So we will use the non-chiral approach, though we cannot rule out the possibility that a simple proof might somehow be possible in the chiral approach.
A drawback of working with non-chiral super Riemann surfaces is that we will have to convert some $\SU(1,1|1)$ formulas into $\OSp(2|2)$ formulas.
This gives formulas that are equivalent but generally more complicated.   That is actually why we did not choose to write the whole article in terms of $\OSp(2|2)$.

As already explained, in either picture based on $\SU(1,1|1)$ or $\OSp(2|2)$, the super upper half plane $\h H$ is defined by ${\mathrm {Im}}\,z>0$, with $z=u/v$.
In section \ref{susysetup}, we will want to know in the $\OSp(2|2)$ picture what should be meant by the conformal boundary of $\h H$.   In fact, the natural conformal boundary is defined
by taking $u,v|\uptheta,\uptheta'$ to all be real.   Note that the real supergroup $\OSp(2|2)$ does act on the locus of real $u,v|\uptheta,\uptheta'$, so the boundary in
this sense is $\OSp(2|2)$-invariant.   Equivalently, after scaling to set $v=1$, the boundary is parametrized by odd and even real variables $x|\uptheta,\uptheta'$.  
We will call this boundary $\partial\h H$.   The reader might be slightly surprised that $\partial\h H$ is of codimension $1|2$, not $1|0$, relative to $\h H$.   This
is actually a common state of affairs in supergeometry: the most natural conformal boundary of a supermanifold
is not necessarily of codimension $1|0$.   For example, in the AdS/CFT correspondence,
in a superspace description, the conformal boundary will frequently have fewer fermionic coordinates than the bulk.

\subsubsection{Converting to  $\OSp(2|2)$}

A first step is to write the $\SU(1,1|1)$ generators of eqns. (\ref{slgenerators}) and (\ref{fermigen}) as $\OSp(2|2)$ generators, that is, as linear transformations
of $u,v|\uptheta,\uptheta'$.   
For bosonic generators, we have
\beq
{\sf e}=\left(\hspace{-1mm}\begin{array}{cc|cc}
0& 1&0 &0\\
 0&0 & 0&0\\
\hline 0& 0 & 0 & 0\\
0 & 0 & 0 & 0\\
\end{array}\hspace{-1mm}\right),
~~{\sf f}=\left(\hspace{-1mm}\begin{array}{cc|cc}
 0 & 0&0 &0\\
 1& 0 & 0& 0\\
\hline 0& 0 &0 & 0\\
 0& 0 &0 & 0\\
\end{array}\hspace{-1mm}\right),
~~{\sf h}=\left(\hspace{-1mm}\begin{array}{cc|cc}
1&0&0 &0\\
0&{\scriptsize -}1 & 0&0\\
\hline
0&0 & 0 &0\\
0&0 &0 &0\\
\end{array}\hspace{-1mm}\right),~~
{\sf z}=\left(\hspace{-1mm}\begin{array}{cc|cc}
0&0&0 &0\\
0&0 & 0&0\\
\hline
0&0 & 0 &1\\
0&0 &-1&0\\
\end{array}\hspace{-1mm}\right)\label{eq:OSPbos}
\eeq
and for fermionic generators, we have
\bea\notag
{\sf q}_1&=&\left(\hspace{-1mm}\begin{array}{cc|cc}
0&0&{\scriptstyle -\frac{1}{2\sqrt{2}}}&{\scriptstyle -\frac{\i}{2\sqrt{2}}}\\
0&0 & 0&0\\
\hline
0&{\scriptstyle \frac{1}{2\sqrt{2}}} & 0 &0\\
0&{\scriptstyle \frac{\i}{2\sqrt{2}}} &0 &0\\
\end{array}\hspace{-1mm}\right),~
{\sf q}_2=\left(\hspace{-1mm}\begin{array}{cc|cc}
0&0&{\scriptstyle \frac{1}{\sqrt{2}}}&{\scriptstyle -\frac{\i}{\sqrt{2}}}\\
0&0 & 0&0\\
\hline
0&{\scriptstyle - \frac{1}{\sqrt{2}}} & 0 &0\\
0&{\scriptstyle \frac{\i}{\sqrt{2}}} &0 &0\\
\end{array}\hspace{-1mm}\right),~~\\ [8pt]
{\sf q}_3&=&\left(\hspace{-1mm}\begin{array}{cc|cc}
0&0&0 &0\\
0&0 & {\scriptstyle \frac{1}{2\sqrt{2}}}&{\scriptstyle \frac{\i}{2\sqrt{2}}}\\
\hline
{\scriptstyle \frac{1}{2\sqrt{2}}}&0 & 0 &0\\
{\scriptstyle \frac{\i}{2\sqrt{2}}}&0 &0 &0\\
\end{array}\hspace{-1mm}\right), ~~~~~
{\sf q}_4=\left(\hspace{-1mm}\begin{array}{cc|cc}
0&0&0 &0\\
0&0 & {\scriptstyle \frac{1}{\sqrt{2}}}&{\scriptstyle- \frac{\i}{\sqrt{2}}}\\
\hline
{\scriptstyle \frac{1}{\sqrt{2}}}&0 & 0 &0\\
{\scriptstyle -\frac{\i}{\sqrt{2}}}&0 &0 &0\\
\end{array}\hspace{-1mm}\right).\label{eq:OSPfer}
\ea

To obtain the kernel that appears  in the recursion relation  requires a knowledge of the measure on the space of flat connections on a three-holed sphere $Y$ .
We determined this in section \ref{sec:GPI}  in the $\SU(1,1|1)$ language, and these results can be translated to $\OSp(2|2)$.  
As a first step, we can rewrite eqn.  \eqref{eq:U0V0} for the monodromies by which we parametrized a flat connection on $Y$ in terms of
algebra generators:
\bea
U_0= e^{\frac{a}{2} {\sf h} + \frac{a}{2\sinh\frac{a}{2}} \kappa {\sf e} + (\pi +\phi_1)\hq {\sf  z} }e^{\psi_2 {\sf q}_2+\psi_4 {\sf q}_4},~~~V_0 =e^{-\frac{b}{2} {\sf h} + \frac{b}{2\sinh\frac{b}{2}}{\sf f} + (\pi + \phi_2) \hq {\sf z}} e^{\psi_1 {\sf q}_1+\psi_3 {\sf q}_3}.
\ea
To write the holonomies as operators on $u,v|\uptheta,\uptheta'$, we now just have to substitute into these formulas the expressions
 \eqref{eq:OSPbos} and \eqref{eq:OSPfer} for the generators.  The expressions for $\kappa$ and the $\uU(1)$ constraint in terms of $a,b|\psi_1,\psi_2,\psi_3,\psi_4$ 
 were derived in section  \ref{sec:GPI}.

In the rest of this derivation, we use conventions similar to those that we used in reviewing Mirzakhani's bosonic derivation.   Thus the boundary lengths of a geodesic three-holed
sphere will be $b,b',b''$ (not $a,b,c$ as hitherto) and the $R$-symmetry twist parameters will be $\phi,\phi',\phi''$ (not $\phi_a,\phi_b,\phi_c$).

\subsubsection{The Supersymmetric Setup}\label{susysetup}

We now consider the kernel that appears in the Mirzakhani-style recursion relation for the $\cN=2$ volumes.   We consider an $\cN=2$  three-holed sphere
$Y$ with geodesic boundaries $\gamma,\gamma',\gamma''$, and boundary   data $\b=(b,\phi)$, $\b'=(b',\phi')$, $\b''=(b'',\phi'')$, and $\OSp(2|2)$ holonomies $U,V,W$.
 We want to find the measure $\h {\sf T}(\b,\b',\b'')$ of
the set $\mathcal S$ consisting of points $p\in \gamma$ such that the geodesic $\ell_p$ orthogonal to $\gamma$ at $p$ reaches $\gamma'$.   
The answer is actually  the obvious superanalog of eqn. (\ref{crossr}).    If $P_0,P_3$ are the endpoints of the $U$-invariant geodesic $\h\gamma\subset \h H$,
and  $P_1,P_2$ are the endpoints of a $V$-invariant geodesic $\h\gamma'\subset \h H$,
then
\be\label{obviousanalog} \h {\sf T}=\log \frac{\langle 0,2\rangle\langle 1,3\rangle}{\langle 0,1\rangle\langle 2,3\rangle},\ee
similarly to eqn. (\ref{crossr}).   The $\N=2$ version of the symbol $\la~,~\ra$ was defined in eqn. (\ref{n2symbol}).

To justify eqn. (\ref{obviousanalog}), we first need to explain what is meant by the terms ``geodesics,'' ``orthogonal,'' and so on.   For $\N=1$ super Riemann surfaces, such
a discussion was given in appendix D of \cite{SW}, leading to the obvious $\N=1$ superanalog of eqns. (\ref{crossr}) and (\ref{obviousanalog}).   A similar discussion applies
for $\N=2$, assuming that one works with non-chiral $\N=2$ super Riemann surfaces, as we will now  explain.

First,  what is meant by a ``geodesic'' in $\h H$?   We recall that the $\N=2$ super upper half plane $\h H$ contains the ordinary upper half plane $H$, embedded
at $\uptheta=\uptheta'=0$.   We want to regard an ordinary geodesic in $H$ as a geodesic in $\h H$.  We also want the notion of a geodesic to be $\OSp(2|2)$-invariant.
So we declare a geodesic in $\h H$  to be any curve -- that is, any submanifold of dimension $1|0$  -- that is equivalent under $\OSp(2|2)$ to a geodesic in $H$.
Note that this definition implies that there is a well-defined length element along any geodesic $\gamma\subset \h H$: we pick an $\OSp(2|2)$ transformation that maps
$\gamma$ to $H$ and then we use the ordinary length element for a geodesic in $H$.   Another consequence of the definition is that a geodesic in $\h H$ has 
endpoints in what we defined in section \ref{nonchiral} as $\partial\h H$, the conformal boundary of $\h H$, parametrized by real $x|\uptheta,\uptheta'$.   Indeed, if $\gamma\subset H$,
then it has endpoints in the conformal boundary of $H$.   These are points of the form $x|0,0$, with real $x$, so they are in $\partial \h H$.  Applying an $\OSp(2|2)$ transformation
and using the fact that $\partial \h H$ is $\OSp(2|2)$-invariant, we learn that any geodesic in $\h H$ has endpoints in $\partial \h H$.  We can also describe this by saying that
any geodesic in $\h H$ has real endpoints (that is, endpoints whose coordinates $x|\uptheta,\uptheta'$ are all real).    \

Now, if $\gamma$ is a geodesic in $\h H$ and $p\in \gamma$, we want to define the geodesic $\ell_p$ that is orthogonal to $\gamma$ at $p$.  If $\gamma\in H$,
we just use the classical definition of the orthogonal geodesic $\ell_p$. Any case can be reduced by this one by an $\OSp(2|2)$ transformation.   We also need to know
what it means to say that a geodesic $\ell_p$ is self-intersecting or intersects another geodesic $\gamma'$.   Since a geodesic is a curve of dimension $1|0$, 
two geodesics in a super Riemann surface of dimension $2|4$ generically do not have a point in common, and indeed asking for two geodesics to have a point in common
 is not the right condition. 
One way to explain the right condition (slightly different from but equivalent to the explanation in \cite{SW} for $\N=1$) 
is to use the fact that we are interested in hyperbolic super Riemann surfaces.  This means
that in discussing whether two geodesics intersect, we can lift the situation to $\h H$ and give a definition there.
We  say that an intersection of two geodesics $\gamma_1,\gamma_2\subset \h H$ consists of a pair of points
$P_1\in\gamma_1, $ $P_2\in\gamma_2$ with $\la 1,2\ra=0$.  Since  there are two
real parameters in the choices of $P_1$ and $P_2$, the complex condition $\la 1,2\ra=0$, equivalent to two real conditions, is a good criterion for intersection of geodesics.

The special case of this that we most need is  the condition that two geodesics $\gamma_1$ and $\gamma_2$ have an endpoint in common.   Suppose that $\gamma_1$ has an endpoint with coordinates $x_1|\uptheta_1,\uptheta'_1$, and $\gamma_2$ has an endpoint  with coordinates $x_2|\uptheta_2,\uptheta'_2$. These coordinates are all real, since the endpoints of any geodesic are real.  The condition under which we say that $\gamma_1$ and $\gamma_2$
have a common endpoint  is that $\la 1,2\ra=0$, again using the pairing (\ref{n2symbol}).    
Since the coordinates are real, this is a single real condition, just as it would be classically.   

Now, going back to the $U$- and $V$-invariant geodesics $\h\gamma,\h\gamma'\subset \h U$, we pick a point
$p\in \h\gamma$ and ask if the orthogonal geodesic $\ell_p$ to $\gamma$ at $p$ has a common endpoint with $\h\gamma'$.   In the bosonic case, the picture corresponding to
this question is drawn
in fig. \ref{fig:recursionkernel}; two points $p$ satisfy the desired condition, one for each endpoint of $\h\gamma'$.   We have included the fermionic variables in such a way
that we did not change the number of equations (the condition $\la 1,2\ra=0$ on the endpoints  is a single real condition, reducing to $x_1=x_2$ in the classical limit)
or unknowns (we defined a geodesic to have dimension $1|0$).    So also supersymmetrically the condition that $\ell_p$ and $\h\gamma'$ have an endpoint in common
 determines two points in
$\h\gamma$.   These points are the endpoints of the set $\mathcal S\subset \h\gamma$ whose length is supposed to be $\h {\sf T}$.   

At this point, then, to compute $\h {\sf T}$ we have to analyze the situation of fig. \ref{fig:recursionkernel} in the supersymmetric case.    It might seem that we have to do a genuinely
new supersymmetric calculation, but actually one can justify the  obvious answer of eqn. (\ref{obviousanalog}) without any really new calculation.
First, we can make an $\OSp(2|2)$ transformation that maps $\h\gamma$ to $H$ and, of course, also acts on $\h\gamma'$.   (We can even pick the $\OSp(2|2)$ transformation
so that $\h\gamma$ is a vertical geodesic in $H$, as pictured in  fig. \ref{fig:recursionkernel}.   But this is not necessary for the argument.)   Then the endpoints $P_0$, $P_3$
of $\h \gamma$ are at $x_0|0,0$ and $x_3|0,0$.   Since $\h\gamma\subset H$, by definition all geodesics orthogonal to $\h\gamma$ are in $H$ also, so for any $p\in\h\gamma$,
an endpoint $Q$ of $\ell_p$ is in $H$, say at $\t x|0,0$. The endpoints $P_1,P_2$ of $\h\gamma'$ are generically not in $H$; they are at, say, $x_1|\uptheta_1,\uptheta'_1$
and $x_2|\uptheta_2,\uptheta_2'$.   But since $Q$ is at $\uptheta=\uptheta'=0$, the conditions $\la 1,Q\ra=0$ and $\la 2,Q\ra=0$ do not depend on the fermionic coordinates of
$P_1,P_2$; those conditions reduce to $y=x_1$ and $y=x_2$.  At this point, the calculation of the length of the set $\mathcal S$ reduces precisely to the calculation that we already
explained in the purely bosonic case, leading to \be\label{interimform} \h {\sf T}=\log \frac{(x_0-x_2)(x_1-x_3)}{(x_0-x_1)(x_2-x_3)},\ee as in eqn. (\ref{crossr}).   

This is the answer in a coordinate system with $\h\gamma\in H$, implying that the points $P_0$, $P_3$ are at $x_0|0,0$ and $x_3|0,0$.   We would like a formula
for $\h{\sf T}$ that does not depend on this condition.   To find such a formula, we just note that the fact that $P_0,P_3$ have vanishing fermionic coordinates implies
that $x_0-x_2=\la 0,2\ra$, $x_1-x_3=\la 1,3\ra$, etc.   So the covariant version of the answer is just eqn. (\ref{obviousanalog}).   

One point that might puzzle the reader is that in analyzing the function $\h {\sf T}$, we made use of the real structure of $\OSp(2|2)\cong \SU(1,1|1)$, but 
 in computing the measure on moduli space
we ignored this
real structure (at least as regards the fermionic moduli).  There is actually no contradiction.   $\SU(1,1|1)$ does have a real structure,
and we are free to use it, if convenient, in computing the function $\h {\sf T}$.   On the other hand, the measure can be analytically continued to complex values of the moduli,
and in computing the measure, we can disregard the real structure if that is convenient.

\subsubsection{Evaluation of the Kernel Appearing in the Recursion}\label{eval}

It remains to reduce eqn. (\ref{obviousanalog}) to an explicit formula.   Unfortunately, this will lead us through some rather complicated formulas, though the simplicity of the final
results suggests that there might be a simpler route to those results.

The first step is to find explicit formulas for the endpoints $P_0,\cdots,P_3$ that appear in this derivation,   In doing so, we will describe these points by homogeneous coordinates
 $u_0,v_0|\uptheta_0,\uptheta_0'$, $u_1,v_1|\uptheta_1,\uptheta_1'$, $u_2,v_2|\uptheta_2,\uptheta_2'$
 and $u_3,v_3|\uptheta_3,\uptheta_3'$. 
 We view the holonomies $U,V$ as $2|2\times 2|2$ matrices 
 acting on $u,v | \uptheta,\uptheta'$. 
 These matrices have two bosonic and two fermionic eigenvectors.   The bosonic eigenvectors of $U$ are the points $P_0,P_3$, and the bosonic eigenvectors of $V$ are the points
 $P_1,P_2$.  
After a detailed calculation, we find the explicit eigenvectors:
\bea \notag
u_0,v_0|\uptheta_0,\uptheta'_0&=&-\frac{\kappa}{2\sinh \frac{b}{2}},1 \Big|\frac{\psi_2+\frac{\kappa}{ 2 \sinh \frac{b}{2}} \psi_4}{\sqrt{2}(1-e^{\i \hq \phi-\frac{b}{2}})} , -\i\frac{\psi_2 + \frac{\kappa}{2 \sinh \frac{b}{2}} \psi_4 }{\sqrt{2}(1-e^{\i \hq \phi-\frac{b}{2}})}, \\ \notag
u_1,v_1|\uptheta_1,\uptheta'_1&=& 0,1 \Big| \frac{\psi_1}{2\sqrt{2}(e^{\frac{b'}{2}-\i \hq \phi'}-1)} ,\i\frac{\psi_1}{2\sqrt{2}(e^{\frac{b'}{2}-\i \hq \phi'}-1)}  ,\\ \notag
u_2,v_2|\uptheta_2,\uptheta'_2&=& -2\sinh \frac{b'}{2},1\Big|\frac{\psi_1 - 2 \sinh \frac{b'}{2} \psi_3}{2\sqrt{2}(e^{-\frac{b'}{2}-\i \hq \phi'}-1)} ,\i \frac{\psi_1 - 2 \sinh \frac{b'}{2} \psi_3}{2\sqrt{2}(e^{-\frac{b'}{2}-\i \hq \phi'}-1)},\\
u_3,v_3|\uptheta_3,\uptheta'_3 &=& 1,0 \Big| \frac{\psi_4}{\sqrt{2}(-1+e^{\frac{b}{2}+\i \hq \phi})},\frac{-\i\psi_4}{\sqrt{2}(-1+e^{\frac{b}{2}+\i \hq \phi})}. \label{longform}
\ea
As explained in section \ref{susysetup}, we now have to evaluate
\beq
\widehat{{\sf T}} = \log \frac{\lb 0,2\rb \lb 1,3\rb }{ \lb 0,1\rb \lb 2,3\rb}=\log\left[ \frac{u_0v_2-u_2v_0-\uptheta_0\uptheta_2-\uptheta'_0\uptheta'_2}{u_0v_1-u_1 v_0-\uptheta_0\uptheta_1-\uptheta'_0\uptheta'_1}\frac{u_1v_3-u_3v_1-\uptheta_1\uptheta_3-\uptheta'_1\uptheta'_3}{u_2v_3-u_3 v_2-\uptheta_2\uptheta_3-\uptheta'_2\uptheta'_3}\right].
\eeq
This function depends on the three geodesic lengths, the two independent $\uU(1)$ holonomies, and the four fermionic moduli.  An explicit formula, using eqn. (\ref{longform})
for the eigenvectors, is
\bea
\widehat{{\sf T}} =\log \left[\frac{-\kappa+2\sinh \frac{b}{2}\, 2 \sinh \frac{b'}{2}+\frac{(\kappa \psi_4 + 2 \sinh \frac{b}{2} \psi_2)(\psi_1 - 2 \sinh \frac{b'}{2} \psi_3)}{2\big(1-e^{-\frac{b}{2}+\i \hq \phi}\big)\big(1-e^{-\frac{b'}{2}-\i \hq \phi'}\big)}}{- \kappa +\frac{(\kappa \psi_4 + 2 \sinh \frac{b}{2} \psi_2)\psi_1}{2\big(1-e^{-\frac{b}{2}+\i \hq \phi}\big)\big(1-e^{\frac{b'}{2}-\i \hq \phi'}\big)}}\frac{1-\frac{\psi_1 \psi_4}{2\big(e^{\frac{b}{2}+\i \hq \phi}-1\big)\big(1-e^{\frac{b'}{2}-\i \hq \phi'}\big)}}{1+\frac{(\psi_1-2\sinh \frac{b'}{2} \psi_3)\psi_4}{2\big(e^{\frac{b}{2}+\i \hq \phi}-1\big)\big(e^{-\frac{b'}{2}-\i \hq \phi'}-1\big)}}\right]\nonumber
\ea
To make the dependence on fermionic moduli explicit, we can use the expansion   $\kappa = \kappa_0 + \sum_{i<j} \kappa_{ij} \psi_i \psi_j + \kappa_4 \psi_1\ldots \psi_4$ presented in \eqref{eq:Kexp}. Then we expand  $\widehat{{\sf T}}$
\beq
\widehat{{\sf T}}=\widehat{{\sf T}}_0 +\sum_{i<j} \widehat{{\sf T}}_{ij} \psi_i\psi_j + \widehat{{\sf T}}_4 \psi_1 \psi_2 \psi_3 \psi_4,
\eeq
Finding the explicit formulas for $\widehat{{\sf T}}_{ij}$ and $\widehat{{\sf T}}_4$ is straightforward but tedious. The function $\widehat{{\sf T}}_0$ is simpler and, importantly, is equal to the bosonic length appearing in the original Mirzakhani recursion
\bea
\widehat{{\sf T}}_0 = {\sf T}_{\rm bos.} = \log \left[\frac{\cosh \frac{b''}{2}+ \cosh \frac{b+b'}{2} }{\cosh \frac{b''}{2}+ \cosh \frac{b-b'}{2} }\right].
\ea
We will not attempt to write down the answers for $\widehat{{\sf T}}_{ij}$ and $\widehat{{\sf T}}_4$, partly because they are long, and also because we ultimately only need the simpler functions that arise after  fermionic integration. 

We are now ready to complete the calculation, and integrate $\h{\sf T}$ over $\mathcal{M}_{0,3}$, the $0|4$ dimensional moduli space of the three-holed sphere.   We define
\beq
{\sf T}({\sf b},{\sf b}',{\sf b}'') = 2\pi \, \int_{\mathcal{M}_{0,3}} \mu_{0,3} \cdot \widehat{{\sf T}} .
\eeq
We included a factor of $2\pi$ on the right hand side to simplify the final form of the recursion. The measure obtained in the previous section is
\beq
\mu_{0,3} = - \frac{2 e^{\i  \hq \phi - \i  \hq \phi'} (\cosh \frac{b''}{2} + \cos  \hq \phi'')}{(2\pi)} (1+K)\delta(\phi+\phi'+\phi''-F) [\d^4 \psi]
\eeq
 When integrating $\mu_{0,3} \cdot \widehat{\sf T}$ over fermionic moduli there are three terms, depending on whether we pick $\widehat{{\sf T}}_0$, $\widehat{{\sf T}}_{ij}$ or $\widehat{{\sf T}}_4$ to absorb the fermionic integrations. The three terms  are given schematically by
\begin{align}\notag
\int [\d^4 \psi]~  \delta(\phi+\phi'+\phi''-F)\,(1+K) \, \widehat{{\sf T}}_0 & \propto \delta''(\sum\phi) \widehat{{\sf T}}_0,\\ \notag
\int [\d^4 \psi] \delta(\phi+\phi'+\phi''-F)\,(1+K) \,\widehat{{\sf T}}_{ij} \psi_i\psi_j& = \delta'(\sum\phi)\, f_{ij}  \widehat{{\sf T}}_{ij}+ \delta(\sum \phi) ~k_{ij}  \widehat{{\sf T}}_{ij},
\\
\int [\d^4 \psi] \delta(\phi+\phi'+\phi''-F)\,(1+K) \,\widehat{{\sf T}}_4 &= \delta(\sum \phi ) \widehat{{\sf T}}_{4}.
\end{align}
The first term involving $\widehat{{\sf T}}_0$ was essentially already computed in section \ref{sec:GPI}. Since $\widehat{{\sf T}}_0$ is already independent of fermionic moduli, the answer is proportional to the quantity $V_{0,3}$, computed in that section. The second term involves the new function $\widehat{{\sf T}}_{ij}$. Surprisingly, we find that $f_{ij}\widehat{\sf T}_{ij}=0$. Finally, we are left with two contributions proportional to $\delta(\phi+\phi'+\phi'')$, from $k_{ij} \widehat{{\sf T}}_{ij}$ and from $\widehat{{\sf T}}_4$. The sum of these  is surprisingly simple: 
\beq
k_{ij} \widehat{{\sf T}}_{ij} + \widehat{{\sf T}}_4 =\frac{e^{\i  \hq \phi' - \i \hq  \phi }}{2(\cosh \frac{b''}{2} + \cos  \hq \phi'')}\Big[  \frac{1}{16 \cosh^2 \frac{b+b'+b''}{4}}-\frac{1}{16 \cosh^2 \frac{b-b'+b''}{4}}+(b''\to-b'')\Big].\nonumber
\eeq  
There might be a more direct route to this relatively simple formula, but this is not obvious.   The final result for $\sf T$ is presented in eqn. (\ref{finalt}) below.

To formulate the recursion relation, we need one more function ${\sf D}({\sf b},{\sf b}',{\sf b}'')$.
This is the integral over the fermionic moduli  of the segment of $\gamma_b$ from which orthogonal geodesics reach neither $\gamma_{b'}$ nor $\gamma_{b''}$. It
can be computed from the following sum rule:
\beq\label{loggo}
2\pi\int_{\mM_{0,3}} \mu_{0,3} \cdot b ={\sf D}({\sf b},{\sf b}',{\sf b}'') + {\sf T}({\sf b},{\sf b}',{\sf b}'') + {\sf T}({\sf b},{\sf b}'',{\sf b}').
\eeq
It is important to formulate this sum rule after integration over moduli, since the way we parametrize the fermionic moduli spoils the symmetry between the $b'$ and $b''$ boundaries. The extra factor of $2\pi$ was introduced for future convenience, like the rescaling of ${\sf T}$. As opposed to the case of $\cN=1$ super Riemann surfaces, the left hand side of
eqn. (\ref{loggo}) does  not vanish and produces an answer proportional to $V_{0,3}$.  The result for  ${\sf D}({\sf b},{\sf b}',{\sf b}'')$ is stated in eqn. (\ref{finald}).

We are now ready to write down the final version of the $\N=2$ recursion relation.

\subsection{Recursion Relation for $\cN=2$ Volumes}\label{sec:RRfinalform}

\subsubsection{Final Evaluation} 

In this section we will summarize the results so far and write down the recursion relation for the Weil-Petersson volumes for $\cN=2$ hyperbolic 
surfaces in a way that is self-contained and can be read independently from the details of the derivation outlined so far.

The computation in  section \ref{eval} leads to explicit formulas for the two kernels ${\sf T}({\sf b},{\sf b}',{\sf b}'')$ and ${\sf D}({\sf b},{\sf b}',{\sf b}'')$ that appear in the recursion relation. 
Here  we remind the reader that ${\sf b} = (b,\phi)$.  The two functions can be expanded
\bea\label{finalt}
{\sf T}({\sf b},{\sf b}',{\sf b}'') &=&-\frac{1}{4\hq^{\,2}}\delta''(\phi+\phi'+\phi''){\sf T}_2(b,b',b'')-\delta(\phi+\phi'+\phi'')  {\sf T}_0(b,b',b'')\\ \label{finald}
{\sf D}({\sf b},{\sf b}',{\sf b}'') &=&-\frac{1}{4\hq^{\,2}}\delta''(\phi+\phi'+\phi'') {\sf D}_2(b,b',b'')-\delta(\phi+\phi'+\phi'')  {\sf D}_0(b,b',b'').
\ea
 The functions ${\sf T}_{0}$, ${\sf T}_2$ and ${\sf D}_{0}$, ${\sf D}_2$ appearing on the right hand side depend  only on the lengths; the subscript indicates the number of derivatives acting on the delta function they multiply. ${\sf T}_{0}$, ${\sf T}_2$ are 
\bea
{\sf T}_2(b,b',b'') &=&\log \left[\frac{\cosh \frac{b''}{2}+\cosh \frac{b+b'}{2} }{\cosh \frac{b''}{2}+ \cosh \frac{b-b'}{2} }\right],\\
{\sf T}_0(b,b',b'') &=& \frac{1}{ 16\cosh^2 \frac{b+b'+b''}{4}}-\frac{1}{16\cosh^2 \frac{b-b'+b''}{4}}+(b''\to - b'').
\ea
${\sf D}_{0}$, ${\sf D}_2$  are obtained through the sum rule 
\bea
{\sf D}_2(b,b',b'') &=&b-{\sf T}_2(b,b',b'') - {\sf T}_2(b,b'',b')  \notag\\
{\sf D}_0(b,b',b'') &=&-{\sf T}_0(b,b',b'') - {\sf T}_0(b,b'',b') 
\ea
that comes from eqn. (\ref{loggo}) by integration over $\mM_{0,3}$.
One can check explicitly that 
\begin{align}{\sf T}_0(b,b',b'')& = \partial_{b'}^2 {\sf T}_2(b,b',b'')\cr {\sf D}_0(b,b',b'')&= \partial_{b'}^2 {\sf D}_2 (b,b',b''). \label{goodrel}\end{align} These relations are  not  accidents and will be crucial to match the $\cN=2$ generalization of the Mirzakhani recursion relation with the matrix ensemble.

We have now all the ingredients we need to obtain the final form of the recursion relation that computes the volumes of moduli spaces of $\mathcal{N}=2$ hyperbolic surfaces\\
\bea
\hspace{-1cm} b V_{g,n+1} ({\sf b}, {\sf B}) &=& \frac{1}{2}(2\pi) \int_0^{\infty}b' \d b'~\int_0^{2\pi}\d \phi' \int_0^{\infty}b'' \d b''~\int_0^{2\pi}\d \phi'' \, {\sf D} ({\sf b},{\sf b}',{\sf b}'')\nn
&&\hspace{-1.5cm}\times \left(V_{g-1,n+2}(\bar{{\sf b}}',\bar{{\sf b}}'',{\sf B}) + \sum_{\rm stable} V_{h_1,|{\sf B}_1|+1} (\bar{{\sf b}}',{\sf B}_1) V_{h_2,|{\sf B}_2|+1} (\bar{{\sf b}}'',{\sf B}_2) \right)\nn
&&\hspace{-1.5cm}+\sum_{k=1}^{|{\sf B}|} \int_0^\infty b' \d b'\int_0^{2\pi} \d \phi' \, \Big[ -\frac{b}{4\hq^{\,2}} \delta''(\phi+\phi'+\phi_k) -{\sf T}({\sf b},{\sf b}',{\sf b}_k) \Big] V_{g,n} (\bar{{\sf b}}',{\sf B} /{\sf b}_k) . \label{eq:N2recrel}
\ea\\
Here ${\sf B}=\{ {\sf b}_1,\ldots, {\sf b}_n\}$ is the set of boundary parameters, and following our notation an overline denote orientation reversal for a given boundary, so that if ${\sf b}=(b,\phi)$ then $\bar{{\sf b}} = (b,-\phi)$.  The coefficients in this relation have precisely the same interpretation
as in the bosonic recursion relation (\ref{eq:Mirzakhanirec}), except that they have been integrated over the fermionic moduli of a three-holed sphere. 
The qualification ``stable'' under the sum has the same meaning it had in the bosonic case reviewed earlier. Integration over the $\uU(1)$ twist parameters will produce a factor of $(2\pi)^2=\int \d\varphi' \d \varphi'' $ in the first line, and of $(2\pi)= \int \d \varphi'$ in the last. We will explain below the relevance of these seemingly innocent factors and the role they play. One of those factors cancels with a similar one in the normalization of $\mu_{0,3}$ and the normalization of the functions ${\sf T}$ and ${\sf D}$. 

As a final ingredient, we need  two initial conditions for this recursion relation.  One initial condition is familiar:
\beq
\label{eqn:V03final2}
V_{0,3}({\sf b}_1, {\sf b}_2, {\sf b}_3) = - \frac{1}{2\pi} \frac{1}{4\hq^{\,2}} \delta''(\phi_1+\phi_2+\phi_3).
\eeq
We also need $V_{1,1}$, which can be easily derived by imitating the derivation of the recursion with a slight modification:\footnote{A Riemann surface of genus 1 with one
hole or puncture has a $\Z_2$ symmetry.   As in \cite{SW}, by $V_{1,1}$, we mean the volume of $\mM_{1,1}$ in the orbifold sense, including a factor of $1/2$ because
every point in $\mM_{1,1}$ has this $\Z_2$ symmetry.   If we do not do this, the statement of the recursion relation (\ref{eq:N2recrel}) would be more awkward.}
\bea
V_{1,1} ({\sf b}) &=& \frac{1}{2b} \int_0^\infty b' \d b' \int_0^{2\pi} \d \phi' \,{\sf D}({\sf b},{\sf b}',\bar{{\sf b}}'),\nn
&=&- (2\pi) \frac{1}{4\hq^{\,2}}\delta''(\phi) \frac{b^2+4\pi^2}{48} - (2\pi) \delta(\phi) \frac{1}{8}
\ea
The recursion relation determines all $V_{g,n}$ in terms of these two.

Eqn. \eqref{eq:N2recrel} is the main result of this section.   As we show in section \ref{equivalence}, it  implies the duality between $\cN=2$ JT gravity and the 
random matrix ensemble.    First, we will present some general properties that follow from the recursion relation.

\subsubsection{$\cN=2$ Volumes at 	``Fixed Charge''}
The volume of the three-holed sphere with geodesic boundaries of fixed $\uU(1)$ holonomies $\phi,\phi',\phi''$ is proportional to $\delta''(\phi+\phi'+\phi'')$. Similarly the kernels appearing in the recursion are sums of delta functions, possibly with derivatives, of a sum of $\uU(1)$ holonomies.  This implies, by induction, that the same is true for the volume of every $\mM_{g,n}$. More precisely, from the recursion relation, one can deduce an expansion 
\beq
V_{g,n}({\sf B}) =(2\pi)^{2g-1} \sum^{2g-2+n}_{m=0} \frac{(-1)^m}{(4\hq^{\,2})^{m}} \delta^{(2m)}(\phi_{{\sf B}}) \, v_{g,n,m}(B),\label{eqn:volformphi}
\eeq
where we introduced some notation to avoid clutter here and later. On the right hand side, $2m$ is the number of derivatives acting on the delta function. 
For  ${\sf B}=\{ {\sf b}_1,\ldots, {\sf b}_n\}$, 
we define $\phi_{\sf B} = \phi_1 +\ldots +\phi_n$. The function $v_{g,n,m}(B)$ depends only on the lengths $B=\{ b_1,\ldots, b_n\}$. The choice of overall normalization in \eqref{eqn:volformphi} is for later convenience.

It is convenient to introduce  ``fixed charge'' volumes.
The simple dependence of $V_{g,n}(\sf B)$ on the holonomies implies the following condition for the Fourier transforms:
\bea\label{usefulfact}
\int \frac{\d^n\phi}{(2\pi)^n}\, e^{\i \sum_{i=1}^n q_i \phi_i }\, V_{g,n}({\sf B}) = 0,~~~{\rm unless}~~q_1 = q_2 \ldots = q_n.
\ea
Note that because $V_{g,n}({\sf B})$ is supported at $\phi_1+\cdots+\phi_n=0$, this integral makes sense and has the stated property as long as all differences
$q_i-q_j$ are integers; it is not necessary for the individual $q_i$ to be integers.
Vanishing of eqn. (\ref{usefulfact}) when the $q_i$ are not all equal is equivalent to the statement that different supermultiplets are statistically independent, to all orders
in the topological expansion.  
To study any one supermultiplet,  we can set $q_1 = \ldots =q_n = q$ and define the ``fixed charge'' volumes
\beq\label{eq:FCV}
V^{(q)}_{g,n}(B) =(2\pi)^{2-2g} \int \frac{\d^n\phi}{(2\pi)^n}\, e^{\i q \phi_{\sf B }}\, V_{g,n}({\sf B}).
\eeq
The normalization will simplify the recursion relation that these volumes satisfy. Here $q$ is not the $\uU(1)$ $R$-charge of a state, but a label of a supermultiplet.
 One can verify (see eqn. (\ref{volcompar})) that when these volumes are glued to trumpets in order to compute the path integral with NAdS boundary conditions, the $R$-charges of the supermultiplet states are $q \pm \hq /2$; therefore $q$ has the same interpretation as in section \ref{sec:N2JTG}.  

In Fourier space,  the ``fixed charge'' volumes are simply polynomials in $q^2$ of the form
\beq
V^{(q)}_{g,n}(B) = \sum^{2g-2+n}_{m=1} \Big(\frac{q^{2}}{4\hq^{\,2}}\Big)^m v_{g,n,m}(b_1,\ldots, b_n).
\eeq
One can easily prove this property by induction on the recursion relation, without going through \eqref{eqn:volformphi}, since the Fourier transforms of the kernels have 
this  property. One can also prove, imitating the proof by Mirzakhani, that the coefficient of a given power of $q^2$ in the fixed charge volumes is a polynomial in the variables $b_i^2$, of a degree we will specify below. 

For future reference in proving the match with the  matrix model, it is useful to translate the recursion for the volumes $V_{g,n}({\sf B})$ into a recursion for the ``fixed charge'' volumes $V_{g,n}^{(q)}(B)$. One can simply apply a Fourier transform on \eqref{eq:N2recrel}. The result is
\bea
b V^{(q)}_{g,n+1} (b, B) &=& \frac{1}{2}\int_0^\infty  b'\text{d}b' ~b''\text{d}b''~\Big(\frac{q^2}{4\hq^{\,2}}\, {\sf D}_2 (b,b',b'') - {\sf D}_0(b,b',b'')\Big) \nonumber\\
&& \hspace{.5cm} \times \left(V^{(q)}_{g-1,n+2}(b',b'',B)+ \sum_{\rm stable} V^{(q)}_{h_1,|B_1|+1} (b',B_1) V^{(q)}_{h_2,|B_2|+1} (b'',B_2) \right)\nn
&&\hspace{-1cm}+\sum_{k=1}^{|B|} \int_0^\infty b'\text{d}b' \Big[ \frac{q^2}{4\hq^{\,2}}( b -{\sf T}_2(b,b',b_k)) +{\sf T}_0(b,b',b_k) \Big] V^{(q)}_{g,n} (b',B /b_k) \label{eq:N2recrelFC}
\ea
This way of presenting the recursion also avoids the cluttering associated to keeping track of the holonomies. The purpose of normalizing the ``fixed charge'' volumes with a factor of $(2\pi)^{2-2g}$ in \eqref{eq:FCV} was to obtain \eqref{eq:N2recrelFC} in this form. The initial conditions for this recursion can be read off from the previous results:
\bea
V_{0,3}^{(q)} =\frac{q^2}{4\hq^{\,2}} \underbrace{1}_{v_{0,3,1}},~~~~V_{1,1}^{(q)} = \frac{q^2}{4\hq^{\,2}} \underbrace{\frac{b^2+4\pi^2}{48}}_{v_{1,1,1}} +\underbrace{(- \frac{1}{8})}_{v_{1,1,0}}.\label{eq:incondfcr}
\ea

\subsubsection{Relation to Bosonic Volumes}

There is a simple property we can notice easily starting from \eqref{eq:N2recrelFC}. Consider the coefficient $v_{g,n,m}$ with the highest number of derivatives, or equivalently the highest power of $q$, with $m=2g-2+n$. This coefficient satisfies a simplified version of the  recursion \eqref{eq:N2recrelFC} in which we keep only the terms proportional to $q^2/4\hq^{\,2}$ in the recursion kernels. Up to a simple rescaling, this
is precisely the same as the bosonic volume recursion of Mirzakhani, given in \eqref{eq:Mirzakhanirec}. A careful analysis of the normalization therefore yields the relation
\bea
v_{g,n,2g-2+n}(B) = V^{\rm bos.}_{g,n}(B).
\ea
This property holds  for the initial conditions given in eqn. \eqref{eq:incondfcr}, and the recursion relation implies that it is true for all $g,n$.
 We can rewrite the same property in terms of the volumes at fixed holonomies as
\bea
V_{g,n}({\sf B}) = (2\pi)^{2g-2} \, V^{\rm bos}_{g,n}(B) \, (-1)^n \frac{2\pi \delta^{(4g-4+2n)}(\phi_{\sf B})}{(4\hq^{\,2})^{2g-2+n}} +\ldots.
\ea  
In bosonic JT gravity coupled to a $\uU(1)$ gauge theory, one would get a similar result, but without the factors $(-1)^n/(4 \hq^{\, 2})^{2g-2+n}$ 
and without the derivatives acting on the delta function. The derivatives of the delta function have their origin in the fermionic moduli, and
the other extra factors are due to choices of normalization in $\cN=2$ JT gravity. 

This relation makes it clear that $v_{g,n,2g-2+n}$ is a polynomial in $b_i^2$ of degree $3g-3+n$, since it follows from Mirzakhani's results. Moreover, one can show using similar techniques that $v_{g,n,2g-2+n-s}$ is a polynomial in $b_i^2$ of degree $3g-3+n-s$. This implies that $v_{g,n,0}$ is a polynomial of degree $g-1$ in $b_i^2$. This is  true even for $g=0$, as $v_{0,n,0}=0$. 

\subsubsection{Some Simple Examples} 
Next we give some concrete simple examples of $\cN=2$ volumes obtained from the recursion relation. 

For any genus $g\geq 1$, there can be terms in the volumes with no derivatives acting on the delta function, since this is true for the seed $V_{1,1}({\sf b})$. That is not so for the case of genus zero, since the seed $V_{0,3}$ has two derivatives. Therefore all volumes at genus zero are sums of expressions with at least two derivatives. We can analytically compute the two-derivative coefficients and verify that  they are non-zero. After performing the integrals over holonomies, the recursion  for $v_{0,n,1}$ reduces to
\beq\label{twoderiv}
b v_{0,n+1,1}(b,B) =\sum_{k=1}^{|B|} \int_0^\infty b'\d b' ~{\sf T}_0(b,b',b_k) ~v_{0,n,1}(b',B/b_k).
\eeq
One can prove by direct evaluation that the following ansatz for the term with two derivatives solves this recursion:
 \beq
 V_{0,n}({\sf B}) =(2\pi)^{-1}(-1)^{n}\frac{(n-1)!}{2} \frac{1}{4\hq^{\,2}}\delta''(\phi_1 + \ldots + \phi_n) + ({\rm higher~derivatives}).
 \eeq
or equivalently $v_{0,n,1} =(-1)^{n+1}\frac{(n-1)!}{2}$. 

To illustrate some of the preceding  statements, we can compute $V_{0,4}^{(q)}$ and $V_{0,5}^{(q)}$:
\bea
V_{0,4}^{(q)}(b_1,b_2,b_3,b_4) &=& \frac{q^4}{4^2\hq^{\,4}}\underbrace{ \frac{4\pi^2 +b_1^2 +b_2^2+b_3^2+b_4^2}{2}}_{v_{0,4,2}=V^{\rm bos.}_{0,4}(B)} +\frac{q^2}{4\hq^{\,2}} \underbrace{(-3)}_{v_{0,4,1}} ,\\
V_{0,5}^{(q)}(b_1,b_2,b_3,b_4,b_5) &=&  \frac{q^{6}}{4^3\hq^{\,3}} V^{\rm bos.}_{0,5}(B) - \frac{q^4}{4^2\hq^{\,4}}\frac{52\pi^2 + 9 \sum_{j=1}^5 b_j^2}{2} + \frac{q^2}{4\hq^{\,2}} 12.
\ea\\
From these expressions, we can recover
 the bosonic volumes and  the two derivative term predicted in eqn. (\ref{twoderiv}).  We see as well that in these examples, the volumes
 are polynomials in the variables $b_i^2$ of the expected degree.

\subsection{Equivalence with Random Matrix Recursion}\label{equivalence}
Now we will finally complete the circle of ideas and show that the $\cN=2$ JT gravity path integral is equal to the result from the random matrix ensemble that we have described. 
We recall eqn. (\ref{eq:Zgnnn}) expressing partition functions in terms of volumes:
\bea
Z_{g,n}= \left[ \prod_{j=1}^n \int_0^\infty (b_j \d b_j) \int_0^{2\pi}(2\pi \d \phi_j) Z_{\rm tr} (\beta_j,\alpha_j;\overline{{\sf b}}_j)\right] V_{g,n}({\sf b}_1,\ldots, {\sf b}_n).\nonumber
\ea
We need a more explicit version of this formula. For this, we use the explicit form  (\ref{trumpet2}) of the $\cN=2$ JT gravity trumpet to rewrite this as
\bea
\hspace{-0.5cm}Z_{g,n}&=& (2\pi)^{n}\sum_{q_1,\ldots,q_n\in \mathbb{Z}+\delta-\frac{1}{2}}\, \prod_{i=1}^n \left[\big(e^{\i\alpha_i (q_i-\frac{\hq}{2})}+e^{\i\alpha_i (q_i +\frac{\hq}{2})}\big) e^{-\beta_i E_0(q_i)}\right]\nonumber\\
&&\hspace{-0.5cm}\times \int_0^\infty \prod_{j=1}^n b_j \d b_j \, Z_{\rm tr}^{\rm bos.}(\beta,b) \int_0^{2\pi} \prod_{j=1}^n\frac{\d \phi_j}{2\pi} e^{\i \phi_j q_j} \,V_{g,n}({\sf b}_1,\ldots,{\sf b}_n),
\ea
where the bosonic trumpet is $Z_{\rm tr}^{\rm bos.}(\beta,b) =e^{-\frac{b^2}{4\beta}}/(2\sqrt{\pi \beta} )$. 
As in the discussion of eqn. (\ref{usefulfact}), the integration over the $\phi_i$ has simple properties because the integrand is supported at $\phi_1+\cdots+\phi_n=0$ and
the differences of the exponents $q_i-q_j$ are integers.
Then we obtain a partition function consistent,  to all orders in $e^{-S_0}$, with statistically independent supermultiplets 
\beq\label{volcompar}
Z_{g,n}= (2\pi)^{2g+n-2}\sum_{q\in \mathbb{Z}+\delta-\frac{1}{2}} \prod_{i=1}^n \Big(e^{\i\alpha_i (q-\frac{\hq}{2})}+e^{\i\alpha_i (q+\frac{\hq}{2})}\Big)  Z_{g,n}^{(q)}(\beta_1,\ldots,\beta_n),
\eeq
where the contribution from each supermultiplet $Z_{g,n}^{(q)}$ is given by\footnote{We can now complete the proof that the number of BPS states does not fluctuate in gravity. Since, as we argued above, $V^{(q)}_{g,n}$ are polynomials in $b_i^2$, if $E_0(q)>0$, the integral over the $b_i$ cannot produce a dependence with $\beta$ that cancels the exponential decay $e^{-\beta E_0(q)}$ at large $\beta$. For the multiplet with $E_0(0)=0$ it would require a pole at $b=0$ to modify the number of BPS states, also ruled out by the fact that $V_{g,n}^{(q)}$ are polynomials. \label{footnoteBPSfluct}}
\beq
Z_{g,n}^{(q)} (\beta_1,\ldots,\beta_n)=  \left[\prod_{j=1}^n  \int_0^\infty b_j \d b_j  \, Z_{\rm tr}^{\rm bos.}(\beta_i,b_i)e^{-\beta_i E_0(q)}\right]V^{(q)}_{g,n}(b_1,\ldots,b_n).\label{Zqg}
\eeq
The factor of $(2\pi)^{2g+n-2}$ can be absorbed by a shift of $S_0$, since the exponent is minus the Euler characteristic of the surface. 
For this reason, we start by focusing on $Z_{g,n}^{(q)}$. 

To show the equivalence with the matrix model recursion, we will proceed in the following way.   Let  $Z_{g,n,\M}^{(q)}$  be the solution of the matrix model recursion of an $(\upalpha_0,\upbeta)=(1,2)$ AZ ensemble with spectral curve $y_q(x)$ and threshold energy $E_0(q)$, both for now generic.\footnote{$\N=2$ JT supergravity has a number of BPS states proportional to $e^{S_0}$. We show in appendix \ref{app:Loop} that the loop equations of an AZ ensemble with measure $(\upalpha,\upbeta)=(1+2\nu,2)$ are independent of $\nu$ when $\nu$ is proportional to $L$ (apart for allowing a pole in the spectral curve at $x=0$). Therefore the loop equations only depend on $\upalpha_0 =\upalpha- \upbeta \nu$ which we take to be one.}  
Similarly, define matrix model ``volumes'':
\beq
Z_{g,n,\M}^{(q)} (\beta_1,\ldots,\beta_n)=  \left[\prod_{j=1}^n  \int_0^\infty b_j \d b_j Z_{\rm tr}^{\rm bos.}(\beta_i,b_i)e^{-\beta_i E_0(q)}\right]V^{(q)}_{g,n,\M}(b_1,\ldots,b_n).\label{Zq}
\eeq
We want to show
that $Z_{g,n,\M}^{(q)}$, with the right choice of spectral curve, obeys the same recursion relation as $Z_{g,n}^{(q)}$.   
 A small generalization of a calculation in \cite{Eynard:2007fi}, which we explain in appendix \ref{app:Loop}, shows that one can rewrite the matrix model loop equations (presented in eqn. \eqref{eq:RRF}) in terms of the matrix model volumes, as
 \bea
b V^{(q)}_{g,n+1,\M} (b, B) &=& \frac{1}{2} \int_0^{\infty} b'\text{d}b' \int_0^{\infty} b''\text{d}b''~D_q(b'+b'',b)\nn
&&\hspace{-0.5cm}\times \left(V^{(q)}_{g-1,n+2,\M}(b',b'',B) + \sum_{\rm stable} V^{(q)}_{h_1,|B_1|+1,\M} (b',B_1) V^{(q)}_{h_2,|B_2|+1,\M} (b'',B_2) \right)\nn
&&\hspace{-0.5cm}+ \frac{1}{2}\sum_{k=1}^{|B|} \int_0^{\infty} b'\text{d}b' \Big(D_q(b'+b_k,b) + D_q(b'-b_k,b) \Big) V^{(q)}_{g,n,\M} (b',B /b_k) \label{eq:RRMMV}
\ea
where the function $D_q(x,y)$ depends on the spectral curve $y_q(x)$ through the following integral transform
\beq\label{dfunction}
D_q(x,y) = -\int_{\varepsilon+\i \mathbb{R}} \frac{\d z}{2\pi \i} \frac{e^{-xz} \sinh (yz) }{z y_q(E_0(q)-z^2)}.
\eeq
The contour of integration is along the imaginary axis with a small positive real part $\varepsilon>0$. This result was deduced in  \cite{Eynard:2007fi}  for a Dyson ensemble with $\upbeta=2$, but is also valid for an $(\upalpha_0,\upbeta)=(1,2)$ AZ ensemble,
since the matrix model recursion is the same.\footnote{The matrix model recursion is different  for ensembles describing theories with time-reversal symmetries.}
The first obvious observation is that the threshold energy has to match the one computed in section \ref{sec:N2JTG} for a supermultiplet with average charge $q$, which is  $E_0(q)=q^2/4\hq^{\,2}$.  We can derive the following identities by summing over residues 
\bea
&&-\int_{\varepsilon+\i \mathbb{R}} \frac{\d z}{2\pi \i} \frac{4\pi }{z \sin 2\pi z} e^{-x z} \sinh (y z) = y + \log  \Big(\frac{\cosh^2\frac{x-y}{4}}{\cosh^2\frac{x+y}{4}} \Big), \\
&&-\int_{\varepsilon+\i \mathbb{R}} \frac{\d z}{2\pi \i} \frac{4\pi z^2}{z \sin 2\pi z} e^{-x z} \sinh (y z) = - \frac{\sinh \frac{x}{2} \sinh \frac{y}{2}}{2(\cosh\frac{x}{2} + \cosh \frac{y}{2})^2}
\ea 
The first and second lines are also equal to 
\bea
&&-\int_{\varepsilon+\i \mathbb{R}} \frac{\d z}{2\pi \i} \frac{4\pi }{z \sin 2\pi z} e^{-x z} \sinh (y z) = y - {\sf T}_2(y,b',x-b') - {\sf T}_2(y,x-b',b'),\\
&&- \int_{\varepsilon+\i \mathbb{R}} \frac{\d z}{2\pi \i} \frac{4\pi z^2}{z \sin 2\pi z} e^{-x z} \sinh (y z) =-    {\sf T}_0(y,b',x-b')-{\sf T}_0(y,x-b',b')
\ea
These relations imply that the recursion relation we derived for the volumes of moduli spaces of $\cN=2$ hyperbolic surfaces, at ``fixed charge'' $q$, is equivalent to the matrix model loop equation for spectral curve $1/y_q(q^2/4\hq^{\,2} - z^2) \sim (q^2/4\hq^{\,2} -z^2)/\sin(2\pi z)$. To verify the claim, insert this spectral curve in \eqref{dfunction}, then write \eqref{eq:RRMMV} explicitly and compare with \eqref{eq:N2recrelFC}. Working out the prefactors, and including the contribution from the overall $(2\pi)^{2g+n-2}$ in \eqref{volcompar}, which further multiplies the spectral curve by a power of $1/2\pi$, we obtain the final version of the spectral curve associated to the matrix model dual to pure $\cN=2$ JT gravity
\beq
y_q(x) =\frac{1}{2\pi} \frac{\sin \Big(2\pi \sqrt{-x + \frac{q^2}{4\hq^{\,2}}}\Big)}{4\pi x}.
\eeq
This is precisely the spectral curve we obtained in section \ref{sec:N2JTG} by performing the gravitational path integral on the disk topology. If we write $x=q^2/4\hq^{\,2}-z^2$ and multiply the spectral curve by $2\pi$, which produces an overall factor of $(2\pi)^{2g+n-2}$, then we obtain $y_q(z) \sim \frac{4\hq^{\,2}}{q^2}\frac{\sin 2\pi z}{4\pi} $ up to a correction of
relative order $1/q$.   This is the spectral curve of bosonic JT gravity, up to a rescaling, giving another explanation of the fact that the ``fixed charge'' volumes are proportional to the bosonic ones at large $q$: $V_{g,n}^{(q)}(B) \sim (q^2/4\hq^{\,2})^{2g+n-2} V^{\rm bos}_{g,n}(B)$.

\subsection{Deformations of $\N=2$ JT gravity}\label{defects}
We have successfully proven, to all orders in the topological expansion, that $\N=2$ JT gravity without additional symmetries is dual to the type of random 
matrix ensemble proposed in section \ref{randommatrix} for $\N=2$ quantum mechanics, namely the supermultiplets of different $R$-charge are statistically independent, and each is
 governed by a $(1,2)$ AZ ensemble.

Such ensembles depend on two choices. The first is the spectrum of BPS states for each $R$-charge. This can be extracted from the disk partition function and, for the ensembles relevant to black holes, the BPS spectrum does not fluctuate (this is explained in section \ref{sec:N2JTG}). The second is the choice of matrix potential that appears in \eqref{roughmeasure}, or 
equivalently the choice of spectral curve.  This is specified by the leading order density of states of non-BPS multiplets in the topological expansion, also computed in section \ref{sec:N2JTG} for pure $\N=2$ JT gravity. In this section, we want to address the question of how to give a gravity interpretation to other choices of BPS spectrum and other matrix potentials in the random matrix ensemble. 

The recipe in gravity to modify the matrix potential was provided in \cite{Maxfield:2020ale} and \cite{Witten:2020wvy} for bosonic JT gravity, and generalized to $\N=1$ JT gravity in \cite{Rosso:2021orf}. Introduce a gas of defects in the gravitational path integral of pure JT gravity. In the $\N=2$ case, the defect species are parametrized by the opening angle $\gamma$, and also a $\uU(1)$$R$-symmetry holonomy $e^{\i \hq \phi}$ around the defect.\footnote{This was referred to as $\alpha$ in \cite{Maxfield:2020ale} and \cite{Witten:2020wvy}, but we avoid this notation here to avoid confusion with the $\uU(1)$$R$-symmetry chemical potential at the NAdS boundaries.} These two parameters can be combined into the following $\SU(1,1|1)$ holonomy matrix around the defect
\bea\label{hyperbolicconj}
U=\left(\begin{array}{cc|c}
e^{\i\hq\phi} \cos \frac{\gamma}{2}&-e^{\i\hq\phi} \sin \frac{\gamma}{2}&0 \\
e^{\i\hq\phi} \sin \frac{\gamma}{2} &e^{\i\hq\phi} \cos \frac{\gamma}{2}&0 \\
\hline
0&0&e^{2\i\hq\phi}  
\end{array}\right).\label{eq:Udef}
\ea 
In the upper left block, we have simply written the product of a rotation and an $R$-symmetry transformation.
This deformation of $\N=2$ JT gravity involves an integral over the location of the defect and a sum over the possible number of them, besides the usual degrees of freedom of pure gravity. For each defect insertion, we add a weighting factor $w$, which is another parameter of the deformation. The monodromy $U$ is conjugate to a diagonal matrix of the form \eqref{eq:Ugeo} with imaginary geodesic length $b= \i \gamma$. Therefore the result for the path integral with such defects can be obtained by analytic continuation of the path integral with geodesic boundaries. More precisely, this is true  when $0<\gamma\leq \pi$; for attempts to extend the results beyond this range, see \cite{Turiaci:2020fjj,Eberhardt:2023rzz}. 

For generic values of $\gamma$ and $\phi$,  the one-defect correction to the disk partition function is given by
\bea
Z_{\rm 1-def}&=&w \sum_{q\in \mathbb{Z} + \delta-\frac{1}{2}} \big( e^{\i \alpha (q-\frac{\hq}{2}) } + e^{\i \alpha (q+\frac{\hq}{2})}\big)  e^{- \i \phi q} e^{-\beta E_0(q)} \frac{1}{2\pi} \frac{ e^{\frac{\gamma^2}{4\beta}}}{2\sqrt{\pi \beta}}\nn
&=&\sum_{q\in \mathbb{Z} + \delta-\frac{1}{2}} \big( e^{\i \alpha (q-\frac{\hq}{2}) } + e^{\i \alpha (q+\frac{\hq}{2})}\big) \int_{E_0(q)}^\infty \d E  e^{-\beta E}  \frac{w e^{-\i \phi q}\cosh\big(\gamma \sqrt{E-E_0(q)}\big)}{8 \pi^2 \sqrt{E-E_0(q)}} .\nonumber
\ea
The first line is the analytic continuation of \eqref{trumpet2} upon $b\to \i \gamma$. In the second line we wrote the defect partition function in a way that makes evident what is the correction to the non-BPS density of states. A first observation is that the threshold energy continues to be $E=E_0(q)$, for each $q$, in the presence of the defect.  To be more exact,
this is true order by order in the defect expansion, but the $1/\sqrt{E-E_0(q)}$ singularity indicates that there is actually a shift of order $w$ in the threshold energy, assuming that
the theory can be described by a matrix integral. A second observation is that the result is complex. It is therefore natural to consider $w$ to be a complex parameter as well, and to introduce for each defect species a  conjugate defect with holonomy $U^{-1}$ and complex conjugate weight. Therefore we include
defect species parametrized by $(\gamma,\phi)$ in pairs, such that we preserve unitarity, and to linear order the correction to $\rho_q(E)$ for $E>E_0(q)$ is
\beq\label{rhoonedd}
\rho_q(E) = \frac{\sinh\big( 2\pi \sqrt{E-E_0(q)}\big)}{8\pi^3 E} + {\rm Re} \left[ w\, e^{-\i \phi q}\frac{\cosh\big(\gamma \sqrt{E-E_0(q)}\big)}{4 \pi^2 \sqrt{E-E_0(q)}}\right] + \mathcal{O}(w^2).
\eeq
We conclude that adding such generic defects modifies the spectral curve, and therefore the matrix potential, without modifying the number of BPS states. We leave for future work to derive the finite $w$ expression for the non-BPS density of states, analogous to the ones derived in \cite{Maxfield:2020ale} and \cite{Witten:2020wvy}. In particular it is important to verify that the resummation over the number of defects can be interpreted in terms of a shift of the threshold energy $E_0(q)$, since otherwise the square root singularity of \eqref{rhoonedd} near threshold is not consistent with a matrix integral.

Instead of considering further the non-BPS sector, in this article we focus on the new element not present in the previous literature: how to modify the number of BPS
states, or in other words  the choice of component of solutions to the constraint $Q^2=0$. It turns out this can also be done with a gas of defects. Even though a generic defect does not contribute to the number of BPS states, supersymmetric defects do.\footnote{An analog for $\N=1$ is that the number of supersymmetric states, which vanishes in the absence
of defects, becomes nonzero in the presence of a gas of Ramond punctures \cite{SW}.}  Locally, a defect preserves supersymmetry whenever the deficit angle is related to the $R$-symmetry holonomy by 
\beq\label{holorel}
\gamma = 2\pi - 2 \hq \,|\phi|.
\eeq 
To write this expression we use that $\phi$ can always be defined to be in the range $-\pi \leq \hq \phi \leq \pi$. When this relation is satisfied, the holonomy \eqref{eq:Udef} commutes with two out of the four fermionic generators. The sign of $\phi$ determines
precisely which fermionic symmetry  is preserved by the defect. The partition function on the disk with such a defect inserted is analyzed in appendix \ref{sec:disktrumpet}. It is not a simple analytic continuation of the  usual trumpet partition function and is instead
\bea
Z_{\rm susy-defect}(\beta,\alpha)&=& w\sum_{n\in \mathbb{Z}} \frac{e^{2\pi \i n\delta} \hq \,\cos\big(\frac{\hq(\alpha+2\pi n)}{2}\big) }{2\pi^3 \big(1-{\rm sgn}(\phi) \frac{\hq (\alpha+2\pi n)}{\pi}\big)} e^{\frac{1}{4\beta}(2\pi- 2\hq|\phi|)^2-\frac{ \hq^{\, 2}}{\beta}(\alpha-\phi+ 2\pi n)^2}\nn
&=&w\sum_{n\in \mathbb{Z}} \frac{e^{2\pi \i n\delta} \hq \,\cos\big(\frac{\hq(\alpha+2\pi n)}{2}\big) }{2\pi^3 \big(1-{\rm sgn}(\phi) \frac{\hq (\alpha+2\pi n)}{\pi}\big)}~~(\beta \to \infty)
\ea
In the second line we take $\beta \to \infty$ and therefore project onto BPS states. Using Poisson resummation one can derive 
\beq
Z_{\rm susy-defect}(\beta\to \infty,\alpha) = w \sum_{k \in \mathbb{Z}+\delta} e^{\i \alpha k} \frac{ e^{ {\rm sgn}(\phi) \i \pi \frac{k}{\hq}}}{4\pi^2 } \Theta(k),
\eeq
where we defined the function 
\be\label{newfn}
 \Theta(k)=\begin{cases} 1 & |k|<\frac{\hq}{2}\\
                                         \frac{1}{2} & |k|=\frac{\hq}{2} \\
                                          0 & |k|>\frac{\hq}{2}. \end{cases}
\ee
Again we see that this correction to the spectrum is complex and therefore for each defect we need to insert an orientation reversed one. The correction to the BPS spectrum is then
\beq\label{eq:BPSssusydef}
N_{\rm BPS}(k) = e^{S_0}\frac{\cos(\frac{\pi k}{\hq})}{4\pi^2} + {\rm Re}\left[ \frac{w\, e^{ {\rm sgn}(\phi) \i \pi \frac{k}{\hq}}}{2\pi^2 } \right] \Theta(k)
\eeq
This equation is exact in both $S_0$ and $w$ as long as $0<\gamma\leq\pi$. For $\gamma$ in this range, configurations either with handles or with more than one defect do not contribute to the BPS spectrum. The reason is that for $\gamma$ in this range, as soon as we have two or more defects or any number of handles there will be a geodesic homotopic to each NAdS boundary. The form of the trumpet partition function guarantees that there will be no contribution to the number of BPS states. 

We can consider multiple supersymmetric defect species, with different values of $\phi$. The contribution to the number of BPS states comes only from either the disk, or a disk with one defect (of any species). Since the one defect contribution to the number of BPS states computed in eqn. \eqref{eq:BPSssusydef} has such a mild dependence on $\phi$, a sum over species will only modify the effective value of $w$.   The answer will still take the form in \eqref{eq:BPSssusydef}, at least as long as all defects satisfy $0<\gamma \leq \pi$.

For the reasons given above, the most general shift in the spectrum of BPS states is a linear combination of $\cos (\frac{\pi k}{\hq})$, associated to the real part of $w$, or $\sin (\frac{\pi k}{\hq})$, associated to the imaginary part of $w$. Only for $w\in \mathbb{R}$ can we preserve charge conjugation invariance, since $\sin(\frac{\pi k}{\hq})$ changes sign under $k\to-k$. Moreover, a contribution $\sin (\frac{\pi k}{\hq})$ produces BPS states at charges $k=\pm \hq/2$. Since the disk predicts a vanishing number of BPS states at these charges, a model with non-real $w$ would seemingly predict a negative number of BPS states for either $k=+\hq/2$ or $-\hq/2$. Such a theory of gravity would then be apparently non-unitary if $\delta=1/2$, so that the spectrum of charges is half-integral.\footnote{A similar issue arises in other examples.  For example, adding a gas of defects to bosonic JT gravity can lead to a negative density of states. On the matrix model side, it has been proposed in \cite{Rosso:2021orf} that this  signals a phase transition.}

Similar to their non-supersymmetric counterpart, the spectrum of non-BPS states is affected by contributions with an arbitrary number of supersymmetric defects, but we will not analyze this problem in this article. 

The spectrum of BPS states that we can achieve with a gas of supersymmetric defects is very restrictive. For example if $w\in \mathbb{R}$, then the correction from a defect has the same dependence on $k$ as the leading answer from the disk. Interestingly, this change in the spectrum does not allow us to construct a theory of gravity with a number of BPS states reproducing the $\N=2$ SYK answer (with the only exception being $\hq=3$.) It would be nice to find a gravitational calculation that reproduces the BPS spectrum of SYK for any\footnote{We
thank J. Maldacena for raising this question.} $\hq$. 

Finally, it is possible to interpret the deformations by a gas of defects as a modification of the dilaton potential of two-dimensional $\N=2$ dilaton-gravity, with a modification that 
depends on both the dilaton and the scalar field coming from the $R$-symmetry $\uU(1)$ $BF$ theory. This approach actually explains why we need to consider defects coming in pairs with opposite orientation: in this way the gravity action is real. Since we did not analyze the explicit form of the $\N=2$ JT gravity action, it is outside of the scope of this article to describe this perspective.

\section{Incorporating Time-Reversal Symmetry}\label{timereversal}

\subsection{Boundary Viewpoint}\label{boundaryview}

In this section, we will incorporate time-reversal symmetry in $\N=2$ matrix models and JT supergravity.

Time-reversal is an anti-unitary transformation that commutes with the Hamiltonian.   But we have to choose whether it will commute or anticommute with the $R$-charge generator
$J$.   The standard convention in particle physics\footnote{The
reason for this convention is that historically, $\tT$ was defined to be an approximate symmetry of ordinary matter (originally, it was believed to be an exact symmetry).  For this
to be the case, additive conserved charges such as baryon and lepton number must be $\tT$-invariant.}
 is that an anti-unitary symmetry that commutes with additive conserved charges as well as with the Hamiltonian is called $\tT$, while an anti-unitary symmetry that commutes with the Hamiltonian but anticommutes with additive conserved charges is called $\CT$.   
So we will consider two cases:
   
(A)   The boundary quantum mechanics system  (or ensemble) may have an antiunitary symmetry $\tT$ that commutes with both $H$ and $J$.  To be a symmetry, $\tT$ will have to conjugate
$Q$ and $Q^\dagger$ to linear combinations of themselves.  Since $[J,Q]=\h q \,Q$, $[J,Q^\dagger]=-\h q \,Q^\dagger$, $\tT$ must conjugate $Q$ to a multiple of itself and likewise
$Q^\dagger$.   After possibly replacing $\tT$ with $\tT e^{\i\gamma J}$ for some $\gamma$, we can assume that $\tT$ commutes with $Q$ and $Q^\dagger$.
So $\tT$ simply commutes with the whole supersymmetry algebra of the boundary theory.

In theories with fermions, one can often define a time-reversal symmetry such that $\tT^2=(-1)^\fF$, the operator that distinguishes bosons and fermions.   In the present context, however,
we have a fermionic operator $Q$ that commutes with $\tT$, and therefore with $\tT^2$.   So necessarily $\tT^2$ commutes with fermionic operators, rather than anticommuting with them.
(This leads in Euclidean signature to what is called a ${\mathrm{pin}}^-$ structure, rather than ${\mathrm{pin}}^+$.)

The operator $\tT^2$ is a unitary operator that commutes with the whole supersymmetry algebra.  
The unitary operators that we know about that commute with the supersymmetry algebra are $c$-numbers $e^{\i a}$, $a\in \R$ and also operators $e^{2\pi \i n J/\h q}$, $n\in\Z$.
 We could consider a boundary theory with an additional unitary symmetry
beyond  the supersymmetry algebra; then there would also be an additional gauge symmetry in bulk.   However, we will avoid introducing additional symmetries.
Given this, $\tT^2$ must be of the form $e^{\i a}e^{2\pi \i n J/\h q}$ for some $a,n$.   
However, $\tT^2$ is further constrained by the fact that it commutes with $\tT$.   The only operators $e^{\i a}e^{2\pi \i n J/\h q}$ that commute with $\tT$ are $\pm 1$.
So the only options are $\tT^2=\pm 1$.

Since $\tT$ commutes with $J$, $\tT$ symmetry places no constraint on the charge defect defined by $J=\delta~{\mathrm{mod}}~\Z$.

(B) The boundary system may have an antiunitary symmetry $\CT$ that commutes with $H$ but anticommutes with $J$.   In this case, $\CT$ must conjugate $Q$ to a multiple of
$Q^\dagger$, and vice-versa.  So  $\CT Q\CT^{-1}= e^{\i\alpha} Q^\dagger$, $\CT Q^\dagger \CT^{-1} =e^{-\i\alpha}Q$,
for some real $\alpha$.   By  replacing $\CT $ with $\CT e^{-\i \alpha J/\h q}$, we can reduce to $\alpha=0$:
\be\label{redrule}\CT Q \CT^{-1}=Q^\dagger,~~\CT Q^\dagger \CT^{-1}=Q.\ee

Again $\CT^2$ commutes with the whole supersymmetry algebra, so if we do not wish to introduce additional symmetries, we must have $\CT^2=e^{\i a} e^{2\pi\i n J/\hq}$ 
for some $a,n$.   The condition that $\CT^2$ commutes with $\CT$ implies that actually $\CT^2=\pm e^{2\pi \i n J/\h q}$ for some $n$.  But we are still free to substitute
$\CT\to\CT e^{-2\pi\i k J/\h q}$, $k\in\Z$, without disturbing eqn. (\ref{redrule}).   By choosing $k$ so that either $2k=n$ or $2k=n+\h q$, we can reduce to $\CT^2=\pm 1$.

Since $J$ is odd under $\CT$, the possible values of $\delta$ in a $\CT$-invariant theory are $0,1/2$.  (In the reduction in the last paragraph, one has to take into account
that if $\delta=1/2$, then $\exp(2\pi\i J)=-1$.)

When the boundary theory has a $\tT$ or $\CT$ symmetry, we expect the two-dimensional bulk supergravity theory to have such a symmetry as well.   In the present article,
we study the bulk theory primarily in Euclidean signature.   In that context, it is more natural to consider an ordinary (linear) reflection symmetry rather than an antilinear time-reversal
symmetry.   So we make use of the $\CPT$ theorem, which tells us that a relativistic theory has a spatial reflection symmetry $\rR$ that commutes with additive conserved
charges such as $J$ if and only if it has a $\CT$ symmetry, and it has a spatial reflection symmetry $\CR$ that anticommutes with additive conserved charges such as $J$
if and only if it has a $\tT$ symmetry.

We will next discuss the  random matrix ensembles that are natural in $\N=2$ supersymmetry extended by $\tT$ or $\CT$ symmetry.
The bulk picture with $\CR$ or $\rR$ symmetry is discussed in section \ref{bulk}.

(A) Since $\tT$ commutes with $J$, it maps the space of BPS states for any given $J$ to itself and places no constraint on the dimensions of these spaces.\footnote{There is one
small exception:
if $\tT^2=-1$, the dimensions of spaces of BPS states must be even, by what is usually called Kramers doubling.    However, this factor of 2
is essentially invisible in a double-scaled model, in which dimensions are proportional to $e^{S_0}$.  There is a similar doubling for the non-BPS supermultiplets.   A random matrix
description of a theory with $\tT^2=-1$ uses an ensemble in which this doubling is built in.}   Likewise, the space of $(k,k+\h q \,)$ representations is $\tT$-invariant,
for each $k$.

It is straightforward to repeat the analysis of sections \ref{randomone} and \ref{randomtwo} to incorporate the $\tT$ symmetry.\footnote{The following amounts mostly to repeating the analysis
in \cite{SW} for the cases that the anomaly coefficient vanishes (no anomaly, $\tT^2=1$) or equals 4 (the anomaly is captured by saying $\tT^2=-1$).}  
 First, consider a single $(k,k+\h q\,)$ multiplet
assuming that the others do not exist.   We have to modify the analysis of section \ref{randomone} by assuming that the matrix $\qQ$ is $\tT$-invariant and similarly
by restricting the symmetry group $G=\uU(L_k)\times \uU(L_{k+\h q})$ to its $\tT$-invariant subgroup.    For $\tT^2=1$, this means that we should assume $\qQ$ to be a real
matrix and take $G={\rm O}(L_k)\times {\rm O}(L_{k+\h q})$.  In other words, $\qQ$ is now a bifundamental of a product of orthogonal, rather than unitary, groups.
 The derivation of eqn. (\ref{azmeasure}) is modified as follows. There is no continuous symmetry rotating the
phase of a given eigenvalue, so $\upalpha=0$ in eqn. (\ref{azmeasure}); and $\w$ is real in eqn. (\ref{matform}), as a result of which the factors $(\lambda_i\pm \lambda_j)^2$
in eqn. (\ref{firstmeasure}) are replaced by $|\lambda_i\pm\lambda_j|$.   The resulting measure is $\prod_i \d\lambda_i \prod_{i<j}|\lambda_i^2-\lambda_j^2|$,
an AZ ensemble with $(\upalpha,\upbeta)=(0,1)$.    In the subsequent derivation
assuming the presence of $\nu$ zero-modes of charge $k$ or $k+\h q$, there are now $\nu$ instead of $2\nu$ additional
symmetry generators for each $\lambda_i$, so the factor $\lambda_i^{2\nu}$ in eqn. (\ref{azmeasuretwo}) is replaced by $\lambda_i^\nu$.   The final result is then
$\prod_i \d\lambda_i\, \lambda_i^\nu\prod_{i<j}|\lambda_i^2-\lambda_j^2|$, an AZ ensemble with $(\upalpha,\upbeta)=(\nu,1)$. 
We can write this as in eqn. (\ref{bonven}), but now with $\upbeta=1$:
\be\label{convagain}\upalpha=\upalpha_0+\upbeta\nu,\hskip1cm (\upalpha_0,\upbeta)=(0,1). \ee
  The derivation in section \ref{randomtwo} to allow
for potential couplings between the different multiplets requires similar small corrections.
The argument $\t\lambda u+\lambda v$ of the delta function
in eqn. (\ref{deltafn}) is now real, so the delta function is an ordinary delta function of a real variable, and also $\w$ is 
now real in eqn. (\ref{symgens}).  These effects cancel as before and the multiplets remain statistically independent. 
Upon allowing for the presence of BPS states,  there are now half as many symmetries mapping a non-vanishing row
or column to a row or column of zeros, so the contribution to $\upalpha$ is half as big as before.  The final prediction is that the $(k,k+\h q\,)$ multiplets are statistically independent 
and each one is described by an AZ multiplet with parameters satisfying (\ref{convagain}). 

In case $\tT^2=-1$, $\qQ$ is a matrix of quaternions and the unitary groups $\uU(L_k)$ and $\uU(L_{k+\h q})$ are replaced by symplectic groups.   In other words,
$\qQ$ is now a bifundamental of a product of symplectic groups.   The group that rotates a single eigenvalue is now $\SU(2)\cong 
\Sp(1)$, with dimension 3, leading to $\upalpha_0=3$,
and the parameter $\w$ in eqn. (\ref{matform}) is now quaternionic, depending on four real parameters and leading to $\upbeta=4$.   After also taking into account that all
parameters are quaternionic and there
are now four generators mapping a nonzero row or column of a bifundamental to a vanishing row or column, one finds that the $(k,k+\h q \,)$ multiplets are statistically
independent and described by an AZ multiplet with $(\upalpha,\upbeta)=(3+4\nu,4)$.
So
\be\label{convthree}\upalpha=\upalpha_0+\upbeta\nu,\hskip1cm (\upalpha_0,\upbeta)=(3,4). \ee

In a bulk description, the difference between a bifundamental of a pair of orthogonal group and a bifundamental of a pair of symplectic groups is a factor of $(-1)^\chi$, that
is a factor of $-1$ for each crosscap.   This is proved in section 4.2.4 of \cite{SW} using the loop equations.

(B) Since $\CT$ anticommutes with $J$, it maps the space of BPS states with $J=k$ to the corresponding space with $J=-k$, placing no constraint on the dimensions $L_k^0$
of the spaces of BPS states except that $L_k^0=L_{-k}^0$ (and in case $\delta=0$ and $\tT^2=-1$, $L_0^0$ must be even by Kramers doubling).   Similarly, the space
of $(k,k+\h q\,)$ representations is mapped by $\CT$ to the space of $(-k-\h q,-k)$ representations.   Thus for generic values of $k$, $\CT$ merely implies that the random matrix
description of the $(k,k+\h q \,)$ states is the same as the random matrix description of the $(-k-\h q,-k)$ states. In terms of the bulk charge $q=k+\hq/2$ introduced in section \ref{sec:N2JTG}, $\CT$ acts as $q\to -q$. 

We learn more only in the case of a supermultiplet with $R$-charges $(-\h q/2,\h q/2)$, or equivalently $q=0$.   As $\h q$ is odd, this 
multiplet only exists if $\delta=1/2$.  For $\delta=1/2$, $\CT$ maps the $(-\h q/2,\h q/2)$ 
states to themselves, and therefore for this multiplet, the random matrix description will be different.   
Since $J$ has eigenvalues $\pm \h q/2$ for these states, it can be put in a block diagonal form 
\be\label{blockformj} J=\frac{\h q}{2}\left(\begin{array}{c|c} 1 & 0 \\ \hline 0 & -1 \end{array} \right). \ee
The blocks are of equal size, as they are exchanged by $\CT$. Since $\CT$ anticommutes with $J$, and $\CT^2=\varepsilon$ where $\varepsilon=\pm 1$,   we can assume  $\CT$ to take the form
\be\label{tform}\CT=\left(\begin{array} {c|c} 0 & 1 \\ \hline \varepsilon & 0 \end{array} \right)\star.\ee
The symmetry group is reduced to the group of unitary transformations that commute with both $J$ and $\CT$.   These are unitary
transformations of the form
\be\label{uform}\U=\left(\begin{array} {c|c} U & 0 \\ \hline 0 & \bar U \end{array} \right),\ee
with $U$ unitary (the adoint, complex conjugate, and transpose of a matrix $M$ will be denoted $M^\dagger, \,\bar M,\, M^\tr$).  
Thus the symmetry group is $\uU(L_{\h q/2})$.  
Because $Q$ and $Q^\dagger$ carry $R$-charges $\h q$ and $-\h q$, they take the general form 
\be\label{qform}  Q=\left(\begin{array} {c|c} 0& M \\ \hline 0 & 0 \end{array} \right),\hskip1cm Q^\dagger=\left(\begin{array} {c|c} 0 & 0 \\ \hline M^\dagger & 0 \end{array} \right).\ee
The condition $\CT Q^\dagger\CT^{-1}= Q$ tells us  that $M=\varepsilon M^\tr$ is  symmetric or antisymmetric depending on whether $\varepsilon=+1$ or $\varepsilon=-1$, 
and the symmetry $Q\to \U Q\U^{-1}$ becomes
$M\to U M U^\tr$.   In other words, $M$ is a complex  second rank tensor of the unitary group $\uU(L_{\h q/2})$, a symmetric tensor if $\varepsilon=1$ and an antisymmetric
one if $\varepsilon=-1$.  

Ensembles of such symmetric or antisymmetric matrices with unitary symmetry also arise in $\N=1$ JT supergravity with anomaly coefficient 2 or 6 mod 8 \cite{SW}.  
The symmetric tensors correspond to an AZ ensemble with $(\upalpha_0,\upbeta)=(1,1)$ and the antisymmetric tensors correspond to an AZ ensemble with $(\upalpha_0,\upbeta)=(1,4)$. 
 This can be shown by arguments similar to those in section \ref{randommatrix}. We consider in turn the two cases:
 
  (1) The canonical form of a symmetric second rank tensor $M$ is $\diag(\lambda_1,\lambda_2,\cdots)$,
 with all $\lambda_j\geq 0$.  For each $j$, the symmetry group ${\rm U}(L_{\h q/2})$ has one generator rotating the phase of $\lambda_j$; this leads to $\upalpha=1$.
For each pair $\lambda_i, \lambda_j$, the symmetry group has two generators rotating the corresponding eigenvectors of $M$ among themselves, as opposed to the four
in eqns. (\ref{matform}), (\ref{matform2}).   This leads to $\upbeta=1$ rather than 2.  So the ensemble is an AZ multiplet with $(\upalpha,\upbeta)=(1,1)$.

(2) The canonical form of an antisymmetric tensor $M$ is a direct sum of
 $2\times 2$ blocks of the form $\begin{pmatrix}0 & \lambda_j\cr -\lambda_j & 0\end{pmatrix}$, with all $\lambda_j\geq 0$.   The unitary group $\uU(L_{\h q/2})$ has for each $j$
 a $\uU(2)$ subgroup acting on this block, of which an ${\mathrm{SU}}(2)$ subgroup leaves $M$ invariant and the one remaining generator rotates the phase of $\lambda_j$, leading to $\upalpha=1$.  By looking at generators that act between a pair of these $2\times 2$ blocks of $M$, one finds
 $\upbeta=4$.  So $(\upalpha,\upbeta)=(1,4)$ for antisymmetric tensors.  
 
 There are no BPS states of charge $\pm \h q/2$, so for these multiplets, there is no
  distinction betweeen $\upalpha$ and $\upalpha_0$.
 
In a bulk description, the difference between a random ensemble of symmetric complex tensors with unitary symmetry and
 a random ensemble of antisymmetric complex tensors with that symmetry consists of a factor $(-1)^\chi$, just like the difference between the two cases in (A).  This is proved using the
 loop equations in section 4.2.4 of \cite{SW}, and it also has a simple proof based on Feynman diagrams explained in appendix B of that paper.

 \subsection{The Cross-Cap With $\rR$ or $\CR$ Symmetry}\label{bulk}
 
In Lorentz signature with coordinates $t=x^0, x=x^1$, consider gamma matrices $\gamma_\mu$, $\mu=0,1$ satisfying $\{\gamma_\mu,\gamma_\nu\}=\pm 2\eta_{\mu\nu}$ with
$\eta_{\mu\nu}=\diag(-1,1)$.   The massless Dirac equation $(\gamma^0\partial_0+\gamma^1\partial_1)\psi=0$ has a time-reversal symmetry
$\tT\psi(t,x)=\gamma_0\psi(-t,x)$ and a reflection symmetry $\rR\psi(t,x) =\gamma_1\psi(t,-x)$.   We observe that $\tT^2=\pm 1$ acting on $\psi$ if and only if $\rR^2=\mp 1$ acting on
$\psi$; the
signs are  determined by the sign in the Clifford algebra.   An operator version of this statement is that\footnote{Here we are considering $\tT^2$ and $\rR^2$ as automorphisms of
the operator algebra. Even if $\tT^2$ or $\rR^2$ is the identity as an automorphism of the operator algebra, it may still act as $-1$ on a quantum Hilbert space, under some
conditions.}       $\tT^2=1$ or $(-1)^\fF$ if and only if $\rR^2=(-1)^\fF$ or 1.
This relationship is actually a general prediction of the $\CPT$ theorem.   

Here we did not incorporate any additive conserved charge.   If one does so, for example by taking $\psi$ to be a complex fermion with a $\uU(1)$ symmetry, one
finds that because time-reversal acts by an antilinear operator while a spatial reflection  acts by a linear one, time-reversal  will commute with the conserved charges if and only if the
reflection anticommutes with them,
and vice-versa.  To see this, expand $\psi $ in Majorana fermions by $\psi=\psi_1+\i \psi_2$.   A time-reversal symmetry can be defined to act by, say,
$\psi_1(t,x)\to \gamma_0 \psi_1(-t,x)$, $\psi_2(t,x)\to \varepsilon \gamma_0\psi_2(-t,x)$, with $\varepsilon=\pm 1$.  Regardless of what signs we pick in the definition of time-reversal,
the reflection symmetry that is related to it by the $\CPT$ theorem will act with the same signs: $\psi_1(t,x)\to \gamma_1 \psi(t,-x)$, $\psi_2(t,x)\to \varepsilon \gamma_1\psi_2(t,-x)$.
Then, as time-reversal is anti-unitary, we see that it maps $\psi$ to $\psi^\dagger=\psi_1-\i\psi_2$ if $\varepsilon=1$, but to $\psi$ if $\varepsilon=-1$.  For a unitary reflection
symmetry, these statements are reversed.   

  So a $\tT$ symmetry satisfying $\tT^2=1$ (or $(-1)^\fF$)  is always accompanied by a $\CR$ symmetry satisfying $\CR^2=(-1)^\fF$ (or 1), and a $\CT$ symmetry satisfying
$\CT^2=1$ (or $(-1)^\fF$) is always accompanied by an $\rR$ symmetry satisfying $\rR^2=(-1)^\fF$ (or 1).
The boundary ensembles discussed in section \ref{boundaryview} have a $\tT$ or $\CT$ symmetry satisfying $\tT^2=1$ or $\CT^2=1$, so they will have a $\CR$ or $\rR$ symmetry
satisfying $\CR^2=(-1)^\fF$ or $\rR^2=(-1)^\fF$.

One more important detail that can be understood from the preceding derivation is the following.   To get $\tT^2=1$ acting on $\psi$, the sign in the Clifford algebra was
$\{\gamma_\mu,\gamma_\nu\}=-2\eta_{\mu\nu}$.   This Clifford algebra has a two-dimensional real representation, so Majorana fermions with $\tT^2=1$ can exist in two
spacetime dimensions.    However, when we Wick rotate to Euclidean signature, where we will perform computations, the Clifford 
algebra becomes $\{\gamma_\mu,\gamma_\nu\}=-2\delta_{\mu\nu}$.   This algebra does not have a two-dimensional real representation, 
but it does have a representation by imaginary $2\times 2$ matrices.   The formula
$\rR\psi(t,x)=\gamma_1\psi(t,-x)$ shows that in acting on fermions $\rR$ (or $\CR$ if conserved charges are incorporated) is proportional to a gamma matrix, so we have to expect
that as a $2\times 2$ matrix acting on fermions, $\rR$ or $\CR$ can be chosen to be imaginary but cannot chosen to be real.   
Since $\gamma_1^2=-1$, $\gamma_1 $ has eigenvalues $\pm \i$
and determinant 1; hence we expect $\rR$ to have determinant 1 as an operator on fermions.  

      \begin{figure}
 \begin{center}
   \includegraphics[width=4in]{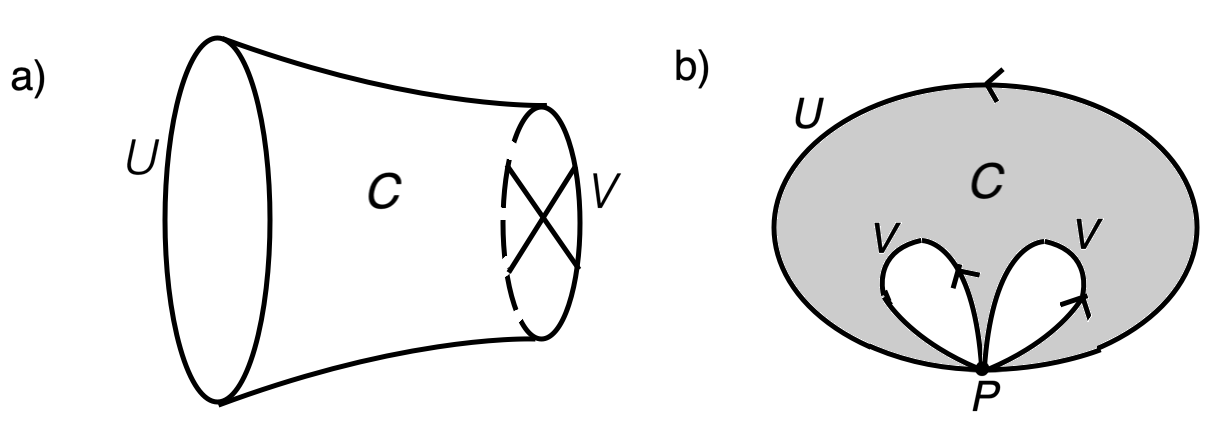}
 \end{center}
\caption{\footnotesize (a) $C$ is a cylinder with opposite points on the right boundary identified (as schematically indicated by the $\Large\times$) to make
an unorientable ``crosscap'' manifold.   After this identification, the boundary on the right becomes a circle of half the original circumference,
and it is no longer a boundary but an interior circle, ultimately an interior geodesic. 
  A flat connection on $C$ with holonomy $U$ around the boundary and holonomy $V$ around the interior geodesic can exist
if and only if $V^2=U$. (b) To calculate the torsion, one can use the very simple cell decomposition of $C$ (or more precisely of a topological space that for these
purposes is equivalent to $C$)
indicated here, with a single vertex $P$, two one-cells labeled $V$ and $U$ (the two one-cells labeled  $V$ that appear in the diagram are identified to make the crosscap), and
one two-cell.   \label{CrossCap}}
\end{figure}

With these preliminaries out of the way, our goal now is to compute the ``torsion'' (or moduli space integration measure) for a basic unorientable two-manifold
 from which all others can be built.
This is the ``crosscap'' manifold (fig. \ref{CrossCap}(a)), which is obtained from a cylinder by identifying opposite points on one boundary, to build an unorientable manifold $C$
with a single boundary circle. 
Gluing this manifold onto a trumpet gives an unorientable asymptotically NAdS spacetime $C'$ with one asymptotic boundary; the path integral
on $C'$ is expected to reproduce the leading correction to the random matrix partition function due to time-reversal symmetry.   The main novel step
in computing this correction is to compute the torsion of a flat connection on $C$.  Once this is known, relatively standard steps similar to those that we have already explained
lead to the prediction for the correction to the partition function.

In computing the torsion, we consider a flat connection on $C$ in which the holonomy around the boundary is $U$ and the holonomy around the ``interior'' circle $\gamma$ where the
identification is made  is $V$.   ($\gamma$ will ultimately be a geodesic in a hyperbolic metric.)   
Flatness of the connection implies that $U=V^2$, because the boundary on the left of fig. \ref{CrossCap}(a) is homotopic to a double cover of  $\gamma$.
We can satisfy these conditions with
\be\label{crosscaprep} 
U=\left(\begin{array}{cc|c}
-e^{\i\h q \phi}e^{\frac{b}{2}}&0&0 \\
0&-e^{\i\h q \phi}e^{-\frac{b}{2}}&0 \\
\hline
0&0&e^{2\i\h q \phi}  
\end{array}\right),\hskip1cm V=\left(\begin{array}{cc|c}
\i e^{\i\frac{\h q \phi}{2}}e^{\frac{b}{4}}&0&0 \\
0&-\i e^{\i\frac{\h q \phi}{2}}e^{-\frac{b}{4}}&0 \\
\hline
0&0&e^{\i\h q \phi}  
\end{array}\right).
\ee
Not coincidentally, this is the same $U$ that we used in eqn. (\ref{eq:Ugeo})  to describe a trumpet with an inner geodesic of 
length $b$ and an $R$-symmetry parameter $\phi$.  Clearly, $V^2=U$,
and if we set $b=\phi=0$ so that $U$ reduces to $\diag(-1,-1|1)$, then $V$ has the properties described earlier for a reflection $\rR$ (acting on the fermions, it is imaginary
with determinant 1).  Replacing $\i$ by $-\i$ in the formula for $V$ would give an equivalent formula, up to a shift of $\phi$ by $2\pi$.   

In eqn. (\ref{crosscaprep}), $V$ is an element of the supergroup $\SU(1,1|1)$, or more exactly its complexification, 
as is appropriate for a model with reflection symmetry $\rR$.  To get $\CR$ symmetry, we have to combine the expression for $V$ with a charge conjugation operation;
to postpone this slight  complication, we consider $\rR$ symmetry first, case (B) in the language of section \ref{boundaryview}.

 In section 3.4.5 of \cite{SW}, using the existence of a very simple cell decomposition of
the crosscap manifold (fig. \ref{CrossCap}(b)), a simple general recipe was described to compute $\tau_C$, the torsion of a flat bundle on $C$.  Here we will just use this
recipe and refer to the paper for an explanation.

For $x\in \mathfrak g$, let $\h V(x)= V x V^{-1}$. Then the torsion of the crosscap manifold $C$ is $\Ber' (\h V-1)$.   Decompose the Lie algebra $\mathfrak g$ of
$\SU(1,1|1)$, as $\mathfrak g^0\oplus \mathfrak g^\perp$, where $\g^0$ is the kernel of $\h V-1$ and $\g^\perp$ is its orthocomplement.   For the case of $\rR$ symmetry, $\mathfrak g^0$ consists of the subalgebra of diagonal matrices
and $\mathfrak g^\perp$ consists of the strictly off-diagonal matrices.   Because this decomposition is $\h V$-invariant, one has a factorization
\be\label{usefuldec} \tau_C=   \tau_C^0\tau_C^\perp,  \ee
with 
\be\label{tordefs}\tau_C^\perp= \Ber' (\h V^\perp-1), \hskip1cm \tau_C^0= \Ber'( \h V^0-1).\ee
Here $\h V^0$ and $\h V^\perp$ are the restrictions of $\h V$ to  to $\mathfrak g^0$ and $\mathfrak g^\perp$.
For the case of $\CR$ symmetry that we consider later, the decomposition $\g=\g^0\oplus \g^\perp$ is slightly different, but it is still $\h V$-invariant, so eqn. (\ref{usefuldec}) still holds.

$\h V^\perp-1$ is invertible and preserves the $\Z_2$-grading of the Lie algebra, with separate blocks $\h V^\perp_+-1$ and $\h V^\perp_{-}-1$
acting on the bosonic and fermionic parts of $\g_\perp$.  So  \be\label{moreber} \Ber'(\h V^\perp-1)=\frac{|\det( \h V^\perp_{+}-1)|}{\det( \h V^\perp_{-}-1)}.\ee
 (See the paragraph after \eqref{facttors} for a definition of $\Ber'$.) 
In the present example, the eigenvalues of $\h V^\perp_{+}-1$  are  $-e^{\pm b/2}-1$, leading to  $|\det\,(\h V^\perp_{+}-1)|=(1+e^{b/2})(1+e^{-b/2})=4\cosh^2 \frac{b}{4}.$ The action on the fermionic generators is
\be\label{vcaction}\h V:\begin{pmatrix} {\sf q}_1 \cr {\sf q}_2 \cr {\sf q}_3 \cr {\sf q}_4\end{pmatrix}\to \begin{pmatrix} \i \, e^{\i \frac{\hq \phi}{2}} e^{\frac{b}{4}} {\sf q}_1 \cr \i \, e^{-\i\frac{\hq \phi}{2}}e^{\frac{b}{4}} {\sf q}_2\cr -\i \, e^{\i \frac{\hq \phi}{2}} e^{-\frac{b}{4}} {\sf q}_3\cr -\i  \, e^{-\i \frac{\hq \phi}{2}}e^{-\frac{b}{4}} {\sf q}_4\end{pmatrix}. \ee
Therefore the eigenvalues of $\h V^\perp_{-}-1$  are $-1\pm \i e^{\pm b/4}e^{-\i \h q \phi/2}$ and $-1\pm \i e^{\pm b/4} e^{\i\h q \phi/2}$, so
$\det \,(\h V^\perp_{-}-1) = 4(\cos \frac{\h q\phi}{2}-\i \sinh \frac{b}{4})^2$.  Combining these results, we get
\be\label{orthotor}\tau_C^\perp=\frac{\cosh^2 \frac{b}{4}}{(\cos \frac{\h q\phi}{2}-\i \sinh \frac{b}{4})^2}. \ee
$\tau_C^0$ can be computed precisely as in the derivation of 
 eqn. (3.49) of \cite{SW}.  It   is a simple ratio of measures for parameters $b,\phi$ that appear in $U$ and $V$ and potential gluing parameters $\varrho,\varphi$
for gluing the left boundary in fig. \ref{CrossCap}(a) onto something else:
\be\label{diagtor}\tau_C^0=\frac{1}{4} \d b\d\phi (\d\varrho \d\varphi)^{-1}. \ee  
The factor of $1/4$ here has the same origin as the $1/2$ in eqn. (3.49) of \cite{SW}: one can write $\tau_C^0=\d \bar b\d\bar\phi (\d\varrho\d\varphi)^{-1}$, where
$\bar b$ is the length of the interior geodesic in the figure and $\bar\phi$ is the $R$-symmetry monodromy around this geodesic; as $b=2\bar b$, $\phi=2\bar\phi$,
this leads to  eqn. (\ref{diagtor}). 
So the crosscap torsion is
\be\label{fulltor}\tau_C=\tau_C^\perp \tau_C^0 =\frac{\cosh^2 \frac{b}{4}}{4(\cos \frac{\h q\phi}{2}-\i \sinh \frac{b}{4})^2} \d b\d\phi (\d\varrho \d\varphi)^{-1}. \ee

To turn this into a prediction that can be compared to random matrix theory, we have to combine this result with the boundary path integral of the Schwarzian mode.
As in eqn. (\ref{tortrumpet}),  the trumpet path integral with boundary conditions natural for the torsion calculation is given by
\be\label{tortrumpetnat} \t Z_{\rm tr}(b,\phi) = Z_{\rm tr}(b,\phi) \frac{\sinh\frac{b}{2}}{\cosh\frac{b}{2}+\cos \h q \phi}, \ee
where $Z_{\rm tr}(b,\phi)$ was computed in \eqref{trumpet2}. A trumpet that ends on a crosscap can be obtained by gluing an ordinary trumpet onto the crosscap manifold of fig. \ref{CrossCap}(a).   The gluing law for the torsion
gives a formula for the path integral that results from this gluing:
\be\label{longformal}Z_{\rm xcap}(b,\phi)= \t Z_{\rm tr}(b,\phi)\d b\d\phi \frac{1}{\tau_S}\tau_C, \ee
Using eqn. (\ref{tortrumpetnat}) for $\t Z_{\rm tr}(b,\phi)$,  eqn. (\ref{eq:torsioncirc}) for $\tau_S$, and eqn. (\ref{fulltor}) for $\tau_C$, we get finally the measure 
for a trumpet that ends on a crosscap:
\be\label{crosspredict} Z_{{\rm xcap},\CT}(b,\phi) =Z_{\rm tr}(b,\phi) \frac{( \cosh \frac{b}{2}+ \cos \h q \phi )\coth\frac{b}{4} }{8(\cos \frac{\h q\phi}{2}-\i \sinh \frac{b}{4})^2} \d b\d\phi, \ee where we make explicit the fact that this is the prediction in a model
with $\CT $ or $\rR$ symmetry.
Since we have already computed $Z_{\rm tr}(b,\phi)$, this result can be compared to the matrix model prediction, as we will do in section \ref{matrixprediction}.

Now we consider the case of $\tT$ or $\CR$ symmetry, case (A) in the language of section \ref{boundaryview}.    This means that we have to include  charge conjugation, which we will
call $\cC$, as part of the monodromy.   As we explain in detail momentarily, $\cC$ is an outer automorphism of $\SU(1,1|1)$, satisfying $\cC^2=1$. It acts trivially on $\SU(1,1)$ but reverses the 
sign of the $R$-charge generator, which
was called $\sf z$ in eqn.  (\ref{slgenerators}).  Roughly, including $\cC$ means replacing the monodromy $V$ by $V_\cC=\cC V$.  However, because the $R$-charge is odd under $\cC$, the conjugacy
class of $V_\cC$ is independent of $\phi$.   To see this, write more explicitly $V_\phi$ for the $\phi$-dependent matrix  $V$ defined in eqn. (\ref{crosscaprep}), and observe that,
 because ${\sf z}$ is odd under $\cC$, we have
$e^{\lambda {\sf z}} \cC V_\phi e^{-\lambda {\sf z}}=\cC e^{-2\lambda {\sf z}} V_\phi=\cC V_{\phi- \lambda/\hq}$.
Hence by conjugation we can set $\phi=0$ in $V$, in which case also $\phi=0$ in $U=V^2$  (concretely, this happens because as the $R$-charge is odd under $\cC$, 
$U=\cC V_\phi \cC V_\phi$ is actually independent of $\phi$).   So we consider the monodromies
\be\label{crmonodromy}U=\left(\begin{array}{cc|c}
-e^{\frac{b}{2}}&0&0 \\
0&-e^{-\frac{b}{2}}&0 \\ 
\hline
0&0& 1
\end{array}\right),\hskip1cm V_\cC=\cC V= \cC \left(\begin{array}{cc|c}
\i e^{\frac{b}{4}}&0&0 \\
0&-\i e^{-\frac{b}{4}}&0 \\
\hline
0&0&1 
\end{array}\right).
\ee
 
$\cC$ cannot be written as a $3\times 3$ matrix, because it is an outer automorphism of the Lie algebra of $\SU(1,1|1)$ that
conjugates the $2|1$-dimensional representation  that we use in writing
most formulas 
into an inequivalent representation, with reversed $R$-charges.   (The formula  $V_\cC =\cC V$ should really be interpreted as a formula in the group or in a $\cC$-invariant
representation such as the adjoint representation, not a formula in $3\times 3$ matrices as written in eqn. (\ref{crmonodromy}).)
$\cC$ acts on the Lie algebra of $\SU(1,1|1)$, in the notation of eqn.     (\ref{slgenerators}), by 
\begin{align}\label{actionc} {\sf q}_1& \leftrightarrow - {\sf q}_2 \cr
                                            {\sf q}_3&\leftrightarrow {\sf q}_4 \cr
                                                  {\sf z}&\leftrightarrow -{\sf z} ,\end{align}
                                                   with ${\sf e},{\sf f},{\sf h}$ invariant.   This defines an automorphism of the Lie algebra, and clearly $\cC^2=1$.   
                                                   
 To compute the torsion, we have to replace $\h V(x)=V x V^{-1}$ by $\h V_\cC(x)=\cC(V x V^{-1})   $ with  $V$ as defined in eqn. (\ref{crmonodromy}).
   One now has $\tau_C^\perp=\Ber'(\h V^\perp_{\cC}-1)$.   
In the previous example, $\mathfrak g^\perp_+$ was spanned by the $\SU(1,1)$ generators ${\sf e,f}$. These generators are $\cC$-invariant,
so acting on them, 
there is no difference between $\h V$ and $\h V_\cC$, and they contribute a factor $4\cosh^2\frac{b}{4}$ to $|\det(\h V_{\cC \perp,+}-1)|$, as before.    
However, as the $R$-charge generator $\sf z$ is $\cC$-odd, it is not in the kernel of  $\h V_\cC-1$ and instead has eigenvalue $-2$.
Hence it contributes a factor of 2 to $|\det\,(\h V^\perp_{\cC,+}-1)|$, which now equals 
$8\cosh^2\frac{b}{4}$.
  Acting on  $\mathfrak g^\perp_-$, $\h V_\cC$ is quite different from $\h V$.   It acts by
\be\label{vcactions}\h V_\cC:\begin{pmatrix} {\sf q}_1 \cr {\sf q}_2 \cr {\sf q}_3 \cr {\sf q}_4\end{pmatrix}\to \begin{pmatrix} -\i e^{b/4} {\sf q}_2 \cr -\i e^{b/4} {\sf q}_1\cr -\i e^{-b/4} {\sf q}_4\cr -\i  e^{-b/4}{\sf q}_3\end{pmatrix}. \ee                                          
From this, one finds $\det(\h V^\perp_{\cC,-}-1)=4\cosh^2\frac{b}{4}=\frac{1}{2}\det(\h V^\perp_{\cC,+}-1)$.  So eqn. (\ref{orthotor}) is replaced by
\be\label{tcreplace}\tau_C^\perp=2. \ee   
What about $\tau_C^0$?     The kernel of $\h V_\cC$ is generated by $\sf h$, and the same derivation as in eqn. (3.49) of \cite{SW} leads to
$\tau_C^0=\frac{1}{2}\d b\d\varrho^{-1}$.    Putting these factors together,
\be\label{combinedtor}\tau_C=\tau_C^\perp \tau_C^0 =\d b \d\varrho^{-1}. \ee
This is a measure for a flat connection on the crosscap manifold, where the monodromy $U$ around the outer circle automatically has $\phi=0$.  

The prediction for the path integral measure for a trumpet ending on a crosscap in a theory with $\tT$ symmetry can be deduced from eqn. (\ref{longformal}) using
this formula for $\tau_C$:
\be\label{crosspredict2} Z_{{\rm xcap},\tT}(b)=Z_{\rm tr}(b,0) \frac{\cosh\frac{b}{4}}{\sinh\frac{b}{4}} \d b \d\varphi. \ee
We have written the left hand side only as a function of $b$, with no $\phi$; and on the right hand side, we have set $\phi=0$ in the trumpet partition function $Z_{\rm tr}(b,\phi)$.

Compared to eqn. (\ref{crosspredict}), we see a notable difference: there is no $\d\phi$, and instead there is a $\d\varphi$.   Technically, the reason that this happened
is that because the $R$-charge generator $\sf z$ is not annihilated by $\h V_\cC$, $\tau_C^0$ did not contain a factor $\d\phi (\d\varphi)^{-1}$ that was present in the previous
analysis.   Eqn. (\ref{crosspredict2}) is telling us that instead of integrating over an $R$-symmetry angle $\phi$ in the monodromy $V$, we have to integrate over an $R$-symmetry
parameter $\varphi$ that appears in gluing the crosscap to the trumpet. 

  The fact that there is no $R$-symmetry angle $\phi$ to integrate over is indeed something we already know, since the monodromies 
(\ref{crmonodromy}) do not depend on any such parameter.
But why are we supposed to integrate over an $R$-symmetry gluing parameter $\varphi$?  The point is that, at the asymptotic boundary of the trumpet (or of any NAdS$_2$
spacetime), gauge transformations are not allowed.    So if $A$ is the $R$-symmetry gauge field, we can define a gauge-invariant parameter $\vartheta=\int_\ell A$, where
the integration path $\ell$ starts on the asymptotic boundary of the trumpet, heads in to the geodesic $\gamma$, winds once around $\gamma$, and then heads back out to
the asymptotic boundary.    In our previous analysis with $\rR$ symmetry rather than $\CR$, the parameter $\vartheta$ would have no gauge-invariant content (beyond what is contained in 
the monodromy $V$ around $\gamma$) because the parts of the integral from the asymptotic boundary to $\gamma$ and back again would cancel.  However, with $\CR$ symmetry,
because the monodromy around $\gamma$ involves the action of $\cC$, under which $A$ is odd, the ``incoming'' and ``outgoing'' parts of the integral add, instead of canceling.
Since these integrals do add, one can identify $\vartheta$ as $2\varphi$, where $\varphi$ is the $R$-symmetry gluing parameter that we have to integrate over, according to
eqn. (\ref{crosspredict2}).  The range of the $\uU(1)$ holonomy parameter $\vartheta$ is $0\leq\vartheta\leq 2\pi$, so the range of $\varphi=\vartheta/2$ is $0\leq \varphi\leq\pi$. 
Integration over $\varphi$ with measure $\d\varphi$ just gives a constant factor $\pi$, since the integrand does not depend on $\varphi$.

Finally, we discuss some topological details that will be important in understanding and completing the calculations in section \ref{matrixprediction}.   
In the case of $\tT$ symmetry, $U$ and $V$ are unique, up to conjugation.   However, with $\CT$ symmetry, $U$ has an arbitrary $R$-symmetry monodromy $e^{\i\hq \phi}$ and
for each choice of $\phi$, there are two possible choices of $V$.   For given $\phi$, each choice of $V$ corresponds to a flat $\uU(1)_R$ bundle $\L\to C$
with holonomy $e^{\i\hq\phi}$ around the boundary of $C$.   The two flat bundles differ by the value of the first Chern class  $c_1(\L)$. This is a relative
first Chern class, because $C$ has a boundary; one cannot say which choice of $V$ corresponds to $c_1(\L)=0$ and which  corresponds to $c_1(\L)\not=0$,
but one can say that they are different.\footnote{The topology is easier to describe if we turn $C$ into a compact manifold without boundary  by gluing on a disc.   In that
way we make a closed, unorientable manifold which is a copy of ${\mathbb {RP}}^2$.  A flat $\uU(1)$ bundle over this manifold is either trivial, or else has holonomy $-1$ around the
generator of $\pi_1({\mathbb{RP}}^2)\cong \Z_2$.   The first Chern class of this nontrivial flat bundle is the nonzero element of $H^2({\mathbb{RP}}^2,\Z)\cong\Z_2$.   This
is a torsion element, so it cannot be detected by curvature.   In general the first Chern class of a line bundle with connection is determined by the curvature of the connection
only modulo torsion.}  $\CT$ invariance implies that
the $\uU(1)_R$ theta-angle has the possible values $0$ and $\pi$.   According to eqn. (\ref{delfo}), $\theta=\pi$ if and only if $\delta=0$.  In the case of a Riemann
surface $C$ without boundary, the meaning of $\theta=\pi$ is that the path integral with a line bundle $\L$ should be weighted by a factor $(-1)^{ c_1(\L)}$.
In the present case, because $C$ has a boundary and $c_1(\L)$ is a relative first Chern class, it is subtle to explain what is meant by $(-1)^{c_1(\L)}.$ 
   A simple statement, however, is that when we multiply $Z_{{\rm xcap},\CT}$ by the trumpet path integral,
which for $\delta=0$ is odd under $\phi\to\phi+2\pi$, the two choices of $V$ will contribute with opposite signs.  This is a manifestation of the fact that $\theta=\pi$ if $\delta=0$.
If $\delta=1/2$, the trumpet is even under $\phi\to \phi+2\pi$, so the two choices of $V$ will contribute with the same sign, in accord with the fact that $\theta=0$.

However, $Z_{{\rm xcap},\CT}(b,\phi)-Z_{{\rm xcap},\CT}(b,\phi+2\pi)$ is imaginary, so if the two choices of $V$ contribute with opposite signs, then 
it seems that the path integral will be imaginary.  Euclidean path integrals can be complex-valued in general, but an imaginary result here will not agree with expectations
from the boundary matrix ensemble.     In fact, a rather subtle additional  factor of
$\pm\i$ appears precisely when $\delta=0$.    A $\CT$-invariant (or $\tT$-invariant) theory with $n$ Majorana fermions all transforming in the same way under $\CT$ (or $\tT$) 
has an anomaly that depends on the value of $n$ mod 8.   The $\N=2$ Schwarzian theory has two Majorana fermions, which transform oppositely under $\tT$, leading to no anomaly,
but transform the same way under $\CT$, so that the $\N=2$ Schwarzian theory with $\CT$ symmetry 
has an anomaly coefficient\footnote{In $\N=1$ JT supergravity, the Schwarzian theory has a single
Majorana fermion, and is anomalous even in the absence of time-reversal symmetry.  This was important in \cite{SW}.  The $\eta$-invariant entered in that analysis
for reasons similar to what we explain here.}  of 2.   If $\delta=1/2$, the boundary theory is anomalous in the same
way and the anomalous Schwarzian theory correctly reproduces the anomaly in the microscopic boundary description,\footnote{We can realize $\delta=1/2$ with an $\N=2$ SYK model
with an odd number $N=2r+1$ of complex fermions.  This is equivalent to $4r+2$ Majorana fermions, so the anomaly coefficient  is $4r+2$, or $\pm 2$ mod 8.   Thus the boundary
anomaly either equals the anomaly of the Schwarzian theory or  differs from it by 4. 
If the two coefficients are equal, the Schwarzian theory simply reproduces the anomaly of the boundary theory.   If they differ by 4, we can compensate
with a bulk factor $(-1)^\chi$, which carries an anomaly coefficient of 4.  Including this factor
is a familiar option and does not affect the fact that the path integral is real.}    but if $\delta=0$, the boundary theory
is anomaly-free and we need to modify the bulk theory to compensate for the anomaly of the $\N=2$ Schwarzian theory.  In appendix \ref{sec:disktrumpet}, a factor in computing
the partition function of the $\N=2$ Schwarzian theory was ${\mathrm{Pf}}(D)$, where ${\mathrm{Pf}}$ is the Pfaffian (the square root of the determinant) and $D$, the rotation generator, can be viewed
as the 1-dimensional Dirac operator
of the two Majorana fermions that are described by the $\N=2$ Schwarzian theory.  This Dirac operator is defined on $S$, the boundary of the spacetime $C'$ obtained
by gluing the crosscap to the trumpet.   A general recipe\footnote{This recipe is suggested by
the Dai-Freed theorem \cite{DF} and was most fully justified from a physical point of view in \cite{WY}.}  to define a potentially anomalous
fermion path integral on a manifold $S=\partial C'$ is to replace ${\mathrm{Pf}}(D)$ with $|{\mathrm{Pf}}(D)|\exp(-\i \pi \eta(D_{C'})/2)$, where $\eta(D_{C'})$ is the Atiyah-Patodi-Singer
eta-invariant of a bulk Dirac operator 
$D_{C'}$ that extends $D$.  In the present case, we can define $D_{C'}$ as the Dirac operator (for some ${\rm pin}^-$ structure)
coupled to a flat gauge field on $C'$ defined by the monodromy $V$ of eqn. (\ref{crosscaprep}) with $\phi$ replaced by $\alpha$.     Any other choice of $D_{C'}$ is equivalent topologically  to this, possibly with a different ${\rm pin}^-$ structure (or equivalently with $\phi=\alpha+2\pi$).   To see the consequences, first note that if $\alpha=0$, then
$D_{C'}$ is the direct sum of two copies of a Dirac operator coupled to a ${\rm pin}^-$ bundle.  Its $\eta$-invariant is then $-1$ or
$1$, depending on the ${\rm pin}^-$ structure, so $\exp(-\i\pi\eta/2)=\pm\i$.  When one varies $\alpha$, because $\eta$ is a topological invariant on a closed manifold of even dimension, 
$\eta(D_{C'})$ is constant except for jumps that precisely ensure that $|{\rm Pf}(D)|e^{-\i\pi \eta(D_{C'})/2}$ varies smoothly with $\alpha$.   So $|{\rm Pf}(D)|e^{-\i\pi \eta(D_{C'})/2}=
\pm\i \,{\rm Pf}(D)$ (with a sign that does not depend on $\alpha$), and the effect of including the  eta-invariant and canceling the anomaly is just to multiply the path integral of the Schwarzian modes, naively
${\rm Pf}(D)$,  by an overall factor $\pm \i$.   The choice of
sign leads  to two bulk theories that differ by a factor of $(-1)^\chi$, dual as usual to two options in the boundary theory.

\subsection{The Matrix Model Prediction}\label{matrixprediction}

We will now compare to matrix model predictions in the two cases of $\tT$ and $\CT$ symmetry.
 
(A) We begin with  ${\sf T}$ symmetry, and consider statistically independent multiplets with $(\upalpha_0,\upbeta)=(0,1)$ or $(3,4)$. The spectral curve of a supermultiplet with charges $(q-\hq/2,q+\hq/2)$ is given by eqn. \eqref{eq:N2sc}. We remind the reader that in gravity it is more convenient to label a multiplet by the bulk charge $q$, which is also equal to the average microscopic $R$-charge of the multiplet. Let us first write down the prediction from $\cN=2$ JT gravity for the partition function of the crosscap more explicitly
\bea
Z_{{\rm xcap},{\sf T}}(\beta,\alpha) &=& \int_0^\infty  \d b \, Z_{\rm tr}(b,0) \pi \coth\Big(\frac{b}{4}\Big)\nn
&=& \sum_{q\in \mathbb{Z}+\delta-\frac{1}{2}} \big( e^{\i \alpha (q-\frac{\hq}{2}) } + e^{\i \alpha (q+\frac{\hq}{2})}\big) \int_0^\infty \d b \, Z_{\rm tr}^{\rm bos.}(\beta,b) V^{(q)}_{{\rm xcap},{\sf T}}(b),
\ea
where we introduced the ``fixed charge'' volume for the crosscap
\beq\label{eq:xcapfc}
V^{(q)}_{{\rm xcap},{\sf T}}(b)=\frac{\coth (\frac{b}{4})}{2}.\hspace{1cm}{\rm Case~A}
\eeq
A factor of $1/(2\pi)$ comes from the normalization of the $\N=2$ trumpet in terms of its bosonic counterpart $Z_{\rm tr}^{\rm bos}$. This definition of the ``fixed charge'' crosscap volume in terms of the partition function is consistent with eqns. \eqref{volcompar} and \eqref{Zqg} specialized to $g=1/2$ and $n=1$. We used a subscript ${\sf T}$ in $V^{(q)}_{{\rm xcap},{\sf T}}$
 to distinguish this crosscap volume from the one in case (B) below. The goal of this section is to reproduce eqn. \eqref{eq:xcapfc} from the matrix model. 

The genus $g=1/2$ correction $R^{(q)}_{\rm xcap}(x)$ to the one boundary resolvent $R(x)$ of a supermultiplet $q$ can be derived from the matrix model loop equation, which we review in appendix \ref{app:Loop}, and is given by 
\beq\label{eq:xcapppp}
R^{(q)}_{\rm xcap}(x) = \int \frac{\d x'}{2\pi \i} \frac{1}{x'-x} \frac{\sqrt{\sigma(x')}}{\sqrt{\sigma(x)}} \left( \frac{(1-\frac{2}{\upbeta})y'(x')}{2y(x')} + \frac{\upalpha_0-1}{2\upbeta x'}\right),
\eeq
and we can use $\sigma(x)=(x-\frac{q^2}{4\hq^{\,2}})$ (this function is introduced in the appendix), since there is a single edge in the double-scaling limit. The integration contour for $x'$ starts as a small circle around $x$ and can be deformed to compute the discontinuity along the cut. The answer for an $(\upalpha_0,\upbeta)=(0,1)$ AZ ensemble with spectral curve \eqref{eq:N2sc} is given by 
\beq
R^{(q)}_{\rm xcap}(x) = \frac{1}{2\sqrt{-x+\frac{q^2}{4\hq^{\,2}}}} \int_{\frac{q^2}{4\hq^{\,2}}}^{\infty} \frac{\sqrt{x'-\frac{q^2}{4\hq^{\,2}}} \, \d x'}{x'-x} \left[-\frac{\coth\big(2\pi \sqrt{x'-\frac{q^2}{4\hq^{\,2}}}\big)}{\sqrt{x'-\frac{q^2}{4\hq^{\,2}}}}  \right]
\eeq
Shifting the integrand $x'\to \frac{q^2}{4\hq^{\,2}} + x'$ and writing $x$ in terms of a new variable $z$ as $x= \frac{q^2}{4\hq^{\,2}} - z^2$, we can put the matrix model prediction for the crosscap in the form
\beq\label{eq:R12A}
R^{(q)}_{\rm xcap}\Big(x= \frac{q^2}{4\hq^{\,2}}-z^2\Big) =- \frac{1}{2z} \int_{0}^{\infty} \frac{\sqrt{x'} \d x'}{x'+z^2} \frac{\coth\big(2\pi \sqrt{x'}\big)}{\sqrt{x'}} 
\eeq
The matrix model prediction of the crosscap volume is given in terms of the resolvent by the integral transform
\bea\label{eq:VolXCAPMM}
V^{(q)}_{{\rm xcap,M}}(b) = \int \frac{\d z}{2\pi \i} e^{b z}(-2 z) R^{(q)}_{\rm xcap}\Big(x= \frac{q^2}{4 \hq^{\,2}}-z^2 \Big),
\ea
where $z$ is integrated in the imaginary direction. To derive this relation, at the level of an individual multiplet, one can combine $Z_{\rm xcap}(\beta) = \int_0^\infty \d b Z_{\rm tr}^{\rm bos.}(\beta,b) e^{-\beta E_0} V_{\rm xcap,M}(b)$ with $R(x) = -  \int_0^\infty \d\beta\, e^{\beta x} Z(\beta)$. Apply first the integral transform to go from the partition function to the resolvent and use that $-\int_0^\infty \d \beta \, e^{\beta x} Z_{\rm tr}^{\rm bos} = e^{-b z}/(-2 z) $, with $x= E_0 - z^2$, to write the resolvent in terms of the volume. Applying an inverse Laplace transform, and using that $E_0= q^2/4 \hq^2$, one arrives at the expression \eqref{eq:VolXCAPMM} given above.

We can insert \eqref{eq:R12A} into the right hand side of \eqref{eq:VolXCAPMM}, to derive $V^{(q)}_{{\rm xcap,M}}(b)$. Even though the integral appearing in \eqref{eq:R12A} diverges for large $x'$, we can obtain a finite answer for the volume by first integrating over $z$ and second over $x'$. We get
\bea\label{eq:xcapTm}
V^{(q)}_{{\rm xcap,M}}(b) &=& \int_{0}^{\infty}\d x' \,\sin\big(b\sqrt{x'}\big)\frac{\coth\big(2\pi \sqrt{x'}\big)}{\sqrt{x'}} =\frac{ \coth (\frac{b}{4})}{2}=V^{(q)}_{{\rm xcap},{\sf T}}(b) .
\ea
This matches precisely the prediction from gravity in \eqref{eq:xcapfc}, as indicated in the right hand side of eqn. \eqref{eq:xcapTm}. The result for the $(\upalpha_0,\upbeta)=(3,4)$ ensemble is the same up to an overall minus sign, which can be reproduced in gravity by including a factor of $(-1)^{n_c}$ in the path integral, where $n_c$ is the number of crosscaps.\\

(B) We now analyze  ${\sf CT}$ symmetry. The gravity calculation performed in the previous section gives the following answer for the crosscap partition function
\bea
Z_{{\rm xcap},{\sf CT}}(\beta,\alpha) &=& \int_0^\infty  \d b \, \int_0^{4\pi} \d \phi \, Z_{\rm tr}(b,\phi) \frac{( \cosh \frac{b}{2}+ \cos \h q \phi )\coth\frac{b}{4} }{8(\cos \frac{\h q\phi}{2}-\i \sinh \frac{b}{4})^2}   ,
\ea
The integrand is precisely \eqref{crosspredict} and the range over the $\uU(1)$ holonomy $\phi$ is from $0$ to $4\pi$ since we need to integrate over all possible holonomies around the crosscap, given in eqn. \eqref{crosscaprep}. 
The integral is not absolutely convergent because of singularities when $b=0$ and $\hq \phi$ is an odd multiple of $\pi$; we  will integrate over $\phi$ before integrating over $b$, a procedure that seems to give sensible results.
Comparing with \eqref{Zqg}, the crosscap volume at ``fixed charge'' is given by 
\beq
V_{{\rm xcap},{\sf CT}}^{(q)}(b) =\int_0^{4\pi} \frac{\d \phi}{2\pi} \, e^{-\i \phi q}\, \frac{( \cosh \frac{b}{2}+ \cos \h q \phi )\coth\frac{b}{4} }{8(\cos \frac{\h q\phi}{2}-\i \sinh \frac{b}{4})^2}
\eeq
A factor of $1/(2\pi)$ comes, again, from the normalization of the $\N=2$ trumpet in terms of $Z_{\rm tr}^{\rm bos.}$. This integral seems complicated but can be computed using residues in a straightforward way. In order to do this, change variables from $\phi$ to $w = e^{\i \frac{\phi}{2}}$ and write the integrand as a meromorphic function in $w$ implementing the following replacements $2\cos \hq \phi \to w^{2\hq}+w^{-2\hq}$, $2\cos \frac{\hq \phi}{2} \to w^{\hq}+w^{-\hq}$ and $e^{-\i \phi q} \to w^{- 2q}$. The corresponding integration contour in the complex $w$-plane is over the unit circle $|w|^2=1$. We can evaluate the integral as the sum over the residues at the poles inside the unit disk in the complex $w$-plane. The integrand has potentially a pole at $w=0$ and $\hq$ poles at $w^{\hq} = - \i e^{-b/4}$ (the poles at $w^{\hq} = \i e^{+b/4}$ are outside of the unit disk and do not contribute to this sum). The final answer is 
\beq
V_{{\rm xcap},{\sf CT}}^{(q)}(b) = \begin{cases}
			\frac{\coth \frac{b}{4}}{2} \delta_{q,0} - e^{-\frac{\i \pi |q|}{\hq}} e^{-\frac{|q|}{2\hq} b}, & \text{if $2q = 0 ~{\rm mod}~ \hq$,}\\
            0, & \text{otherwise}
		 \end{cases}
		 \label{CTxcapv}
\eeq
As a preliminary observation, charge conjugation acts as $q\to -q$ and we see indeed that the crosscap volume depends only on $|q|$. We remind the reader that a ${\sf CT}$ symmetry is only possible if $\delta = 0$ or $1/2$. We separately analyze each case below.

\paragraph{$\delta = 1/2$:} Let us consider first the anomalous theory. In this case the spectrum of $R$-charges $k$ is half-integral and therefore the bulk charge of a supermultiplet $q=k\pm \hq/2$ is an integer. The random matrix ensemble analysis implies that for all $q>0$ we should study an AZ ensemble with $(\upalpha_0,\upbeta)=(1,2)$ while for $q=0$ we should study the $(\upalpha,\upbeta)=(1,1)$ or $(1,4)$ ensembles, depending on the sign of ${\sf CT}^2$. Let us therefore begin with $q=0$, corresponding to a supermultiplet with $R$-charges $(-\hq/2,\hq/2)$. In this case the crosscap volume given in eqn. \eqref{CTxcapv} becomes
\beq
V_{{\rm xcap},{\sf CT}}^{(0)}(b) = \frac{\coth \frac{b}{4}}{2} - 1
\eeq
Next compare this with the matrix model crosscap in the $(\upalpha,\upbeta)=(1,1)$ AZ ensemble. The answer for $(1,4)$ is the same up to an overall minus sign that in gravity can be interpreted as adding a factor $(-1)^{n_c}$ in the path integral, where $n_c$ is the number of crosscaps. The matrix model prediction to the crosscap contribution to the resolvent, using the spectral curve given in eqn. \eqref{eq:N2sc} and using the loop equation \eqref{eq:xcapppp}, is given by 
\bea
R^{(0)}_{\rm xcap}(x=-z^2) &=&  \int \frac{\d x'}{2\pi \i} \frac{1}{x'-x} \frac{\sqrt{\sigma(x')}}{\sqrt{\sigma(x)}} \left( \frac{(1-\frac{2}{\upbeta})y'(x')}{2y(x')} + \frac{\upalpha_0-1}{2\upbeta x'}\right)\nonumber\\
 &=&\frac{1}{2z} \int_{0}^{\infty} \frac{\sqrt{x'} \d x'}{x'+z^2} \left[-\frac{\coth\big(2\pi \sqrt{x'}\big)}{\sqrt{x'}}  + \frac{1}{\pi x'}\right],
\ea
where $\sigma(x)= x$ in this case. We remind the reader in deriving this equation that for the $q=0$ supermultiplet, the gap vanishes $E_0(q=0)=0$ and there are no BPS states with charges $\pm \hq/2$. Since there is no pole in the spectral curve for this multiplet at $x=0$ we also do not need to add any logarithmic correction to the matrix potential. We can use the relation between the matrix model prediction for the volume and the matrix model resolvent to verify
\bea
V^{(0)}_{{\rm xcap,M}} (b) &=& \int \frac{\d z}{2\pi \i} e^{b z} (-2z) R^{(0)}_{\rm xcap}(-z^2) \notag \\ \notag
&=& \int_{0}^{\infty}\d x' ~\sin\big(b\sqrt{x'}\big)\left[\frac{\coth\big(2\pi \sqrt{x'}\big)}{\sqrt{x'}}- \frac{1}{\pi x'} \right]\\
&=&\frac{\coth \frac{b}{4}}{2}  -1=V_{{\rm xcap},{\sf CT}}^{(0)}(b) .
\ea
We therefore obtain a match between gravity and the matrix ensemble proposed at the beginning of this section for the $q=0$ sector. 

Next we analyze the other multiplets with $q\neq 0$. An $(\upalpha_0,\upbeta)=(1,2)$ AZ ensemble predicts a vanishing crosscap contribution to the resolvent. In gravity we obtained that the crosscap volume vanishes for most charges except when $2q=0~{\rm mod}~\hq$, which in the $\delta=1/2$ theory reduces to $q = 0 ~{\rm mod}~\hq$. The crosscap volume in this case is 
\beq\label{eqCTqn0}
V_{{\rm xcap},{\sf CT}}^{(q)}(b) = - (-1)^q e^{-\frac{|q|}{2\hq} b}.
\eeq
How can we make sense of this non-zero result? It is customarily assumed that, when taking the large $L$ limit of a matrix integral, we keep fixed the matrix potential or equivalently the spectral curve $y(x)$. Nevertheless, one is free to include small corrections to the spectral curve (or the matrix potential) of order $1/L$.  This will contribute to the resolvent
at the same order in $e^{-S_0}$ as a crosscap.   This possibility has
not been necessary in previous discussions of two-dimensional gravity and matrix models, but it does seem to be necessary in $\N=2$ JT supergravity with $\CT$ symmetry.

Let us write $y(x) = y_0(x) + e^{-S_0} y_{1/2}(x) + \ldots $ and compute the correction $y_{1/2}(x)$ needed to reproduce \eqref{eqCTqn0}. We can use the crosscap volume from gravity to evaluate the change in the partition function
\beq
Z^{(q)}_{\rm xcap} (\beta) =  \int_0^\infty \d b\, \frac{e^{-\frac{b^2}{4\beta}}}{2\sqrt{\pi \beta}} e^{-\beta \frac{q^2}{4 \hq^2}} \, (-1)^{q+1} e^{-\frac{|q|b}{2\hq}} .
\eeq
The correction to the resolvent is given by
\beq\label{eq:Rxcapqnq0}
R_{\rm xcap}^{(q)}(x) = - \int_0^\infty \d \beta \, e^{\beta x} \, Z^{(q)}_{\rm xcap} (\beta) =  \frac{(-1)^q}{2z(z+\frac{|q|}{2\hq})},~~~{\rm where}~~x=\frac{q^2}{4\hq^2}-z^2.
\eeq
The difference between $R(x)$ and $y(x)$ is $V'(x)/2$, which can be an arbitrary analytic function of $x$ up to the possible presence of a simple pole at $x=0$. But
$y$ is odd under the hyperelliptic involution $z\leftrightarrow -z$.   Therefore we can read off the change in
$y$ from the odd piece of \eqref{eq:Rxcapqnq0}:
\beq\label{xcapcorry}
y_{1/2}(x) =(-1)^q \frac{|q|}{2\hq}\, \frac{1}{2 x z}.
\eeq
The first check of our proposal is that  $y_{1/2}(x)$ has the right analytic behavior to be interpreted as a correction to the spectral curve. Following \cite{Maxfield:2020ale} and \cite{Witten:2020wvy}, the pole in $z$ at $z=0$ indicates this is the first term in a series that resums into a shift of the ground state energy of the supermultiplet 
\beq\label{shiftE0}
E_0(q) \to E_0(q) + e^{-S_0}(-1)^q \frac{2\pi |q|}{\hq} + \mathcal{O}(e^{-2S_0}).
\eeq
This statement, and the precise prefactors, can be verified easily by expanding the spectral curve \eqref{eq:N2sc} under a small variation of $E_0$ near $z=0$. In addition,
 the correction to the spectral curve \eqref{xcapcorry} has a pole at $x=0$ indicating the need for the correction to the matrix potential  to have a $\log x$ term as well.   
 A singularity of $y$ other than the
$1/z$ and $1/x$ singularities that we have actually found would not be consistent with interpreting $y_{1/2}(x)$ as a correction to the spectral curve.

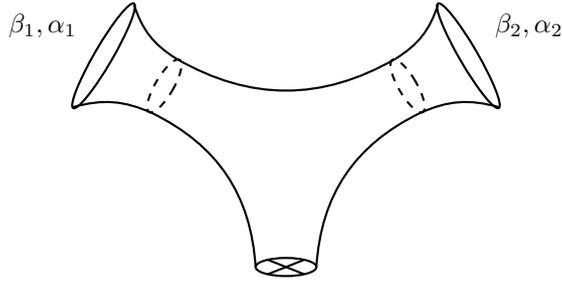
\begin{figure}[t!]
\begin{center}
\begin{tikzpicture}[scale=0.8, baseline={([yshift=-0.1cm]current bounding box.center)}]
\draw[thick] (0,-1) ellipse (.5 and .15);%Boundary
\draw[thick, rotate around={-30:(-3,2.5)}] (-3,2.5) ellipse (.15 and 1);%Boundary
\draw[thick, rotate around={30:(3,2.5)}] (3,2.5) ellipse (.15 and 1);%Boundary
 \draw[thick, bend right=30] (-1.8,2.45) to (1.8,2.45);
  \draw[thick, bend left=20] (-1.8,2.45) to (-2.5,3.35);
  \draw[thick, bend right=20] (1.8,2.45) to (2.5,3.35);
  \draw[thick, bend right=20] (-2.2,1.55) to (-3.5,1.65);
  \draw[thick, bend left=20] (2.2,1.55) to (3.5,1.65);
 \draw[thick, bend left=30] (-2.2,1.55) to (-0.5,-1);
  \draw[thick, bend right=30] (2.2,1.55) to (0.5,-1);
  \draw[thick] (-.3,-1-.12) to (.3,-1+.12);
   \draw[thick] (.3,-1-.12) to (-.3,-1+.12);
  %\draw[thick, bend right=30] (0.5,-1) to (1,-2);
  \draw[thick, dashed, rotate around={-30:(-2,2)}] (-2,2) ellipse (.15 and .5);%Geodesic
\draw[thick, dashed, rotate around={30:(2,2)}] (2,2) ellipse (.15 and .5);%Geodesic
  \node at (-4,3) {\small $\beta_1,\alpha_1$};
   \node at (4,3) {\small $\beta_2,\alpha_2$};
    \end{tikzpicture} 
    \end{center}
    \caption{\footnotesize Contribution to the connected two-boundary path integral with a crosscap in the interior. The two NAdS boundaries are labeled by their inverse temperature and $R$-charge chemical potential, while the crosscap is indicated with a cross. The geodesic dividing the NAdS boundaries from the internal surface are denoted by dashed lines.}
    \label{fig:2bdyxcap}
    \end{figure}

This interpretation of $y_{1/2}$  has non-trivial implications for higher order corrections to the resolvent, and we will not attempt a full analysis here.  But we will verify our proposal in  the first nontrivial case. The leading order, connected, two-boundary partition function of the matrix integral of an AZ ensemble in the double scaling limit is universal and depends solely on $E_0$, the ground state energy, 
\beq
Z_{0,2}=\frac{\sqrt{\beta_1\beta_2}}{2\pi(\beta_1+\beta_2)} e^{-(\beta_1+\beta_2)E_0}.
\eeq 
Therefore a change in $y$ will influence $Z_{0,2}$ only through the shift in $E_0$. 
With this in mind, consider the topological expansion of the two-boundary partition function as $Z_2= Z_{0,2}+e^{-S_0}Z_{\frac{1}{2},2}+{\mathcal O}(e^{-2S_0})$.  If the $e^{-S_0}Z_{\frac{1}{2},2}$ contribution
reflects a correction to $y$, then we can predict $Z_{\frac{1}{2},2}$ in terms of
the shift of the ground state energy computed in \eqref{shiftE0}.   In terms of the fixed charge partition function, the prediction is 
\beq\label{refback}
Z^{(q)}_{\frac{1}{2},2} (\beta_1,\beta_2) = -(-1)^q \frac{2\pi |q|}{\hq}\frac{\sqrt{\beta_1\beta_2}}{2\pi} e^{-(\beta_1 + \beta_2) E_0(q)}.
\eeq 
We can reproduce this prediction directly from gravity. For this purpose, we consider the path integral on a hyperbolic cylinder with a crosscap in the interior, see fig. \ref{fig:2bdyxcap}. We first introduce $V_{\frac{1}{2},2}({\sf b}_1,{\sf b}_2)$ as the path integral of $\N=2$ JT gravity on the sphere with two geodesic boundaries and one crosscap. This quantity is easy to compute combining the rules given in section \ref{sec:GPI} and section \ref{bulk}, since it does not involve a non-trivial action of the mapping class group. The answer is 
\bea
V_{\frac{1}{2},2}({\sf b}_1,{\sf b}_2) &=& \int_0^{4\pi} \d \phi_3 \,\int_0^{\infty} \d b_3 V_{0,3}({\sf b}_1, {\sf b}_2, \overline{{\sf b}}_3)\frac{( \cosh \frac{b_3}{2}+ \cos \h q \phi_3 )\coth\frac{b_3}{4} }{8(\cos \frac{\h q\phi_3}{2}-\i \sinh \frac{b_3}{4})^2}    \nonumber \\
&=&\sum_{q} \frac{1}{2\pi}\frac{q^2}{4\hq^2}\, e^{\i q \phi_1 + \i q \phi_2} \int_0^{\infty} \d b_3 \int_0^{4\pi} \frac{\d \phi_3}{2\pi} \, e^{-\i \phi_3 q}\, \frac{( \cosh \frac{b_3}{2}+ \cos \h q \phi_3 )\coth\frac{b_3}{4} }{8(\cos \frac{\h q\phi_3}{2}-\i \sinh \frac{b_3}{4})^2}\nonumber\\
&=& - \sum_{q}  \frac{(-1)^{q}}{2\pi} \frac{q^2}{4\hq^2}\, e^{\i q \phi_1 + \i q \phi_2} \int_0^{\infty} \d b_3   e^{-\frac{|q|}{2\hq} b} \nonumber\\
&=& - \sum_{q}  \frac{(-1)^{q}}{2\pi} \frac{|q|}{2\hq}\, e^{\i q \phi_1 + \i q \phi_2} 
\ea
In the first line, we start with a three-holed sphere and glue a crosscap to one hole, labeled $3$. We included an orientation reversal of  the crosscap $R$-symmetry holonomy with respect to the three-holed sphere as indicated by $\overline{{\sf b}}_3 = (b_3, -\phi_3)$. In the second line we used \eqref{eqn:V03final} and wrote the delta function on holonomies as a sum over charges $q$
\beq
-\delta''(\phi_1 + \phi_2 - \phi_3) = \frac{1}{2\pi} \sum_{q\in\mathbb{Z}} q^2 e^{\i q \phi_1 + \i q \phi_2 - \i q \phi_3}.
\eeq
which allowed us to combine a factor of $e^{-\i \phi_3 q}$ with the crosscap contribution into the same integral we already computed above in \eqref{CTxcapv}. Finally we performed the $\phi_3$ integral. In the last two lines we did not include the contribution of \eqref{CTxcapv} with $q=0$ since, even though the $b_3$ integral is divergent, it is multiplied by a factor of $q^2$ and therefore automatically vanishes. (This is consistent with the fact that the $q=0$ multiplet does not require any correction to the spectral curve.) Finally, we can glue this volume to the two external trumpets and obtain the correction to the two-boundary partition function in a sector of ``fixed charge'' $q$
\bea
Z^{(q)}_{\frac{1}{2},2}(\beta_1,\beta_2) &=&- (-1)^q \int_0^\infty b_1 \d b_1 Z_{\rm tr}^{\rm bos.}(b_1,\beta_1)\int_0^\infty b_2 \d b_2 Z_{\rm tr}^{\rm bos.}(b_2,\beta_2) (2\pi) \frac{|q|}{2\hq}\\
&=& -(-1)^q \frac{\sqrt{\beta_1\beta_2}}{\pi} (2\pi) \frac{|q|}{2\hq}=- (-1)^q \frac{|q|}{\hq}\sqrt{\beta_1\beta_2}
\ea
This answer obtained from the gravitational path integral, in the second line, matches precisely the matrix model prediction (\ref{refback}) in the presence of a correction to the spectral curve, verifying our proposal. 

Even though we have found a matrix integral that reproduces the gravitational path integral to this order in the topological expansion, we should stress that we do not have an understanding of why in the presence of $\CT$ symmetry, and not otherwise, such a modification of the spectral curve is necessary.

\paragraph{$\delta = 0$:} Finally we analyze the non-anomalous theory with $\delta=0$. Since the spectrum of $R$-charges $k$ is integral, the spectrum of bulk charges $q=k\pm\hq/2$ will be half-integral. Again we find from \eqref{CTxcapv} that most multiplets have a vanishing contribution to the crosscap except when $2 q = 0~{\rm mod}~\hq$. In those cases, the crosscap volume has the same $b$-dependence as considered above and we can again interpret it as a shift of the spectral curve. The check for the cylinder with a crosscap works as well.  The only new feature in this case is that the ``fixed charge'' crosscap volume is imaginary since it is proportional to $e^{-\frac{\i \pi |q|}{\hq}} = \i (-1)^{q+\frac{\hq}{2}} $ with half-integer $q$.  To get a real answer, we
need the factor of $\pm \i$, coming from the eta-invariant, that was explained at the end of section \ref{bulk}.

There is a subtlety here concerning the supermultiplets $(-\hq,0)$ and $(0,\hq\,)$ that contain states of $R$-charge zero.   While a generic supermultiplet in the presence of
$\CT$ symmetry is described by an AZ
multiplet with $(\upalpha_0,\upbeta)=(1,2)$, the $(-\hq,0)$ and $(0,\hq\,)$ multiplets are different because they both contain states of vanishing charge, and states of vanishing charge
are mapped to themselves
by $\CT$.   An analysis similar to what we have described before for in other examples shows that these multiplets are naturally described by ensembles with 
 $(\upalpha_0,\upbeta)=(-1,2)$. With an analytic potential, such an ensemble would produce a non-vanishing crosscap correction. As discussed in appendix \ref{app:Loop}, we
  can add a $\log x$ term to the matrix potential that shifts the effective value of $\upalpha_0$.  After such a shift, we can make the measure for the singular values of $Q_{0}$ and $Q_{-\hq}$ take the same form as an $(\upalpha_0,\upbeta)=(+1,2)$ ensemble, therefore removing the contribution to the crosscap coming from the loop equations. After doing this, we can again add a small correction to the spectral curve in a way that reproduces the non-vanishing crosscap volume computed from gravity. 

Finally, we note that in both the $\tT$-invariant case and the $\CT$-invariant case, the $b$ integral is divergent at $b=0$, which means that it is probably difficult to
extend the comparison between the matrix model and the crosscap to higher orders\footnote{In the bosonic case, the divergence can be regulated in gravity by working with the $(2,p)$ minimal string which becomes JT gravity only in the large $p$ limit \cite{SSS, SW, Mertens:2020hbs}. It is reasonable to expect that if a similar duality between matrix integrals and the non-critical $\N=2$ string exists, the crosscap would again be finite.}.   In the case of $\CT$ symmetry, the divergence is only at $q=0$ (assuming that one
integrates over $\phi$ before integrating over $b$, as we have done), so for other values
of $q$, there appears to be no difficulty in computing in higher orders.   However, this is precisely the case in which we have found that the matrix potential has higher order
corrections.  We do not know if further corrections are needed in going to higher orders, so we do not have clear predictions about the higher order amplitudes.

\section{$\mathcal{N}=4$ JT Gravity Black Hole}\label{N4JT}

In this section we present some partial results regarding the study of pure $\cN=4$ JT gravity including a sum over topologies.

\def\gl{{\rm gl}}
\def\SO{{\mathrm{SO}}}
\subsection{The $\cN=4$ JT Black Hole Spectrum} 
The general
structure of $\N=4$ JT supergravity is similar to what we have already discussed for $\N=2$.
 First, the action has, besides a topological term proportional to $S_0$, a bulk action that can be written as a topological $BF$ theory with gauge supergroup $G=\PSU(1,1|2)$.  A
 maximal bosonic subgroup of $G$ is  ${\rm SL}(2,\mathbb{R}) \times \SU(2)$. The first factor describes gravity while the second  is the gauge
 group of  an $\SU(2)$ $BF$ theory. We denote generators of this group as $J^1, J^2, J^3$. For asymptotically NAdS boundary conditions, boundary terms are required; these are simpler to write in a second order formulation.   The bosonic fields in bulk are a graviton and
 dilaton along with the $\SU(2)$ gauge fields.   The gravitino is now in the fundamental of $\SU(2)$. 
 Because the group $\PSU(1,1|2)$ has an $\SU(2)$ group of outer automorphisms, the $BF$ theory of $\PSU(1,1|2)$ has $\SU(2)$ as a global symmetry group.\footnote{If one gauges
 this global $\SU(2)$, one gets a $BF$ theory of the supergroup $D(2,1;\alpha)$, potentially related to a near extremal limit of black holes that have AdS$_3$ throats with large $\cN=4$
 superconformal symmetry.}  To distinguish the two $\SU(2)$'s, we refer to the one that is gauged as the $R$-symmetry group $\SU(2)_R$, and the other one we call
 $\SU(2)_{\gl}$.

The disk partition function for $\N=4$ again reduces to the path integral of a boundary reparametrization mode, the $\N=4$ Schwarzian mode. We work at inverse temperature $\beta$ and we turn on a chemical potential for the Cartan of $\SU(2)$,  parametrizing the fugacity by $e^{\i \alpha J^3}$. This problem was studied in \cite{Heydeman:2020hhw}. The boundary theory governs reparametrizations of the $\cN=4$ supercircle $(t,\theta,\bar{\theta})$ where $\theta$ is a complex $\SU(2)$ doublet and $\bar{\theta}$ its complex conjugate. The grand canonical partition function is found by imposing the following twisted conditions 
\bea
(t,\theta, \bar{\theta}) \to \big(t+\beta, - e^{\i \alpha J^3}  \theta, - e^{- \i \alpha J^3} \bar{ \theta}\big).
\ea
Since $\theta$ is in the fundamental representation, we can write $J^3=Z/2$ where $Z={\rm diag}(1,-1)$ is a Pauli matrix.  Thus $\alpha$  has  periodicity $\alpha\cong\alpha+4\pi$
as well as the $\SU(2)$ Weyl invariance, which acts by $\alpha\to -\alpha$.   The final result for the partition function is
\beq\label{diskN4}
Z_{\rm disk} =e^{S_0} \sum_{n\in \mathbb{Z}} \beta \frac{ \cot(\frac{\alpha}{4})(\alpha+4\pi n)}{\pi^4(1-\frac{1}{4\pi^2}(\alpha+4\pi n)^2)^2} \, e^{\frac{\pi^2 }{\beta}\big(1-\frac{1}{4\pi^2}(\alpha+4\pi n)^2\big)}\,.
\eeq
The derivation is outlined in appendix \ref{sec:disktrumpet}.   

With NAdS boundary conditions, the boundary quantum mechanics has four supercharges $Q_{A\dot A}$, $A,\dot A=1,2$ transforming as a bifundamental
of $\SU(2)_R$ (acting on the $A$ index) and $\SU(2)_\gl$ (acting on $\dot A$) and satisfying 
\be\label{fouralg} \{Q_{A\dot A},Q_{B\dot B}\}=\varepsilon_{AB}\varepsilon_{\dot{A} \dot{B}} H, \ee
with $H$ the Hamiltonian. Here $\varepsilon_{AB}$ and $\varepsilon_{\dot{A}\dot{B}}$ are antisymmetric tensors with $\varepsilon_{12}=\varepsilon_{\dot 1\dot 2}=1$.
Alternatively, since $(\SU(2)_R\times \SU(2)_\gl)/\Z_2\cong \SO(4)$ (where here $\Z_2$ is the product of the centers of the two $\SU(2)$'s), and a bifundamental
of $\SU(2)_R\times \SU(2)_\gl$ is a vector of $\SO(4)$, the supersymmetry algebra contains four hermitian supercharges $Q_I$, $I=1,\cdots,4$ transforming as a vector of $\SO(4)$,
and satisfying
\be\label{ikpo} \{Q_I,Q_J\}=2\delta_{IJ} H .\ee

In quantum mechanics with $\N=4$ supersymmetry, a  BPS multiplet  consists  of a  single  irreducible representation of $\SU(2)_R$ of any spin $J$, while  an irreducible 
non-BPS multiplets decomposes under $\SU(2)_R$ as $J \oplus (J-1/2) \oplus (J-1/2) \oplus (J-1)$ with some $J\geq 1/2$
(for $J=1/2$, the multiplet instead reduces to  $0\oplus 0\oplus 1/2$). The partition function \eqref{diskN4} can be expanded in such multiplets as 
\beq\label{fourdisc}
\hspace{-0.2cm}Z_{\rm disk} = e^{S_0} +\sum_{J\geq \frac{1}{2}} \big( \chi_J(\alpha) + 2 \chi_{J-\frac{1}{2}}(\alpha) + \chi_{J-1}(\alpha) \big)  \int_{J^2}^\infty \d E  e^{-\beta E} \frac{e^{S_0}J\, \sinh \big( 2\pi \sqrt{E-J^2}\big)}{\pi^2 E^2} .
\eeq
where $\chi_J(\alpha) = \frac{\sin{((2J+1)\alpha/2)}}{\sin (\alpha/2)}$ is the character of a spin $J$ representation of $\SU(2)$, and the gap in the spectrum is given by $E_0(J)=J^2$ for a supermultiplet with maximum spin $J$. 
(For $J=1/2$, since formally $\chi_{-1/2}(\alpha)=0$, the expression $ \chi_J(\alpha) + 2 \chi_{J-\frac{1}{2}}(\alpha) + \chi_{J-1}(\alpha)$ for the character of a non-BPS multiplet
reduces to $\chi_{1/2}(\alpha)+2\chi_0(\alpha)$, which is the character of the exceptional multiplet $J=1/2$.)   Since $J\geq 1/2$ in all cases,
  the energy gap $J^2$  is  non-vanishing for all non-BPS multiplets. 
  In contrast to the case of $\N=2$, there is no multiplet with a vanishing gap.
The BPS states are described by the term $e^{S_0}$ in eqn. (\ref{fourdisc}), and since this term is independent of $\alpha$, they all have spin 0.

The spectrum extracted from this partition function is very simple. 
In addition to $e^{S_0}$ BPS states, all of spin 0, there is for each $J\geq 1/2$ a continuous spectrum of non-BPS supermultiplets with highest spin $J$.
The density of states is 
\bea
\rho_J(E) =\begin{cases} e^{S_0} \frac{J \sinh \big(2\pi\sqrt{E-J^2}\big)}{\pi^2 E^2}& E\geq J^2\cr 
 0 & E<J^2.\end{cases}       \ea
To obtain this answer from a matrix model, the spectral curve should be 
\bea
y_J(x) = \frac{J\sin \big( 2\pi \sqrt{-x+J^2}\big)}{\pi x^2}.
\ea
The behavior of the spectrum near threshold is the typical square root behavior $y_J(J^2 + \varepsilon) \sim \sqrt{\varepsilon}$ of a Dyson ensemble in random matrix theory,
or an AZ ensemble in which the threshold energy is positive. The behavior at $x=0$ merits discussion. Naively $ y_J(x)$  has a double pole at $x=0$:
\bea
y_J(x) = \frac{J \sin(2\pi J)}{\pi x^2} - \frac{\cos(2\pi J)}{ x} +({\rm analytic~in~}x).
\ea
The double-pole would be problematic since it would imply a highly divergent term in the matrix potential at $x=0$ that would spoil the derivation of the loop equations in their usual form. This is explained in appendix \ref{app:Loop}. Fortunately, since $2J$ is an integer the coefficient of the double-pole vanishes: $\sin(2\pi J)=0$. The coefficient of the simple pole is non-zero and equal to $(-1)^{2J+1}$.   Thus the residue of the simple pole is positive or negative for half-integer or integer $J$.
A simple pole  is consistent with the loop equations in the presence of a logarithmic term in the potential, as discussed in the appendix.

$\N=4$ JT supergravity does not have some of the generalizations that exist for $\N=2$.  There is no analog of the parameter $\hq$,
 since the $R$-symmetry group $\SU(2)_R$ is simply-connected. For essentially the same reason, there is no analog of the parameter $\delta$.

\subsection{The Two-Boundary Wormhole} 
Having obtained the spectrum and the multiplet structure, the next question is whether different multiplets are statistically independent or not. 
The most basic test of this question comes from the amplitude for a cylinder, the leading connected amplitude with two NAdS boundaries.
The path integral on the cylinder reduces to an integral over the moduli space of flat $\PSU(1,1|2)$ connections, modulo gauge transformations. These are labeled by the supergroup holonomy around the non-contractible circle. In order to restrict to gauge connections that have an interpretation in hyperbolic geometry we restrict to holonomies with the property that the $\SU(1,1)$ block is in the hyperbolic conjugacy class. We assume no other restriction. This holonomy is parametrized by a geodesic length $b$, describing the conjugacy class of the $\SU(1,1)$ block, and an $\SU(2)$ holonomy  conjugate to $e^{\i \phi J^3}$, with parameter $\phi$, 
where in the fundamental representation $J^3=Z/2$ with $Z$ the diagonal Pauli matrix. Like $\alpha$, $\phi$ has a periodicity $\phi\to\phi+4\pi$ and a Weyl symmetry
$\phi\to -\phi$.   For concreteness we take $\phi>0$. There is also a Schwarzian reparametrization mode on each boundary.  The path integral of the reparametrization mode along each boundary is given by 
\beq\label{N4trumpet1}
Z_{\rm tr}(\beta,\alpha; b,\phi) = \sum_{n\in \mathbb{Z}}  \frac{\cot(\frac{\alpha}{4})}{4\pi \beta} e^{-\frac{1}{4\beta}b^2} \Big( e^{-\frac{1}{4\beta}(\alpha-\phi+4\pi n)^2}-e^{-\frac{1}{4\beta}(\alpha+\phi+4\pi n)^2}\Big) \,.
\eeq
We review the derivation in appendix \ref{sec:disktrumpet}. Using Poisson resummation, this expression can be written as a sum over representations:
\beq\label{N4trumpet2}
Z_{\rm tr}(\beta,\alpha; b,\phi) = \sum_{J\geq \frac{1}{2}} \left( \chi_J(\alpha) + 2 \chi_{J-\frac{1}{2}}(\alpha) + \chi_{J-1}(\alpha) \right) 2 \sin(J\phi)   \, \frac{1}{4\pi}\frac{e^{-\beta J^2-\frac{b^2}{4\beta}}}{2\sqrt{\pi \beta}}.
\eeq
The $\beta$ dependence is the same as that of the trumpet in bosonic JT ($Z_{\rm tr}^{\rm bos.}(\beta,b) =e^{-\frac{b^2}{4\beta}}/(2\sqrt{\pi \beta} )$) up to a shift of the group state energy to $E_0(J)=J^2$. 

We can now compute the cylinder partition function by gluing two of these trumpets along a common geodesic boundary. The final answer is 
\beq
Z_{0,2}(\beta_1, \alpha_1; \beta_2,\alpha_2) = 2\pi \int_0^\infty b\d b \int_0^{4\pi} \d \phi \, Z_{\rm tr} (\beta_1,\alpha_1; b,\phi) Z_{\rm tr} (\beta_2,\alpha_2;b,-\phi).
\eeq
The factor of $2\pi$ comes from the integral over the twist parametrizing the maximal torus of $\SU(2)$, including a factor of $1/2$ because of the Weyl symmetry. (In any case, we should emphasize we do not claim to properly understand the absolute normalization of the trumpet.) After gluing the two trumpets we obtain
\bea
Z_{0,2}(\beta_1, \alpha_1; \beta_2,\alpha_2) &=& \sum_{J_1,J_2\geq \frac{1}{2}} \left( \chi_{J_1}(\alpha_1) +2 \chi_{J_1-\frac{1}{2}}(\alpha_1) + \chi_{J_1-1}(\alpha_1) \right) \nonumber\\
&&\times\left( \chi_{J_2}(\alpha_2) +2 \chi_{J_2-\frac{1}{2}}(\alpha_2) + \chi_{J_2-1}(\alpha_2) \right)\int_0^{4\pi} \frac{\d \phi}{8\pi}(2 \sin J_1\phi)(2\sin J_2\phi) \nn
&&\times  e^{-\beta_1 J_1^2 - \beta_2 J_2^2} \int_0^\infty b\d b \, \frac{e^{-\frac{b^2}{4\beta_1}-\frac{b^2}{4\beta_2}}}{4 \pi \sqrt{ \beta_1\beta_2}}.
\ea
The integral over the geodesic length $b$ gives the same as before, since the trumpet is again proportional to the bosonic one. Integration over $\phi$ imposes that $J_1=J_2$. Therefore again we can only have the same multiplet propagating on each boundary to get a non-zero answer. The final result is then
\beq
Z_{0,2}(\beta_1, \alpha_1; \beta_2,\alpha_2)  = \sum_{J\geq \frac{1}{2}} \Big[ \prod_{i=1,2} \big( \chi_J(\alpha_i) +2\chi_{J-\frac{1}{2}}(\alpha_i) + \chi_{J-1}(\alpha_i) \big) \Big] \frac{\sqrt{\beta_1 \beta_2} \, e^{-(\beta_1+\beta_2) J^2}}{2\pi(\beta_1+\beta_2)}.
\eeq
The multiplets are statistically independent, and the temperature dependence matches the expectations from a double-scaled matrix integral with a gap  $E_0(J)= J^2$.   This is
encouraging for a possible random matrix description.

\subsection{$\N=4$ Random Matrix Theory}\label{randomfour}

Finally, in this section we will explore aspects  of $\N=4$ random matrix theory.   We will consider the case that only $J=1/2$ multiplets (multiplets that decompose under
$\SU(2)_R$ as $0\oplus 0\oplus 1/2$) are present.
By similar logic to that of section \ref{randomtwo}, we will argue that it is natural  for random $\N=4$ supersymmetry with all multiplets of $J=1/2$ 
to be described by an AZ ensemble with $(\upalpha,\upbeta)=(1,2)$, the familiar ensemble that is important in $\N=1$ and $\N=2$ JT supergravity.  We expect the same to be true
for multiplets of any $J$, but it would be technically more complicated to demonstrate this.

The $\N=4$ algebra in the $\SO(4)$ language of eqn. (\ref{ikpo}) has an obvious resemblance  to a Clifford algebra.  
That fact enables an easy construction of the $J=1/2$ representation, from which all other representations are easily constructed.   Consider four gamma matrices
$\gamma_I$, $I=1,\cdots ,4$, satisfying $\{\gamma_I,\gamma_J\}=2\delta_{IJ}$, transforming as a vector of $\SO(4)$ (or equivalently a 
bifundamental of $\SU(2)_R\times \SU(2)_\gl$) and acting in the usual way on a four-dimensional
Hilbert space $\H_0$. As usual we let $\gamma_5=\gamma_1\gamma_2\gamma_3\gamma_4$ be the product of the four $\gamma_I$.  
Then, we get an irreducible representation of $\N=4$ global supersymmetry by taking
\be\label{zero} Q_I=\lambda\gamma_I,~~~~H=\lambda^2,\ee
with a nonzero constant $\lambda$.   To make the $Q_I$ hermitian, $\lambda$ must be real, and by possibly conjugating with $\gamma_5$, we can reduce to $\lambda>0$.
A familiar fact about spinors is that $\H_0$ transforms under $\SU(2)_R\times \SU(2)_\gl$ as $(1/2,0)\oplus (0,1/2)$.   So we can describe $\H_0$ as $\S_{1/2}\oplus \S'_{1/2}$,
where $\S_{1/2}$ is a two-dimensional vector space in the spin 1/2 representation of $\SU(2)_R$, and $\S'_{1/2}$ is a two-dimensional vector space in the spin 1/2 representation 
of $\SU(2)_\gl$.    However, since $\SU(2)_\gl$ is not part of the supersymmetry
algebra, we often want to describe $\H_0$ just as a representation of $\SU(2)_R$.  From that point of view, $\S'_{1/2}\cong \C^2$ is just a two-dimensional vector space on which
$\SU(2)_R$ acts trivially, so $\H_0=\C^2\oplus \S_{1/2}$.

We can now also easily describe an arbitrary irreducible representation of the $\N=4$ algebra.  For $J$ a positive integer or half-integer, let $\H_J$ be a Hilbert space of dimension
$2J+1$ on which $\SU(2)_R$ acts as spin $J$.   Let $\H=\H_0\otimes \H_{J-1/2}$.  Setting $Q_I=\lambda \gamma_I\otimes 1$, we get an irreducible representation of the $\N=4$
algebra.  $\H$  transforms under $\SU(2)_R$ as $(J-1)\oplus (J-1/2)\oplus (J-1/2)\oplus J$, and we label this representation by the highest spin it contains,
namely $J$.     We note that $\SU(2)_\gl$ actually acts on
this representation via its action on $\H_0$, transforming the supersymmetry generators in the usual way.  So even though it is not part of the 
supersymmetry algebra, $\SU(2)_\gl$ acts on every irreducible representation of this algebra, and therefore on every representation.\footnote{In general, an outer automorphism of
a group $G$ will either act in a given representation $W$, or conjugate $W$ to some inequivalent representation.   In the present case, since the group $\SU(2)_\gl$ is connected,
there is no way for it to act nontrivially on the discrete label $J\in \{1/2,1,3/2,\cdots\}$ of a representation and instead it 
acts in each representation.}  This fact turns out to be helpful in understanding the random matrix theory.   

Here, however, as already noted,
we will restrict ourselves to considering a random realization of the $\N=4$ algebra that is isomorphic to the direct sum of some number $L$ of copies of the basic $J=1/2$
representation.    For this, we consider a bosonic Hilbert space $\H_b\cong \C^{2L}$ and a fermionic Hilbert space $\H_f=\S_{1/2}\otimes \H_f'$ with $\H_f'\cong \C^L$.
 Thus, $\SU(2)_R$ acts trivially on $\H_b$, and $\H_f$ transforms as
the direct sum of $L$ copies of $\S_{1/2}$.   $\SU(2)_R$ also has an expected action on $Q_I$ coming from 
the fact that they transform as $(1/2,1/2)$ of $\SU(2)_R\times \SU(2)_\gl$.
Then we look for a set of four operators $Q_I:\H_b\leftrightarrow \H_f$ that obey the algebra (\ref{fouralg}) and intertwine with the action
of $\SU(2)_R$, acting on the four $Q_I$ and on the Hilbert space $\H=\H_b\oplus \H_f$ in the expected fashion.   

The symmetry group of this problem is simply the group of unitary transformations of $\H$ that commute with $\SU(2)_R$.   This is $\uU(2L)\times \uU(L)$, where the first factor
acts on $\H_b$ and the second on $\H_f'$.    

Given our knowledge of the classification of representations of the $\N=4$ algebra, we can use the group $\uU(2L)\times \uU(L)$ to put
any set of $Q_I$ that generate such a representation in a canonical form.   A representation of the $\N=4$ algebra 
on $\H_b\oplus \H_f=\C^{2L}\oplus S_{1/2}\otimes \C^L$ is 
generically\footnote{The reason that this is only generically the case is that a $J=1/2$ representation with $H=\lambda^2$ 
degenerates to a collection of BPS states if $\lambda=0$.}  the direct sum of $L$ copies of the $J=1/2$ representation.   In a single copy of that representation, we have as before
$\H_f=\S_{1/2}$
and $\H_b=\S'_{1/2}$, and we can assume that  $Q_I=\lambda\gamma_I$, with the $\gamma_I$ being generators of a Clifford
algebra on $\H_b\oplus\H_f$.  To get $L$ copies of this representation, we introduce an $L$-dimensional vector space $V\cong \C^L$ with trivial action of $\SU(2)_R$, and then take
$\H_b=S'_{1/2}\otimes V$, $\H_f=S_{1/2}\otimes V$, and 
\be\label{qvac} Q_I=\gamma_I\otimes M, \ee
where $M$ is a hermitian matrix acting on $V$.  By a unitary transformation, we can put $M$ in a diagonal form $M={\rm diag}(\lambda_1,\lambda_2,\cdots,\lambda_L)$, and,
since it is possible to make a $\gamma_5$ transformation separately for each ``eigenvalue'' $\lambda_i$, we can assume that $\lambda_i\geq 0$ for all $i$.

We have already determined that the symmetry group in this problem, taking into account the $\SU(2)_R$ symmetry,
 is $K=\uU(2L)\times \uU(L)$, with real dimension $5L^2$.   We will now show that, given the constraint of
$\SU(2)_R$-invariance, to specify $Q_I$ requires $8L^2$ real parameters.   And we will show that the supersymmetry algebra imposes $3L^2$ real constraints on the $Q_I$.
One might therefore expect that the moduli space of representations of the supersymmetry algebra with the assumed dimension and action of $\SU(2)_R$ would have
dimension $8L^2-5L^2-3L^2=0$.   Instead, by explicit construction in eqn. (\ref{qvac}), we see that this moduli space has dimension $L$, parametrized by $\lambda_1,\lambda_2,\cdots,
\lambda_L$.   What is happening is that the $5L^2$ generators of $K$ are not all effective at reducing the moduli space dimension, because an $L$-dimensional subalgebra,
consisting of the matrices on $V$ that commute with $M$, leaves fixed $Q_I$ in eqn. (\ref{qvac}).   Hence the dimension of the moduli space
is actually $8L^2-(5L^2-L)-3L^2 = L$.   

This counting shows that the $3L^2 $ constraints that come from the supersymmetry algebra are all effective at reducing the dimension of the moduli space.
So they are all  independent; there are no relations between these constraints.
In section \ref{randomtwo}, we found that in $\N=2$ random matrix theory, there are relations between the constraints as soon as 
at least  four consecutive values of the $R$-charge are present. 
For $\N=4$, it turns out that there are relations between the constraints 
 for all non-BPS representations with $J>1/2$, even if only one type of representation is present.\footnote{In fact, there are not only relations between the constraints;
 there are relations between the relations, and higher order relations between those relations, and so on, with a number of steps of order $J$.}   That is the main reason
that here we will analyze the random matrix theory only for $J=1/2$; the case of higher $J$ is technically more complicated.

It remains to verify  that after taking into account the action of $\SU(2)_R$, there are $8L^2$ parameters in the $Q_I$, and the supersymmetry algebra places
$3L^2$ relations among those parameters.  

To count parameters in $Q_I$, we note first that we can write $Q_I=\Q_I+\Q_I^\dagger$,  where $\Q_I$ maps $\H_f$ to $\H_b$ and $\Q_I^\dagger$ is the adjoint map
from $\H_b$ to $\H_f$.   $\Q_I$ is constrained by $\SU(2)_R$ invariance, but it does not satisfy any condition of hermiticity (and there is no constraint associated to $\SU(2)_\gl$,
since $\SU(2)_\gl$ invariance is not assumed).    At this point, it is convenient to think of $\Q_I$ as a bifundamental of $\SU(2)_R\times \SU(2)_\gl$ rather than a vector
of $\SO(4)$.   Four operators $\Q_{A\dot A}:\S_{1/2}\otimes \C^L\to \C^{2L}$ that transform with spin 1/2 under $\SU(2)_R$ are equivalent to a pair
of operators $\X_{\dot A}:\C^L\to \C^{2L}$ that obey no particular constraint.  
The two operators $\X_{\dot A}$ depend on a total of $2 \cdot L \cdot 2L=4L^2$ complex parameters, or $8L^2$ real parameters.   This is the number of real parameters
involved in the choice of the $Q_I$, after imposing the action of $\SU(2)_R$, but without imposing the constraint that the $Q_I$ satisfy the $\N=4$ algebra.

The number of parameters involved in imposing the $\N=4$ algebra can be determined similarly.
{\it A priori}, $\{Q_I,Q_J\}$ transforms as $(0,0)\oplus (1,1)$ under $\SU(2)_R\times \SU(2)_\gl$.  The $(0,0)$ part is the Hamiltonian, and the supersymmetry constraint says
that the $(1,1)$ part vanishes.    Of course, $\{Q_I,Q_J\}$ maps $\H_b$ to $\H_b$ and $\H_f$ to $\H_f$.   An operator that transforms as spin 1 under $\SU(2)_R$ cannot
map $\H_b$ nontrivially to itself, since $\H_b$ transforms with spin 0 under $\SU(2)_R$.   Since $\H_f$ transforms under $\SU(2)_R$ with spin 1/2, and spin 1/2 arises in 
the decomposition $1/2\otimes 1 =1/2\oplus 3/2$, it is possible for a spin 1 operator
to map $\H_f$ to itself.  A set of three hermitian operators mapping $\H_f=\S_{1/2}\otimes \C^L$ to itself and transforming with spin 1 under $\SU(2)_R$ are determined by a single hermitian operator
on $\C^L$, and thus depends on $L^2$ real parameters.   The $(1,1)$ part of $\{Q_I,Q_J\}$ consists of 9 hermitian operators on $\H_f$, making up three vectors of $\SU(2)_R$.
This is equivalent to three hermitian operators on $\C^L$, so the number of real parameters is $3L^2$, as claimed earlier.    These parameters transform under $\SU(2)_\gl$ as $L^2$ copies
of the spin 1 representation.

We still have  to determine the measure for integration over the ``eigenvalues'' $\lambda_i$.   By definition, when we do this, we are trying to compute the measure on the
space of $Q_I$'s that do satisfy the constraints of $\N=4$ supersymmetry  (modulo equivalence generated by the group $K$).    As noted earlier, every solution of the constraints
is invariant under $\SU(2)_\gl$, acting via a suitable subgroup of $K$.   In computing the measure, we have to consider (in first order) perturbations of the $Q_I$ that do
not necessarily preserve the $\SU(2)_\gl$ symmetry, but $\SU(2)_\gl$ is still very useful in organizing and constraining the calculation.

As in eqn. (\ref{qvac}), when the constraints are satisfied, $\H_b$ and $\H_f$ have decompositions $\H_b=\S'_{1/2}\otimes \C^L$, $\H_f=\S_{1/2}\otimes \t \C^L$,
where $\SU(2)_\gl$ acts on $\S'_{1/2}$ (in the spin 1/2 representation) and not on anything else.   Here $\C^L$ and $\t \C^L$ are two different vector spaces, both
of dimension $L$; to emphasize that they are not the same, we now denote the second one as $\t\C^L$.   With this decomposition, it is now clear that the subgroup of
$K=\uU(2L)\times \uU(L)$ that commutes with $\SU(2)_\gl$ is $K'=\uU(L)\times \uU(L)$, where the two factors act respectively on $\C^L$ and $\t\C^L$.
The group $K'$ has dimension $2L^2$, while $K$ has dimension $5L^2$.   The part of the Lie algebra of $K$ that is orthogonal to the Lie algebra of $K'$
consists of $L^2$ copies of the spin 1 representation of $\SU(2)_\gl$.   (These constraints come from operators transforming as $(1,1)$ under $\SU(2)_R\times \SU(2)_\gl$
and acting on an $\SU(2)_\gl$-invariant Hilbert space $\H_f$, so they all transform with spin 1 under $\SU(2)_\gl$.)   

It turns out that the effect on the measure of $3L^2$ symmetry generators transforming at $L^2$ copies of spin 1 of $\SU(2)_\gl$ cancels the effect on the
measure of $3L^2$ constraints transforming in the same way.  Before demonstrating this cancellation, we will first argue that if it does occur,
then a random realization of the $\N=4$ algebra with the representation content that we have assumed is described by an AZ ensemble with $(\upalpha,\upbeta)=(1,2)$.

We simply write down the general form of the four hermitian operators $Q_I$, assuming that they intertwine with an action of $\SU(2)_R\times \SU(2)_\gl$, not
just $\SU(2)_R$ as required by the supersymmetry algebra:
\be\label{genformodd} Q_I=\gamma_I \frac{1+\gamma_5}{2} M +\gamma_I \frac{1-\gamma_5}{2} M^\dagger. \ee 
Here $M$ is a completely unconstrained linear operator mapping $\C^L$ to $\t\C^L$, and $M^\dagger $ is its adjoint acting in the opposite direction.
In other words, $M$ is a bifundamental of $K'=\uU(L)\times \uU(L)$.   As we explained in section \ref{randomone}, a bifundamental of $\uU(L)\times \uU(L)$ is 
naturally associated in random matrix theory to an AZ ensemble with $(\upalpha,\upbeta)=(1,2)$.   

So all that remains is to show that there is no net effect on the measure of the symmetry generators and constraints that transform as spin 1 of $\SU(2)_\gl$.  
The analysis can be simplified by using the residual symmetries.   After putting $Q_I$ in the canonical form (\ref{qvac}) with a diagonal matrix
$M={\rm diag}(\lambda_1,\cdots,\lambda_L)$,  there is  a residual $\uU(1)^L$ symmetry of diagonal matrices that commute with $M$.
A cancellation between symmetries and constraints occurs separately for each $\uU(1)^L$ representation.   Symmetries and perturbations of $L$ that are $\uU(1)$-invariant
preserve the diagonal structure of $Q_I$, and do not ``mix'' with the off-diagonal symmetries and perturbations.   For each pair of distinct eigenvalues $\lambda_i,\lambda_j$, the $ij$ matrix
elements of a symmetry generator or a perturbation of the $Q_I$ transform in a $\uU(1)^L$ representation that is different for each pair $i,j$.   This implies that one can
analyze the $ij$ matrix elements for given $i,j$ independently of the others.

To understand the cancellation that occurs for diagonal matrix elements, we do not really need to do a calculation.   Diagonal symmetry generators and perturbations of the $Q_I$
preserve the diagonal structure, and as long as this diagonal structure is preserved, the different diagonal matrix elements do not influence each other.
Therefore, to study a diagonal matrix element, it suffices to take $L=1$.
To understand why a cancellation must occur in this case, all we need to know is that for $L=1$,  the $Q_I$ are proportional to $\lambda$, the canonical form being
$Q_I=\lambda\gamma_I$.   This means that the vector fields corresponding to the generators of $K$ that have spin 1 under $\SU(2)_\gl$ are all proportional to $\lambda$,
so each of these generators produces a factor $\lambda$ in the measure.   On the other hand, for any first order variation $\delta Q_I$ of the $Q_I$, the constraints
give delta functions that assert the vanishing of the $(1,1)$ part of $\{\delta Q_I,Q_J\}+\{Q_I,\delta Q_J\}$.   This quantity is linear in the $Q_I$, so it is linear in $\lambda$.
So the constraints give delta functions of the general form $\delta(\lambda f)$, where $f$ is independent of $\lambda$.   As $\delta(\lambda f)=\frac{1}{\lambda}\delta(f)$,
each constraint gives a factor $1/\lambda$, canceling the factors that come from the symmetry generators.

To study off-diagonal symmetries and perturbations, it suffices to consider the case of $L=2$ with $M={\rm diag}(\lambda_1,\lambda_2)$.      
The generators of $\SO(4)$ acting on $\H_0=\S'_{1/2}\otimes \S_{1/2}$ are the antihermitian matrices $\sigma_{IJ}=\frac{1}{2}[\gamma_I,\gamma_J]$.   With an appropriate
orientation convention, the matrices on $\H_0$ that transform as spin 1 of $\SU(2)_\gl$ -- that is, the $\SU(2)_\gl$ generators -- are $\zeta^{IJ}\sigma_{IJ}$, where $\zeta_{IJ}=\frac{1}{2}
\epsilon_{IJKL}\zeta^{KL}$ is self-dual ($\epsilon_{IJKL}$ is the antisymmetric tensor with $\epsilon_{1234}=1$).    Thus, with 
\be\label{exac} Q_I=\gamma_I\otimes \begin{pmatrix} \lambda_1 & 0 \cr 0 & \lambda_2\end{pmatrix}, \ee
we can consider the off-diagonal symmetry generators
\be\label{nxac} r(\zeta) = \zeta^{IJ}\sigma_{IJ} \otimes \begin{pmatrix}0 & 1\cr 0 & 0\end{pmatrix}-{\rm h.c.} \ee
The induced variation of $Q_I$ is
\be\label{nec}\delta Q_I = [r(\zeta),Q_I]= -2\zeta_{IJ}\gamma^J \begin{pmatrix} 0 & \lambda_1+\lambda_2\cr 0 & 0 \end{pmatrix} +2\zeta_{IJ}\gamma^J\gamma_5 \begin{pmatrix}
  0 & \lambda_1-\lambda_2\cr 0 & 0 \end{pmatrix}+{\rm h.c.} \ee
In the norm squared of $\delta Q_I$ (namely $\Tr\,\delta Q_I^2$), there is no cross term between the parts of $\delta Q_I$ with and without a factor of $\gamma_5$.
So the norm squared is a multiple of $$(\zeta^{IJ}\bar\zeta_{IJ})\left( (\lambda_1-\lambda_2)^2+(\lambda_1+\lambda_2)^2\right)=2\zeta^{IJ}\bar\zeta_{IJ} (\lambda_1^2+\lambda_2^2).$$  Hence,
 each of the six independent real components of $\zeta_{IJ}$ contributes a factor $(\lambda_1^2+\lambda_2^2)^{1/2}$ to the measure, giving an overall factor $(\lambda_1^2+\lambda_2^2)^3$.
 
 However, the corresponding matrix elements of the constraints cancel this factor.   A normalized off-diagonal perturbation $\delta' Q_I$ in the same $2\times 2$ block that
 is orthogonal to $\delta Q_I$ is
 \be\label{mxec}\delta' Q_I=\frac{1}{\sqrt{\lambda_1^2+\lambda_2^2} }\left(\zeta_{IJ}\gamma^J \begin{pmatrix}0& \lambda_1-\lambda_2\cr 0&0\end{pmatrix} +\zeta_{IJ}\gamma^J\gamma_5 \begin{pmatrix} 0 & \lambda_1+\lambda_2\cr 0&0\end{pmatrix} \right)
 + {\rm h.c.}. \ee
 We then find that
 \be\label{zxec}\{Q_I,\delta' Q_J\}+\{Q_J,\delta' Q_I\}=  2\left(\sigma_{IK}\zeta_{JK}+\sigma_{JK}\zeta_{IK}\right)\begin{pmatrix} 0 & \sqrt{\lambda_1^2+\lambda_2^2}\cr
                                                                                                                                                                                                           0 & 0 \end{pmatrix}+{\rm h.c.} \ee
For each independent real component of $\zeta_{IJ}$, we impose the vanishing of $\{Q_I,\delta' Q_J\}+\{Q_J,\delta' Q_I\}$ as a constraint.   In each case, the delta
function that imposes this constraint has an argument that is a multiple of $\sqrt{\lambda_1^2+\lambda_2^2}$.   The resulting factor in the measure
is $1/(\lambda_1^2+\lambda_2^2)^3$, canceling the measure factor that came from the corresponding symmetry generators.

\vskip1cm
 \noindent {\it {Acknowledgements}}  
We thank M. Heydeman and D. Stanford for discussion. GJT is supported by the Institute for Advanced Study and the NSF under Grant  PHY-2207584, and by the Sivian Fund. Research of EW supported in part by NSF Grant PHY-2207584.

 \appendix

\section{The Disk and the Trumpet}\label{sec:disktrumpet}

In this appendix we explain the derivation of the disk and trumpet partition functions for $\cN=2$ and $\cN=4$ JT supergravity. 

In both cases, there are subtleties concerning the overall normalization, which depend on the normalization that one assumes for zero-mode wavefunctions.   

In the case of the disk, the overall normalization is inessential, in the sense that a change in this normalization is equivalent to a shift in the ground state
entropy $S_0$.   However, if one does change this normalization, one must compensate by renormalizing the path integral of a three-holed sphere by an inverse factor.

In the case of the trumpet, since the trumpet has an inner boundary and is always glued to something else,  it is not meaningful to specify the normalization 
of the partition function without also specifying the measure that will be used in gluing.   In the body of this paper, we have normalized the gluing measure as $\d b\d \varrho \d\phi\d\varphi$.

In practice, we will use normalizations that have been justified in related problems by comparing to the symplectic form of the bulk moduli space in bosonic JT gravity \cite{SSS} 
or by comparing a localization procedure to canonical quantization in some other cases, for example in \cite{Kapec:2019ecr}.    However, the most conclusive justification for the 
normalizations that we will assume is that with these normalizations, $\cN=2$ JT supergravity agrees with matrix model predictions, as found in the body of the 
article.\footnote{It would presumably be possible to get a more complete understanding of the normalizations
by computing the precise relationship between the symplectic measure and the torsion for $\N=2$ super Riemann surfaces.  It is expected
that they are equivalent, in the sense that the symplectic measure and the torsion differ by a factor $e^{c\chi}$, where $\chi$ is the Euler characteristic of a surface, and $c$ is
a constant.  If this relative normalization were known, some considerations of \cite{SSS} could be adapted to the $\N=2$ case.}   

In all cases, the  path integrals that we will evaluate are one-loop exact, meaning that they can be evaluated by summing over fixed points of a rotation symmetry,
with each fixed point weighted by a Gaussian integral.   A conventional approach to the one-loop calculation around a critical point 
is to expand the action, in the present context a Schwarzian action,
 to get the appropriate quadratic action, find the measure -- a symplectic measure if the localization is justified using the symplectic structure of the phase space \cite{Stanford:2017thb} --
  and then  do a Gaussian integral.   However, the localization procedure implies a shortcut which
we will use repeatedly: the output of the Gaussian integral is $1/\sqrt{\det' D}$ (or the inverse of this for fermions), where $D$ is the operator that generates the rotation symmetry
of the fixed point, and the notation $\det'$ means that modes of $D$ that can be generated by symmetries of the disk or trumpet should be discarded.   We note that the modes
that are discarded are zero-modes in the sense that they do not appear in the action or in the symplectic form, but they are in general not zero-modes of $D$.  So in general a few
 eigenvalues of  $D$ have to be omitted by hand.  We call the eigenvalues of $D$ ``rotation eigenvalues.'' 

\subsection*{$\cN=2$ JT supergravity}

\paragraph{The trumpet} We follow the notation of section \ref{sec:N2JTG}. We consider first  the $n=0$ fixed point 
\beq
y(t,\theta) = \frac{2\pi}{\beta} t, ~~~\psi(t,\theta)= e^{\i \hq (\phi-\alpha)t/\beta} \theta.
\eeq
The general answer can be obtained by a shift $\alpha \to \alpha+2\pi n$. The action of this fixed point is easy to evaluate. Since the fermions are turned off, we can restrict the $\cN=2$ Schwarzian to its bosonic subsector. It was found in \cite{FGMS,Stanford:2017thb, Forste:2017kwy, Forste:2017apw} that a reparametrization $(t,\theta) \to (y(t), e^{\i \sigma(t)} \theta)$ on the trumpet has an action 
\beq
S = - \frac{1}{2}\int_0^{\beta} \d t \Big[ \Big\{\tanh \Big(\frac{b\, y}{4\pi}\Big) ,t\Big\}- 2 (\partial_t \sigma)^2 \Big] = \frac{b^2}{4 \beta} + \frac{\hq^{\, 2}}{\beta}(\alpha - \phi)^2.
\eeq
The operator that generates the rotation symmetry of this fixed point is 
\bea
D = \frac{\partial}{\partial t} - \frac{2\pi}{\beta} \frac{\partial}{\partial y} - \frac{\hq (\phi-\alpha)}{\beta}\, \i (\psi \partial_\psi - \bar{\psi} \partial_{\bar{\psi}}).
\ea
We first consider the reparametrization mode and expand $y(t,\theta) = 2\pi t/\beta + \delta y(t)$ with $\delta y(t+\beta)=\delta y(t)$. The eigenvalue equation $\i D \delta y = \lambda \delta y $ reduces to $\i \partial_t \delta y = \lambda \delta y$; the eigenvalues are $\lambda_m=2\pi m /\beta$ with $m\in \mathbb{Z}$. The trumpet has a bosonic $\uU(1)$ isometry  $y \to y +{\rm const}$ that corresponds to $m=0$ and has to be removed from the spectrum of fluctuations. The contribution of the reparametrization mode to the one-loop determinant of the trumpet is
\beq
Z_{\rm 1-Loop,~T}^{\rm Rep.} =\prod_{m\geq 1}\frac{4\pi }{\lambda_m} =  \prod_{m\geq 1}\frac{ 2\beta}{m} = \frac{1}{2\sqrt{\pi \beta}}.
\eeq
To extract $\sqrt{\det' D}$, we took the product of eigenvalues over $m>0$ only, after discarding the zero-mode at $m=0$.
 The final result has been obtained using zeta-function regularization.  In zeta-function regularization, replacing $\lambda_m$ by $\lambda_m/4\pi$ has the effect of
 multiplying the determinant by a constant factor.    The normalization we chose is the same as in \cite{SSS,SW}.
 
Next we consider the rotation eigenvalues coming from the phase mode $\psi \to e^{ \i \delta \sigma(t) } \psi $. On such modes the eigenvalue equation for $D$ becomes $\i \partial_t \delta \sigma = \lambda \delta \sigma  $. The zero-mode of $\delta \sigma$ comes from a $\uU(1)$ symmetry of the trumpet and should be discarded.
 The final result is proportional to $ \prod_{m\geq 1} 2\beta/m$. The normalization is a bit subtle. One can work with either an ``ultralocal'' measure, which gives a result consistent with the canonical quantization of the phase mode, or a symplectic measure appropriate to the BF theory origin of this mode. It was shown in appendix A of \cite{Kapec:2019ecr} that the partition function with the symplectic measure is equal to the partition function with the ultralocal measure divided by the volume of the group, in this case $2\pi$. This leads to the answer 
\beq
Z_{\rm 1-Loop,~T}^{\uU(1)} =2\hq\prod_{m\geq 1}\frac{ 2\beta}{ m} = \frac{\hq}{\sqrt{\pi \beta}}.
\eeq
The one-loop determinant corresponding to the ultralocal measure is then $\hq\sqrt{4\pi/\beta}$, and we can check this is consistent with canonical quantization. The partition function of the phase mode alone would be, taking $\phi=0$ for simplicity, $\sum_{n\in\mathbb{Z}}  \frac{\hq \sqrt{4\pi}}{\sqrt{\beta}} e^{-\frac{\hq^2(\alpha + 2\pi n)^2}{\beta}}= \sum_{k\in \mathbb{Z}} e^{\i \alpha k} e^{-\beta\frac{ k^2}{4\hq^{\, 2}}}$, where we used Poisson resummation to write the right hand side as a sum over states with different charges and energies. 

Next we analyze the rotation eigenvalues of the fermionic fields. We write a generic fermionic perturbation around the fixed point as $\psi \to \psi + \delta \psi$ with $\delta\psi (t) = e^{\i \hq \phi t/\beta} \chi(t)$, and $\chi(t+\beta)=-\chi(t)$. The eigenvalue equation $\i D\delta \psi = \lambda \delta \psi$ then reduces to 
\beq
\Big( \i \frac{\partial}{\partial t} - \frac{\hq \, \alpha}{\beta} \Big) \chi = \lambda \chi
\eeq
and therefore the eigenvalues are $ (2\pi m+\hq\,\alpha)/\beta$ with $m \in \mathbb{Z}+1/2$. Importantly, the answer is independent of the holonomy around the geodesic and depends only on the chemical potential at the NAdS boundary. The regularized version of this determinant is 
\beq\label{fermpfaff}
Z_{\rm 1-Loop,~T}^{\rm ferm.} = \prod_{m \in \mathbb{Z}+ \frac{1}{2}} \frac{m+\frac{\hq\,\alpha}{2\pi}}{2\beta} = 2 \cos \Big( \frac{\hq \alpha}{2} \Big),
\eeq
which makes sense  since this is the partition function of a complex fermion of charges $\pm \hq/2$ in quantum mechanics.   In this example, there are no zero-modes,
so there is no subtlety with the normalization.
 
We can put the three contributions together with the classical action and write the result for the $n=0$ fixed point:
 \beq
  \frac{\hq \, \cos(\frac{\hq\alpha}{2})}{\pi \beta} e^{-\frac{b^2}{4\beta}-\frac{\hq^{\, 2}}{\beta}(\alpha-\phi)^2}.
 \eeq
Shifting $\alpha \to \alpha + 2\pi n $ gives the final answer presented in \eqref{trumpet}. Deriving \eqref{trumpet2} from \eqref{trumpet} is straightforward. To do this first rewrite \eqref{trumpet} as 
\bea
Z_{\rm tr}(\beta,\alpha; b,\phi)&=& Z_{\rm tr}^{\rm bos.}(\beta,b) 2\cos\big(\frac{\hq\alpha}{2}\big) \sum_{n\in \mathbb{Z}}  \frac{\hq \, e^{2\pi\i n(\delta+1/2)} }{\sqrt{\pi \beta}}e^{-\frac{\hq^{\, 2}}{\beta}(\alpha-\phi+2\pi n)^2} .
\ea
Since $\hq$ is an odd integer, we have substituted $\cos\big(\pi \hq (\frac{\alpha}{2\pi}+n)\big)=e^{\i \pi n} \cos(\frac{\hq\alpha}{2})$, to simplify the $n$-dependence in the sum. Now we can apply Poisson resummation on the final term 
\beq
\sum_{n\in \mathbb{Z}}  \frac{\hq e^{2\pi\i n(\delta+1/2) }  }{\sqrt{\pi \beta}} e^{-\frac{\hq^{\, 2}}{\beta}(\alpha-\phi+2\pi n)^2} = \frac{1}{2\pi}\sum_{q \in \mathbb{Z} + \delta-\frac{1}{2}}  e^{\i \alpha q} e^{-\i \phi q} e^{-\beta \frac{q^2}{4\hq^{\,2}}}
\eeq
This immediately leads to \eqref{trumpet2} after multiplication by  $2\cos(\frac{\hq\alpha}{2})$.

\paragraph{The disk} Next we consider the evaluation of the disk partition function. The calculation is very similar to the trumpet and we only point out the differences. The $n=0$ fixed point is given by $y=2\pi t/\beta$ and $\psi = e^{- \i \hq \alpha t/\beta} \theta$ and the bosonic part of the action is now instead 
\beq
S = - \frac{1}{2}\int_0^{\beta} \d t \Big[ \Big\{\tan \Big(\frac{y}{2}\Big) ,t\Big\}+ 2 (\partial_t \sigma)^2 \Big] = - \frac{\pi^2}{\beta} - \frac{\hq^{\,2}}{\beta}\alpha^2. 
\eeq
We analyze next the rotation eigenvalues and start with the reparametrization mode. In the case of the trumpet we restricted to $m\geq1$ since $m=0$ corresponds to the $\uU(1)$ isometry of the geometry. The hyperbolic disk has a full ${\rm SL}(2,\mathbb{R})$ isometry, generated by the modes with $m=0,\pm 1$. This leads to a modified one-loop determinant
\beq
Z_{\rm 1-Loop,~D}^{\rm Rep.} = \prod_{m\geq 2}\frac{ 2\beta}{m} = \frac{1}{4\sqrt{\pi }\beta^{3/2}}.
\eeq
The one-loop determinant of the phase mode is unchanged and gives $Z_{\rm 1-Loop,~D}^{\uU(1)} = \hq /\sqrt{\pi \beta}$. Finally the $\N=2$ hyperbolic disk has four real fermionic isometries, corresponding to complex modes with rotation eigenvalues labeled by $m=\pm 1/2$. These are the fermionic generators of $\SU(1,1|1)$. The fermionic one-loop determinant is
\beq
Z_{\rm 1-Loop,~D}^{\rm Ferm.}= \prod_{m\neq \pm \frac{1}{2}} \frac{m+\frac{\hq \alpha}{2\pi}}{2\beta} = \frac{16 \beta^2}{1-\frac{\hq^{\, 2} \alpha^2}{\pi^2}}\, 2 \cos \Big(\frac{\hq\alpha}{2}\Big).
\eeq  (In this case, we extracted the square root by considering just one of the two fermions, and instead allowed both signs of $m$.)
We can now put the ingredients together. We also multiply the disk partition function by $1/(4\pi)^2$ which we can think of as a shift in the action by a term proportional to the Euler characteristic. The normalization of the disk partition function correlates with the normalization used in section \ref{sec:GPI} for the torsion of a three-holed sphere.
 With these choices, the final answer is 
\beq
\frac{\hq \cos\big(\frac{\hq\alpha}{2}\big)}{2\pi^3\big(1-\frac{\hq^{\, 2} \alpha^2}{\pi^2}\big)} e^{\frac{\pi^2}{\beta}\big(1-\frac{\hq^{\, 2} \alpha^2}{\pi^2}\big)},
\eeq
and the disk partition function is obtained by summing over shifts $\alpha \to \alpha + 2\pi n $, reproducing the result in section \ref{sec:N2JTG}. This result was first derived in \cite{Stanford:2017thb}.

We now explain how to obtain the spectrum \eqref{diskdos} from the partition function. First, use $2\cos \pi \hq (n+\frac{\alpha}{2\pi})=(-1)^{n}2 \cos(\frac{\hq\alpha}{2})$ to write the partition function as 
\beq
Z_{\rm disk}(\beta,\alpha) = 2\cos\big(\frac{\hq\alpha}{2}\big) \sum_{n\in \mathbb{Z}} \frac{\hq \, e^{2\pi\i n(\delta+1/2)} }{4\pi^3 (1-\frac{\hq^{\, 2} (\alpha+2\pi n)^2}{\pi^2})} e^{\frac{\pi^2}{\beta}(1-\frac{\hq^{\, 2} (\alpha+2\pi n )^2}{\pi^2})},
\eeq
Let us begin by finding the BPS states. These are the states with zero energy and therefore they are the only contribution that survives a zero temperature $\beta \to \infty$ limit. The classical action is proportional to the temperature so in this limit we can evaluate it to zero
\beq
Z_{\rm disk}(\beta\to\infty,\alpha) = 2\cos\big(\frac{\hq\alpha}{2}\big) \sum_{n\in \mathbb{Z}}  \frac{\hq \, e^{2\pi\i n(\delta+1/2)} }{4\pi^3 (1-\frac{\hq^{\, 2} (\alpha+2\pi n)^2}{\pi^2})},
\eeq
This is not in a useful form, in particular we would like to extract the number of BPS states for each charge $k$. To do this we apply Poisson resummation,
\beq\label{eq:bpsapps}
 \sum_{n\in \mathbb{Z}} \frac{e^{     2\pi\i n(\delta+1/2)}2 \hq \,\cos\big(\frac{\hq\alpha}{2}\big) }{4\pi^3 (1-\frac{\hq^{\, 2} (\alpha+2\pi n)^2}{\pi^2})} = \sum_{k \in \mathbb{Z}+\delta, |k|<\frac{\hq}{2}} \, e^{\i \alpha k}\,\frac{\cos(\frac{\pi k}{\hq})}{4\pi^2},
\eeq
by taking the Fourier transform of the summand in the left hand side with respect to $n$, using the poles coming from the denominator. We now derive the density  of non-BPS states.  In order to do this it is convenient to take a derivative with respect to $\beta$ 
\bea
-\partial_\beta Z_{\rm disk} &=&  2\cos\big(\frac{\hq\alpha}{2}\big) \sum_{n\in \mathbb{Z}} e^{ 2\pi\i n(\delta+1/2)} \frac{\hq }{4\pi \beta^2} e^{\frac{\pi^2}{\beta} (1-\frac{\hq^{\, 2} (\alpha+2\pi n )^2}{\pi^2})},\\
&=&    2\cos\big(\frac{\hq\alpha}{2}\big) \sum_{q\in \mathbb{Z}+\delta-\frac{1}{2}} \frac{e^{\i \alpha q}}{2\pi} \frac{e^{\frac{\pi^2 }{\beta}}}{4 \sqrt{\pi }\beta^{3/2}} e^{-\beta E_0(q)} ,\\
&=&   \sum_{q\in \mathbb{Z}+\delta-\frac{1}{2}} \frac{e^{\i \alpha (q-\frac{\hq}{2}) } + e^{\i \alpha (q+\frac{\hq}{2})}}{2\pi}  \int_0^\infty \d s \frac{2 s \sinh 2\pi s}{4\pi^2} e^{- \beta (E_0(q)+s^2)} , \\
&=&   \sum_{q\in \mathbb{Z}+\delta-\frac{1}{2}} \frac{e^{\i \alpha (q-\frac{\hq}{2}) } + e^{\i \alpha (q+\frac{\hq}{2})}}{2\pi}  \int_{E_0(q)}^\infty \d E \,\frac{ \sinh 2\pi \sqrt{E-E_0(q)}}{4\pi^2} e^{- \beta E} .
\ea
In the second line, we apply Poisson resummation to rewrite the sum over fixed points as a sum over charges. Since the derivative with respect to $\beta$ removed the denominator that came from the fermion zero-modes, this Poisson resummation is straightforward.  Motivated by the fact that the temperature dependence in the second line is that of
bosonic JT gravity, in the third line we rewrote this expression as an integral over a parameter $s$.
   In the fourth line, we changed integration variables from $s$ to $E=E_0+s^2$. 

It is straightforward to integrate the final line in the formula above with respect to $\beta$. The integration produces simply a factor of $1/E$ inside the integral over energies. We conclude 
\beq
Z_{\rm disk} = Z_{\rm disk}(\beta=\infty) + \sum_{q\in \mathbb{Z}+\delta-\frac{1}{2}} (e^{\i \alpha (q-\frac{\hq}{2}) } + e^{\i \alpha (q+\frac{\hq}{2})})\int_{E_0(q)}^\infty \d E \frac{ \sinh 2\pi \sqrt{E-E_0(q)}}{8\pi^3E} e^{- \beta E} .
\eeq
The integral on the right vanishes for $\beta\to\infty$ in all cases; it vanishes exponentially fast if $E_0(q)>0$, and as $1/\sqrt \beta$ if $E_0(q)=0$.
 We already computed the integration constant $Z_{\rm disk} (\beta=\infty)$ in \eqref{eq:bpsapps}, since it corresponds to the BPS  contribution. This leads to \eqref{diskdos}.

\paragraph{Schwarzian Theory from Virasoro} We briefly show how to reproduce these results from a canonical quantization approach. We can view the Schwarzian path integral arising from the disk as computing a classical partition function of a system whose phase space is infinite-dimensional and parametrized by $\cN=2$ superconformal transformations modulo $\SU(1,1|1)$. This is a super Virasoro coadjoint orbit, and an extension of \cite{Witten:1987ty,Alekseev:1988ce}, see also \cite{Mertens:2017mtv}, would imply that the quantization of this phase space is given by the vacuum representation of the $\cN=2$ Virasoro algebra with central charge $c=3 \widehat{c}$. The quantum partition function is the character of this representation. The classical limit of the quantum result, meaning the large $\widehat{c}$ limit of the Virasoro character, reproduces the integral over the classical phase space, i.e. the Schwarzian path integral. 

The $\N=2$ Virasoro vacuum character on a torus with modulus $q=e^{2\pi \i \tau}$ and $\uU(1)$ fugacity $y=e^{2\pi \i z}$ is given by \cite{Eguchi:2003ik}
\beq
{\rm Tr}_{\rm vac} \left[ q^{L_0} y^{J_0}\right] = q^{-\frac{\widehat{c}-1}{8}} \frac{1-q}{(1+yq^{1/2})(1+y^{-1}q^{1/2})} \frac{\theta_3 (\tau, z) }{\eta(\tau)^3},~~~~Z_{\rm vac}(q,y) = e^{\frac{\i \pi \widehat{c} }{\tau}(z-\frac{1}{2})^2} {\rm Tr} \left[ q^{L_0} y^{J_0}\right]
 \eeq
The trace is over the vacuum representation, but to compare with the Schwarzian partition function we need to include an elementary exponential factor  to go from a sum over states to a path integral $Z_{\rm vac}$; see section 5 of \cite{Kraus:2006wn} for a general discussion on the origin of this factor. Finally, both in integrating over the coadjoint orbit 
(to perform the Schwarzian path integral) and in quantizing it, one has to sum over the same fixed points.   In quantizing the orbit, the different fixed points give representations of
the $\N=2$ super Virasoro algebra that 
differ by integer spectral flow \cite{Eguchi:2003ik}, and summing over them gives a representation of the $\N=2$ super Virasoro group, not just the corresponding Lie algebra.

The torus is a product of two circles. As explained in \cite{Mertens:2017mtv}, to relate the coadjoint orbit quantization on the torus  in the large $\widehat{c}$ limit to the Schwarzian path integral, we need to send the size of one of the circles to zero so that the modular parameter $\tau$ vanishes as $\tau \sim 1/\widehat{c}$. In this limit, the Schwarzian mode lives on the circle that remains of finite size. More precisely, the relation between the 2d CFT parameters and the Schwarzian ones is 
\beq
\tau =\i  \frac{2\pi}{\beta} \frac{2}{\widehat{c}-1},~~{\rm and}~~z= \frac{1}{2}+\frac{\hq\,\alpha}{2\pi} \tau,~~~~~\widehat{c} \to \infty.
\eeq
The half-integer shift in $z$ appears so that, for $\alpha=0$, the fermions are periodic around the circle which disappears in the Schwarzian limit. Otherwise, there would be no fermions in the remaining large circle where the Schwarzian mode lives, and one would reproduce instead the partition function of a bosonic theory. 

The  large $\widehat{c}$ limit of the following quantities will be useful:
\bea
1-q \to 2 \pi \i \tau,~~1+y^{\pm1} q^{1/2} \to -\i \pi \tau \Big( 1\pm \frac{\hq\,\alpha}{\pi}\Big),~~\frac{\theta_{3}(\tau,z)}{\eta(\tau)^3} \to (-\i \tau) 2 \cos \Big(\frac{\hq\,\alpha }{2}\Big).
\ea
Using these relations, the semiclassical limit of the $\N=2$ Virasoro vacuum character is 
\beq
e^{\frac{\i \pi \widehat{c} }{\tau} (z-\frac{1}{2})^2}q^{-\frac{\widehat{c}-1}{8}} \frac{1-q}{(1+yq^{1/2})(1+y^{-1}q^{1/2})} \frac{\theta_3 (\tau, z) }{\eta(\tau)^3} \to \frac{4}{\pi} \frac{\cos (\frac{\hq\,\alpha}{2})}{1-\frac{\hq^{\,2} \alpha^2}{\pi^2}} \,e^{\frac{\pi^2}{\beta}(1-\frac{\hq^{\, 2}\alpha^2}{\pi^2})}.
\eeq
This reproduces the disk partition function for the $n=0$ fixed point. The sum over $n$ in the Schwarzian limit corresponds to the integral spectral flow of \cite{Eguchi:2003ik} ensuring that the $\uU(1)$ symmetry is compact. 

We can repeat this analysis for the trumpet. This case corresponds to a different coadjoint orbit of the $\N=2$ Virasoro group, where we only mod out by $\uU(1)\times \uU(1)$ generated by the $R$-symmetry and time translations. The Hilbert space that quantizes this phase space is identified with a general non-degenerate representation of the Virasoro group labeled by scaling dimension $h$ and $\uU(1)$ charge ${\sf q}$. These parameters correspond to the length of the geodesic and the $R$-charge holonomy of the Schwarzian trumpet, through the relations 
\beq
h = \frac{\widehat{c}-1}{8} + \frac{\widehat{c}-1}{8} \frac{b^2}{4\pi^2}+\frac{{\sf q}^2}{2(\widehat{c}-1)} ,~~~{\sf q} =- \frac{\widehat{c} -1}{2\pi} \,\hq\,\phi .
\eeq
The opposite sign between ${\sf q}$ and $\phi$ is chosen to match the conventions in the Virasoro algebra with the ones used in JT gravity. These identifications are the same one obtains when considering the bosonic Virasoro times Kac-Moody group. The relevant super Virasoro character is 
\beq
{\rm Tr}_{(h,{\sf q})} \left[ q^{L_0} y^{J_0}\right] = q^{h-\frac{\widehat{c}-1}{8}}y^{\sf q}\,  \frac{\theta_3 (\tau, z) }{\eta(\tau)^3}
\eeq
The Schwarzian limit of the path integral version gives, up to an unimportant overall phase,
\beq
e^{\frac{\i \pi \widehat{c}}{\tau} (z-\frac{1}{2})^2}q^{h-\frac{\widehat{c}-1}{8}}y^{\sf q}\,  \frac{\theta_3 (\tau, z) }{\eta(\tau)^3} \to 
\frac{ 2 \cos(\frac{\hq \alpha}{2})}{\beta} e^{-\frac{b^2}{4\beta} -\frac{\hq^{\, 2}}{\beta}(\alpha - \phi)^2}.
\eeq
This again reproduces precisely the answer coming from localization, after summing over integral spectral flow.

\paragraph{Supersymmetric Defect} We finish by considering the path integral of the boundary graviton on the hyperbolic disk with a defect. If the deficit angle is bigger than $\pi$, we can obtain the partition function by analytically continuing the trumpet by $(b,\phi) \to (\i \gamma, \phi)$. The opening angle around the defect is $\gamma$ in these conventions, and
 $e^{\i \hq\,\phi}$ is the holonomy around the defect. When $\gamma$ and $\phi$ are arbitrary, the answer is the same as the trumpet.

When $\gamma = 2\pi -2\,\hq \,|\phi|$ the defect is supersymmetric. In terms of the connection with Virasoro, these defects correspond to chiral primary fields with scaling dimension related to the charge as $h=|{\sf q}|/2$. Picking ${\sf q}$ positive ($\phi$ negative) leads to $\gamma = 2\pi +2 \hq \phi$ while negative ${\sf q}$ ($\phi$ positive) leads to $\gamma = 2\pi - 2\hq \phi$. To simplify these and some expressions below, when considering these defects we take $\phi$ to be in the range $-\pi \leq \hq\phi \leq + \pi$.

The classical action is a smooth function of $\gamma,\phi$, and nothing special happens when the defect becomes supersymmetric.  The same is true for most of the rotation eigenvalues,
but one fermionic eigenvalue, with  $m=+1/2$ or $-1/2$ depending on the sign of $\phi$, is now associated to a symmetry of the defect geometry and should be
discarded. Something similar was analyzed for $\N=4$ JT gravity defects in \cite{Iliesiu:2021are}. Taking this into account,  the final answer for the partition function is
\beq
Z_{\rm susy-defect}(\beta,\alpha)= w\sum_{n\in \mathbb{Z}} \frac{e^{2\pi \i n(\delta+1/2) }2 \hq \,\cos\big(\frac{\hq\alpha}{2}\big) }{4\pi^3 \big(1- {\rm sgn}(\phi) \frac{\hq (\alpha+2\pi n)}{\pi}\big)} e^{\frac{1}{4\beta}(2\pi-2 \hq| \phi|)^2-\frac{ \hq^{\, 2}}{\beta}(\alpha-\phi+ 2\pi n)^2}.
\eeq
We added a weighting factor $w$ that comes with each defect insertion. This expression can also be reproduced as a limit of a Virasoro character.

\subsection*{$\cN=4$ JT supergravity} 

The previous analysis can be repeated for $\N=4$ JT gravity, with some modifications due to the $\SU(2)$ $R$-symmetry. We begin again with the trumpet. We refer to the $\SU(2)$ holonomy around the geodesic as $U$. In a $2\times 2$ representation, this $\SU(2)$ matrix can be diagonalized to $e^{\i \phi Z/2}$, where $Z={\rm diag}(1,-1)$. We can parametrize the $\N=4$ cylinder with coordinates $(x,\psi,\bar{\psi})$, with $\psi=(\psi^1,\psi^2)$ in the fundamental of $\SU(2)$.  The choice of holonomy translates into the identification
\beq
(x,\psi,\bar{\psi} ) \cong (x+2\pi , - U \psi, - U^{-1} \bar{\psi}).
\eeq
Similar to the conventions of section \ref{sec:N2JTG}, we take the geodesic boundary at ${\rm Im}\,x=0$ and the NAdS boundary to be at $x=\i T + y$ with real $y$. The geometry of the trumpet then induces the following identification on the boundary:
\beq
(y,\psi,\bar{\psi} ) \cong (y+2\pi , - U \psi, - U^{-1} \bar{\psi}).
\eeq
The boundary quantum mechanics is formulated on a supercircle with coordinates $(t,\theta,\bar{\theta})$, where again $\theta$ is in the fundamental representation. The equivalence relation that corresponds to computing the partition function with fugacity $V\in \SU(2)$ is 
\beq
(t, \theta,\bar{\theta}) \cong (t+\beta, -V \theta, - V^{-1} \bar{\theta}).
\eeq
Following the notation in the main text, we  label $V$ by its eigenvalues $e^{\i \alpha Z/2}$. Fixed points take the general form
\beq
y(t,\theta)= \frac{2\pi}{\beta} t,~~~~\psi(t,\theta) = g(t) \theta,~~~~g:S^1 \to \SU(2).
\eeq
For this to satisfy the appropriate periodicity conditions, $g(t)$ must satisfy 
\beq
g(t+\beta) = U g(t) V^{-1}.
\eeq
$U$ and $V$ can be diagonalized by separate unitary transformations of $\psi$ and $\theta$.   After doing so, the fixed point can be written, in analogy to the $\N=2$ case, as 
\beq\label{app:eq:fp}
\psi^1(t,\theta) = e^{\i \frac{(\phi-\alpha) t}{2 \beta}} \theta^1,~~~\psi^2(t,\theta) = e^{-\i \frac{(\phi-\alpha) t}{2 \beta}} \theta^2,
\eeq 
where we wrote $\psi$ and $\theta$ in components to be more explicit.

As in $\N=2$ we have a family of fixed points obtained by a shift $\alpha \to \alpha + 4 \pi n$ with $n\in \mathbb{Z}$.   Because the diagonalization of $U$ or $V$ is unique
only up to the action of the Weyl group, which exchanges the eigenvalues, there is also a second family of fixed points obtained by $\alpha\to-\alpha$.
 For concreteness we focus first on the choice in \eqref{app:eq:fp}.

The classical action is easy to obtain. In the basis that diagonalizes $U$ and $V$, the bosonic sector of the Schwarzian action is essentially the same as for $\N=2$.  Taking account of the normalization used in defining $\alpha$ and $\phi$, the action is
\beq
S = \frac{b^2}{4\beta} + \frac{1}{4\beta}(\alpha-\phi)^2.
\eeq
Next, we need to compute the  eigenvalues of the generator 
\beq
D = \frac{\partial}{\partial t} - \frac{2\pi}{\beta} \frac{\partial}{\partial y} - \frac{\i (\phi-\alpha)}{\beta} (\psi^1 \partial_{\psi^1} -\psi^2 \partial_{\psi^2} -\bar{\psi}^1 \partial_{\bar{\psi}^1} + \bar{\psi}^2 \partial_{\bar{\psi}^2})
\eeq
of the rotation symmetry of the fixed point.
The eigenvalues for  the bosonic reparametrization mode are the same as $\N=2$ or $\N=0$, so  again $Z_{\rm 1-Loop,~T}^{\rm Rep.} = \frac{1}{2\sqrt{\pi \beta}}$. The fermionic rotation eigenvalues are also straightforward: we essentially have two copies of the fermions that we had for $\N=2$, leading to
\beq
Z_{\rm 1-Loop,~T}^{\rm Ferm.} = 4 \cos^2\big(\frac{\alpha}{4}\big).
\eeq
What remain are the modes associated to time-dependent $\SU(2)$ transformations. The fluctuations in the Cartan of $\SU(2)$ are again precisely the same as the phase mode of $\N=2$. The rotation eigenvalues for off-diagonal generators are 
\be\label{thoseq}\lambda_{m,\pm} =\pm \frac{2\pi m+\alpha}{\beta},\hskip1cm m\in \Z.\ee
This leads, with zeta-function regularization, to $\sqrt{\det D}=\sin\frac{\alpha}{2}$. 
The overall normalization can be fixed
using the relation, described in appendix A of  \cite{Kapec:2019ecr}, between the one-loop determinant derived from the symplectic measure, appropriate for this calculation, and the one-loop determinant derived from the ultralocal measure which is equivalent to canonical quantization. The final result is
\beq
~~Z_{\rm 1-Loop,~T}^{\SU(2)} %=\frac{1}{4\pi} \frac{\sqrt{\pi}}{2 \sqrt{\beta}} \frac{1}{\sin(\frac{\alpha}{2})}
=\frac{1}{8 \sqrt{\pi\beta}} \frac{1}{\sin(\frac{\alpha}{2})}.
\eeq
The path integral around one fixed point is therefore 
\beq
\frac{\cot(\frac{\alpha}{4})}{4\pi \beta} e^{-\frac{b^2}{4\beta} - \frac{(\alpha-\phi)^2}{4\beta}}.
\eeq
To obtain the final answer given in \eqref{N4trumpet1} we need to sum over $\alpha \to \alpha + 4\pi \mathbb{Z}$ and add another contribution with $\alpha \to -\alpha$. Since $\cot(-\frac{\alpha}{4}) = -\cot(\frac{\alpha}{4})$ there is a relative minus sign between the two. 

We now explain how to re-express the  result  \eqref{N4trumpet2}  in terms of a sum over $\N=4$ supermultiplets. We begin with an identity derived from Poisson resummation:
\beq\label{tempo}
\sum_{n\in \mathbb{Z}}  \frac{\sqrt{\pi}}{2\sqrt{\beta}\sin(\frac{\alpha}{2})} \Big( e^{-\frac{1}{4\beta}(\alpha-\phi+4\pi n)^2}-e^{-\frac{1}{4\beta}(\alpha+\phi+4\pi n)^2}\Big) = \sum_{J\geq \frac{1}{2}} \chi_{J-\frac{1}{2}}(\alpha) \sin(J\phi) e^{-\beta J^2}\,.
\eeq  Here $\chi_J(\alpha)=\frac{\sin((J+1/2)\alpha)}{\sin(\alpha/2)}$ is the character of the spin $J$ representation of $\SU(2)$.
To get the $\N=4$ trumpet path integral, the expression in eqn. (\ref{tempo}) has to be multiplied by the bosonic JT gravity trumpet, and also by a factor of $4 \cos^2 \alpha/2$. The latter turns the $\SU(2)$ character into a sum over $\N=4$ multiplets, since 
\beq
4\cos^2\big(\frac{\alpha}{4}\big) \,\chi_{J-\frac{1}{2}}(\alpha) = \chi_J(\alpha) + 2 \chi_{J-\frac{1}{2}}(\alpha) + \chi_{J-1}(\alpha).
\eeq
Putting all this together, we arrive at \eqref{N4trumpet2}. 

Next we consider the disk path integral and point out the main differences with the trumpet. The fixed points and action are the same as the trumpet but with $\phi=0$. A new feature is that we do not need to sum over the action of the Weyl group and there is a unique family of saddles related by $\alpha \to \alpha + 4\pi \mathbb{Z}$. The bosonic reparametrization one-loop determinant is the same as, for example, the $\N=2$ disk computed earlier in this section. The fermionic one-loop determinant is essentially two copies of the $\N=2$ disk one, so it is proportional to
\beq
\frac{4 \cos^2 (\frac{\alpha}{4})}{(1-\frac{\alpha^2}{4\pi^2})^2}.
\eeq
The denominator comes from the eight fermionic isometries of the $\N=4$ disk which make up the supergroup $\PSU(1,1|2)$. We have two complex fermions each with frequency $+1/2$ and $-1/2$ in time. Finally, for the $\SU(2)$ mode, the rotation eigenvalues are the same as before, but the $m=0$ modes in eqn. (\ref{thoseq}) are now generated by a symmetry of the bulk
and so should be omitted.  This multiplies $\sqrt{\det'\,D}$ by $\beta/\alpha$, so $1/\sqrt{\det' D}$ is now  $\alpha/\beta^{3/2}\sin\frac{\alpha}{2}$, where the overall coefficient
can be determined from \cite{picken1989propagator}.  Putting the pieces together, we obtain 
\beq
\beta \frac{\cot(\frac{\alpha}{4}) \alpha }{\pi^4(1-\frac{\alpha^2}{4\pi^2})^2} e^{\frac{\pi^2}{\beta}(1-\frac{\alpha^2}{4\pi^2})}
\eeq 
A sum over shifts of $\alpha$ gives the result \eqref{diskN4} for the disk partition function. We also choose the normalization of $S_0$ to match the normalization in \eqref{diskN4}. 

The evaluation of the density of states works similarly to the procedure we outlined for $\N=2$ JT gravity. First we determine the BPS states. To do this, we take $\beta \to \infty$. The new feature is the factor of $\beta$ remaining in the one-loop determinant. This appears to give a divergent answer, but the coefficient vanishes 
\beq
\sum_{n\in\mathbb{Z}} \beta \frac{(\alpha+ 4\pi n)}{(1-\frac{1}{4\pi^2}(\alpha+4\pi n)^2)^2} = 0.
\eeq
This implies we should not set $1/\beta=0$ in the action and instead Taylor expand in $1/\beta$. The linear term cancels the factor of $\beta$ in the one-loop determinant and now the prefactor is
\beq
Z(\beta \to \infty )  \to e^{S_0} \sum_{n\in\mathbb{Z}} \frac{\cot(\frac{\alpha}{4})(\alpha+4\pi n)}{\pi^2 (1-\frac{(\alpha+4\pi n)^2}{4\pi^2})} = e^{S_0}.
\eeq  Since this is independent of $\alpha$, there are BPS states only at $J=0$.
Once we have the number of BPS states, we can use the same trick as before. We take two derivatives $\partial^2_\beta  Z_{\rm disk}$. The two derivatives acting on the classical term will cancel the denominator from the fermion zero-modes. After that we can easily apply Poisson resummation, as  for the trumpet. Finally we integrate over $\beta$ twice to get back $Z_{\rm disk}$ and use the BPS states as boundary condition at $1/\beta=0$. 

One can reproduce these results again from taking a large $c$ limit of $\N=4$ Virasoro characters (the vacuum for the disk, generic for the trumpet). In this case, the relevant torus partition functions were computed by Eguchi and Taormina in \cite{Eguchi:1987sm,Eguchi:1987wf,Eguchi:1988af,Eguchi:2008ct}.

\section{Revised Loop Equations and Topological Recursion}\label{app:Loop}
In this section we analyze the loop equations of the AZ $(\upalpha,\upbeta)$ ensembles, following the presentation in \cite{SW}. We focus on the new features that appear relative
to bosonic JT gravity and $\N=1$ JT supergravity.

A first new feature is that in $\N=2$ and $\N=4$ JT supergravity, the number of BPS states, extracted from the disk partition function, is proportional to $e^{S_0}$. In contrast, in  $\cN=1$ JT supergravity, the number of supersymmetric  states does not scale with $e^{S_0}$ and vanishes in
the absence  of Ramond punctures. A second and related new feature is that we need to allow a logarithmic singularity in the matrix potential. We will see that even though the two effects are physically distinct, they can be analyzed together.  

 We have already seen in sections \ref{randommatrix} and \ref{sec:N2JTG} that different multiplets are statistically independent. Therefore we can focus on any one of them separately. 
Consider a single supermultiplet of charges $(k,k+\hq\,)$ with $L$ nonzero matrix eigenvalues $\lambda_i$ and  $\nu$ BPS states.   In the case of an $(\upalpha,\upbeta)$ AZ ensemble, the
measure in integration over the $\lambda_i$ is 
\bea\label{mmeasure}
 \prod_i\d\lambda_i\prod_{i<j} |\lambda_i^2 - \lambda_j^2|^{\upbeta} \, \prod_{i=1}^L \lambda_i^{\upalpha}\, e^{- L \frac{\upbeta}{2} V(\lambda_i^2)}.
\ea
To maintain the standard structure of topological recursion in matrix integrals, $V$ should be independent of\footnote{We have actually had to relax this condition
in section \ref{matrixprediction} in the discussion of $\CT$-invariant models, but it is a useful condition in the rest of the article.}   $L$.   From   eqn. (\ref{mmeasure}), we see immediately that a shift in $\upalpha$ by a multiple
of $L$ can be absorbed in an $L$-independent logarithmic correction to $V$.   In sections \ref{randomtwo} and \ref{boundaryview}, we have found in all cases that
$\upalpha=\upalpha_0+\upbeta\nu$, where $\upalpha_0$ is the value of $\upalpha$ in an ensemble of  some given type in the absence of BPS states, and $\nu$ is the number of
BPS states.   Moreover, in $\N=2$ and $\N=4$ JT supergravity, $\nu$ is always a multiple of $e^{S_0}$ and thus, in the matrix model, it is a multiple of $L$.  
So the $\upbeta\nu$ term in $\upalpha=\upalpha_0+\upbeta\nu$ can be eliminated by adding to $V(x)$ an $L$-independent term $\upgamma \log x$, $\upgamma=\nu/L$.
The upshot is that we can reduce to the case of only considering
 $(\upalpha_0,\upbeta)$ ensembles  if we allow the matrix potential to have a logarithmic singularity at $x=0$:
 \be\label{logterm} V(x) = \upgamma \log x + \tilde{V}(x),\ee where $\tilde{V}(x)$ is an analytic function.\footnote{For  discussion of matrix models with such logarithmic potentials, see \cite{Srednicki:1992sp,Penner1,Penner2,Ambjorn:1994bp}. } 
 One of the reasons that this makes sense in our application is that in all cases except the multiplet of charges $(-\h q/2,\h q/2)$, the threshold energy $E_0$ is positive and (at least
 to all orders in the topological expansion) the support of 
 the matrix eigenvalues does not reach the logarithmic singularity at $x=0$; on the other hand, the $(-\h q/2,\h q/2)$ multiplet has no BPS states and can be described by an analytic
 potential.

As usual, we define the resolvent 
\bea
R(x) = \sum_{j=1}^L \frac{1}{x-\lambda_i^2},~~~~R(x_1,\ldots,x_n) = R(x_1)\ldots R(x_n).
\ea 
The sum runs over the generically nonzero matrix eigenvalues, not the BPS states.
The loop equations, introduced in \cite{Migdal:1975zg} and put in their optimal form in \cite{Eynard:2004mh}, are derived from the identity
\bea
&&\left( (1-\frac{2}{\upbeta})\partial_x + \frac{\upalpha_0-1}{\upbeta x} - L V'(x) \right) \langle R(x, I )\rangle_c + \langle R(x,x,I) \rangle_c+ \sum_{J } \langle R(x,J) \rangle_c \langle R(x,I/J) \rangle_c \nn
&&+ \frac{2}{\upbeta x} \sum_{k=1}^n \partial_{x_k} \frac{x\langle R(x,I/x_k)\rangle_c - x_k \langle R(I)\rangle_c}{x-x_k} =-\frac{L}{x}\langle P(x;I) \rangle.\label{eq:RRRRR}
\ea
See section 4 of \cite{SW} for a derivation. We follow here the notation of that reference. We denote $I=\{ x_1,\ldots,x_n\}$. The quantity appearing on the right hand side is 
\bea
P(x;I) = \sum_{a=1}^L \frac{xV'(x)-\lambda_a^2 V'(\lambda_a^2)}{x-\lambda_a^2}\, R(I).
\ea
The connected resolvent has an 't Hooft expansion at large $L$ given by 
\bea
\langle R(I) \rangle_c \cong \sum_{g=0,\frac{1}{2},1,\ldots}  \frac{R_{g,n}(I)}{L^{2g+n-2}},
\ea
and inserting this expansion in \eqref{eq:RRRRR} gives a recursion relation for the genus $g$ correlators $R_{g,n}(I)$. The derivation of the loop equations relies on the fact that the function $P(x;I)$ is analytic in $x$. This is obvious if the potential is analytic but  remains true in the presence of a logarithmic term. To see this, notice that $x \partial_x( \log x) = 1$ and therefore the two terms in the numerator coming from the logarithm cancel. 
 
We start with the order $L^2$ term in the loop equation. It is given by 
\bea
R_{0,1}(x)^2 - V'(x) R_{0,1}(x) = \frac{1}{x}( {\rm analytic~in~}x).
\ea
Introducing $y(x) = R_{0,1}(x) - V'(x)/2$, this equation can be rewritten as 
\bea
y(x)^2 = \frac{\upgamma^2}{4x^2} + \frac{1}{x}({\rm analytic~in~}x),
\ea  an equation that defines the spectral curve -- a double cover of the $x$-plane.
Thus  a logarithmic term in the potential (or  a number of BPS states proportional to $L$) leads to a pole in $y(x)$ at $x=0$.  The sign of the residue of the pole is fixed by demanding that the resolvent $R_{0,1}(x)$ does not have a pole, since this would imply an unprotected exact zero-energy supermultiplet, which would be unnatural. Therefore we always have opposite poles in $V'(x)$ and $y(x)$.

In a single cut AZ matrix model 
with a possible logarithmic term in the potential, apart from a possible pole at $x=0$, $y$ has a cut on an interval $(a_-,a_+)$, $0\leq a_-<a_+$.
In our applications to\footnote{In $\N=4$ JT supergravity, we have not attempted to incorporate $\tT$ or $\CT$ symmetry, so we  have only encountered $(1,2)$ ensembles, always
with $a_->0$ and with BPS states only at $J=0$.} $\N=2$ JT supergravity,   $(a_-,a_+)=(E_0(q),\infty)$, with $E_0(q)$ the threshold energy of the given multiplet with $R$-charges $q\pm \hq/2$.
The only case with $E_0(q)=0$ is the multiplet with charges $\pm \hq/2$.   In this case, there are no BPS states and no pole in $y$, so this multiplet is described by  the standard $(1,2)$ ensemble considered in \cite{SW} with square matrices and a regular potential.
In the rest of this discussion we consider the other cases, which all have  $a_-=E_0(q) > 0$. 
When $|q|< \hq/2$, the pole in $y_q(x)$ matches with the number of BPS states at charge $q-\hq/2$ or $q+\hq/2$, consistent with an interpretation in terms of rectangular matrices and no
logarithmic term in the potential.
When $|q|>\hq/2$ there are no BPS states and therefore the supercharge is a square matrix, but the pole remains in $y_q(x)$. 
It is interpreted in terms of a logarithmic term in the matrix potential for that multiplet.

We consider now $I=\{x_1\}$ and $g=0$. The loop equation to order $L$ becomes 
\bea
2 y(x) R_{0,2}(x,x_1) + \frac{2}{\upbeta} \frac{R_{0,1}(x)}{(x-x_1)^2} = \frac{1}{x}({\rm analytic~in~}x).
\ea
On the right hand side, we can replace $R_{0,1}(x)$ by $y(x)$, since the difference, proportional to $V'(x)$ times a function analytic in the neighborhood of the cut, has the same form as the right hand side. Now we can multiply this equation by $\sqrt{\sigma(x)}/(x y(x))$, where $\sigma(x)=(x-a_-)(x-a_+)$. This ratio is analytic in $x$ in a neighborhood of the cut, 
  since the  branch cuts in $y(x)$ at $a_\pm$ are removed by the numerator.
It has
no pole at $x=0$, since $y$ has a pole there. Following the derivation in section 4 of \cite{SW}, this leads to 
\bea
R_{0,2}(x_1,x_2) = \frac{1}{\upbeta(x_1-x_2)^2} \left( \frac{x_1 x_2 - \frac{a_+ + a_-}{2} (x_1+x_2) + a_+ a_-}{\sqrt{\sigma(x_1) } \sqrt{\sigma(x_2)}} -1\right).
\ea
This is the same as the answer for a Dyson $\upbeta$-ensemble, or for an AZ ensemble with either an analytic potential or $\nu$ of order one.  A possible
logarithmic term in the potential has not affected $R_{0,2}(x_1,x_2)$.

We can now move on and look at the loop equation with $I=\emptyset$ to order $L$. In this case we get 
\bea
\Big( \Big(1-\frac{2}{\upbeta}\Big) \partial_x + \frac{\upalpha_0-1}{\upbeta x} \Big) R_{0,1} (x) + 2 y(x) R_{\frac{1}{2},1}(x) = \frac{1}{x}({\rm analytic~in~}x).
\ea
In section \ref{timereversal} we referred to $R_{\frac{1}{2},1}(x)$ as $R_{\rm xcap}(x)$. So far the answer is the same as in \cite{SW}. It is convenient, especially in the double scaling limit, to rewrite the loop equations in terms of $y(x)$ only. At this point we get new terms 
\beq
\Big( \Big(1-\frac{2}{\upbeta}\Big) \partial_x + \frac{\upalpha_0-1}{\upbeta x} \Big) y (x) + 2 y(x) R_{\frac{1}{2},1}(x) =- \Big( \Big(1-\frac{2}{\upbeta}\Big) \partial_x + \frac{\upalpha_0-1}{\upbeta x} \Big) \frac{V' (x)}{2}+ \frac{1}{x}({\rm analytic~in~}x).
\eeq
For an analytic potential, the  terms on the right hand side can be combined as an analytic function divided by $x$, leading to the derivation in \cite{SW}.
Allowing for a logarithmic term in the potential, the general result is 
\beq
\Big( \Big(1-\frac{2}{\upbeta}\Big) \partial_x + \frac{\upalpha_0-1}{\upbeta x} \Big) y (x) + 2 y(x) R_{\frac{1}{2},1}(x) =\frac{\upgamma(\upbeta-\upalpha_0-1)}{2 \upbeta x^2}+ \frac{1}{x}({\rm analytic~in~}x).
\eeq
For generic $\upalpha_0,\upbeta$, and $\upgamma$, the $1/x^2$ term on the right hand side of this equation will modify the analysis.   However,
in all cases encountered in the present article, either $\upgamma=0$ or $\upbeta-\upalpha_0-1=0$.
Specifically, we encounter $(1,2)$, $(0,1)$, and $(3,4)$ ensembles with various values of $\upgamma$, and $(1,1)$ and $(1,4)$ ensembles with $\upgamma=0$.
Given this, the double pole drops out and  a simple exercise in complex analysis and integration similar to the analysis in \cite{SW} gives the crosscap prediction
\bea
R_{\frac{1}{2},1}(x) = \int_{\mathcal{C}} \frac{\d x'}{2\pi \i} \frac{1}{x'-x} \frac{\sqrt{\sigma(x')}}{\sqrt{\sigma(x)}} \left( \frac{(1-\frac{2}{\upbeta})y'(x)}{2 y(x')} + \frac{\upalpha_0-1}{2\upbeta x'}\right).
\ea
 The contour $\mathcal{C}$ surrounds the cut of the spectral curve counterclockwise in the $x$ plane. We use this expression in section \ref{timereversal} to match with the gravity result for the crosscap. 

We are ready to consider the most general case for the loop equations of the AZ ensemble with either $\nu$ proportional to $L$ or a logarithmic term in the potential. We expand \eqref{eq:RRRRR} to order $L^{2-2g-n}$. The result is 
\bea
2 x y(x) R_{g,n+1}(x,I) + x F_{g,n}(x,I) = ({\rm analytic ~ in ~} x),
\ea
where 
\bea
F_{g,n}(x,I) &=& \Big((1-\frac{2}{\upbeta})\partial_x + \frac{\upalpha_0-1}{\upbeta x} \Big)R_{g-\frac{1}{2},n+1}(x,I)  \nn
&&+ R_{g-1,n+2}(x,x,I) + \sum_{\rm stable} R_{h,|J|+1} (x,J) R_{g-h,|I/J|+1}(x,I/J)\nn
&&+2 \sum_{k=1}^n \Big( R_{0,2}(x,x_k) + \frac{1}{\upbeta} \frac{1}{(x-x_k)^2} \Big) R_{g,n}(x,I/x_k).
\ea
The recursion relation one derives from this is
\bea
R_{g,n+1}(x,I) = \int_{\mathcal{C}} \frac{\d x'}{2\pi \i} \frac{1}{x'-x} \frac{\sqrt{\sigma(x')}}{\sqrt{\sigma(x)}} \frac{F_{g,n}(x',I)}{2 y(x')}. \label{eq:RRF}
\ea
This is exactly the same as the one derived in \cite{SW}. The only new feature is that now $y(x)$ is allowed to have a pole at $x=0$.   For $(\upalpha_0,\upbeta)=(1,2)$, the recursion relation becomes identical to the one of the Dyson $\upbeta=2$ ensemble. We used this fact in section \ref{sec:RR}.    When $V'(x)$ has a pole at $x=0$ (or $\nu\propto L$), getting a recursion relation of the
same form as if $V'(x)$ is regular and there are no BPS states has 
depended on the condition 
$\upbeta-\upalpha_0-1=0$, which made it possible at an initial stage of the expansion to replace $R_{0,1}(x)$ with $y(x)$ without disturbing the structure of the topological recursion.  
Conveniently this condition is satisfied in our applications.

In conclusion, 
allowing for a simple pole in $V'(x)$ or a number of BPS states of order $e^{S_0}$ has very little effect  on the structure of the topological recursion, 
at least if $\upbeta-\upalpha_0-1=0$.  What happens if we have instead a double pole in $V'(x)$? In this case the analysis is modified. To begin with, the function $P(x;I)$ would 
not be analytic in $x$ anymore: it would itself have a pole. This would potentially have an effect at all orders in the genus expansion. It is therefore a nice feature that a potential 
double pole in the spectral curve of $\cN=4$ JT gravity, which
appears if the formula for $y_J(x)$ is analytically continued away from integer values of  $2J$, actually vanishes when $2J$ is an integer. This is of course the only 
relevant case to consider since $J$ is the spin of a representation of $\SU(2)$.

\section*{Relation to Mirzakhani's topological recursion}
Finally, in the rest of this appendix we show that the loop equations \eqref{eq:RRF} imply the topological recursion for the matrix model volumes written in \eqref{eq:RRMMV}. This has been done before for the specific spectral curve associated to bosonic JT gravity \cite{Eynard:2007fi} and also $\N=1$ JT supergravity \cite{SW}. Here we generalize this derivation to an arbitrary spectral curve. This helps streamline the proof, given in section \ref{equivalence}, that the $\N=2$ volumes introduced in this paper satisfy the matrix model loop equations, but might be also of use in other applications. 

Let us begin by rewriting the loop equations in the following way. It is convenient, instead of working with the resolvent argument $x$, to introduce a variable $z$ such that $x=E_0-z^2$, where $E_0$ is the end-point of the density of states. Then we can rewrite \eqref{eq:RRF} as an integral over $z$ 
\beq
R_{g,n+1}(E_0-z^2,I) =- \int_{\varepsilon+ \i \mathbb{R}} \frac{\d z'}{2\pi \i} \frac{z'{}^2}{z'{}^2-z^2}  \frac{F_{g,n}(E_0-z'{}^2,I)}{ z y(E_0-z'{}^2)}. \label{eq:RRF2}
\eeq
We remind the reader that the $x'$ contour of integration was around the branch-cut of $y(x)$ in the counterclockwise direction. This corresponds, in the $z$ plane, to a contour along the imaginary axis with a small positive real part $\varepsilon>0$ as indicated in eqn. \eqref{eq:RRF2}. In the case of random matrix ensembles without time reversal the kernel $F_{g,n}(x,I)$ simplifies
\bea\label{Fgnor}
(\upalpha_0,\upbeta)=(1,2):~~~F_{g,n}(x,I) &=& R_{g-1,n+2}(x,x,I) + \sum_{\rm stable} R_{h,|J|+1} (x,J) R_{g-h,|I/J|+1}(x,I/J)\nn
&&+2 \sum_{k=1}^n \Big( R_{0,2}(x,x_k) + \frac{1}{2} \frac{1}{(x-x_k)^2} \Big) R_{g,n}(x,I/x_k).
\ea
The volumes computed through the matrix model observables are defined by the integral transform
\beq\label{VintofR}
V_{g,n,{\rm M}}(B) = \int_{\varepsilon+ \i \mathbb{R}} R_{g,n}(E_0-z_1^2,\ldots,E_0-z_n^2) \prod_{j=1}^n \frac{\d z_j}{2\pi \i} \frac{-2z_j}{b_j}e^{b_j z_j}.
\eeq
This relation is the same as \eqref{Zq}, but defined in terms of resolvents instead. The inverse relation will also be needed
\beq\label{RintofV}
R_{g,n}(E_0-z_1^2,\ldots, E_0-z_n^2) = \int_0^\infty V_{g,n,{\rm M}}(B) \prod_{j=1}^n b_j \d b_j \frac{e^{-b_j z_j}}{-2z_j}.
\eeq
We are now ready to derive \eqref{eq:RRMMV}. The procedure is the following. Begin by writing the volume $V_{g,n+1,M}(b,B)$ as an integral transform of $R_{g,n+1}(x,I)$ using eqn. \eqref{VintofR}. Then use the loop equations to replace $R_{g,n+1}(x,I)$ in the right hand side in terms of other resolvents using \eqref{eq:RRF2} and \eqref{Fgnor}. Next, we can write the resolvents on the right hand side of the loop equations as integrals over the volumes using eqn. \eqref{RintofV}. We focus on the first term in eqn. \eqref{Fgnor} first. We obtain
\bea
bV_{g,n+1,{\rm M}}(b,B) &\supset&b \int_{\varepsilon'+\i \mathbb{R}} \frac{\d z}{2\pi \i} \frac{2z}{b} e^{b z} \int_{\varepsilon+\i \mathbb{R}} \frac{\d z'}{2\pi \i} \frac{z'{}^2}{z'{}^2-z^2} \frac{1}{z y(E_0-z'{}^2)}\nonumber\\
&&\times \int_0^\infty b' \d b' \int_0^\infty b'' \d b'' V_{g-1,n+2,{\rm M}}(b',b'',B) \frac{e^{-b'z'}}{2z'} \frac{e^{-b'' z'}}{2z'}\nonumber\\
&\supset& \frac{1}{2} \int_0^\infty b' \d b' \int_0^\infty b'' \d b'' V_{g-1,n+2,{\rm M}}(b',b'',B)   \int_{\varepsilon+\i \mathbb{R}} \frac{\d z'}{2\pi \i}  \frac{e^{-(b'+b'')z'}}{y(E_0-z'{}^2)}\int_{\varepsilon'+\i \mathbb{R}} \frac{\d z}{2\pi \i}\frac{e^{b z}}{z'{}^2-z^2}\nonumber\\
&\supset& \frac{1}{2} \int_0^\infty b' \d b' \int_0^\infty b'' \d b'' V_{g-1,n+2,{\rm M}}(b',b'',B)  \int_{\varepsilon+\i \mathbb{R}} \frac{\d z'}{2\pi \i}  \frac{e^{-(b'+b'')z'}}{y(E_0-z'{}^2)} \left( - \frac{\sinh (b z')}{z'}\right) \nonumber\\
&\supset& \frac{1}{2} \int_0^\infty b' \d b' \int_0^\infty b'' \d b'' V_{g-1,n+2,{\rm M}}(b',b'',B) D(b'+b'',b)\label{eq:trmmterm1}
\ea
where $\varepsilon' > \varepsilon>0$. In the first step, first two lines, we replaced explicitly the first term of \eqref{Fgnor}. The Laplace and inverse Laplace transforms required in order to go from resolvent to volume cancel each other for all arguments inside the set $B$, related to the set $I$ in the resolvent. We also multiply by a factor of $b$ for later convenience. In the second step, third line, we change the order of integration and leave the $b'$ and $b''$ integrals to be done last. The answer already takes the same form as the first term on the right hand side of \eqref{eq:RRMMV}, with a kernel obtained from evaluating the $z$ and $z'$ integrals. We can change the order of integration and perform the $z$ integral first, picking up the residues at $z=\pm z'$. The answer is shown in parenthesis in the fourth line. In the fifth line we write the term in its final form introducing 
\beq\label{eq:KernelRRMM}
D(x,y) = - \int_{\varepsilon+\i \mathbb{R}} \frac{\d z}{2\pi \i} \frac{e^{-x z} \sinh(y z)}{z y(E_0-z^2)}.
\eeq
This is precisely the result quoted in eqn. \eqref{dfunction}. The exact same manipulations apply to the second term in the right hand side of \eqref{Fgnor}. We are left to consider the final term in the second line of \eqref{Fgnor}, which has a slightly different structure
\bea
bV_{g,n+1,{\rm M}}(b,B) &\supset& \sum_{k=1}^n \int \frac{\d z}{2\pi \i}  e^{b z} \int \frac{\d z'}{2\pi \i} \frac{z'{}}{z'{}^2-z^2} \int_0^\infty b' \d b' \frac{e^{-b' z'}}{y(E_0-z'{}^2)}\nonumber\\
&&\times  V_{g,n,{\rm M}}(b',B/b_k) \underbrace{\int \frac{\d z_k}{2\pi \i}  \frac{2 z_ke^{b_k z_k}}{b_k} \left( \frac{1}{2 z' z_k(z'+z_k)^2}+\frac{1}{ (z'{}^2-z_k^2)^2}\right)}_{=\frac{2 \cosh (b_k z')}{2z'}}\nonumber\\
&\supset& \sum_{k=1}^n \int_0^\infty b' \d b' V_{g,n,{\rm M}}(b',B/b_k)\sum_{\pm} \int \frac{\d z'}{2\pi \i} \frac{e^{-(b' \pm b_k)z'}}{2y(E_0-z'{}^2)} \underbrace{\int \frac{\d z}{2\pi \i}   \frac{e^{b z}}{z'{}^2-z^2}}_{= -\frac{\sinh(b z')}{z'}}  \nonumber\\
&\supset&\frac{1}{2} \sum_{k=1}^n \int_0^\infty b' \d b' V_{g,n,{\rm M}}(b',B/b_k)(D(b'+b_k,b)+D(b'-b_k,b))\label{eq:trmmterm3}
\ea
In the first step, first and second line, we perform the integral transformations to turn the loop equation into a statement for the volumes. The main difference now is that since the term in the kernel $F_{g,n}$ we are considering involves $R_{g,n}(x,I/x_k)$ we need to keep explicitly the integration over $z_k$. The integral over $z_k$ can be done via residues and gives a sum of two terms proportional to $\cosh(b_k z') = \sum_{\pm} e^{\pm b_k z'}$. In the third line we integrate first over $z$ and in the fourth line we recognize the sum of the same kernel $D(x,y)$ as before evaluated at $x=b'+b_k$ and $x=b'-b_k$. 

Putting \eqref{eq:trmmterm1} (and a similar contribution from the second term of \eqref{Fgnor}) and \eqref{eq:trmmterm3} together, we reproduce the full answer \eqref{eq:RRMMV} with the kernel \eqref{eq:KernelRRMM}. The result is valid for any spectral curve. Moreover, since the loop equations for the Dyson ensemble $\upbeta=2$ and the AZ ensemble $(\upalpha_0,\upbeta)=(1,2)$ are the same, it is valid in both cases (the only difference being that for the AZ ensemble $E_0$ is strictly non-negative and if $E_0=0$ then $y\sim 1/z$ near $z=0$ while for the Dyson ensemble $y \sim z$ near $z=0$ always). For example, for $E_0=0$ and $y(x) = \frac{\sin(2\pi \sqrt{-x})}{4\pi}$ we reproduce the bosonic version of Mirzakhani's topological recursion.

\section{Holonomy Constraint for Three-Holed Sphere}\label{sec:constn2}
In this appendix, we fill in details that were omitted in section 
\ref{sec:3hs}, by computing the fermionic contributions to the $\uU(1)$ holonomy constraints for the three-holed sphere and  to the parameter $\kappa$, as a function of other moduli.
We assume a three-holed sphere with two boundary holonomies conjugate to $U_0$ and $V_0$ defined in \eqref{eq:U0V0}.  The third holonomy is therefore conjugate
to $W_0=V_0^{-1}U_0^{-1}$.

We need to compute the three eigenvalues of $W_0$.   
Because $W_0$ is unimodular, its eigenvalues can be expressed in terms of two parameters
$c$ and $\phi_c$, as in eqn. \eqref{eq:U0V0}.
To determine these parameters, it is enough to find two equations that they satisfy. Such
equations are  provided by  conjugation invariants, such as the supertraces ${\rm Str}\,W_0$ and ${\rm Str}\,W_0^{-1}$, which can be compared directly from what
one computes from the diagonal form \eqref{eq:U0V0} of $W_0$.
 To simplify some expressions, we introduce $H_1 = - e^{\i \hq\phi_a}$, $H_2 = -e^{\i \hq\phi_b}$ and $H_3 = - e^{\i \hq\phi_c}$. The supertrace of $W$ is given by 
\beq
{\rm Str}\, W = \frac{2 \cosh \frac{a-b}{2}}{H_1 H_2} +\frac{\kappa}{H_1 H_2}- \frac{1}{H_1 ^2 H_2^2} +  \frac{\psi_2 (\psi_1 -e^{\frac{b}{2}} \psi_3) e^{-\frac{b}{2}} \psi_1 \psi_4 }{H_1^2 H_2} .
\eeq
The supertrace of $W^{-1} = U V$ is 
\beq
{\rm Str}\, W^{-1} = H_1H_2 2 \cosh \frac{a-b}{2} +H_1H_2 \kappa- H_1^2H_2^2 +  H_1H_2^2[ \psi_3 (e^{\frac{a}{2}} \psi_2 +\kappa  \psi_4) +e^{-\frac{a}{2}} \psi_1 \psi_4].
\eeq
In terms of $c$ and $\phi_c$, the supertraces should be 
\bea
{\rm Str}\, W &=&H_3 2 \cosh \Big(\frac{c}{2}\Big) - H_3^2.  \\
{\rm Str} \, W^{-1} &=&H_3 2 \cosh\Big( \frac{c}{2}\Big) -H_3^2.
\ea
We now have enough equations to determine $H_3$ and $\kappa$. Due to the fermionic nature of $\psi$, these parameters can be written in the form 
\begin{equation}
\kappa = \kappa_0 + \sum_{i<j} \kappa_{ij} \psi_i \psi_j + \kappa_4 \psi_1 \psi_2 \psi_3 \psi_4,~~~\phi_a +\phi_b + \phi_c=\sum_{i<j} f_{ij} \psi_i \psi_j + f_4 \psi_1\psi_2\psi_3\psi_4,
\eeq
where $\kappa_0\equiv - 2 \cosh \frac{c}{2} - 2 \cosh \frac{a-b}{2} $. The goal here is to compute the coefficients $\kappa_{0},\kappa_{ij},\kappa_{4}$ and $f_{0},f_{ij},f_{4}$ explicitly. They are functions of $\phi_{a,b}$ and the geodesic lengths only. When the fermionic moduli vanish, the $H_i$ obey the bosonic constraint $H_1 H_2 H_3  = 1$.

 To begin the derivation, we retain $\kappa$ as an independent variable, though eventually we will want to express it in terms of lengths and holonomies.  From the equations for the supertraces of $W$ and $W^{-1}$, we can deduce 
\beq
H_1^2 H_2^2 H_3^2 = \frac{2H_1H_2 \cosh \frac{a-b}{2} + H_1 H_2 \kappa - H_1^2 H_2^2 -1 + H_2 \psi_1 \psi_2 + e^{-\frac{b}{2}} H_2 (-\psi_1 \psi_4 + e^{b} \psi_2 \psi_3)}{2H_1H_2 \cosh \frac{a-b}{2} + H_1 H_2 \kappa - H_1^2 H_2^2 -1 -  H_1 H_2^2  (\kappa \psi_3 \psi_4 + e^{-\frac{a}{2}}\psi_1 \psi_4 - e^{\frac{a}{2}} \psi_2 \psi_3 )}
\eeq
Let us define now the combination $X\equiv 2H_1H_2 \cosh \frac{a-b}{2} + H_1 H_2 \kappa - H_1^2 H_2^2 -1$ to shorten some equations that will appear next, such that 
\beq
H_1^2 H_2^2 H_3^2 = \frac{X + H_2 \psi_1 \psi_2 + e^{-\frac{b}{2}} H_2 (-\psi_1 \psi_4 + e^{b} \psi_2 \psi_3)}{X -  H_1 H_2^2  (\kappa \psi_3 \psi_4 + e^{-\frac{a}{2}}\psi_1 \psi_4 - e^{\frac{a}{2}} \psi_2 \psi_3 )}
\eeq
 We denote by $(H_1^2 H_2^2 H_3^2)_n$ the term with $n$ fermions (when the formula is written in terms of $\kappa$). Then
\beq
(H_1^2 H_2^2 H_3^2)_0=1
\eeq
\beq
(H_1^2 H_2^2 H_3^2)_2= \frac{H_2}{X} \psi_1 \psi_2  +\frac{H_2^2 H_1 e^{-\frac{a}{2}} -H_2 e^{-\frac{b}{2}}}{X} \psi_1 \psi_4 -\frac{H_2^2 H_1 e^{\frac{a}{2}} -H_2 e^{\frac{b}{2}}}{X} \psi_2 \psi_3  +\frac{H_2^2 H_1 \kappa }{X} \psi_3 \psi_4
\eeq
\beq
(H_1^2 H_2^2 H_3^2)_4=\frac{H_2^3 H_1 (\kappa + 2 \cosh\frac{a-b}{2}- 2 H_1H_2) }{X^2}\psi_1 \psi_2 \psi_3 \psi_4 
\eeq
Consider the fact that we know the bosonic solution for $\kappa= - 2 \cosh \frac{c}{2} - 2 \cosh \frac{a-b}{2} + \mathcal{O}(\psi^2) + \mathcal{O}(\psi^4) $. We can see what happens when $\kappa$ is expanded in fermionic moduli. First, the term with no fermions is always $H_1^2H_2^2H_3^2=1$. To get the terms with two fermions we can simply replace $\kappa= - 2 \cosh \frac{c}{2} - 2 \cosh \frac{a-b}{2} $ in $(H_1^2 H_2^2 H_3^2)_2$. To get the term with four fermions we do the same replacement in $(H_1^2 H_2^2 H_3^2)_4$, but we also need to include $\psi^2$ corrections to $(H_1^2 H_2^2 H_3^2)_2$ which contribute to quartic order in fermions. This last step is the main thing left to do. 

To find $\kappa$ to order $\psi^2$ can be done in the following way. Consider one of the supertrace equations $H_3 2 \cosh \frac{c}{2} - H_3^2 = \frac{2 \cosh \frac{a-b}{2}}{H_1 H_2}-\frac{1}{H_1^2 H_2^2} + \frac{\kappa}{H_1 H_2} + \mathcal{O}(\psi^2)$, where the precise form of the $\psi^2$ terms can be found above. We just argued that we know $H_3$ to order $\psi^2$. Then we can replace it in the left hand side of this formula to find  $\kappa$ to order $\psi^2$. Once we know $\kappa$ to order $\psi^2$, we can compute $H_3$ to order $\psi^4$, as explained in the previous paragraph. Knowing $H_3$ to order $\psi^4$ we can replace it again in the same supertrace equation, and matching terms with four fermions obtain $\kappa_4$. Following this procedure we find
\bea
&& \kappa_{12} =-\frac{H_2(H_1H_2+\cosh \frac{c}{2})}{H_1H_2 2\cosh \frac{c}{2} + H_1^2H_2^2+1 },\\
&&\kappa_{14}= \frac{H_2(e^{-\frac{b}{2}}(H_1H_2+\cosh \frac{c}{2})+e^{-\frac{a}{2}}(1+H_1H_2\cosh \frac{c}{2}))}{H_1H_2 2\cosh \frac{c}{2} + H_1^2H_2^2+1 }\\
&& \kappa_{23} =-\frac{H_2(e^{\frac{b}{2}}(H_1H_2+\cosh \frac{c}{2})+e^{\frac{a}{2}}(1+H_1H_2\cosh \frac{c}{2}))}{H_1H_2 2\cosh \frac{c}{2} + H_1^2H_2^2+1 } , \\
&& \kappa_{34} =- \frac{2H_2 (1+ H_1 H_2 \cosh \frac{c}{2})(\cosh \frac{c}{2} + \cosh \frac{a-b}{2})}{H_1H_2 2\cosh \frac{c}{2} + H_1^2H_2^2+1} .
\ea
The final $\kappa_4$ is for now undetermined but we will be able to compute it below, after using this result to fully determine $H_3$. Define for now 
\beq
H_1^2H_2^2H_3^2=1 + \sum_{i<j} \hat{f}_{ij} \psi_i \psi_j + \hat{f}_4 \psi_1\psi_2\psi_3\psi_4,
\eeq
Then as we explained, to get $ \hat{f}_{ij}$ we can simply replace $\kappa \to \kappa_0$ while to get $ \hat{f}_4$ we need to include the dependence with the $\kappa_{ij}$ terms. The final answer is 
\bea
&& \hat{f}_{12} = -\frac{H_2}{H_1H_2 2\cosh \frac{c}{2} + H_1^2H_2^2+1},~~~ \hat{f}_{14}=\frac{H_2(e^{-\frac{b}{2}}-e^{-\frac{a}{2}} H_1 H_2)}{H_1H_2 2\cosh \frac{c}{2} + H_1^2H_2^2+1},\\
&&  \hat{f}_{23} =\frac{H_2(-e^{\frac{b}{2}}+e^{\frac{a}{2}} H_1 H_2)}{H_1H_2 2\cosh \frac{c}{2} + H_1^2H_2^2+1} ,~~~ \hat{f}_{34}=\frac{2H_1H_2^2(\cosh \frac{c}{2} + \cosh \frac{a-b}{2})}{H_1H_2 2\cosh \frac{c}{2} + H_1^2H_2^2+1},\\
&&  \hat{f}_4=- \frac{H_1 H_2^3(\cosh \frac{c}{2} + H_1 H_2)}{(H_1H_2 2\cosh \frac{c}{2} + H_1^2H_2^2+1)^2}.\label{eq:Fexpapp}
\ea
For the reasons explained above, knowing $ \hat{f}_4$ allows us to compute $\kappa_4$. The final answer is 
\beq
\kappa_4 = \frac{H_2^2(\cosh \frac{c}{2} + 2 H_1 H_2 + H_1^2 H_2^2 \cosh \frac{c}{2})}{2(H_1H_2 2\cosh \frac{c}{2} + H_1^2H_2^2+1)^2}
\eeq
This completes the derivation of the fermionic contribution to $\kappa$. After writing it in the original variables it becomes the eqn. \eqref{eq:Kexp} quoted in the main text. In terms of the holonomy constraint, the final step is to rewrite it in terms of $\phi$ instead of $H$. A simple Taylor expansion gives 
\bea
\hq(\phi_a+\phi_b+\phi_c) &=& \frac{1}{2}\log\left( 1 + \sum_{i<j} \hat{f}_{ij} \psi_i \psi_j +  \hat{f}_4 \psi_1\psi_2\psi_3\psi_4 \right) \\
&=&  \sum_{i<j}\hq f_{ij} \psi_i \psi_j + \hq f_4 \psi_1\psi_2\psi_3\psi_4.
\ea 
where $f_{ij}={\scriptstyle \frac{1}{2\i \hq}}\hat{f}_{ij}$ and $f_4 = {\scriptstyle \frac{1}{2\i \hq}}(\hat{f}_4 - \hat{f}_{12}\hat{f}_{34} - \hat{f}_{14} \hat{f}_{23})$. After using \eqref{eq:Fexpapp} and simplifying we obtain \eqref{eq:Fexp}.

\section{JT gravity with $\uU(1)$ gauge field}\label{app:JTU1bos}

In analyzing the torsion  of a three-holed sphere in section \ref{torthree}, we encountered a subtlety in the treatment of the $\uU(1))$ $R$-symmetry.
We encountered in eqn. (\ref{answer}) an uncanceled factor $1/[\d r_z]$, which we interpreted as an instruction to divide by the volume of the $R$-symmetry group,
with the volume form $[\d r_z]$.

Here we will verify that  this prescription gives the correct answer in the much simpler case of  bosonic JT gravity coupled to a $\uU(1)$ gauge field, related to $BF$ theory of the
 group $G={\rm PSL}(2,\mathbb{R}) \times \uU(1)$.  The partition function in this case can  
 be obtained by other means. Consider a flat $G$-bundle on a three-holed sphere, with two holonomies being $U_0 =\pm \Big(\hspace{-1mm}\begin{array}{cc}
 e^{a/2} & \kappa \\
0& e^{-a/2}
\end{array}\hspace{-1mm}\Big) \otimes e^{\i \phi_a} $ and $V_0 = \pm \Big(\hspace{-1mm}\begin{array}{cc}
 e^{-b/2} &0 \\
 1& e^{b/2}
\end{array}\hspace{-1mm}\Big)\otimes e^{\i \phi_b}$. The torsion of the three-holed sphere, according to our prescription, is given by 
\bea
\tau = \frac{1}{2\pi} \, 8 \sinh \frac{a}{2} \sinh \frac{b}{2}\sinh \frac{c}{2}\, \delta(\phi_a+\phi_b+\phi_c) \, \d a\, \d b \, \d c \, \d \phi_a \, \d \phi_b \, \d \phi_c,
\ea
where, following our prescription, we have replaced $1/[\d r_z]$ with $1/2\pi$, where $2\pi$ is the volume of $\uU(1)$ assuming that the measure is $\d r_z$.The third holonomy $\phi_c$ can be  integrated in using the constraint  $1=\int \d \phi_c\, \delta(\phi_a + \phi_b+\phi_c)$. The torsion of a circle is 
\bea
\tau_S = 4 \sinh^2 \frac{a}{2} \, \d a \, (\d \varrho)^{-1} \, \d \phi \, (\d \varphi)^{-1}.
\ea
When we glue along a circle, the measure is $\d a \, \d \phi \, \d \varrho \, \d \varphi$. The $\uU(1)$ twist parameter $\varphi$ is a modulus of a flat $\UU(1)$ connection,
though as $\uU(1)$ is abelian, nothing in the computation is being twisted.
 Integrating over the twist parameter gives a measure  $ a \d a \, 2\pi \d \phi$. Normalizing by the torsion of the circle and fixing the boundary moduli, the measure of the three-holed sphere is
\bea
\mu_{0,3} = \frac{1}{2\pi} \delta(\phi_a+\phi_b+\phi_c).
\ea
We derive a recursion relation for the volumes of hyperbolic surfaces with a flat $\uU(1)$ connection, adapting the procedure we used for $\N=2$ JT supergravity. We obtain
\bea
b V_{g,n+1} ({\sf b}, {\sf B}) &=& \frac{1}{2}(2\pi) \int_0^\infty  b'\text{d}b' \, \int_0^{2\pi} \d \phi' \, \int_0^\infty b''\text{d}b'' \int_0^{2\pi} \d \phi''\, \delta(\phi+\phi'+\phi'')\mathsf{D}_2 (b,b',b'')\nn
&&\hspace{-5mm} \times \left(V_{g-1,n+2}(\bar{{\sf b}}',\bar{{\sf b}}'',{\sf B}) + \sum_{\rm stable} V_{h_1,|B_1|+1} (\bar{{\sf b}}',{\sf B}_1) V_{h_2,|B_2|+1} (\bar{{\sf b}}'',{\sf B}_2) \right)\nn
&&\hspace{-5mm}+\sum_{k=1}^{n} \int_0^{\infty} b'\text{d}b' \int_0^{2\pi} \d \phi' \delta(\phi+\phi'+\phi_k)\Big(b -\mathsf{T}_2(b,b',b_k) \Big) V_{g,n} (\bar{{\sf b}}',{\sf B} /{\sf b}_k) .
\ea
Integration over the $\uU(1)$ twist parameters produces a factor of $(2\pi)^2 = \int \d \varphi' \d \varphi''$ in the first line and a factor of $(2\pi)= \int \d \varphi'$ in the third line. One of those factors of each line cancels with a similar one in the normalization of $\mu_{0,3}$ which comes from our treatment of $\d r_z$. The initial conditions for this recursion are straightforward to obtain using the volumes for a three-holed sphere and for a torus with one boundary. The solution to this recursion is
\beq
V_{g,n}({\sf B}) = (2\pi)^{2g-1} \delta( \phi_1 + \ldots + \phi_{n}) V_g^{\rm bos.}(B),
\eeq
where $2\pi = {\rm Vol}(U(1))$ and $V_{g,n}^{\rm bos.}(B)$ is the bosonic volumes obtained by the recursion relation of Mirzakhani. We can compute the ``fixed charge'' volumes defined through the Fourier transform $V_{g,n}^{(q)}(B) = \int \frac{\d^n \phi}{(2\pi)^n} e^{\i q \sum_i \phi_i} V_{g,n}({\sf B})$. The explicit expression is given by 
\bea
V_{g,n}^{(q)}(B) = (2\pi)^{2g-2} \, V_{g,n}^{\rm bos.}(B).
\ea
The contribution $(2\pi)^{2g-2}$ is consistent with eqn. (21) of \cite{Kapec:2019ecr} derived independently using $BF$ theory techniques.

  \bibliographystyle{JHEP}
\bibliography{bib}

\providecommand{\href}[2]{#2}\begingroup\raggedright\begin{thebibliography}{10}

\bibitem{Almheiri:2020cfm}
A.~Almheiri, T.~Hartman, J.~Maldacena, E.~Shaghoulian and A.~Tajdini,
  \emph{{The entropy of Hawking radiation}},
  \href{https://doi.org/10.1103/RevModPhys.93.035002}{\emph{Rev. Mod. Phys.}
  {\bfseries 93} (2021) 035002}
  [\href{https://arxiv.org/abs/2006.06872}{{\ttfamily 2006.06872}}].

\bibitem{Gibbons:1976ue}
G.~W. Gibbons and S.~W. Hawking, \emph{{Action Integrals and Partition
  Functions in Quantum Gravity}},
  \href{https://doi.org/10.1103/PhysRevD.15.2752}{\emph{Phys. Rev. D}
  {\bfseries 15} (1977) 2752}.

\bibitem{Maldacena:2001kr}
J.~M. Maldacena, \emph{{Eternal black holes in anti-de Sitter}},
  \href{https://doi.org/10.1088/1126-6708/2003/04/021}{\emph{JHEP} {\bfseries
  04} (2003) 021} [\href{https://arxiv.org/abs/hep-th/0106112}{{\ttfamily
  hep-th/0106112}}].

\bibitem{Cotler:2016fpe}
J.~S. Cotler, G.~Gur-Ari, M.~Hanada, J.~Polchinski, P.~Saad, S.~H. Shenker
  et~al., \emph{{Black Holes and Random Matrices}},
  \href{https://doi.org/10.1007/JHEP05(2017)118}{\emph{JHEP} {\bfseries 05}
  (2017) 118} [\href{https://arxiv.org/abs/1611.04650}{{\ttfamily
  1611.04650}}].

\bibitem{Saad:2018bqo}
P.~Saad, S.~H. Shenker and D.~Stanford, \emph{{A semiclassical ramp in SYK and
  in gravity}},  \href{https://arxiv.org/abs/1806.06840}{{\ttfamily
  1806.06840}}.

\bibitem{SSS}
P.~Saad, S.~H. Shenker and D.~Stanford, \emph{{JT gravity as a matrix
  integral}},  \href{https://arxiv.org/abs/1903.11115}{{\ttfamily 1903.11115}}.

\bibitem{Jackiw:1984je}
R.~Jackiw, \emph{{Lower Dimensional Gravity}},
  \href{https://doi.org/10.1016/0550-3213(85)90448-1}{\emph{Nucl. Phys.}
  {\bfseries B252} (1985) 343}.

\bibitem{teitelboim1983gravitation}
C.~Teitelboim, \emph{Gravitation and hamiltonian structure in two spacetime
  dimensions}, {\emph{Physics Letters B} {\bfseries 126} (1983) 41}.

\bibitem{Almheiri:2014cka}
A.~Almheiri and J.~Polchinski, \emph{{Models of AdS$_{2}$ backreaction and
  holography}}, \href{https://doi.org/10.1007/JHEP11(2015)014}{\emph{JHEP}
  {\bfseries 11} (2015) 014} [\href{https://arxiv.org/abs/1402.6334}{{\ttfamily
  1402.6334}}].

\bibitem{Jensen:2016pah}
K.~Jensen, \emph{{Chaos in AdS$_2$ Holography}},
  \href{https://doi.org/10.1103/PhysRevLett.117.111601}{\emph{Phys. Rev. Lett.}
  {\bfseries 117} (2016) 111601}
  [\href{https://arxiv.org/abs/1605.06098}{{\ttfamily 1605.06098}}].

\bibitem{Maldacena:2016upp}
J.~Maldacena, D.~Stanford and Z.~Yang, \emph{{Conformal symmetry and its
  breaking in two dimensional Nearly Anti-de-Sitter space}},
  \href{https://doi.org/10.1093/ptep/ptw124}{\emph{PTEP} {\bfseries 2016}
  (2016) 12C104} [\href{https://arxiv.org/abs/1606.01857}{{\ttfamily
  1606.01857}}].

\bibitem{Engelsoy:2016xyb}
J.~Engelsöy, T.~G. Mertens and H.~Verlinde, \emph{{An investigation of
  AdS$_{2}$ backreaction and holography}},
  \href{https://doi.org/10.1007/JHEP07(2016)139}{\emph{JHEP} {\bfseries 07}
  (2016) 139} [\href{https://arxiv.org/abs/1606.03438}{{\ttfamily
  1606.03438}}].

\bibitem{Mertens:2022irh}
T.~G. Mertens and G.~J. Turiaci, \emph{{Solvable Models of Quantum Black Holes:
  A Review on Jackiw-Teitelboim Gravity}},
  \href{https://arxiv.org/abs/2210.10846}{{\ttfamily 2210.10846}}.

\bibitem{SW}
D.~Stanford and E.~Witten, \emph{{JT gravity and the ensembles of random matrix
  theory}}, \href{https://doi.org/10.4310/ATMP.2020.v24.n6.a4}{\emph{Adv.
  Theor. Math. Phys.} {\bfseries 24} (2020) 1475}
  [\href{https://arxiv.org/abs/1907.03363}{{\ttfamily 1907.03363}}].

\bibitem{AZ}
A.~Altland and M.~R. Zirnbauer, \emph{{Nonstandard symmetry classes in
  mesoscopic normal-superconducting hybrid structures}},
  \href{https://doi.org/10.1103/PhysRevB.55.1142}{\emph{Phys. Rev. B}
  {\bfseries 55} (1997) 1142}
  [\href{https://arxiv.org/abs/cond-mat/9602137}{{\ttfamily
  cond-mat/9602137}}].

\bibitem{Johnson:2019eik}
C.~V. Johnson, \emph{{Nonperturbative Jackiw-Teitelboim gravity}},
  \href{https://doi.org/10.1103/PhysRevD.101.106023}{\emph{Phys. Rev. D}
  {\bfseries 101} (2020) 106023}
  [\href{https://arxiv.org/abs/1912.03637}{{\ttfamily 1912.03637}}].

\bibitem{Mertens:2020hbs}
T.~G. Mertens and G.~J. Turiaci, \emph{{Liouville quantum gravity --
  holography, JT and matrices}},
  \href{https://doi.org/10.1007/JHEP01(2021)073}{\emph{JHEP} {\bfseries 01}
  (2021) 073} [\href{https://arxiv.org/abs/2006.07072}{{\ttfamily
  2006.07072}}].

\bibitem{Maldacena:2023acv}
J.~Maldacena, \emph{{A simple quantum system that describes a black hole}},
  \href{https://arxiv.org/abs/2303.11534}{{\ttfamily 2303.11534}}.

\bibitem{Mohaupt:2000mj}
T.~Mohaupt, \emph{{Black hole entropy, special geometry and strings}},
  \href{https://doi.org/10.1002/1521-3978(200102)49:1/3<3::AID-PROP3>3.0.CO;2-#}{\emph{Fortsch.
  Phys.} {\bfseries 49} (2001) 3}
  [\href{https://arxiv.org/abs/hep-th/0007195}{{\ttfamily hep-th/0007195}}].

\bibitem{Cabo-Bizet:2018ehj}
A.~Cabo-Bizet, D.~Cassani, D.~Martelli and S.~Murthy, \emph{{Microscopic origin
  of the Bekenstein-Hawking entropy of supersymmetric AdS$_{5}$ black holes}},
  \href{https://doi.org/10.1007/JHEP10(2019)062}{\emph{JHEP} {\bfseries 10}
  (2019) 062} [\href{https://arxiv.org/abs/1810.11442}{{\ttfamily
  1810.11442}}].

\bibitem{Choi:2018hmj}
S.~Choi, J.~Kim, S.~Kim and J.~Nahmgoong, \emph{{Large AdS black holes from
  QFT}},  \href{https://arxiv.org/abs/1810.12067}{{\ttfamily 1810.12067}}.

\bibitem{Benini:2018ywd}
F.~Benini and E.~Milan, \emph{{Black Holes in 4D $\mathcal{N}$=4
  Super-Yang-Mills Field Theory}},
  \href{https://doi.org/10.1103/PhysRevX.10.021037}{\emph{Phys. Rev. X}
  {\bfseries 10} (2020) 021037}
  [\href{https://arxiv.org/abs/1812.09613}{{\ttfamily 1812.09613}}].

\bibitem{Boruch:2022tno}
J.~Boruch, M.~T. Heydeman, L.~V. Iliesiu and G.~J. Turiaci, \emph{{BPS and
  near-BPS black holes in $AdS_5$ and their spectrum in $\mathcal{N}=4$ SYM}},
  \href{https://arxiv.org/abs/2203.01331}{{\ttfamily 2203.01331}}.

\bibitem{Heydeman:2020hhw}
M.~Heydeman, L.~V. Iliesiu, G.~J. Turiaci and W.~Zhao, \emph{{The statistical
  mechanics of near-BPS black holes}},
  \href{https://doi.org/10.1088/1751-8121/ac3be9}{\emph{J. Phys. A} {\bfseries
  55} (2022) 014004} [\href{https://arxiv.org/abs/2011.01953}{{\ttfamily
  2011.01953}}].

\bibitem{Lin:2022zxd}
H.~W. Lin, J.~Maldacena, L.~Rozenberg and J.~Shan, \emph{{Looking at
  supersymmetric black holes for a very long time}},
  \href{https://doi.org/10.21468/SciPostPhys.14.5.128}{\emph{SciPost Phys.}
  {\bfseries 14} (2023) 128}
  [\href{https://arxiv.org/abs/2207.00408}{{\ttfamily 2207.00408}}].

\bibitem{Lin:2022rzw}
H.~W. Lin, J.~Maldacena, L.~Rozenberg and J.~Shan, \emph{{Holography for people
  with no time}},  \href{https://arxiv.org/abs/2207.00407}{{\ttfamily
  2207.00407}}.

\bibitem{CraneRabin}
L.~Crane and J.~M. Rabin, \emph{$\text{Super Riemann surfaces: Uniformization
  and Teichm{\"u}ller theory}$},
  \href{https://doi.org/10.1007/BF01223239}{\emph{Communications in
  Mathematical Physics} {\bfseries 113} (1988) 601}.

\bibitem{Cohn:1986wn}
J.~D. Cohn, \emph{{N=2 Super-Riemann Surfaces}},
  \href{https://doi.org/10.1016/0550-3213(87)90039-3}{\emph{Nucl. Phys. B}
  {\bfseries 284} (1987) 349}.

\bibitem{Ip:2016ojn}
I.~C.~H. Ip, R.~C. Penner and A.~M. Zeitlin, \emph{{$\mathcal{N}=2$
  $\text{Super-Teichm\"uller theory}$}},
  \href{https://doi.org/10.1016/j.aim.2018.08.001}{\emph{Adv. Math.} {\bfseries
  336} (2018) 409} [\href{https://arxiv.org/abs/1605.08094}{{\ttfamily
  1605.08094}}].

\bibitem{Kanazawa:2017dpd}
T.~Kanazawa and T.~Wettig, \emph{{Complete random matrix classification of SYK
  models with $\mathcal{N}=0$, $1$ and $2$ supersymmetry}},
  \href{https://doi.org/10.1007/JHEP09(2017)050}{\emph{JHEP} {\bfseries 09}
  (2017) 050} [\href{https://arxiv.org/abs/1706.03044}{{\ttfamily
  1706.03044}}].

\bibitem{VZ}
J.~J.~M. Verbaarschot and I.~Zahed, \emph{{Spectral density of the QCD Dirac
  operator near zero virtuality}},
  \href{https://doi.org/10.1103/PhysRevLett.70.3852}{\emph{Phys. Rev. Lett.}
  {\bfseries 70} (1993) 3852}
  [\href{https://arxiv.org/abs/hep-th/9303012}{{\ttfamily hep-th/9303012}}].

\bibitem{Anderson:1990nw}
A.~Anderson, R.~C. Myers and V.~Periwal, \emph{{Complex random surfaces}},
  \href{https://doi.org/10.1016/0370-2693(91)90401-B}{\emph{Phys. Lett. B}
  {\bfseries 254} (1991) 89}.

\bibitem{Anderson:1991ku}
A.~Anderson, R.~C. Myers and V.~Periwal, \emph{{Branched polymers from a double
  scaling limit of matrix models}},
  \href{https://doi.org/10.1016/0550-3213(91)90411-P}{\emph{Nucl. Phys. B}
  {\bfseries 360} (1991) 463}.

\bibitem{Dalley:1991qg}
S.~Dalley, C.~V. Johnson and T.~R. Morris, \emph{{Multicritical complex matrix
  models and nonperturbative 2-D quantum gravity}},
  \href{https://doi.org/10.1016/0550-3213(92)90217-Y}{\emph{Nucl. Phys. B}
  {\bfseries 368} (1992) 625}.

\bibitem{Dalley:1991vr}
S.~Dalley, C.~V. Johnson and T.~R. Morris, \emph{{Nonperturbative
  two-dimensional quantum gravity}},
  \href{https://doi.org/10.1016/0550-3213(92)90218-Z}{\emph{Nucl. Phys. B}
  {\bfseries 368} (1992) 655}.

\bibitem{Stanford:2017thb}
D.~Stanford and E.~Witten, \emph{{Fermionic Localization of the Schwarzian
  Theory}}, \href{https://doi.org/10.1007/JHEP10(2017)008}{\emph{JHEP}
  {\bfseries 10} (2017) 008}
  [\href{https://arxiv.org/abs/1703.04612}{{\ttfamily 1703.04612}}].

\bibitem{FGMS}
W.~Fu, D.~Gaiotto, J.~Maldacena and S.~Sachdev, \emph{{Supersymmetric
  Sachdev-Ye-Kitaev models}},
  \href{https://doi.org/10.1103/PhysRevD.95.026009}{\emph{Phys. Rev. D}
  {\bfseries 95} (2017) 026009}
  [\href{https://arxiv.org/abs/1610.08917}{{\ttfamily 1610.08917}}].

\bibitem{Heydeman:2022lse}
M.~Heydeman, G.~J. Turiaci and W.~Zhao, \emph{{Phases of $ \mathcal{N} $ = 2
  Sachdev-Ye-Kitaev models}},
  \href{https://doi.org/10.1007/JHEP01(2023)098}{\emph{JHEP} {\bfseries 01}
  (2023) 098} [\href{https://arxiv.org/abs/2206.14900}{{\ttfamily
  2206.14900}}].

\bibitem{OganesyanHuse}
V.~Oganesyan and D.~A. Huse, \emph{Localization of interacting fermions at high
  temperature}, \href{https://doi.org/10.1103/PhysRevB.75.155111}{\emph{Phys.
  Rev. B} {\bfseries 75} (2007) 155111}.

\bibitem{ABGR}
Y.~Y. Atas, E.~Bogomolny, O.~Giraud and G.~Roux, \emph{Distribution of the
  ratio of consecutive level spacings in random matrix ensembles},
  \href{https://doi.org/10.1103/PhysRevLett.110.084101}{\emph{Phys. Rev. Lett.}
  {\bfseries 110} (2013) 084101}.

\bibitem{Forste:2017kwy}
S.~Forste and I.~Golla, \emph{{Nearly AdS$_2$ sugra and the super-Schwarzian}},
  \href{https://doi.org/10.1016/j.physletb.2017.05.039}{\emph{Phys. Lett. B}
  {\bfseries 771} (2017) 157}
  [\href{https://arxiv.org/abs/1703.10969}{{\ttfamily 1703.10969}}].

\bibitem{Forste:2017apw}
S.~F\"orste, J.~Kames-King and M.~Wiesner, \emph{{Towards the Holographic Dual
  of N = 2 SYK}}, \href{https://doi.org/10.1007/JHEP03(2018)028}{\emph{JHEP}
  {\bfseries 03} (2018) 028}
  [\href{https://arxiv.org/abs/1712.07398}{{\ttfamily 1712.07398}}].

\bibitem{Mertens:2017mtv}
T.~G. Mertens, G.~J. Turiaci and H.~L. Verlinde, \emph{{Solving the Schwarzian
  via the Conformal Bootstrap}},
  \href{https://doi.org/10.1007/JHEP08(2017)136}{\emph{JHEP} {\bfseries 08}
  (2017) 136} [\href{https://arxiv.org/abs/1705.08408}{{\ttfamily
  1705.08408}}].

\bibitem{Cardenas:2018krd}
M.~C\'ardenas, O.~Fuentealba, H.~A. Gonz\'alez, D.~Grumiller, C.~Valc\'arcel
  and D.~Vassilevich, \emph{{Boundary theories for dilaton supergravity in
  2D}}, \href{https://doi.org/10.1007/JHEP11(2018)077}{\emph{JHEP} {\bfseries
  11} (2018) 077} [\href{https://arxiv.org/abs/1809.07208}{{\ttfamily
  1809.07208}}].

\bibitem{Forste:2020xwx}
S.~F\"orste, A.~Gerhardus and J.~Kames-King, \emph{{Supersymmetric black holes
  and the SJT/nSCFT$_{1}$ correspondence}},
  \href{https://doi.org/10.1007/JHEP01(2021)186}{\emph{JHEP} {\bfseries 01}
  (2021) 186} [\href{https://arxiv.org/abs/2007.12393}{{\ttfamily
  2007.12393}}].

\bibitem{Kapec:2019ecr}
D.~Kapec, R.~Mahajan and D.~Stanford, \emph{{Matrix ensembles with global
  symmetries and \textquoteright{}t Hooft anomalies from 2d gauge theory}},
  \href{https://doi.org/10.1007/JHEP04(2020)186}{\emph{JHEP} {\bfseries 04}
  (2020) 186} [\href{https://arxiv.org/abs/1912.12285}{{\ttfamily
  1912.12285}}].

\bibitem{RaySinger}
D.~Ray and I.~Singer, \emph{{R Torsion and the Laplacian on Riemannian
  manifolds}}, \href{https://doi.org/10.1016/0001-8708(71)90045-4}{\emph{Adv.
  Math.} {\bfseries 7} (1971) 145}.

\bibitem{Schwarz:1978cn}
A.~S. Schwarz, \emph{{The Partition Function of Degenerate Quadratic Functional
  and Ray-Singer Invariants}},
  \href{https://doi.org/10.1007/BF00406412}{\emph{Lett. Math. Phys.} {\bfseries
  2} (1978) 247}.

\bibitem{Reidemeister}
K.~Reidemeister, \emph{{Homotopieringe und Linsenraume}},
  \href{https://doi.org/10.1007/BF02940717}{\emph{Abhandlungen aus dem
  Mathematischen Seminar der Universitat Hamburg} {\bfseries 11} (1935) }.

\bibitem{Cheeger}
J.~Cheeger, \emph{Analytic torsion and the heat equation}, {\emph{Annals of
  Mathematics} {\bfseries 109} (1979) 259}.

\bibitem{Muller1}
W.~Muller, \emph{{Analytic torsion and R-torsion of riemannian manifolds}},
  {\emph{Advances in Mathematics} {\bfseries 28} 233}.

\bibitem{Muller2}
W.~Muller, \emph{{Analytic torsion and R-torsion for unimodular
  representations}}, {\emph{Journal of the American Mathematical Society}
  {\bfseries 6} 721}.

\bibitem{BismutZhang}
J.-M. Bismut and W.~Zhang, \emph{{An extension of a theorem by Cheeger and
  Muller}}, {\emph{Asterisque} {\bfseries 205} 235}.

\bibitem{Witten:2012bg}
E.~Witten, \emph{{Notes On Supermanifolds and Integration}},
  \href{https://doi.org/10.4310/PAMQ.2019.v15.n1.a1}{\emph{Pure Appl. Math.
  Quart.} {\bfseries 15} (2019) 3}
  [\href{https://arxiv.org/abs/1209.2199}{{\ttfamily 1209.2199}}].

\bibitem{Witten:1991we}
E.~Witten, \emph{{On quantum gauge theories in two-dimensions}},
  \href{https://doi.org/10.1007/BF02100009}{\emph{Commun. Math. Phys.}
  {\bfseries 141} (1991) 153}.

\bibitem{mirzakhani2007simple}
M.~Mirzakhani, \emph{Simple geodesics and weil-petersson volumes of moduli
  spaces of bordered riemann surfaces}, {\emph{Inventiones mathematicae}
  {\bfseries 167} (2007) 179}.

\bibitem{Mirzakhani:2006eta}
M.~Mirzakhani, \emph{{Weil-Petersson volumes and intersection theory on the
  moduli space of curves}},
  \href{https://doi.org/10.1090/S0894-0347-06-00526-1}{\emph{J. Am. Math. Soc.}
  {\bfseries 20} (2007) 1}.

\bibitem{norbury2023enumerative}
P.~Norbury, \emph{Enumerative geometry via the moduli space of super riemann
  surfaces},  \href{https://arxiv.org/abs/2005.04378}{{\ttfamily 2005.04378}}.

\bibitem{Witten:2012ga}
E.~Witten, \emph{{Notes On Super Riemann Surfaces And Their Moduli}},
  \href{https://doi.org/10.4310/PAMQ.2019.v15.n1.a2}{\emph{Pure Appl. Math.
  Quart.} {\bfseries 15} (2019) 57}
  [\href{https://arxiv.org/abs/1209.2459}{{\ttfamily 1209.2459}}].

\bibitem{Eynard:2007fi}
B.~Eynard and N.~Orantin, \emph{{Weil-Petersson volume of moduli spaces,
  Mirzakhani's recursion and matrix models}},
  \href{https://arxiv.org/abs/0705.3600}{{\ttfamily 0705.3600}}.

\bibitem{Maxfield:2020ale}
H.~Maxfield and G.~J. Turiaci, \emph{{The path integral of 3D gravity near
  extremality; or, JT gravity with defects as a matrix integral}},
  \href{https://doi.org/10.1007/JHEP01(2021)118}{\emph{JHEP} {\bfseries 01}
  (2021) 118} [\href{https://arxiv.org/abs/2006.11317}{{\ttfamily
  2006.11317}}].

\bibitem{Witten:2020wvy}
E.~Witten, \emph{{Matrix Models and Deformations of JT Gravity}},
  \href{https://doi.org/10.1098/rspa.2020.0582}{\emph{Proc. Roy. Soc. Lond. A}
  {\bfseries 476} (2020) 20200582}
  [\href{https://arxiv.org/abs/2006.13414}{{\ttfamily 2006.13414}}].

\bibitem{Rosso:2021orf}
F.~Rosso and G.~J. Turiaci, \emph{{Phase transitions for deformations of JT
  supergravity and matrix models}},
  \href{https://doi.org/10.1007/JHEP02(2022)187}{\emph{JHEP} {\bfseries 02}
  (2022) 187} [\href{https://arxiv.org/abs/2111.09330}{{\ttfamily
  2111.09330}}].

\bibitem{Turiaci:2020fjj}
G.~J. Turiaci, M.~Usatyuk and W.~W. Weng, \emph{{2D dilaton-gravity,
  deformations of the minimal string, and matrix models}},
  \href{https://doi.org/10.1088/1361-6382/ac25df}{\emph{Class. Quant. Grav.}
  {\bfseries 38} (2021) 204001}
  [\href{https://arxiv.org/abs/2011.06038}{{\ttfamily 2011.06038}}].

\bibitem{Eberhardt:2023rzz}
L.~Eberhardt and G.~J. Turiaci, \emph{{2D dilaton gravity and the
  Weil-Petersson volumes with conical defects}},
  \href{https://arxiv.org/abs/2304.14948}{{\ttfamily 2304.14948}}.

\bibitem{DF}
X.-z. Dai and D.~S. Freed, \emph{{eta invariants and determinant lines}},
  \href{https://doi.org/10.1063/1.530747}{\emph{J. Math. Phys.} {\bfseries 35}
  (1994) 5155} [\href{https://arxiv.org/abs/hep-th/9405012}{{\ttfamily
  hep-th/9405012}}].

\bibitem{WY}
E.~Witten and K.~Yonekura, \emph{{Anomaly Inflow and the $\eta$-Invariant}},
  in \emph{{The Shoucheng Zhang Memorial Workshop}}, 9, 2019,
  \href{https://arxiv.org/abs/1909.08775}{{\ttfamily 1909.08775}}.

\bibitem{Witten:1987ty}
E.~Witten, \emph{{Coadjoint Orbits of the Virasoro Group}},
  \href{https://doi.org/10.1007/BF01218287}{\emph{Commun. Math. Phys.}
  {\bfseries 114} (1988) 1}.

\bibitem{Alekseev:1988ce}
A.~Alekseev and S.~L. Shatashvili, \emph{{Path Integral Quantization of the
  Coadjoint Orbits of the Virasoro Group and 2D Gravity}},
  \href{https://doi.org/10.1016/0550-3213(89)90130-2}{\emph{Nucl. Phys. B}
  {\bfseries 323} (1989) 719}.

\bibitem{Eguchi:2003ik}
T.~Eguchi and Y.~Sugawara, \emph{{Modular bootstrap for boundary N = 2
  Liouville theory}},
  \href{https://doi.org/10.1088/1126-6708/2004/01/025}{\emph{JHEP} {\bfseries
  01} (2004) 025} [\href{https://arxiv.org/abs/hep-th/0311141}{{\ttfamily
  hep-th/0311141}}].

\bibitem{Kraus:2006wn}
P.~Kraus, \emph{{Lectures on black holes and the AdS(3) / CFT(2)
  correspondence}}, {\emph{Lect. Notes Phys.} {\bfseries 755} (2008) 193}
  [\href{https://arxiv.org/abs/hep-th/0609074}{{\ttfamily hep-th/0609074}}].

\bibitem{Iliesiu:2021are}
L.~V. Iliesiu, M.~Kologlu and G.~J. Turiaci, \emph{{Supersymmetric indices
  factorize}}, \href{https://doi.org/10.1007/JHEP05(2023)032}{\emph{JHEP}
  {\bfseries 05} (2023) 032}
  [\href{https://arxiv.org/abs/2107.09062}{{\ttfamily 2107.09062}}].

\bibitem{picken1989propagator}
R.~Picken, \emph{{The propagator for quantum mechanics on a group manifold from
  an infinite-dimensional analogue of the Duistermaat-Heckman integration
  formula }}, {\emph{Journal of Physics A: Mathematical and General} {\bfseries
  22} (1989) 2285}.

\bibitem{Eguchi:1987sm}
T.~Eguchi and A.~Taormina, \emph{{Unitary Representations of $N=4$
  Superconformal Algebra}},
  \href{https://doi.org/10.1016/0370-2693(87)91679-0}{\emph{Phys. Lett.}
  {\bfseries B196} (1987) 75}.

\bibitem{Eguchi:1987wf}
T.~Eguchi and A.~Taormina, \emph{{Character Formulas for the $N=4$
  Superconformal Algebra}},
  \href{https://doi.org/10.1016/0370-2693(88)90778-2}{\emph{Phys. Lett.}
  {\bfseries B200} (1988) 315}.

\bibitem{Eguchi:1988af}
T.~Eguchi and A.~Taormina, \emph{{On the Unitary Representations of $N=2$ and
  $N=4$ Superconformal Algebras}},
  \href{https://doi.org/10.1016/0370-2693(88)90360-7}{\emph{Phys. Lett.}
  {\bfseries B210} (1988) 125}.

\bibitem{Eguchi:2008ct}
T.~Eguchi, Y.~Sugawara and A.~Taormina, \emph{{Modular Forms and Elliptic
  Genera for ALE Spaces}},  in \emph{{Workshop on Exploration of New Structures
  and Natural Constructions in Mathematical Physics: On the Occasion of
  Professor Akhiro Tsuchiya's Retirement}}, 3, 2008,
  \href{https://arxiv.org/abs/0803.0377}{{\ttfamily 0803.0377}}.

\bibitem{Srednicki:1992sp}
M.~Srednicki, \emph{{A New construction of the Penner model}},
  \href{https://doi.org/10.1142/S0217732392004237}{\emph{Mod. Phys. Lett. A}
  {\bfseries 7} (1992) 2857}
  [\href{https://arxiv.org/abs/hep-th/9206085}{{\ttfamily hep-th/9206085}}].

\bibitem{Penner1}
R.~C. Penner, \emph{{The moduli space of a punctured surface and perturbative
  series}}, {\emph{Bull. Am. Math. Soc.} {\bfseries 15} (1986) 73}.

\bibitem{Penner2}
R.~C. Penner, \emph{{Perturbative series and the moduli space of Riemann
  surfaces}}, {\emph{J. Diff. Geom.} {\bfseries 27} (1988) 35}.

\bibitem{Ambjorn:1994bp}
J.~Ambjorn, C.~F. Kristjansen and Y.~Makeenko, \emph{{Generalized Penner models
  to all genera}}, \href{https://doi.org/10.1103/PhysRevD.50.5193}{\emph{Phys.
  Rev. D} {\bfseries 50} (1994) 5193}
  [\href{https://arxiv.org/abs/hep-th/9403024}{{\ttfamily hep-th/9403024}}].

\bibitem{Migdal:1975zg}
A.~A. Migdal, \emph{{Recursion Equations in Gauge Theories}}, {\emph{Sov. Phys.
  JETP} {\bfseries 42} (1975) 413}.

\bibitem{Eynard:2004mh}
B.~Eynard, \emph{{Topological expansion for the 1-Hermitian matrix model
  correlation functions}},
  \href{https://doi.org/10.1088/1126-6708/2004/11/031}{\emph{JHEP} {\bfseries
  11} (2004) 031} [\href{https://arxiv.org/abs/hep-th/0407261}{{\ttfamily
  hep-th/0407261}}].

\end{thebibliography}\endgroup
\end{document}